\documentclass[a4paper,11pt]{article}

\usepackage{jcappub} 

\usepackage[T1]{fontenc} 

\usepackage{graphicx}
\usepackage{subcaption}
\usepackage{booktabs, multirow}
\usepackage[normalem]{ulem}

\title{\boldmath Periodic orbits and their gravitational waves in EMRIs: supermassive black hole affected by galactic dark matter halos}

\author[a]{Guo-He Li,}
\author[b]{Chen-Kai Qiao,}
\author[a]{Jun Tao}

\affiliation[a]{College of Physics, Sichuan University, Chengdu, 610065, China}
\affiliation[b]{College of Physical Science and New Energy, Chongqing University of Technology, Chongqing, 400054, China}

\emailAdd{liguohe@stu.scu.edu.cn}
\emailAdd{chenkaiqiao@cqut.edu.cn}
\emailAdd{taojun@scu.edu.cn}

\abstract{Periodic orbits exhibiting zoom-whirl behavior have become attractive topics for studying particle dynamics and gravitational wave emission in extreme-mass-ratio inspirals (EMRIs). This study systematically investigates periodic orbits around black holes and their gravitational wave radiation in three dark matter halo environments: NFW, Beta, and Moore models. The dark matter distribution in these models can be effectively incorporated using two parameters --- the dark matter characteristic mass and halo characteristic radius. Our results reveal that for a larger dark matter mass and a smaller characteristic radius, the shapes of the periodic orbits and the corresponding gravitational waveforms show more significant deviations from the Schwarzschild case. As the halo characteristic radius increases, the orbital shapes and waveform characteristics gradually converge with the Schwarzschild black hole results. Our results also suggest that the NFW and Beta models produce nearly indistinguishable results, while the Moore model shows distinct signatures compared with Beta/NFW models. Furthermore, calculations of characteristic strains in frequency spectra show that gravitational-wave signals associated with these periodic orbits lie above the sensitivity curves of LISA, TianQin, and Taiji, indicating their detectability in future space-based gravitational wave observatories. These findings deepen our understanding of dark matter halo effects on periodic motions and gravitational wave signatures.
	
\ \ 
	
dark matter theory; astrophysical black holes}

\begin{document}
\maketitle
\flushbottom

\section{Introduction}

A substantial number of independent astrophysical observations have been made, including cosmic microwave background radiation~\cite{WMAP:2010sfg,Planck:2013oqw,Planck:2013pxb,Planck:2018vyg}, galactic rotation curves~\cite{Rubin:1970zza,Corbelli:1999af}, and gravitational lensing, which collectively indicate that our universe is dominated by dark matter (DM) and dark energy. Precise measurements reveal that approximately 26.8\% of the total mass of the universe is dark matter, 68.3\% is dark energy, and the familiar baryonic matter constitutes only 4.9\% of our universe \cite{Planck:2013pxb,Planck:2013oqw,Planck:2018vyg}. It has been demonstrated that dark matter and dark energy are not only crucial for the evolution of the universe, but also play a pivotal role in the formation and development of galaxies and galaxy clusters~\cite{Blumenthal:1984bp,Reddick:2012qy,Markevitch:2003at,Rubin:1970zza,Sofue:2000jx}. These mysterious elements of our universe significantly influence the structure of galaxies and the orbits of particles and stars in galaxies.  

In a large number of galaxies, dark matter is known to form a halo structure around the supermassive black hole (SMBH)~\cite{Cooray:2002dia,Wang:2019ftp}. The mass of these dark matter halos spans multiple orders of magnitude, and their density distribution characteristics are crucial for understanding the formation and evolution of galaxies~\cite{Bhattacharya:2011vr}. Through extensive researches using numerical simulations and astrophysical observations, scholars have developed various theoretical models to describe the density profiles of dark matter halos, such as the Navarro-Frenk-White (NFW) model~\cite{Navarro:1995iw,Navarro:1996gj}, Beta model~\cite{Navarro:1994hi,Cavaliere:1976tx}, Moore model~\cite{Moore:1999gc}, Burkert model~\cite{Salucci:2000ps}, Einasto model~\cite{Dutton:2014xda,Graham:2005xx,Navarro:2008kc}, Dehnen-type Model~\cite{Dehnen:1993uh} and Brownstein model~\cite{Brownstein:2009gz}. These models have been demonstrated to possess significant value and have practical applications in phenomenological studies, providing critical constraints for both indirect and direct detection of dark matter~\cite{Cuoco:2016eej,Iocco:2015xga,Pato:2015tja,Salucci:2018hqu}.

The motion of test particles (or small celestial bodies) in a gravitational field serves as a pivotal approach to probe the properties of gravitational fields. Bound precession orbits and circular orbits play a foundational role in the classical tests of gravitational theories, the long-term stability and evolution of gravitational systems (especially strong-field regimes), and gravitational wave emission~\cite{Poisson:1993vp,Cutler:1993vq,Apostolatos:1993nu,Dai:2023cft,GRAVITY:2020gka,Cardoso:2014sna,Cunha:2022gde,Guo:2024cts,Mummery:2022ana}. However, the motion of small celestial bodies and test particles in strong gravitational fields may also exhibit another important motion pattern --- periodic orbits~\cite{Levin:2008mq,Levin:2008ci,Grossman:2008yk,Misra:2010pu}. Periodic orbits have gradually become an inspiring research topic in recent years. They have been extensively studied in quantum-corrected spacetime under quantum gravity theory~\cite{Tu:2023xab,Yang:2024lmj,Jiang:2024cpe,Al-Badawi:2025yum,Huang:2025vpi,Deng:2020yfm}, hairy black holes~\cite{Zhao:2024exh,Lin:2023rmo,Meng:2024cnq,Deng:2025wzz}, and other spacetime backgrounds~\cite{Liu:2018vea,Junior:2024tmi,Zhou:2020zys,Wei:2019zdf,Babar:2017gsg,Yang:2024cnd,Meng:2025kzx,Gu:2024dna,Li:2025sfe,Wang:2025hla,Wang:2025wob,Zare:2025aek,Alloqulov:2025bxh,Alloqulov:2025ucf,Tan:2024hzw,Haroon:2025rzx}. 

Periodic orbits exist universally in various gravitational systems. These orbits form closed trajectories in the gravitational field with perfect time periodicity, and they are conventionally classified by three integers $(z~w~v)$ ~\cite{Levin:2008mq}. Specifically, $z$ denotes the number of leaves in a full periodic period (with greater $z$ values indicating increased trajectory complexity), $w$ represents the count of additional whirls the particle performs during its outward drift toward the apoapsis, and $v$ characterizes the behavior of the subsequent vertex the particle encounters after leaving the initial vertex (initial apoapsis)~\cite{Levin:2008mq,Levin:2008ci,Grossman:2008yk}. Their precession behavior, orbital shape and gravitational wave signatures of periodic orbits are all inherited in these integers. For instance, the precession behavior of periodic orbits can be quantitatively characterized by a precession parameter $q=\frac{\Delta\phi}{2\pi}-1$. For a closed periodic trajectory, this parameter must be a rational number determined by three integers $q=w+\frac{v}{z}$. Studying these periodic orbits holds profound theoretical significance for understanding more general bound orbits. Since any irrational number can be infinitely approximated by a sequence of rational numbers, more complex bound orbits in gravitational systems characterized by irrational precession parameters can always be approached by a sequence of nearby periodic orbits with rational $q$.

The zoom-whirl features of periodic orbits can be directly reflected in the gravitational waveform and frequency spectrum~\cite{Yang:2024lmj,Haroon:2025rzx}, which may provide potential observational signatures during the observation of extreme-mass-ratio inspirals (EMRIs) through next-generation space-based detectors. The EMRIs is typically characterized by a compact celestial object spiraling towards a supermassive black hole (SMBH)~\cite{Glampedakis:2005hs,Hughes:2000ssa}. These systems are identified as primary targets for the next generation of space-based detectors, including LISA~\cite{LISA:2017pwj,LISA:2022yao}, TianQin~\cite{TianQin:2015yph,Liu:2020eko}, Taiji~\cite{Hu:2017mde,Gong:2021gvw}, and DECIGO~\cite{Musha:2017usi}. In the observations of EMRIs, the zoom-whirl features of periodic orbits produce unique fingerprints that distinguish them from other bound orbits, both in the waveforms and the frequency spectra of gravitational wave signals. In the time domain, the resulting waveforms are characterized by strong, high-frequency emissions during the whirl phase, alternating with extended low-amplitude intervals during the zoom phase \cite{Yang:2024lmj,Haroon:2025rzx,Zhao:2024exh}. In the frequency domain, because of the periodicity of orbits, the corresponding gravitational waves exhibit discrete spectra in the frequency domain \cite{Agrawal:2026rwu,Xamidov:2026kqs,Ahmed:2025azu,Alloqulov:2025bxh,Yang:2024lmj,Haroon:2025rzx}. Analyzing their discrete frequency spectra not only reveals the characteristics of periodic orbits in gravitational fields, but also provides means for probing the background spacetime geometry and gravitational theories. Comparing these spectral features against the sensitivity curves of space-based detectors allows for an assessment of the detectability for these periodic orbits. Additionally, the application of the Fisher matrix analysis in gravitational waves enables more thorough explorations of these periodic orbits and the physical parameters in gravitational systems.

Recently, theoretical studies have examined the influences of dark matter on the physical processes in the vicinity of SMBHs. These studies involve a wide range of topics and physical processes, including accretion~\cite{Heydari-Fard:2022xhr,Saurabh:2020zqg}, gravitational lensing~\cite{Qiao:2024ehj,vanUitert:2012bj,Atamurotov:2021hoq,Pantig:2022toh,Pantig:2020odu,Qiao:2022nic,Liu:2023xtb}, circular geodesics and black hole shadow ~\cite{Konoplya:2022hbl,Das:2020yxw,Rayimbaev:2021kjs,Jusufi:2020cpn,Filho:2023abd,Yang:2023tip,Pantig:2022sjb,Ovgun:2025bol}, quasi-normal mode and gravitational waves~\cite{Cardoso:2021wlq,Liu:2023vno,Liu:2024xcd,Li:2025qtb,Zhao:2023itk,Zhou:2024vhk,Ashoorioon:2025ezk,Cole:2022yzw}. The complex interactions between dark matter environments and celestial objects can produce notable influences on the gravitational wave signals emitted from EMRI systems and other binary black hole (BBH) systems ~\cite{Kavanagh:2020cfn,Speeney:2022ryg,Duque:2023seg,Karydas:2024fcn,Kavanagh:2024lgq,Vicente:2025gsg,Karydas:2025bkj}. However, the studies on periodic orbits around SMBH in dark matter environments remain in a very preliminary stage, which are usually concentrated on a single dark matter model \cite{Alloqulov:2025ucf,Haroon:2025rzx,Tan:2024hzw}. The comparison of periodic orbits from different dark matter distributions is still absent. The specific impact of dark matter mass and density profiles in halo structures (described by different halo models) on the periodic orbits around SMBHs, as well as the subsequent influence on the generated gravitational wave signals, remain unresolved and challenging issues. These questions are of significant importance for the interpretation of gravitational wave signals in EMRIs in gravitational wave astronomy.

Inspired from the aforementioned motivations, this study focuses on the properties of periodic orbits around supermassive black holes in dark matter halo environments. This work enables us to give a comprehensive analysis on dark matter influences from different halo models. Specifically, we select three typical dark matter halo models (NFW, Beta, and Moore), whose validity has been tested by theoretical investigations and astrophysical observations in galaxies \cite{Sofue:2020rnl}. The characteristics of test particles' periodic orbits calculated within these halo models under different dark matter masses $k$ and halo scales $h$ are analyzed through numerical calculations. Meanwhile, we investigate the gravitational wave signals generated by periodic orbits in EMRIs influenced by these dark matter models, clarifying the impacts of different dark matter halos on the shape of periodic orbits and gravitational waveforms. Finally, to give an assessment of the detectability of these orbits in next-generation space-based observations, we study the characteristics of periodic orbits' gravitational wave in the frequency spectra, and compare them with the detectors' sensitivities of LISA, TianQin and Taiji. The analysis presented in this work would be helpful for gaining a deeper understanding of how different dark matter density distributions precisely affect particle orbits around SMBHs, and it may provide potential applications for capturing observable signatures in future space-based gravitational wave detections.

The structure of this paper is as follows: In Section~\ref{s2}, we review the spacetime metrics generated by supermassive black holes in dark matter halo environments. We further derive the equations of motion and effective potential for massive test particles, and develop the framework to compute the orbital precession angle $q$ and  particle orbits. In Section~\ref{s3}, we compare and analyze the periodic orbits of black holes in the backgrounds of different dark matter halos. The influence of dark matter halo parameters on the periodic orbits is emphasised in this section. In Section~\ref{s4}, we explore the gravitational waveforms generated by these periodic orbits. In Section~\ref{s5}, we study the gravitational wave characteristics in frequency spectra and evaluate the detectability of these periodic orbits in the context of future space-based gravitational wave observatories. Finally, Section~\ref{s6} contains the conclusions and discussions of our work. Throughout the paper, we use the geometric unit system with $G = c = 1$.

\section{Supermassive Black Holes in Dark Matter Halos}\label{s2}

To investigate the orbital dynamics and gravitational wave emission characteristics of particles around supermassive black holes in different dark matter halo environments, it is necessary to derive the spacetime metric of the system. Therefore, this section begins by introducing the spacetime metrics for SMBHs enclosed by different dark matter halo distributions. Subsequently, based on the obtained spacetime metric, the effective potential is given to study the orbital motions of particles. Finally,to characterize the periodic orbits around the black hole, we introduce the concept of the relative value of precession angle (labeled by $q$). 

\subsection{Spacetime Metric}
In most galaxies, the dark matter distribution can be described by spherically symmetric halo models. Established astrophysical frameworks for describing these dark matter halo structures in our galaxy and external spiral galaxies include the NFW, Einasto, Beta, Burkert, Brownstein, Dehnen and Moore models, and most of them favor the spherically symmetric dark matter distributions.~\cite{Cavaliere:1976tx,Navarro:1994hi,Navarro:1995iw,Navarro:1996gj,Burkert:1995yz,Moore:1997sg,Moore:1999gc,Brownstein:2009gz,Dehnen:1993uh,Dutton:2014xda,Graham:2005xx,Navarro:2008kc}.
Under such circumstance, our analysis focuses specifically on static and spherically symmetric black hole solutions surrounded by dark matter halos, with the spacetime metric given by,
\begin{equation}
	\mathrm{d}s^2=-f(r)\mathrm{d}\mathrm{t}^2+f(r)^{-1}\mathrm{d}\mathrm{r}^2+r^2\mathrm{d}\theta^2+r^2\mathrm{s}in^2\theta\mathrm{d}\phi^2.\label{e1}
\end{equation}
From the gravitational geodesic equations,for particles moving in the equatorial plane of spherically symmetric spacetime, rotational velocity is determined by the metric function $f(r)$ through~\cite{Xu:2018wow,Matos:2000ki},
\begin{equation}
	v_{tg}{}^2(r)=\frac{r}{\sqrt{f(r)}}\cdot\frac{d\sqrt{f(r)}}{dr}=\frac{r(\mathrm{dln}\sqrt{f(r)})}{dr}.\label{e2}
\end{equation}
Based on the rotational velocity relationship in Eq.~(\ref{e2}), the dark matter halo contribution to metric function can be obtained by solving the differential equation,
\begin{equation}
	f_{DM}(r)=\exp\left[2\int\frac{v_{tg}{}^2(r)}{r}\mathrm{d}r\right].\label{e3}
\end{equation}
On the other hand, according to astrophysical constraints, the rotational velocity of stars in galactic environments follows $v_{tg}^2(r) \approx M(r)/r$, with the total mass $M(r)$ including all constitutes in galaxies. Concretively, the cumulative mass function for the dark matter halo is expressed as,
\begin{equation}
	M_{DM}(r)=4\pi\int_0^r\rho(r^{\prime})r^{\prime2}dr^{\prime},
\end{equation}
where $\rho(r^{\prime})$ represents the dark matter density function, whose specific forms are provided by established astrophysical halo models. 

This study utilizes several renewed spherically symmetric dark matter profiles, including the NFW, Beta, and Moore models. The dark matter distributions in these models can be described by analytical expressions~\cite{Cavaliere:1976tx,Navarro:1995iw,Navarro:1996gj,Moore:1999gc},
\begin{subequations}
	\begin{align}
		\rho_{\text{NFW}}(x) &= \frac{\rho_0}{x(1 + x^2)}, \label{eq:nfw} \\
		\rho_{\text{Beta}}(x) &= \frac{\rho_0}{(1 + x^2)^{3/2}}, \label{eq:beta} \\
		\rho_{\text{Moore}}(x) &= \frac{\rho_0}{x^{3/2}(1 + x^{3/2})}, \label{eq:moore}
	\end{align}
\end{subequations}
where $x = r/h$, with $\rho_0$ and $h$ representing the characteristic density and characteristic radius (or characteristic scale) of dark matter halo respectively. The effective spacetime metric for supermassive black holes enclosed by dark matter halos is obtained by combining the dark matter density profile $\rho$ and black hole mass $M$ in the Einstein field equations \cite{Xu:2018wow}. After some simplification, the metric can be decomposed as $f(r) = f_{\text{DM}}(r) - \frac{2M}{r}$, where $f_{\text{DM}}(r)$ represents the dark matter contribution (see Eq.~(\ref{e3})) and $-\frac{2M}{r}$ accounts for the central black hole's gravitational effect. For different dark matter models, the resulting metric functions take the following forms ((the derivations of metric functions can be found in reference~\cite{Qiao:2024ehj}),
\begin{subequations}
	\begin{align}
		&f_{\text{NFW}}(r) = (1 + x)^{-\frac{8\pi k}{r}} - \frac{2M}{r}, \label{e6a} \\
		&f_{\text{Beta}}(r)= e^{-\frac{8\pi k}{r}\sinh^{-1}x} - \frac{2M}{r}, \label{e6b} \\
		&f_{\mathrm{Moore}}(r)=e^{\frac{16\pi k}{\sqrt{3}h}\arctan\frac{2\sqrt{x}-1}{\sqrt{3}}}\cdot(1+x^{3/2})^{-\frac{16\pi k}{3r}}\cdot\left(\frac{1+x-\sqrt{x}}{1+x+2\sqrt{x}}\right)^{-\frac{8\pi k}{3h}}-\frac{2M}{r}. \label{e6c}
	\end{align}
\end{subequations}
The parameter $M$ represents the supermassive black hole's mass, while $k = \rho_0 \cdot h^3$ provides an estimate of the dark matter mass. All of the above models revert to the Schwarzschild metric when \( k = 0 \) or \( h \to \infty \).

In most galaxies, the total mass of dark matter, the scale of the dark matter halo, and the event horizon of the SMBH in the galactic center usually follow a hierarchical relation (in geometric units): $ M \ll k \ll h $. The supermassive black hole at the center of the Milky Way galaxy is Sgr A*, with a mass of $ M = 4.3 \times 10^6 \, M_\odot $ and a dark matter halo's characteristic radius of approximately $ h = 10.94 \, \text{kpc} $, which results in $ k \approx 10^3 M $ and $ h \approx 10^{10} M $~\cite{Kafle:2014xfa,Lin:2019yux,Junior:2023xgl}. For the Virgo galaxy (M87), its central supermassive black hole has a mass of $ M = 6.5 \times 10^9 \, M_\odot$, with a dark matter halo's characteristic radius $ h = 91.2 \, \text{kpc} $ and characteristic density $ \rho_0 = 6.9 \times 10^6 \, M_\odot/\text{kpc}^3 $~\cite{EventHorizonTelescope:2021bee,Lu:2023bbn,Jusufi:2019nrn}. Dark matter parameters in Virgo galaxy satisfy $ k \approx 10^3 M $ and $ h \approx 10^8 M $. In addition, there exist more massive black holes, such as Ton 618, whose mass is estimated to be $ M = 6.6 \times 10^{10} \, M_\odot $, with a dark matter halo's characteristic density of $ 1.4 \times 10^7 \, M_\odot/\text{kpc}^3 $ and a characteristic scale of $ 500 \, \text{kpc} $~\cite{Shemmer:2004ph}; the dark matter halo parameters satisfy $ k \approx 10^4 M $ and $ h \approx 10^8 M $. In the rest of this work, we will vary the dark matter mass parameter $k$ from $10^3\sim10^4 M$ and the dark matter halo scale $h$ from $10^7 M \sim 10^{10} M$ to highlight the effects of dark matter on periodic orbits in EMRI systems.

\subsection{Effective Potential, ISCO and MBO}\label{sub2}

Once the spacetime metric is obtained, the characteristics of particle motions in this spacetime can be further analyzed, which needs the concept of effective potential. For a massive test particle orbiting the  black hole, we examine the dynamics through its Lagrangian,
\begin{equation}
	\mathcal{L} = \frac{m}{2}g_{\mu\nu}\dot{x}^{\mu}\dot{x}^{\nu},\label{e7}
\end{equation}
where dot denotes differentiation with respect to proper time, and $m$ represents the test particle's mass. Without loss of generality, we can set $m=1$ and introduce the generalized momentum per unit mass,
\begin{equation}
	p_{\mu} = \frac{\partial\mathcal{L}}{\partial\dot{x}^{\mu}} = g_{\mu\nu}\dot{x}^{\nu},\label{e8}
\end{equation}
Substituting Eq.~(\ref{e8}) into Eq.~(\ref{e1}) yields the equations of motion of the particle. For a static and spherically symmetric black hole, the conserved quantities of the system --- energy $E$ and angular momentum $L$ --- are related to $p_t$ and $p_\phi$.
\begin{subequations}
	\begin{align}
		p_{t}&=-f(r)\dot{t}=-E,\label{e9a}\\
		p_{\phi}&=r^{2}\sin^{2}\theta\dot{\phi}=L,\label{e9b}\\
		p_{r}&=f(r)^{-1}\dot{r},\label{e9c}\\
		p_{\theta}&=r^{2}\dot{\theta}.\label{e9d}
	\end{align}
\end{subequations}
It can also be derived from Eqs.~(\ref{e9a}) and (\ref{e9b}) as follows,
\begin{subequations}
	\begin{align}
		\dot{t} &= \frac{E}{f(r)}, \label{e10a} \\
		\dot{\phi} &= \frac{L}{r^2\sin^2\theta}. \label{e10b}
	\end{align}
\end{subequations}
Because of the spherical symmetry, we can always constrain the particle orbit to the equatorial plane (\(\theta = \pi/2\) and \(\dot{\theta} = 0\)). For a particle following a timelike geodesic in a gravitational field, its four-velocity $\dot{x}^\mu$ satisfies the normalization condition,
\begin{equation}
	g_{\mu\nu}\dot{x}^\mu\dot{x}^\nu = -1.\label{e15}
\end{equation}
Utilizing the relation between conserved quantities and four-velocity in Eqs.~(\ref{e10a}), (\ref{e10b}), the normalization of four-velocity Eq.~(\ref{e15}) reduces to the following radial orbit equation,
\begin{equation}
	\dot{r}^2 + V_{\text{eff}} = E^2,\label{e17}
\end{equation}
and $V_{\mathrm{eff}}$ is the effective potential of the test particle
\begin{equation}
	V_{\mathrm{eff}}=f(r)\left(1+\frac{L^2}{r^2}\right)\label{e18}
\end{equation}
Evidently, when particle escapes to infinity, as \(r \to \infty\), we have \(\lim_{r \to \infty} V_{\mathrm{eff}}= 1\). 

\begin{figure}
	\centering 
	\begin{subfigure}{0.45\textwidth}
		\includegraphics[width=7cm,height=6cm]{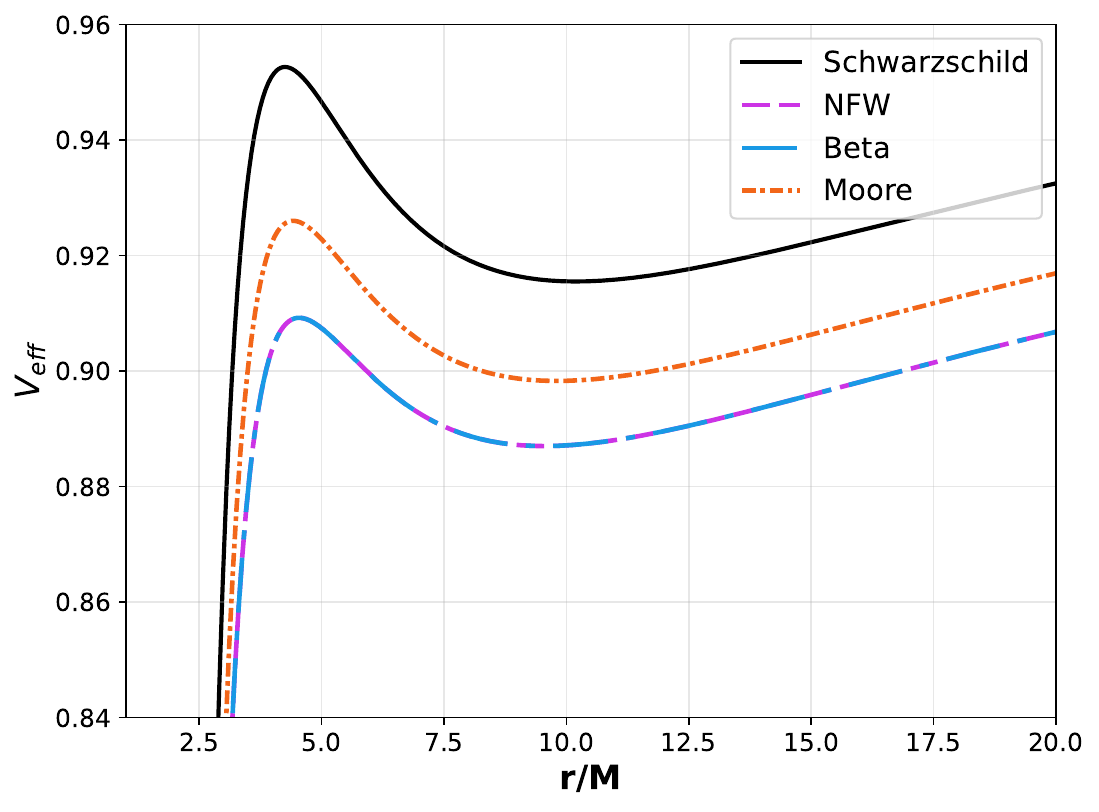}
		\caption{}
		\label{Veff_1}
	\end{subfigure}
	\hfill 
	\begin{subfigure}{0.45\textwidth}
		\includegraphics[width=7cm,height=6cm]{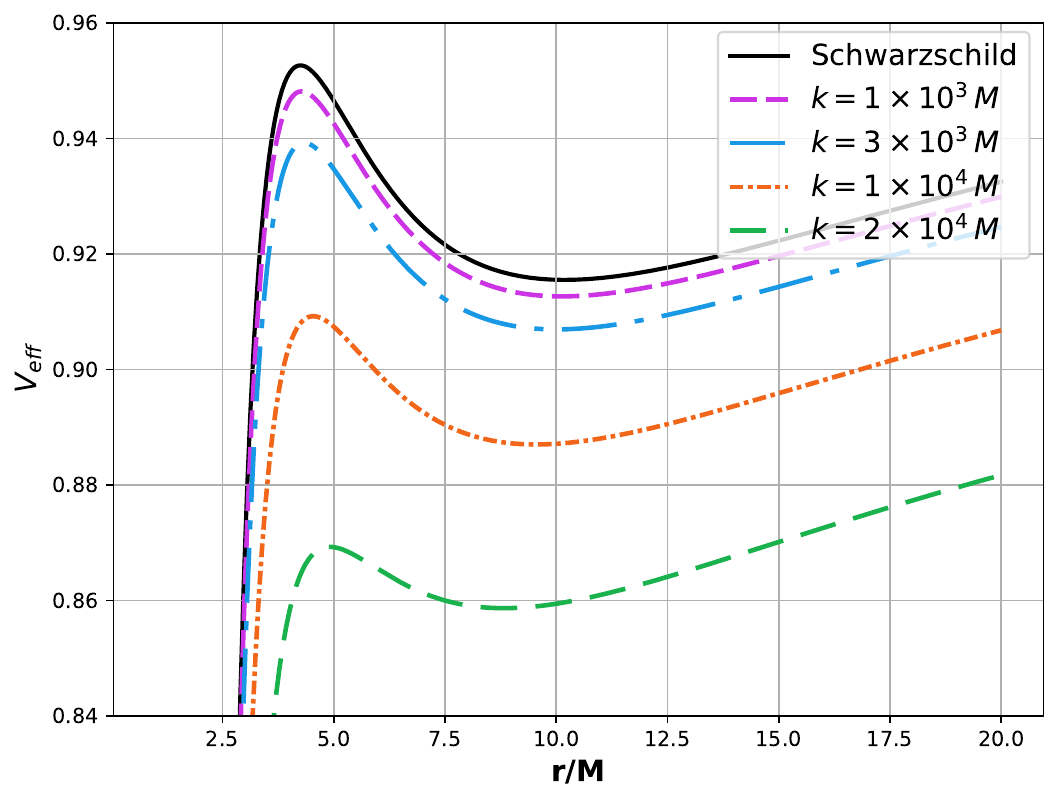}
		\caption{}
		\label{Veff_2}
	\end{subfigure}
	\caption{The influence of dark matter on the effective potential of particles. (a) The left subplot highlights the effects of different dark matter halo models, with \( k = 10^4 M \). (b) The right subplot emphasizes the role of the dark matter mass \( k \). In both subplots, the particle angular momentum is set to \( L = 3.8 \), and the dark matter halo characteristic radius is \( h = 10^7 M \). The black curve corresponds to the case without dark matter (i.e., the Schwarzschild black hole).}
	\label{Veff}
\end{figure}

To show the dark matter halo influences on effective potential, we give an illustration of effective potential in Fig.~\ref{Veff}. Fig.~\ref{Veff_1} exhibits the effective potentials of particles influenced by different dark matter halo models for the same dark matter parameters and orbital angular momentum. Fig.~\ref{Veff_2} exhibits the effective potentials affected by dark matter characteristic mass $k$ for the same dark matter density profile (taking the NFW model as an example; results from other halo models have a similar tendency). It can be seen that the effective potential exhibits two extrema. The minimum value of the effective potential corresponds to stable circular orbits, while the maximum value corresponds to unstable circular orbits. It is particularly noteworthy that the curves for NFW and Beta cases overlap in Fig.~\ref{Veff_1}, while the Moore model results can be distinguished from NFW/Beta. Moreover, the existence of dark matter halos reduces the extrema of the effective potential. It can also be observed from Fig.~\ref{Veff_2} that the larger dark matter mass causes a reduction in the extrema of the effective potential.

In the exploration of particle motion and gravitational field dynamics, several critical orbits are fundamental to understanding and distinguishing between bound and scattering orbits near black holes. In particular, the marginally bound orbit (MBO) and the innermost stable circular orbit (ISCO) are two representative examples of such critical orbits, which significantly expands our understanding of the particles behavior under strong gravitational fields. The MBO represents the critical threshold separating bound and unbound particle trajectories. At this orbit, particles possess precisely the energy required to escape to infinity with zero kinetic energy. The corresponding conditions of the MBO are
\begin{equation}
	V_{\text{eff}} = E = 1, \quad \frac{dV_{\text{eff}}}{dr} = 0. 
\end{equation}
The innermost stable circular orbit (ISCO) marks the inner boundary of stable circular motion. Particles orbiting within this radius would inevitably undergo a rapid decrease of orbital radius and directly fall into the black hole, due to the instability of circular orbits very close to the event horizon (with $r_H < r <r_{ISCO}$ or $L<L_{ISCO}$). The conditions that determine the ISCO are as follows,
\begin{equation}
	\dot{r} = 0, \quad \frac{dV_{\text{eff}}}{dr} = 0, \quad \frac{d^2V_{\text{eff}}}{dr^2} = 0.
\end{equation}

\begin{figure}
	\centering 
	\begin{subfigure}{0.475\textwidth}
		\includegraphics[width=7cm,height=6cm]{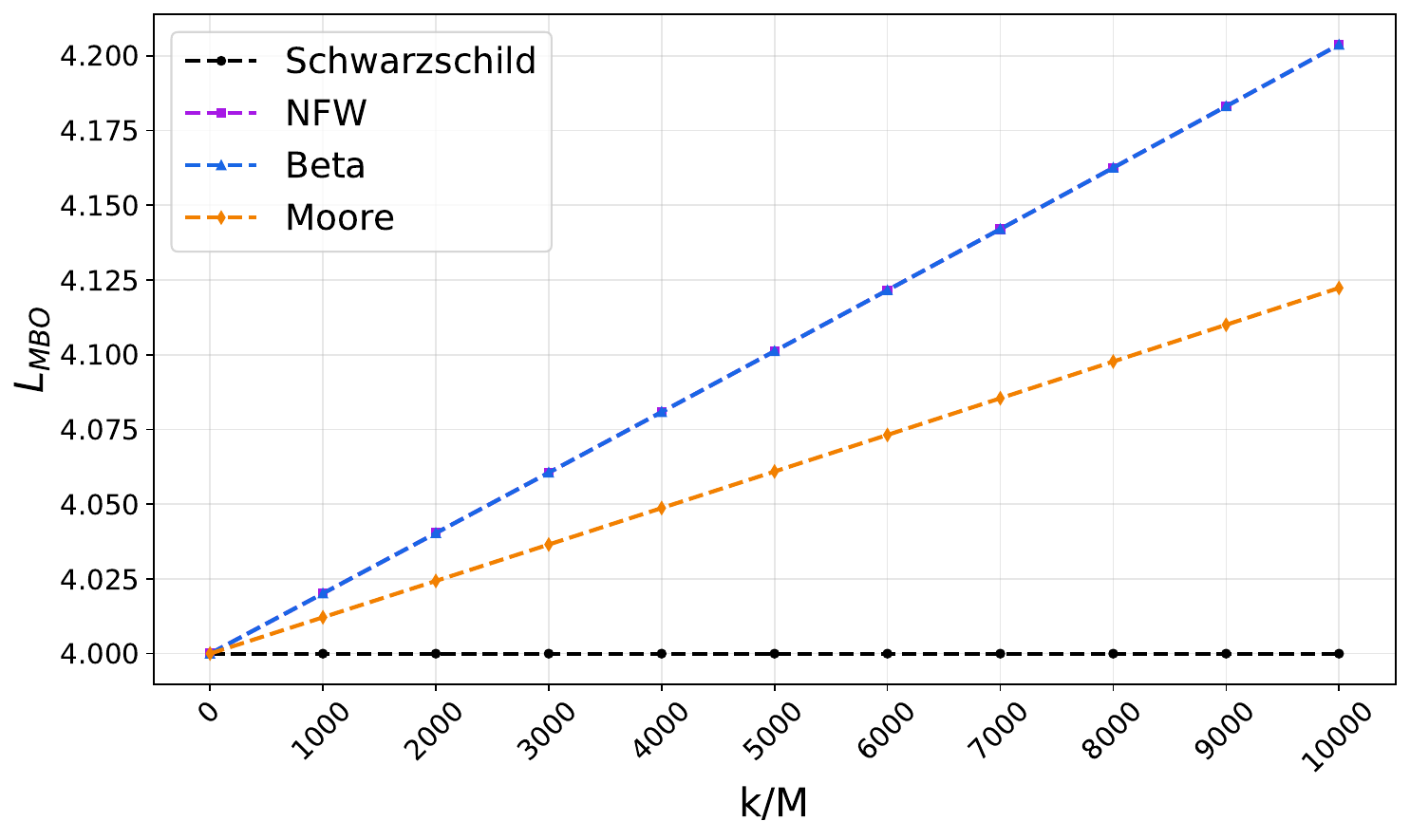}
		\caption{}
		\label{L_1}
	\end{subfigure}
	\hfill 
	\begin{subfigure}{0.475\textwidth}
		\includegraphics[width=7cm,height=6cm]{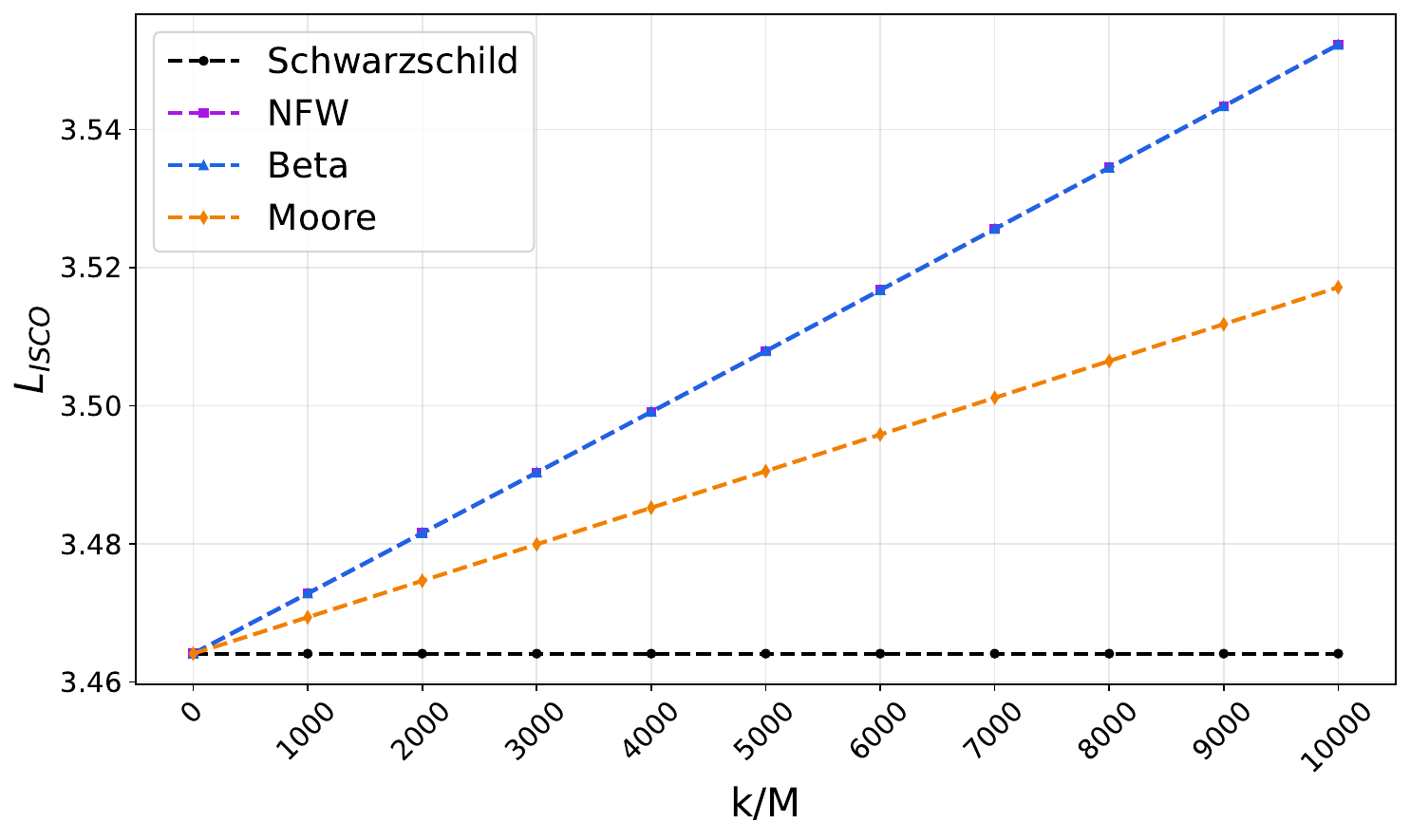}
		\caption{}
		\label{L_2}
	\end{subfigure}
	\caption{The orbital angular momentum of MBO and ISCO affected by dark matter mass $k$ in different dark matter halo models (with characteristic radius  \( h = 10^7 M \)). (a) Marginally bound orbit (MBO); (b) Innermost stable circular orbit (ISCO). The black curve represents the case of a Schwarzschild black hole without dark matter.}
	\label{L}
\end{figure}

To investigate the influence of dark matter halos on test particles' orbital angular momentum, we analyzed the effects of the dark matter mass $k$ on the angular momentum of two critical orbits --- MBO and ISCO --- using different dark matter halo models, as shown in Fig.~\ref{L_1} and~\ref{L_2}. For both MBO and ISCO orbits, a similar trend is observed: the presence of dark matter halos increases the orbital angular momentum of these orbits. As the dark matter mass parameter \( k \) increases, both angular momenta \( L_{\text{MBO}} \) and \( L_{\text{ISCO}} \) rise accordingly, with the curves for the NFW and Beta models nearly overlapping. For the same dark matter mass parameter \( k \), the \( L_{\text{MBO}} \) and \( L_{\text{ISCO}} \) values of the NFW and Beta models are higher than those of the Moore model. This indicates that under identical dark matter mass and halo scale, the gravitational effects exerted by the NFW and Beta  are stronger than those of the Moore.

In this paper, we mainly focus on the properties of periodic orbits around black holes embedded in different dark matter halos, which requires,
\begin{equation}
	L_{\text{ISCO}} \leq L \quad \text{and} \quad E_{\text{ISCO}} \leq E \leq E_{\text{MBO}} = 1. \label{e20} 
\end{equation}
This inequality delineates the permissible range for bound orbital motion. Violations of this inequality  result in either the capture of particles into black hole or the escape of particles into infinity.

\subsection{Periodic Orbits and Their Precession Parameter $q$}

Periodic orbits and their precession behavior constitute important aspects of particle bound orbital motions in curved spacetime, offering crucial insights into the understanding of gravitational fields. In a strong gravitational field, one of the most important phenomena associated with bound orbits is orbital precession—a perturbation effect where the orbit's orientation gradually shifts over successive periods, causing the periastron to advance around the central body. A periodic orbit is defined as a trajectory that returns to its initial position and velocity after a finite time interval, completing a closed path in phase space. The orbital precession angle for periodic orbits must take discrete values (described with three integers), otherwise the test particle could not return to its initial position after a time period.

\begin{figure}
	\centering 
	\begin{subfigure}{0.475\textwidth}
		\includegraphics[width=7cm,height=6cm]{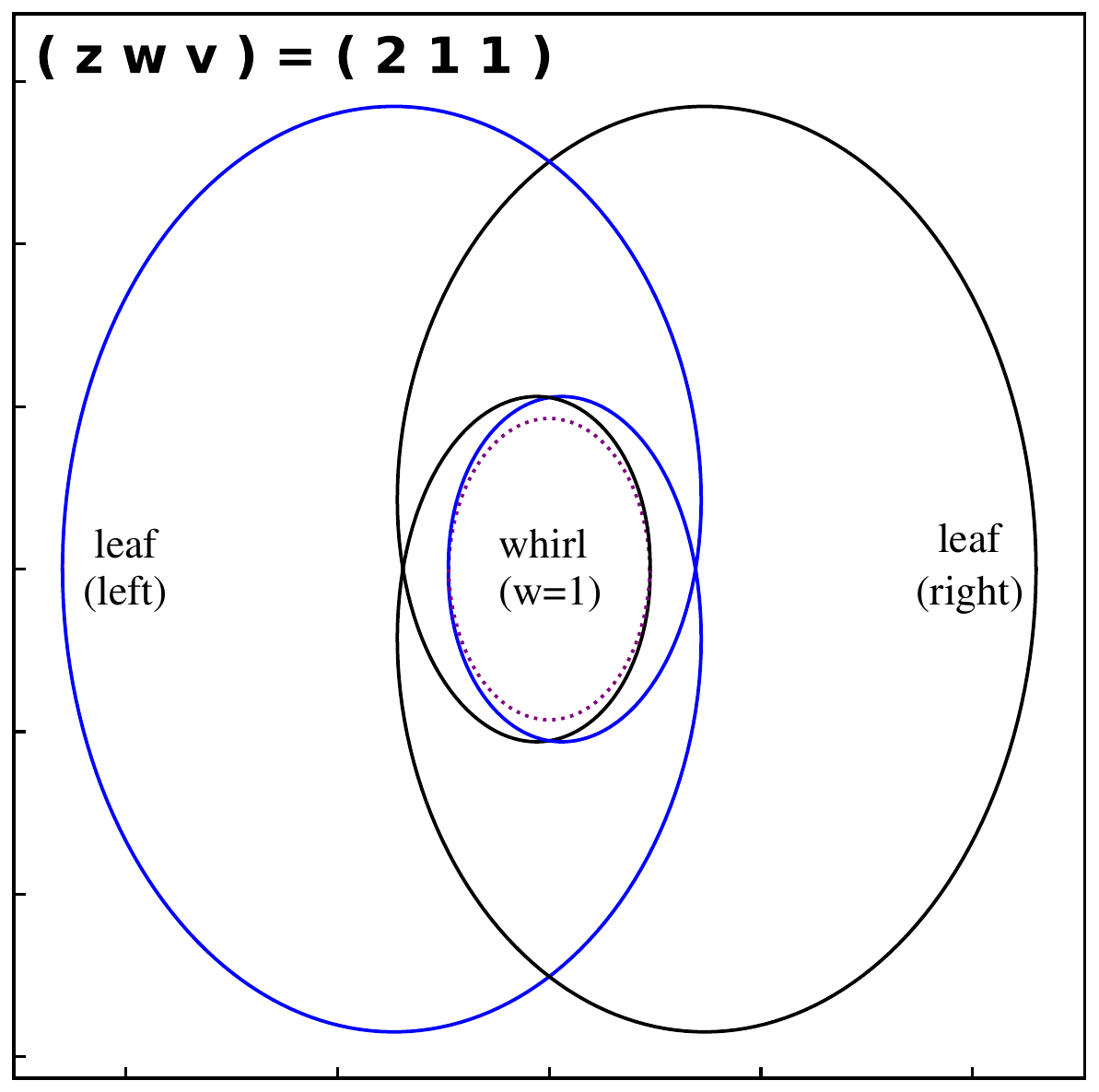}
		\caption{}
		\label{( 2 1 1 )}
	\end{subfigure}
	\begin{subfigure}{0.475\textwidth}
		\includegraphics[width=7cm,height=6cm]{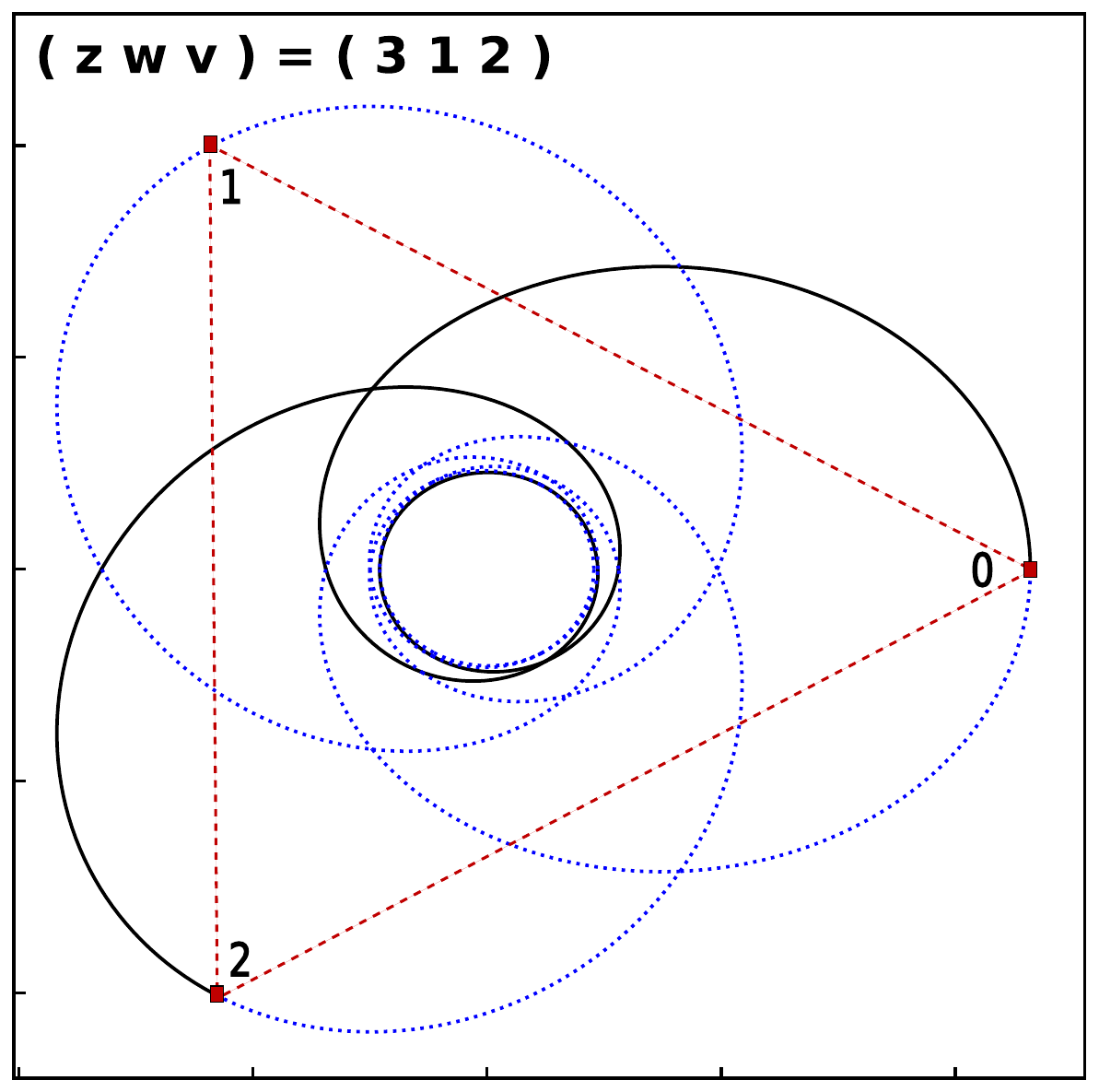}
		\caption{}
		\label{( 3 1 2 )}
	\end{subfigure}
	\caption{ Schematic of periodic orbits. (a) The $(z~w~v) = (2~1~1)$ orbit, where z = 2 is visualized by the blue and black curves (representing 2 leaves in one full orbital period), and w = 1 is characterized by the purple dotted curve (indicating 1 additional whirl in the inner part of periodic orbit during particles drift outward to apoapsis). (b) The $(z~w~v)$ = (3 1 2) orbit illustrating the vertex labeling scheme: vertices are numbered 0, 1, 2 along successive apastra, where v = 2 indicates the orbit skips one vertex and moves to the vertex 2 (at the next apoapsis) after leaving the initial apoapsis at vertex 0.}
	\label{( z w v )}
\end{figure}

To quantitatively characterize periodic orbits and their precession behavior, we introduce a rational parameter $q$, 
\begin{equation}
	q = \frac{\Delta\phi}{2\pi} - 1 = w + \frac{v}{z}.\label{e24}
\end{equation}
Here, $\Delta\phi$ denotes the precession angle --- the total azimuthal angular change during one complete radial oscillation from periastron $r_1$ to apoastron $r_2$ and back. The physical significance of the frequency ratio $\frac{\Delta\phi}{2\pi}$ becomes apparent when we consider truly periodic orbits; for such orbits, this ratio must be a rational number~\cite{Levin:2008mq,Levin:2008ci}. In this expression, the configurations $(z~ w~ v)$ represent the zoom, whirl, and vertex numbers, respectively, which characterize the geometric properties of the periodic orbit. \(z\) denotes the number of leaves in a full periodic orbit trajectory, with greater \(z\) values indicating increased trajectory complexity; \(w\) represents the count of additional whirls the particle performs during its outward drift toward the apoapsis; and \(v\) characterizes the behavior of the subsequent vertex the particle encounters after leaving the initial vertex (apoapsis)~\cite{Levin:2008mq,Levin:2008ci,Grossman:2008yk}. Two illustrative examples of the orbit with configurations \((z~ w~ v) = (2~1~1)\) and \((z~ w~ v) = (3~1~2)\) are presented in Fig.~\ref{( z w v )}.

The precession angle can be computed through direct integration over one complete radial cycle~\cite{Haroon:2025rzx},

\begin{equation}
	\Delta\phi = \oint d\phi = \int_{r_{1}}^{r_{2}}\frac{d\phi}{dr}dr+\int_{r_{2}}^{r_{1}}\frac{d\phi}{dr}dr.\label{e22}
\end{equation}
To evaluate this integral, we need to express $\frac{d\phi}{dr}$ in terms of the orbital parameters $(E,L)$. By deriving from the equations of motion (Eqs.~(\ref{e10b}) and (\ref{e17})), we obtain,

\begin{equation}
	\frac{d\phi}{dr}=\frac{d\phi}{d\tau}\cdot\frac{d\tau}{dr}=\pm \frac{L}{r^2 \sqrt{E^2 - f(r) \left(1 + \frac{L^2}{r^2}\right)}}.\label{e23}
\end{equation}
The sign of $\frac{d\phi}{dr}$ depends on the direction of radial motion: it is positive during the outward journey from periastron $r_1$ to apoastron $r_2$ (where $\dot{r} > 0$), and negative during the inward return from $r_2$ to $r_1$ (where $\dot{r} < 0$).
Substituting Eq.~(\ref{e23}) into Eq.~(\ref{e22}) and utilizing the symmetry of the radial motion, the total precession angle becomes,

\begin{equation}
	\Delta\phi = 2\int_{r_1}^{r_2} \frac{L}{r^2 \sqrt{E^2 - f(r) \left(1 + \frac{L^2}{r^2}\right)}} dr.
\end{equation}
Finally, combining this result with the definition in Eq.~(\ref{e24}), we arrive at the feasible integral expression for the precession parameter:

\begin{equation}
	q= \frac{1}{\pi} \int_{r_1}^{r_2} \frac{L}{r^2 \sqrt{E^2 - f(r) \left(1 + \frac{L^2}{r^2}\right)}} dr - 1. \label{e25}
\end{equation}
The integral formula provides a direct way to compute the precession of periodic orbits. It can be seen that the precession paramater \( q \) depends on the energy \( E \), angular momentum \( L \), and the metric function \( f(r) \), which varies with the dark matter halo background surrounding the SMBH, as detailed in Eqs.~(\ref{e6a})-(\ref{e6c}).

For a given periodic orbit, once the configurations \((z~ w~v)\) and $(E,L)$ are determined, the corresponding particle trajectory \(r(\phi)\) can be obtained by solving the following differential equation:
\begin{equation}
	\left(\frac{dr}{d\phi}\right)^2 = \frac{r^4\left[E^2 - f(r)\left(1 + \frac{L^2}{r^2}\right)\right]}{L^2}
\end{equation}
By differentiating both sides of this equation with respect to \(\phi\), a second-order ordinary differential equation for $r(\phi)$ can be derived,
\begin{equation}
	\frac{d^2 r}{d\phi^2} = -\frac{r}{2L^2} \times \left[r\left(L^2 \frac{f(r)}{dr} + r^2  \frac{f(r)}{dr} - 4E^2 r\right) + 2f(r)\left(L^2 + 2r^2\right)\right]
\end{equation}
Using $x = r(\phi)\cos\phi$ and $y = r(\phi)\sin\phi$, we can visualize the orbit in the Cartesian coordinate system.

\section{Effects of Dark Matter Halos on Periodic Orbits}\label{s3}

In this section, we present numerical results on periodic orbits, focusing on the dark matter halo effects on the shape of these orbits. For a given periodic orbit characterized by a group of zoom, whirl, vertex configurations \((z~w~v)\), its shape and trajectory are determined by the conserved energy $E$ and orbital angular momentum $L$ of a massive particle. A proper choice of \((E, L)\) in their permissible ranges, which are prescribed by Eq.~(\ref{e20}), could result in a periodic orbit around the SMBH with integers \((z~w~v)\). To systematically investigate the effects of dark matter halos on the precession angle and periodic orbits' shape, we can either maintain the orbital angular momentum $L$ and change the energy (which leads to a series \(E_{(z~w~v)}\)), or maintain the energy as constant and vary the angular momentum (which results in a series \(L_{(z~w~v)}\)). In this section, we maintain the angular momentum to give comparison of periodic orbits calculated from different dark matter halo models, while relegating the case of selected energy $E$ to Appendix~\ref{a1}. Specifically, we let the angular momentum $L$ satisfy the relation
\begin{equation}
	L = L_{\text{ISCO}} + \varepsilon(L_{\text{MBO}} - L_{\text{ISCO}}), \label{e28}
\end{equation}
where \(\varepsilon \in [0, 1]\), such that the permissible range in Eq.~(\ref{e20}) is automatically satisfied. For the sake of simplicity, we choose \(\varepsilon = 0.5\) for angular momentum in this sections \ref{s3}-\ref{s5}.

Furthermore, as demonstrated by Eqs.~(\ref{e6a})-(\ref{e6c}), the spacetime metric of the gravitational field is determined by the dark matter mass $k$ and the halo characteristic radius parameter $h$. The geometric shapes of periodic orbits around SMBH could be inevitably affected by these parameters.  Therefore, in the remaining part of this work, we primarily examine two scenarios: A. the effect of dark matter mass; B. the impact of dark matter halo scale.

\subsection{The Effect of Dark Matter Mass on Periodic Orbits}\label{s3_1}

\begin{figure}[b]
	\centering 
	\begin{subfigure}{0.3\textwidth}
		\includegraphics[width=\linewidth]{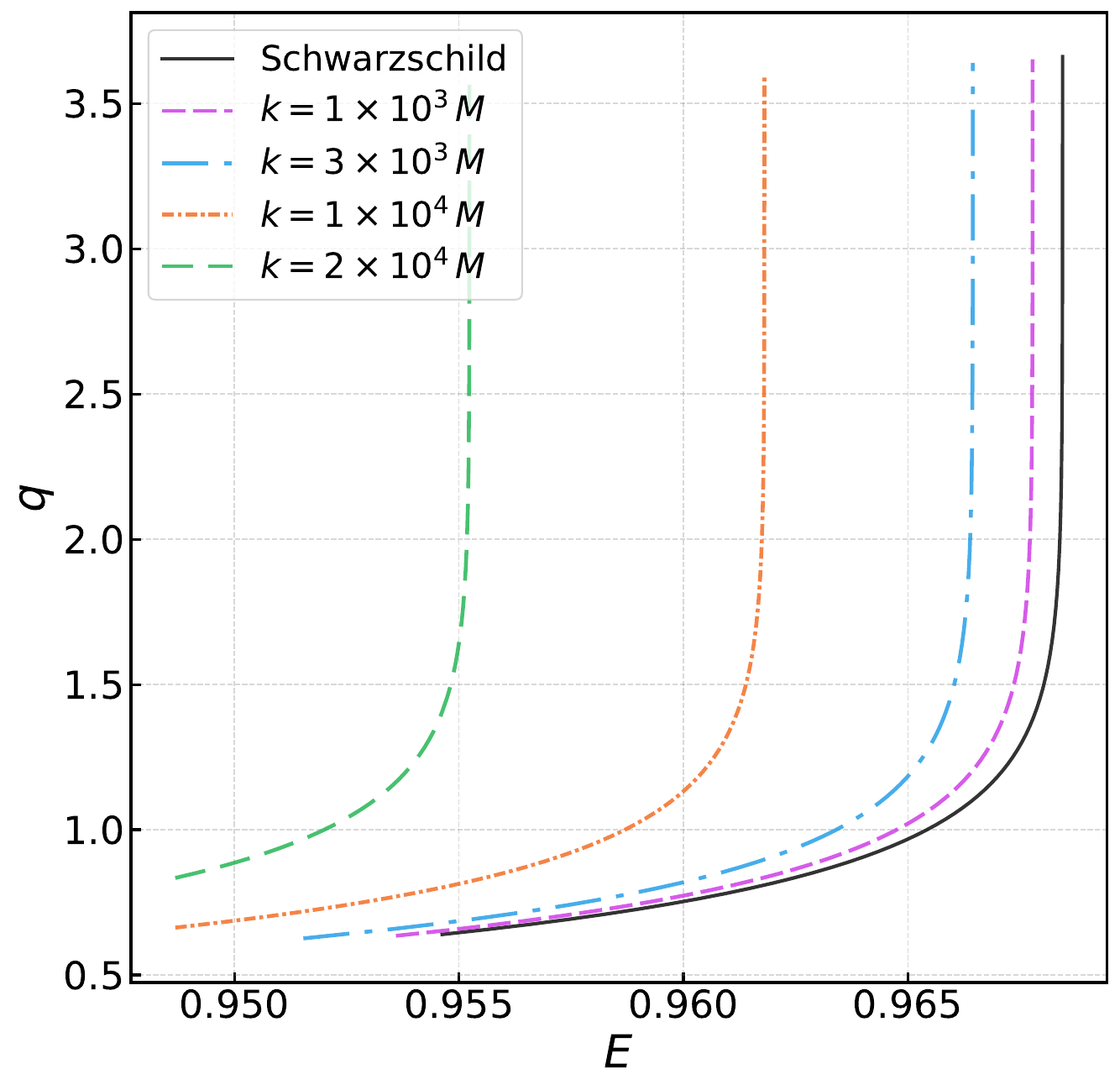}
		\caption{NFW}
	\end{subfigure}
	\hfill 
	\begin{subfigure}{0.3\textwidth}
		\includegraphics[width=\linewidth]{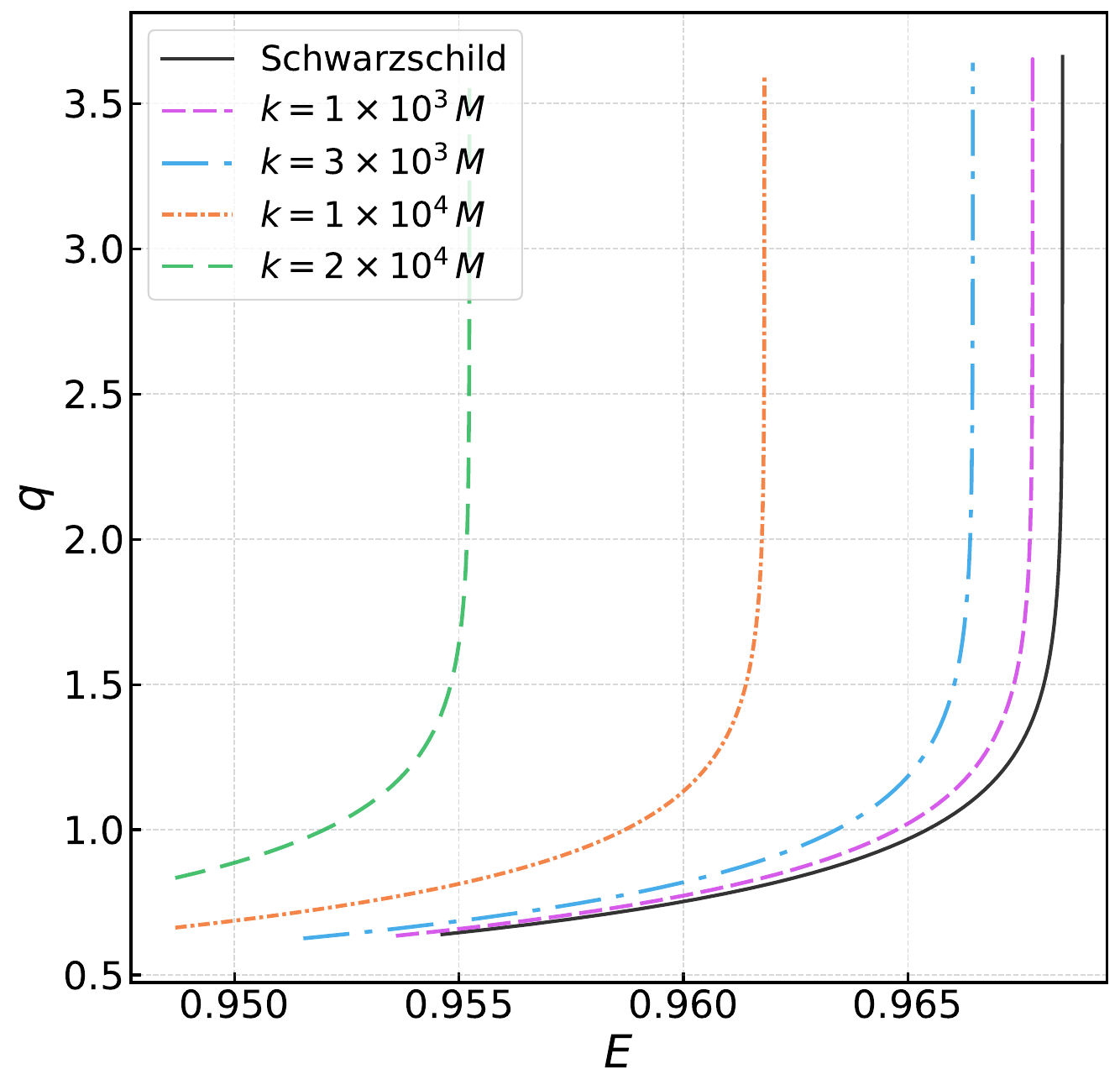}
		\caption{Beta}
	\end{subfigure}
	\hfill
	\begin{subfigure}{0.3\textwidth}
		\includegraphics[width=\linewidth]{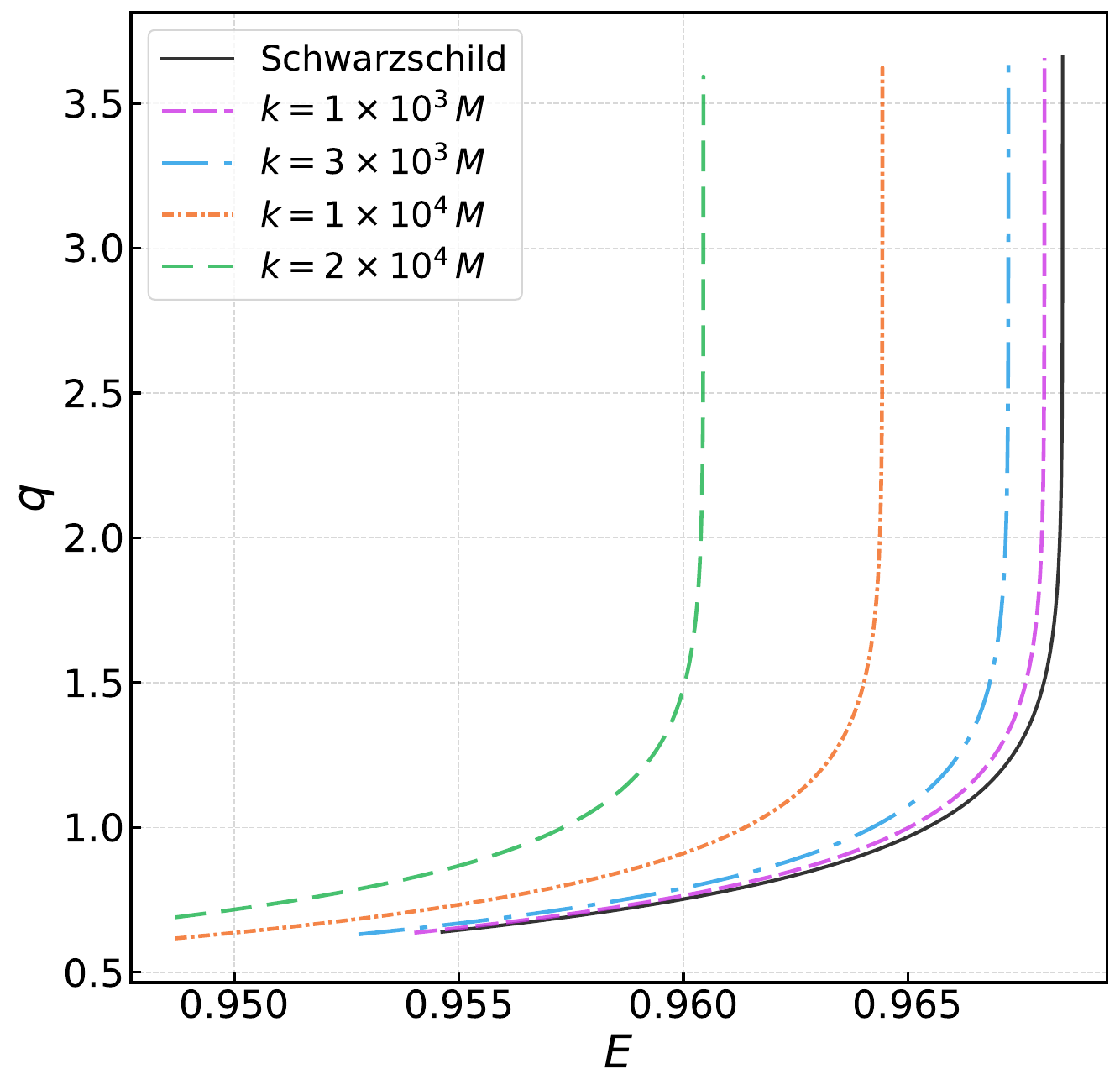}
		\caption{Moore}
	\end{subfigure}
	
	\caption{Precession angle parameter $q$ changes with orbital energy. The dark matter halo scale is fixed at $h=10^7M$, and the effects of different dark matter masses $k$ (ranging from $1\times10^3M \sim 2\times10^4M$) on the precession angles are compared within each halo model: (a) NFW model; (b) Beta model; (c) Moore model. The Schwarzschild black hole results (dashed black lines) serve as reference baselines.}
	\label{dif_k_q_models}
\end{figure}

\begin{figure}
	\centering 
	\begin{subfigure}{0.22\textwidth}
	\includegraphics[width=\linewidth]{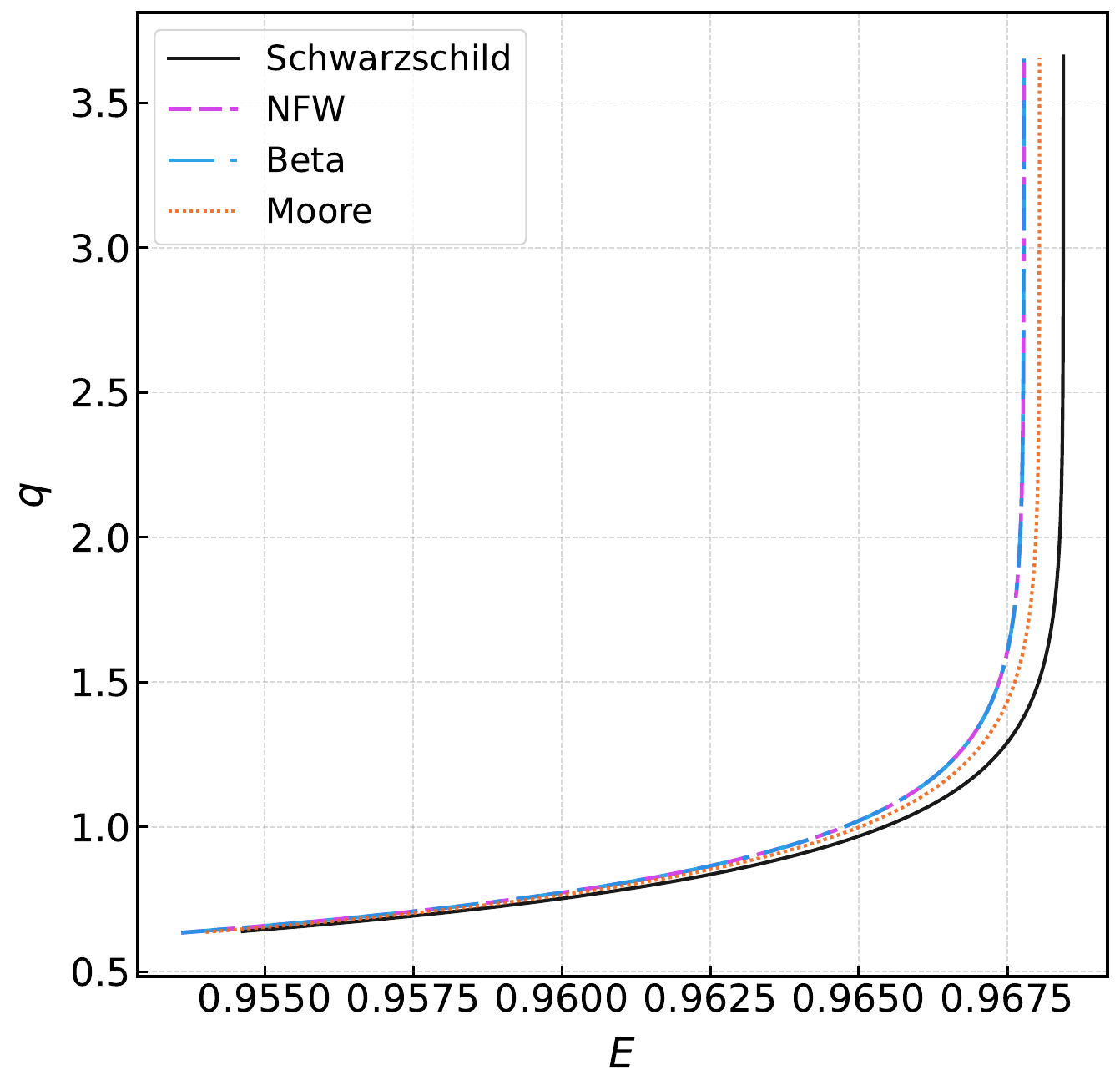}
	\caption{$k=1 \times 10^3\,M$}
	\label{subfig:k1e3}
	\end{subfigure}
	\hfill 
	\begin{subfigure}{0.22\textwidth}
	\includegraphics[width=\linewidth]{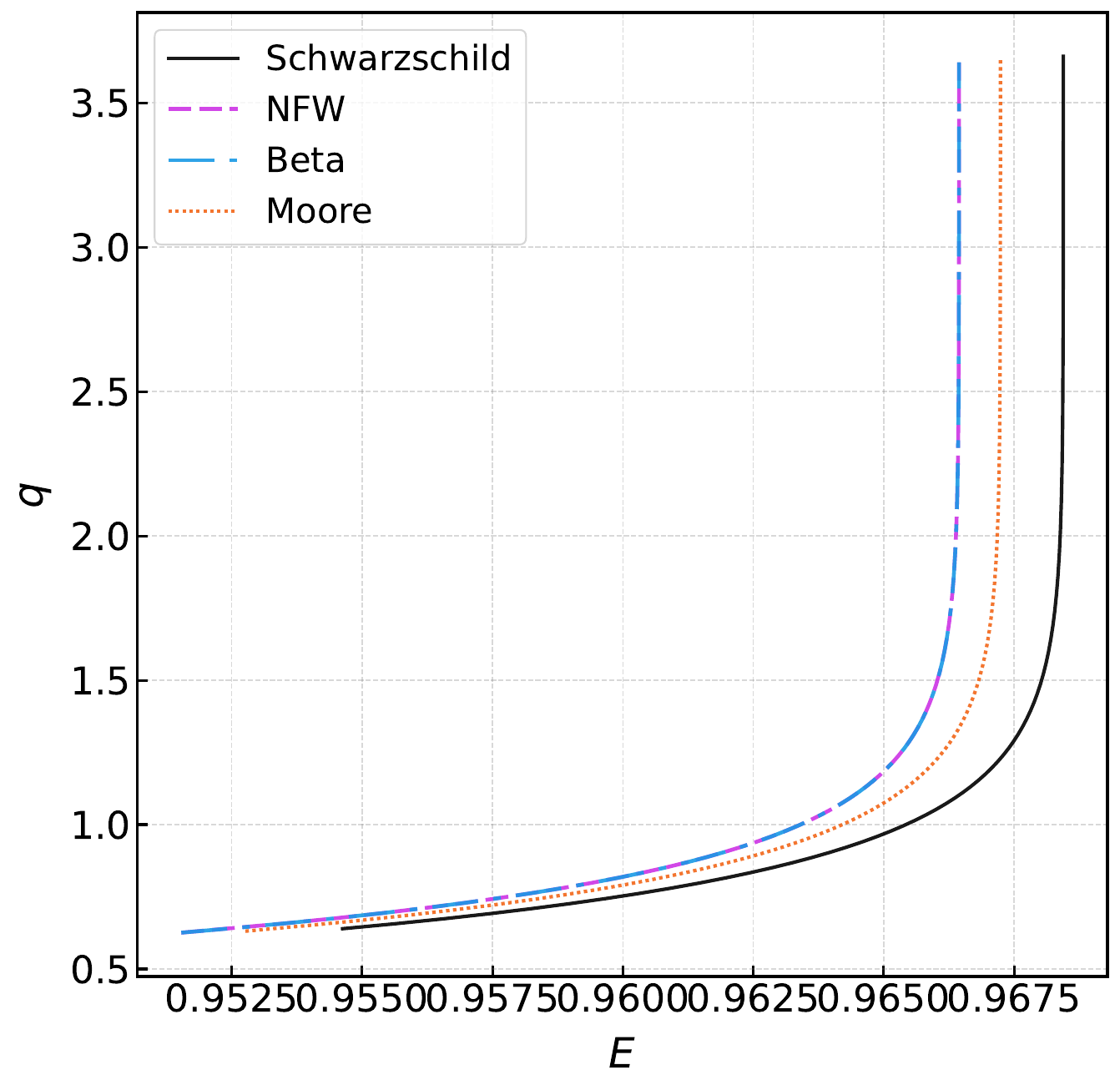}
	\caption{$k=3 \times 10^3\,M$}
	\label{subfig:k3e3}
	\end{subfigure} 
	\hfill 
	\begin{subfigure}{0.22\textwidth}
	\includegraphics[width=\linewidth]{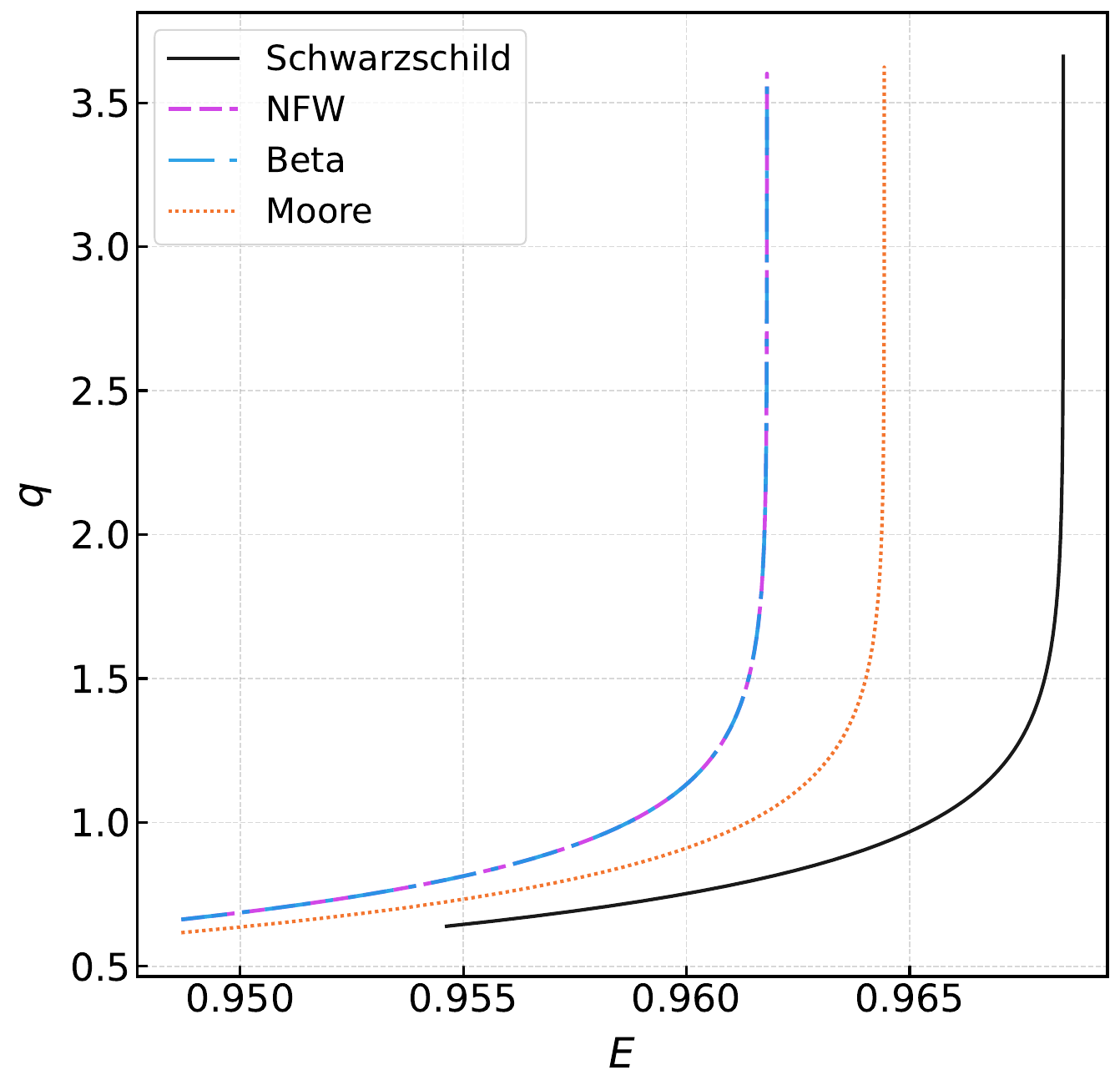}
	\caption{$k=1 \times 10^4\,M$}
	\label{subfig:k1e4}
	\end{subfigure}
	\hfill 
	\begin{subfigure}{0.22\textwidth}
	\includegraphics[width=\linewidth]{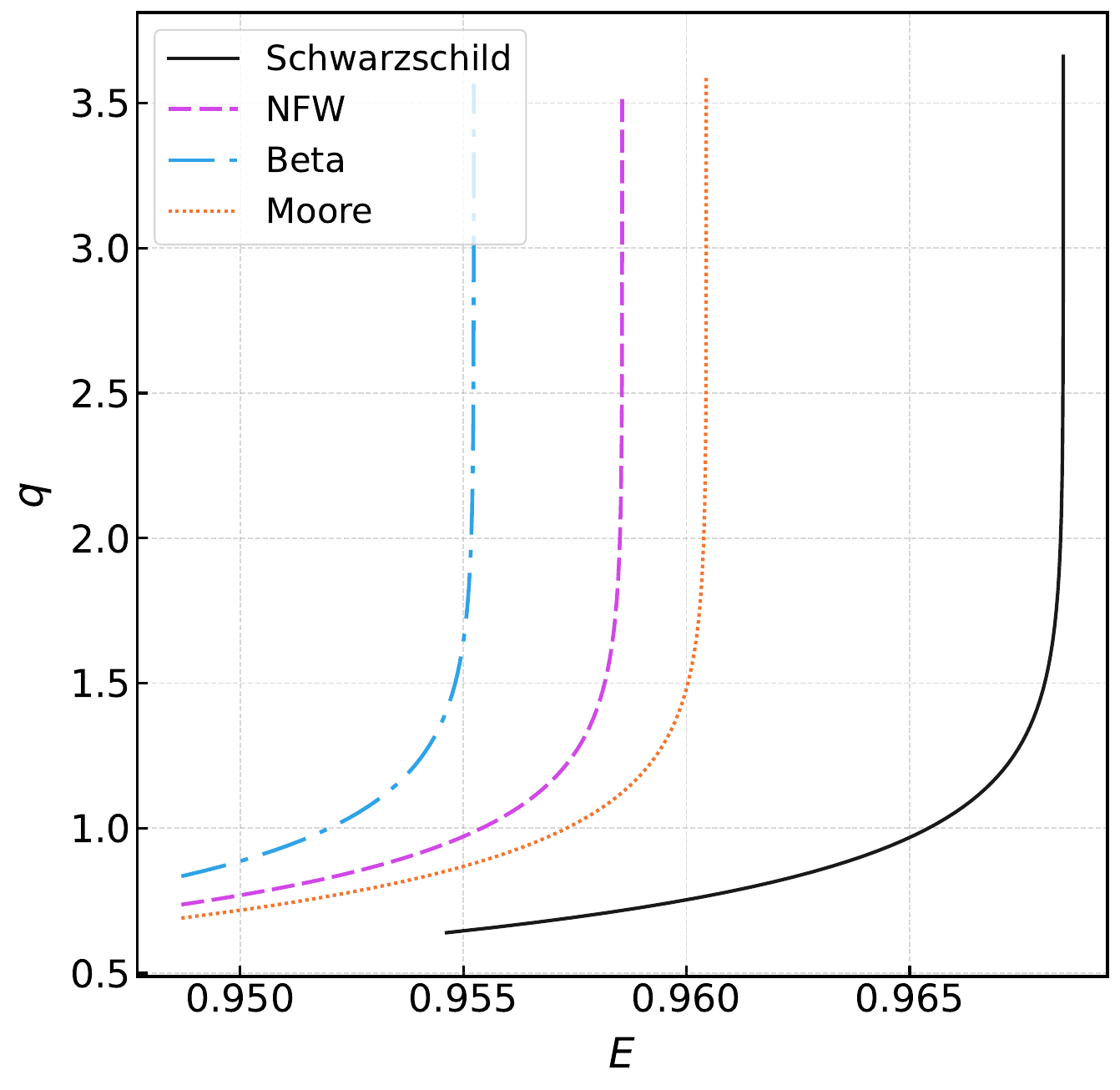}
	\caption{$k=2 \times 10^4\,M$}
	\label{subfig:k2e4}
	\end{subfigure}
	
	\caption{Comparison of precession angles for different dark matter halo models: the dark matter halo scale is fixed at \(h=10^7M\), and the subfigures illustrate results under four distinct dark matter masses, including (a) \(k=1\times10^3M\); (b) \(k=3\times10^3M\); (c) \(k=1\times10^4M\); and (d) \(k=2\times10^4M\).
	}
	\label{dif_k_q}
\end{figure}

In the study of dark matter effects on periodic orbits (which is conducted under a selected angular momentum $L = L_{\text{ISCO}} + \frac{1}{2}(L_{\text{MBO}} - L_{\text{ISCO}})$), the primary step is to determine the characteristic energy \( E_{(z~w~ v)} \).The accurate value of this energy can be calculated from the numerical results of the precession angle parameter \( q \) varying with \( E \), as is derived from Eq.~(\ref{e25}).  Fig.~\ref{dif_k_q_models} exhibits the variation patterns of precession angles under three different dark matter halo models (NFW, Beta, and Moore) with different dark matter masses \( k \), where the dark matter halo scale is fixed at \( h=10^7 M \). Fig.~\ref{dif_k_q} further compares the precession angle among different models under fixed dark matter mass, and halo scale. The energy \( E_{(z~w~v)} \) corresponding to periodic orbits can be determined via the intersection points of the curves with $q = w + v/z$, and the obtained precise values of $E_{(z~w~v)}$ is detailed in Appendix~\ref{a2}.

An important feature is observed from Fig.~\ref{dif_k_q_models}: under the same precession angle \( q \), the larger the dark matter mass \( k \), the lower the energy \( E \) required to achieve that precession angle. This phenomenon indicates that an increase in dark matter mass significantly reduces the energy \( E_{(z~w~v)} \) of specific periodic orbit patterns. The comparative analysis in Fig.~\ref{dif_k_q} further reveals substantial differences among different dark matter halo models. Under the same precession angle \(q\), the required orbital energies \(E_{(z~ w~v)}\) in the NFW and Beta models are generally lower than those in Moore model, caused by contributions of different dark matte density distributions to gravitational field. Notably, among the dark matter masses examined in our 
study --- \(k = 1\times10^{3}M\), \(3\times10^{3}M\), \(1\times10^{4}M\), and \(2\times10^{4}M\) --- the curves of the NFW and Beta models almost overlap, making them difficult to distinguish. This suggests that the NFW and Beta density profiles produce very similar effects on orbital precession.

\begin{figure}
	\centering 
	
	\begin{subfigure}{\textwidth}
	\includegraphics[width=0.23\linewidth]{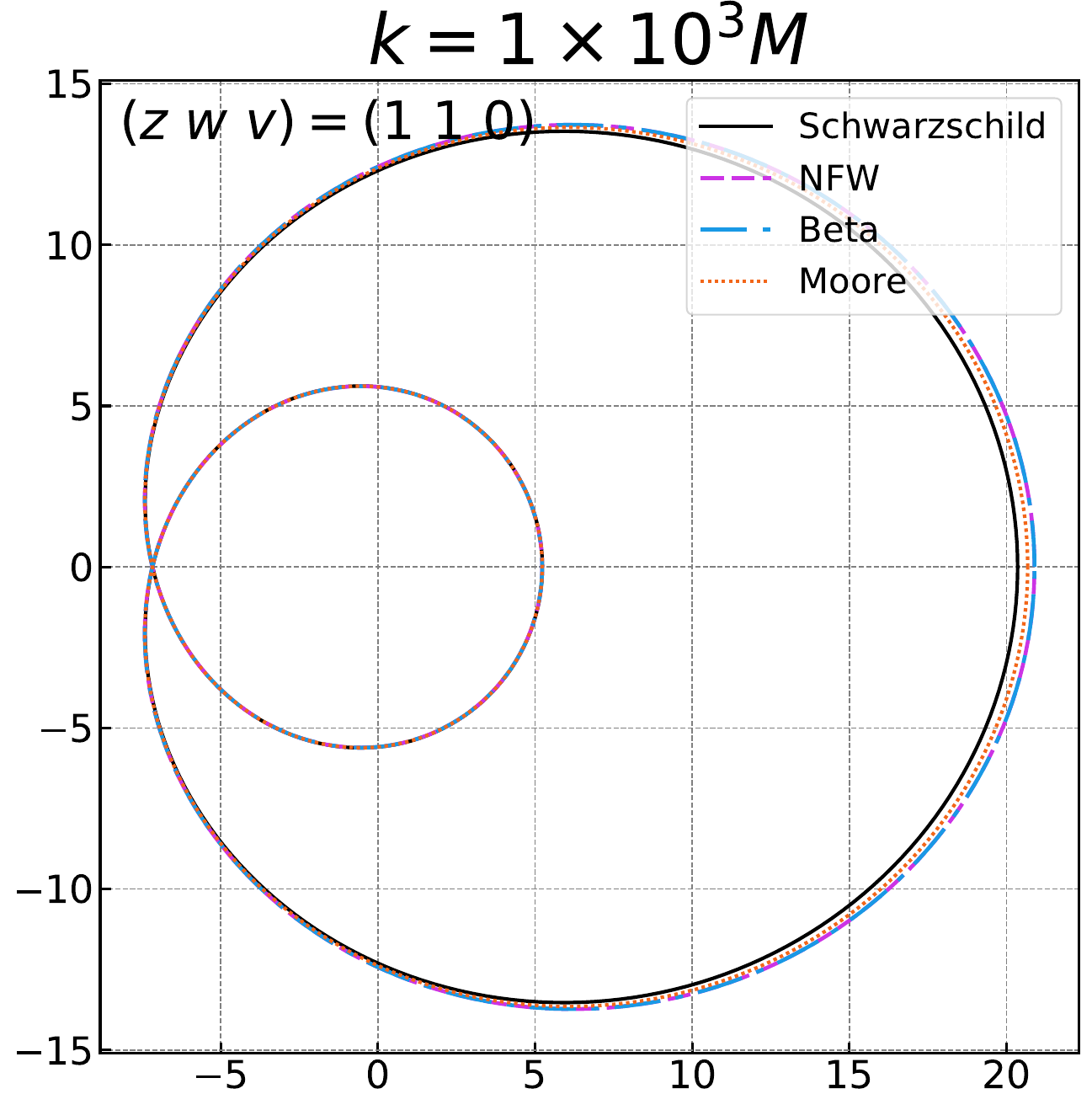}
	\includegraphics[width=0.23\linewidth]{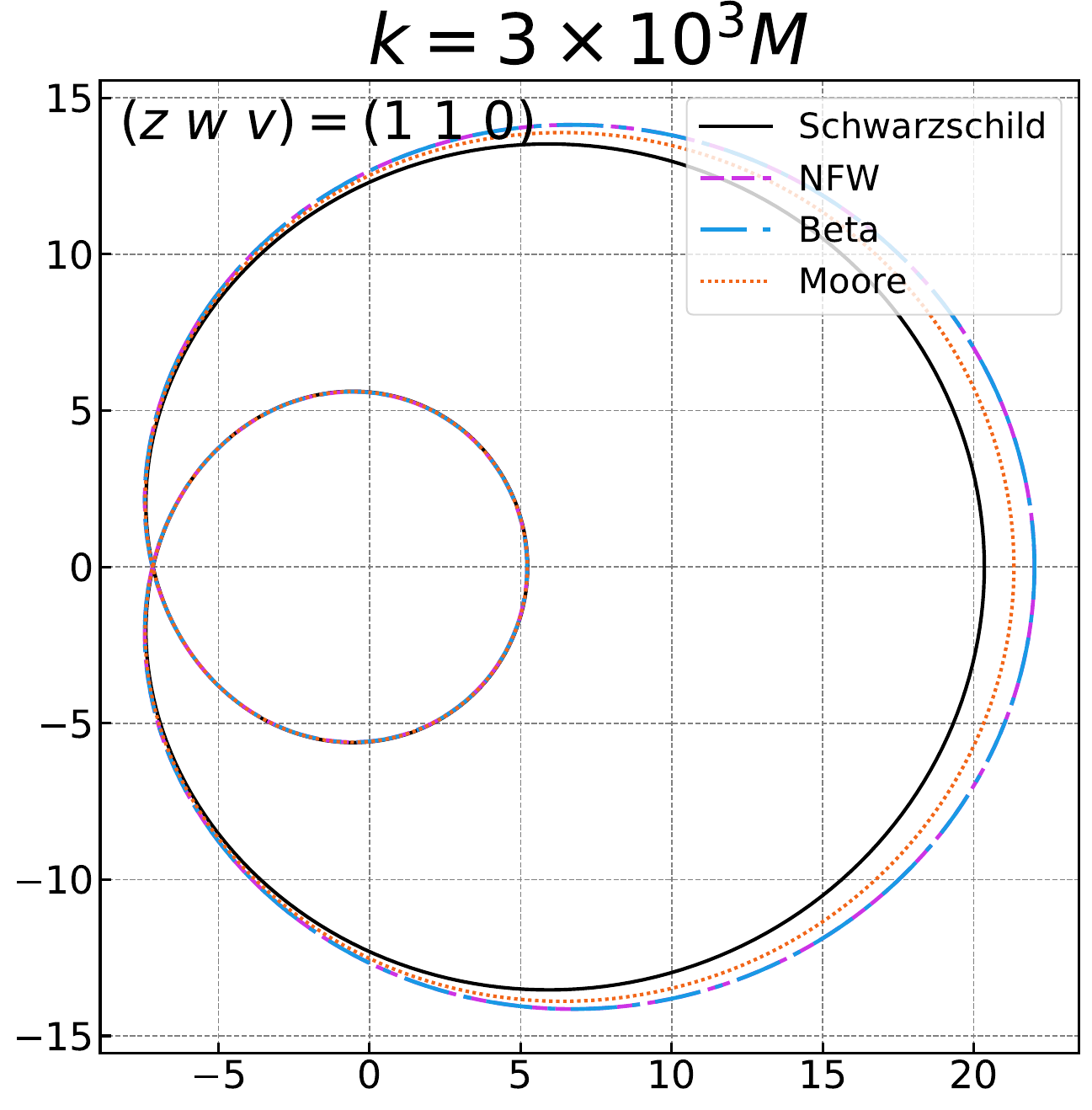}
	\includegraphics[width=0.23\linewidth]{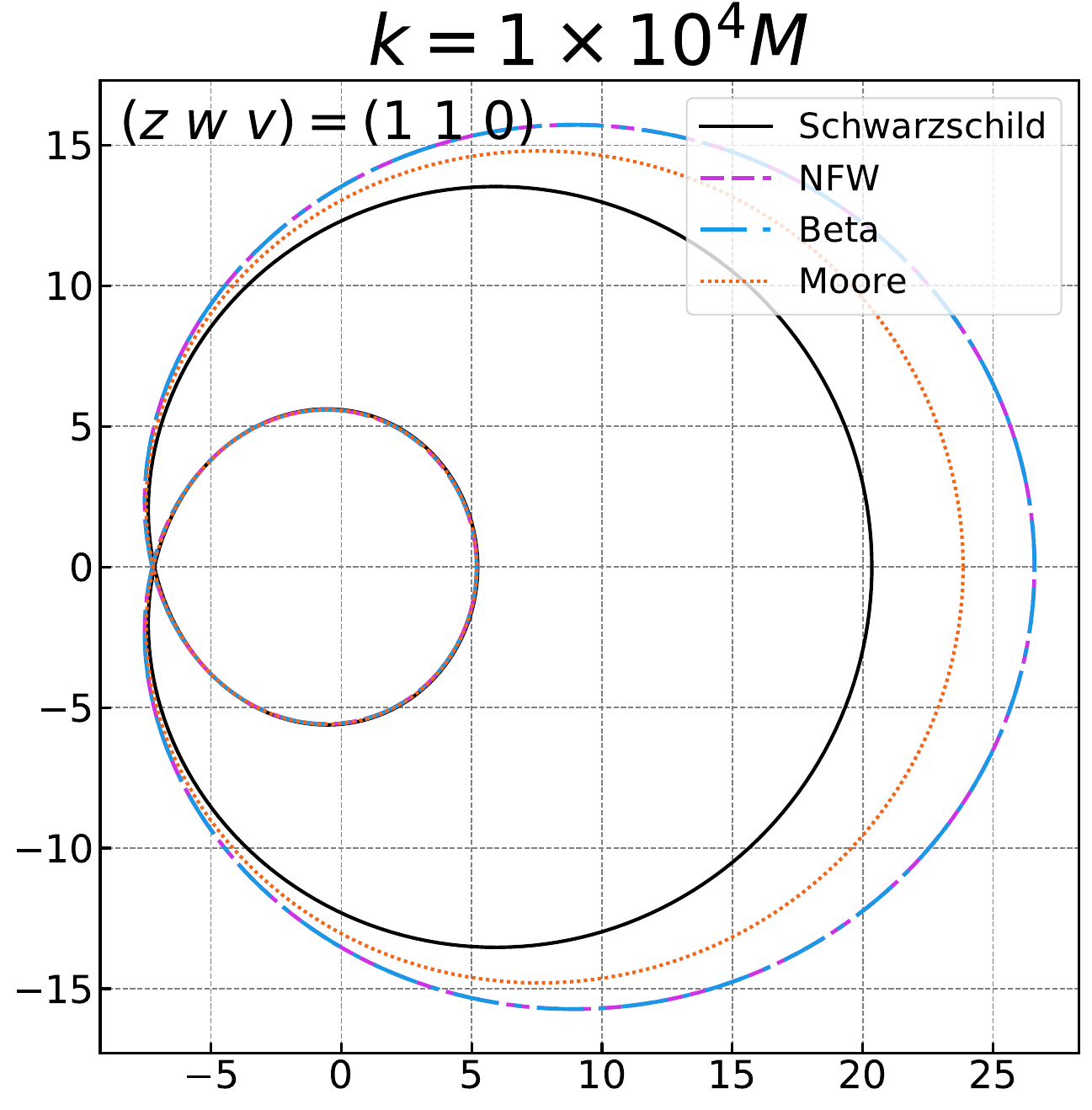}
	\includegraphics[width=0.23\linewidth]{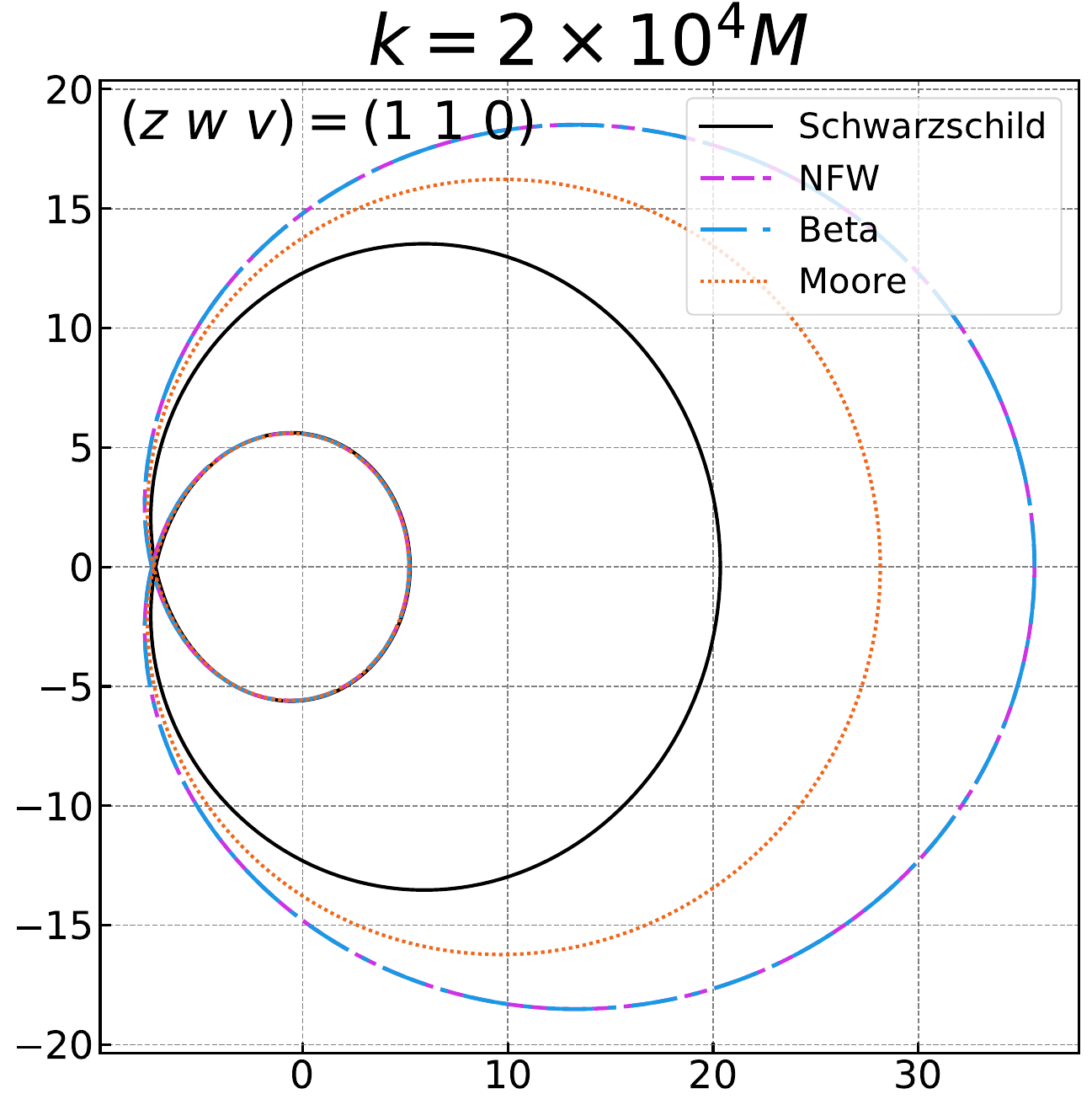}
	\end{subfigure}
	
	\vspace{0.4cm} 
	
	\begin{subfigure}{\textwidth}
	\includegraphics[width=0.23\linewidth]{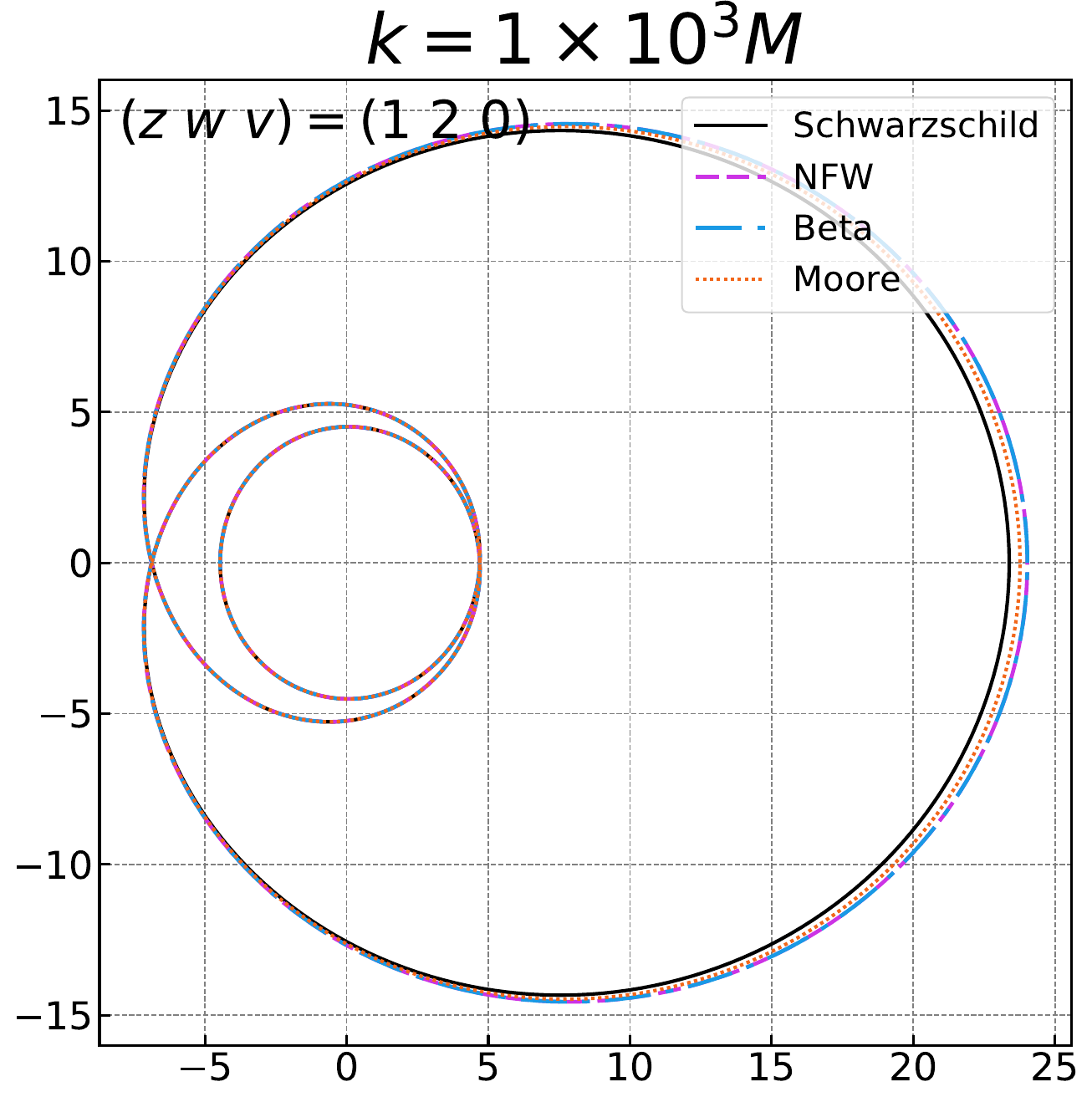}
	\includegraphics[width=0.23\linewidth]{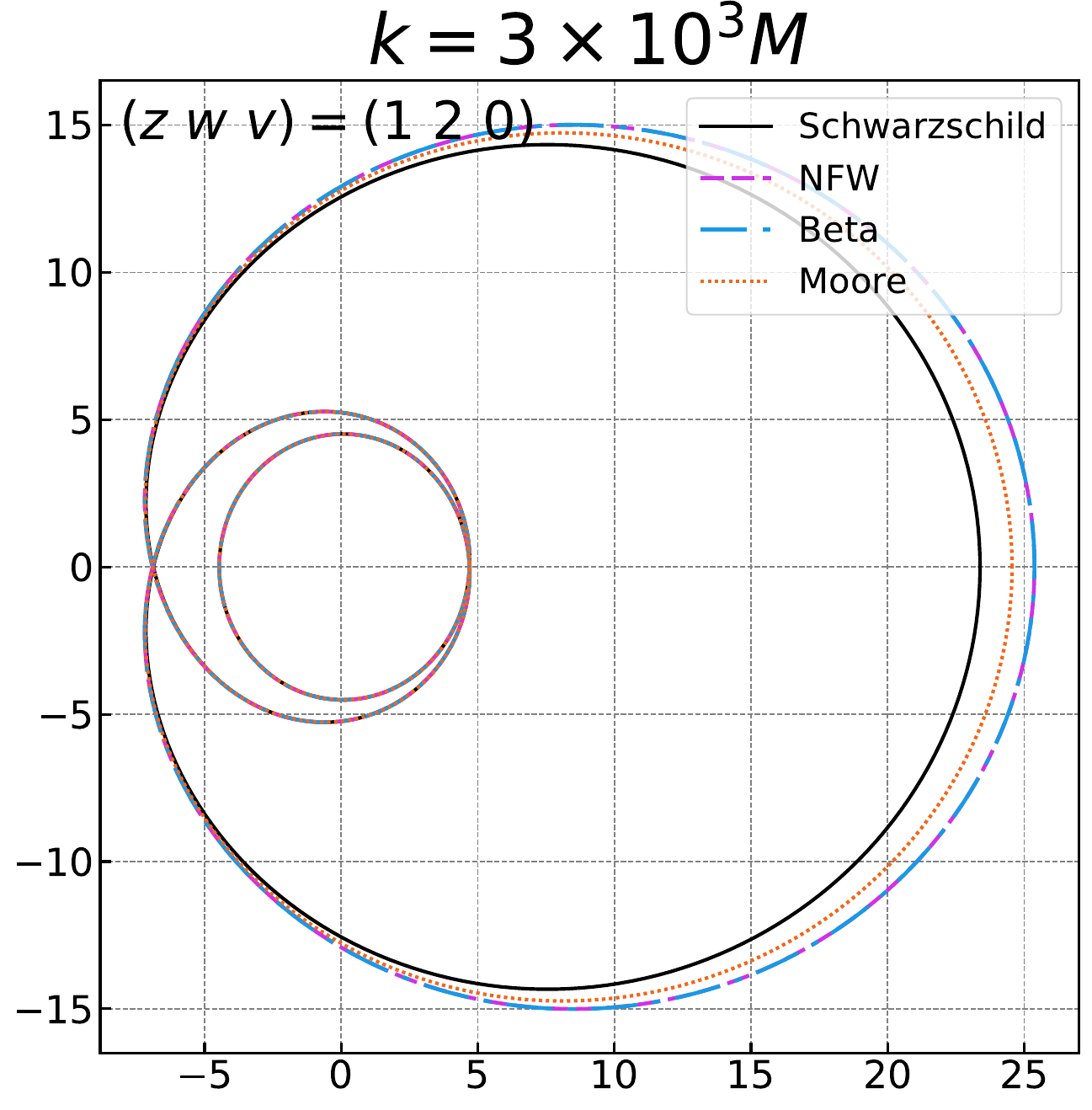}
	\includegraphics[width=0.23\linewidth]{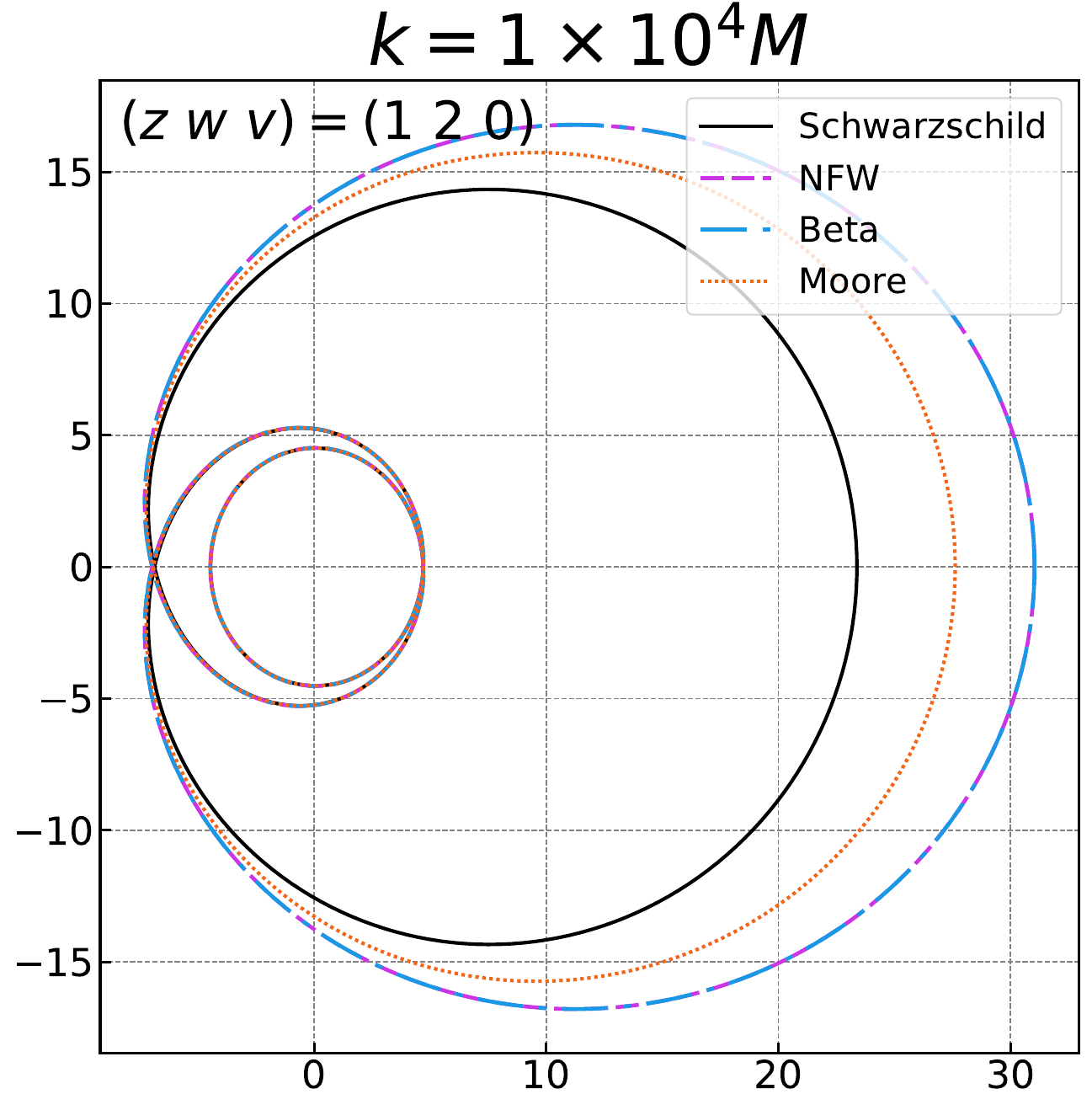}
	\includegraphics[width=0.23\linewidth]{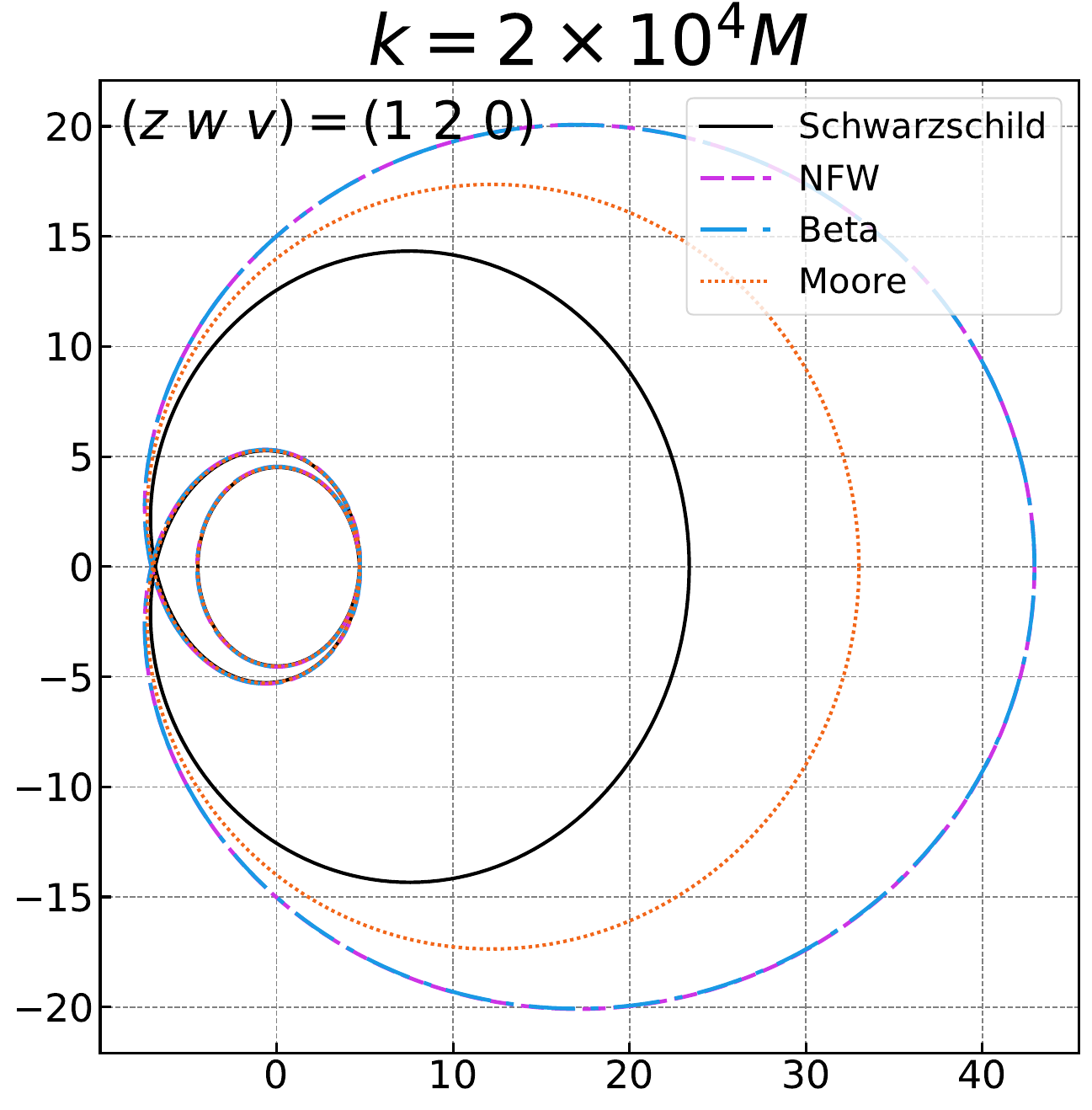}
	\end{subfigure}
	
	\vspace{0.4cm}
	
	\begin{subfigure}{\textwidth}
	\includegraphics[width=0.23\linewidth]{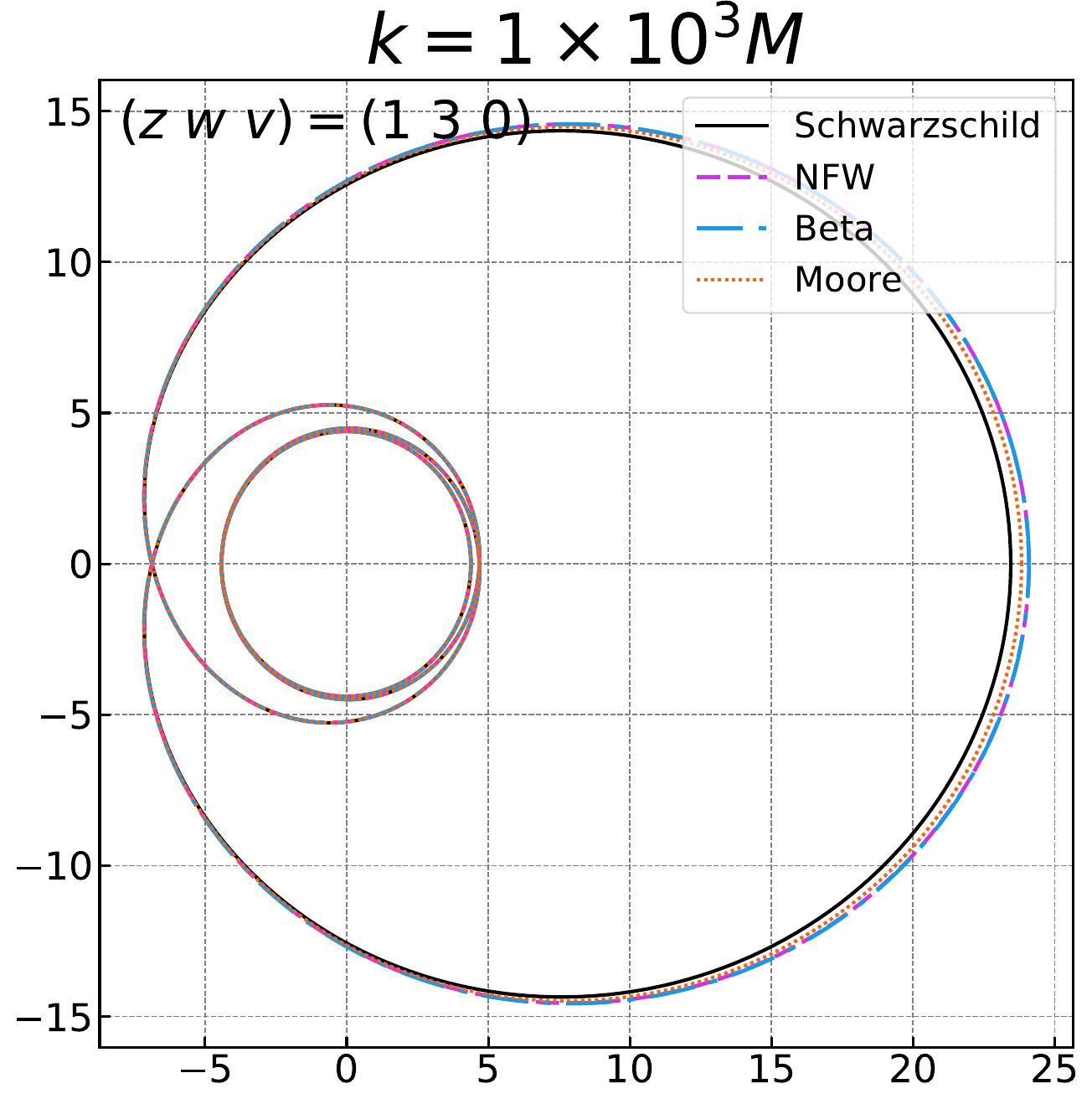}
	\includegraphics[width=0.23\linewidth]{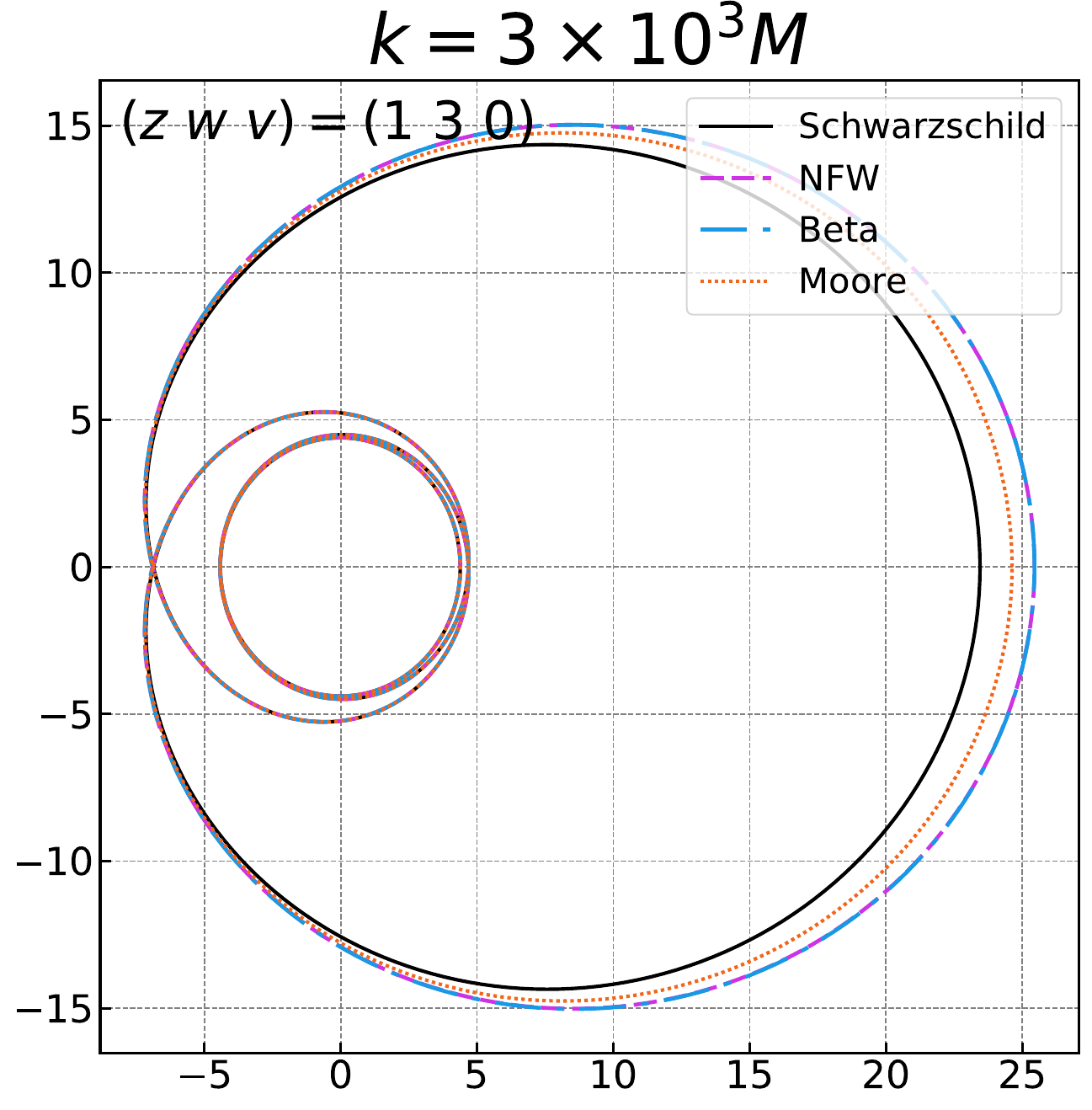}
	\includegraphics[width=0.23\linewidth]{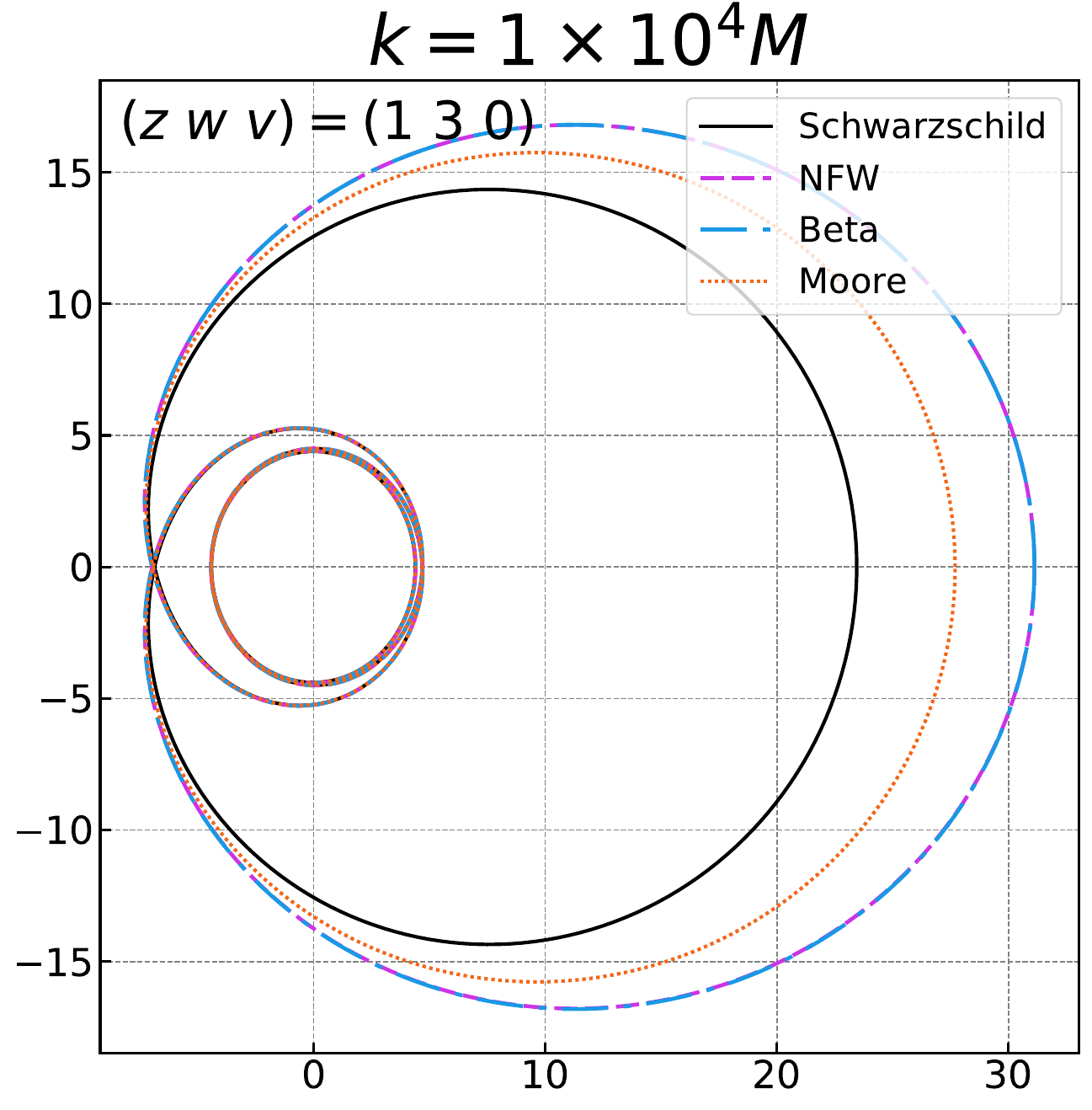}
	\includegraphics[width=0.23\linewidth]{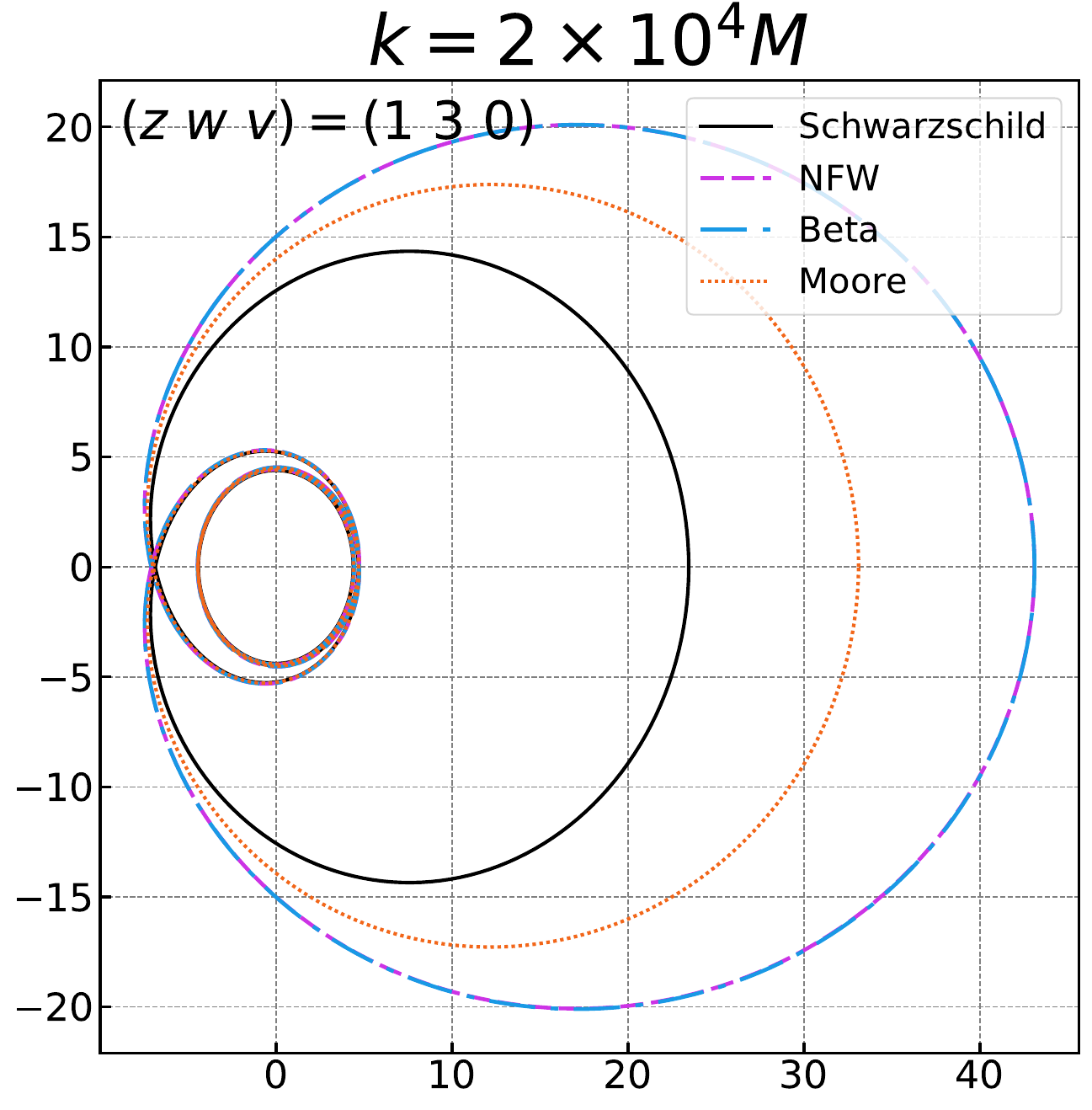}
	\end{subfigure}
	
	\vspace{0.4cm}
	
	\begin{subfigure}{\textwidth}
	\includegraphics[width=0.23\linewidth]{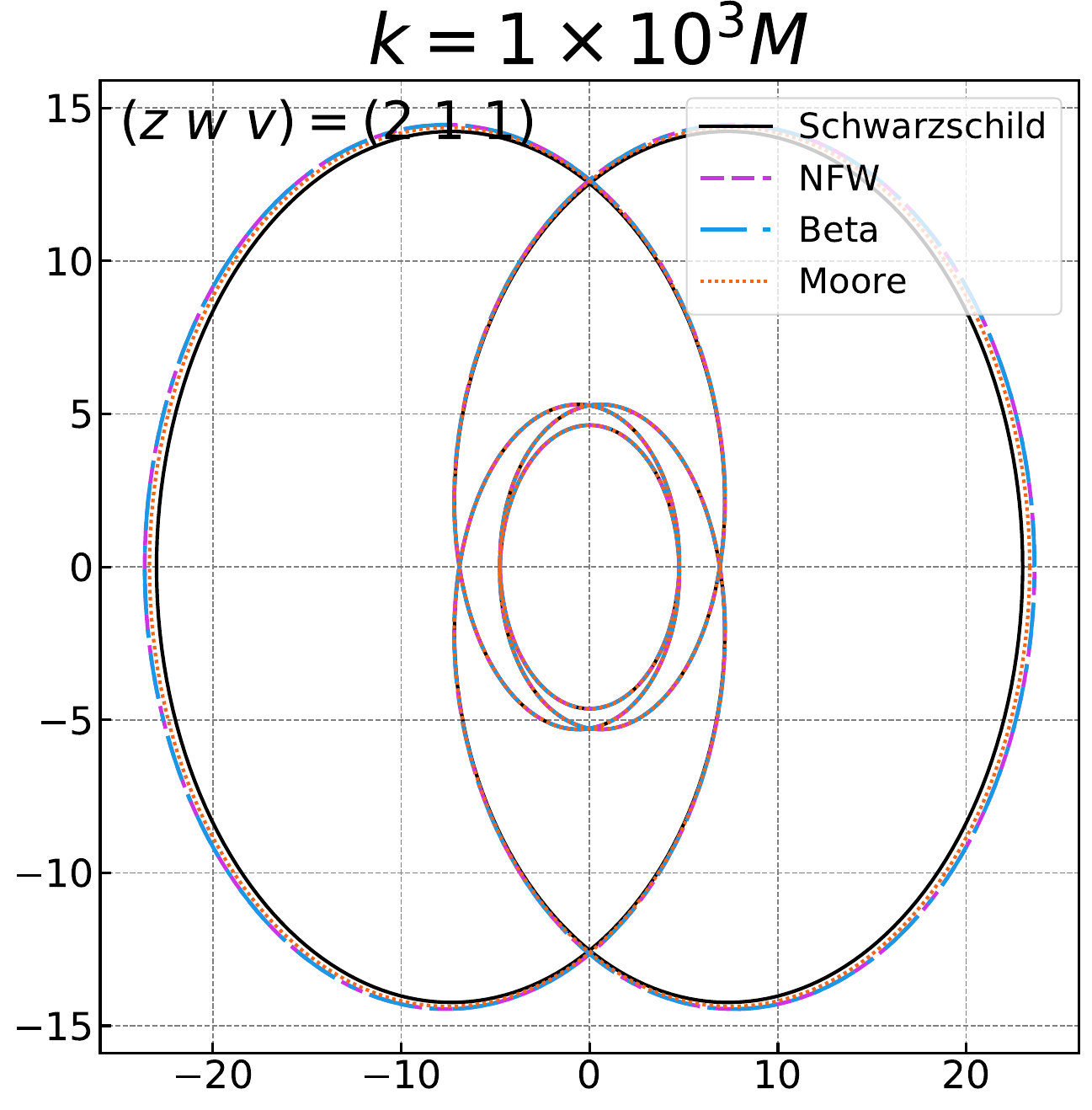}
	\includegraphics[width=0.23\linewidth]{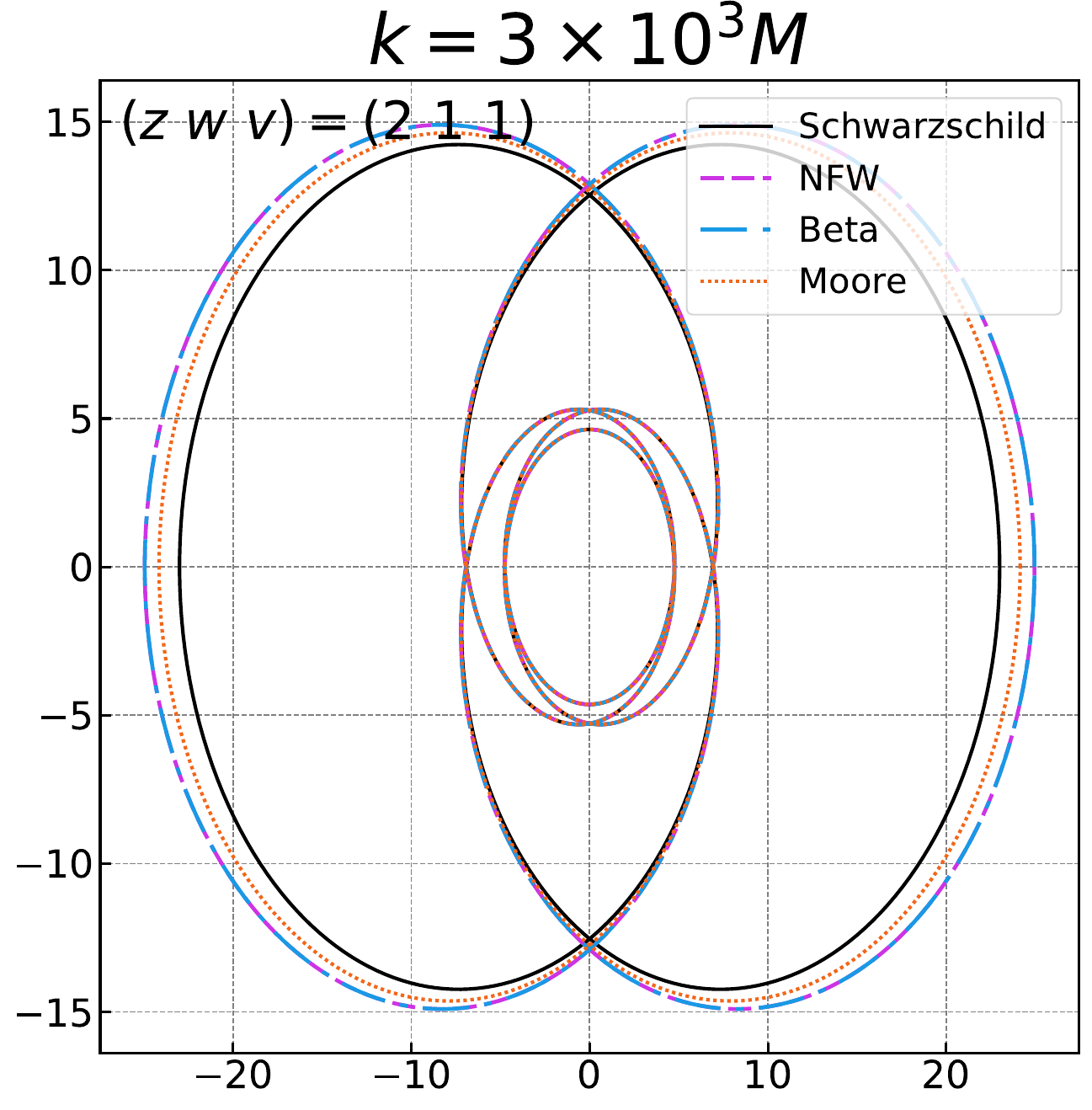}
	\includegraphics[width=0.23\linewidth]{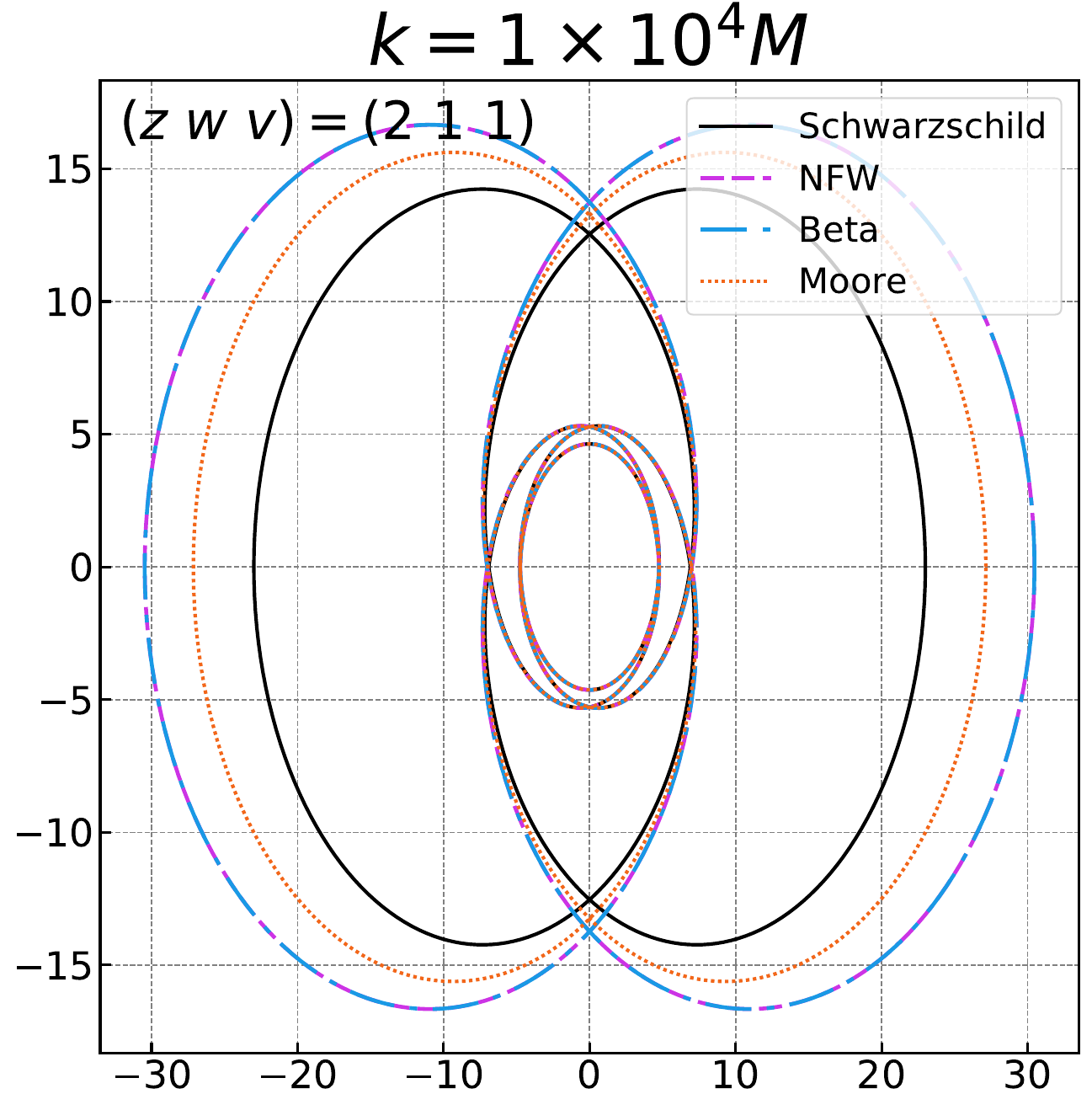}
	\includegraphics[width=0.23\linewidth]{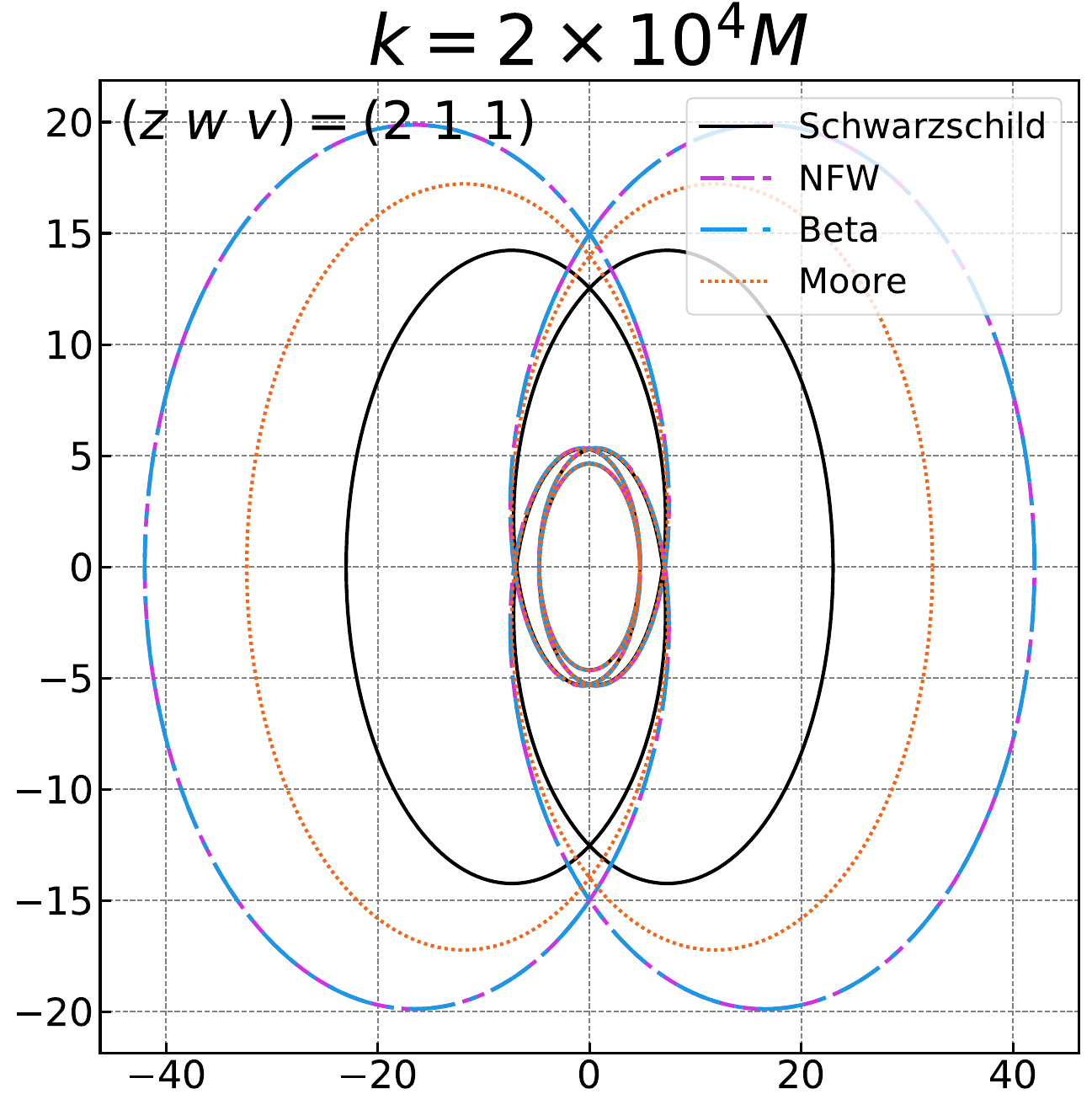}
	\end{subfigure}
	
	\vspace{0.4cm}
	
	\begin{subfigure}{\textwidth}
		\includegraphics[width=0.23\linewidth]{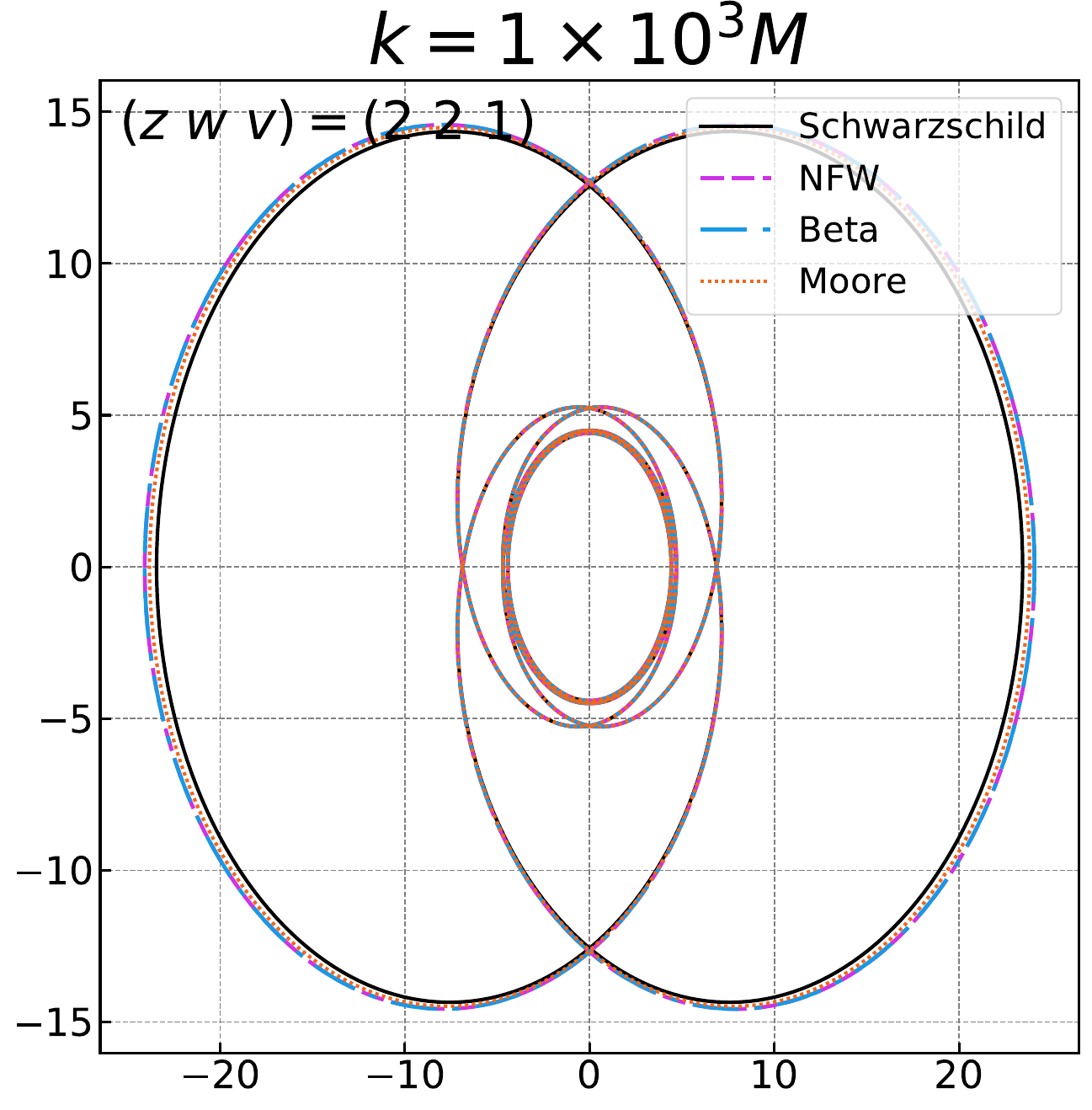}
		\includegraphics[width=0.23\linewidth]{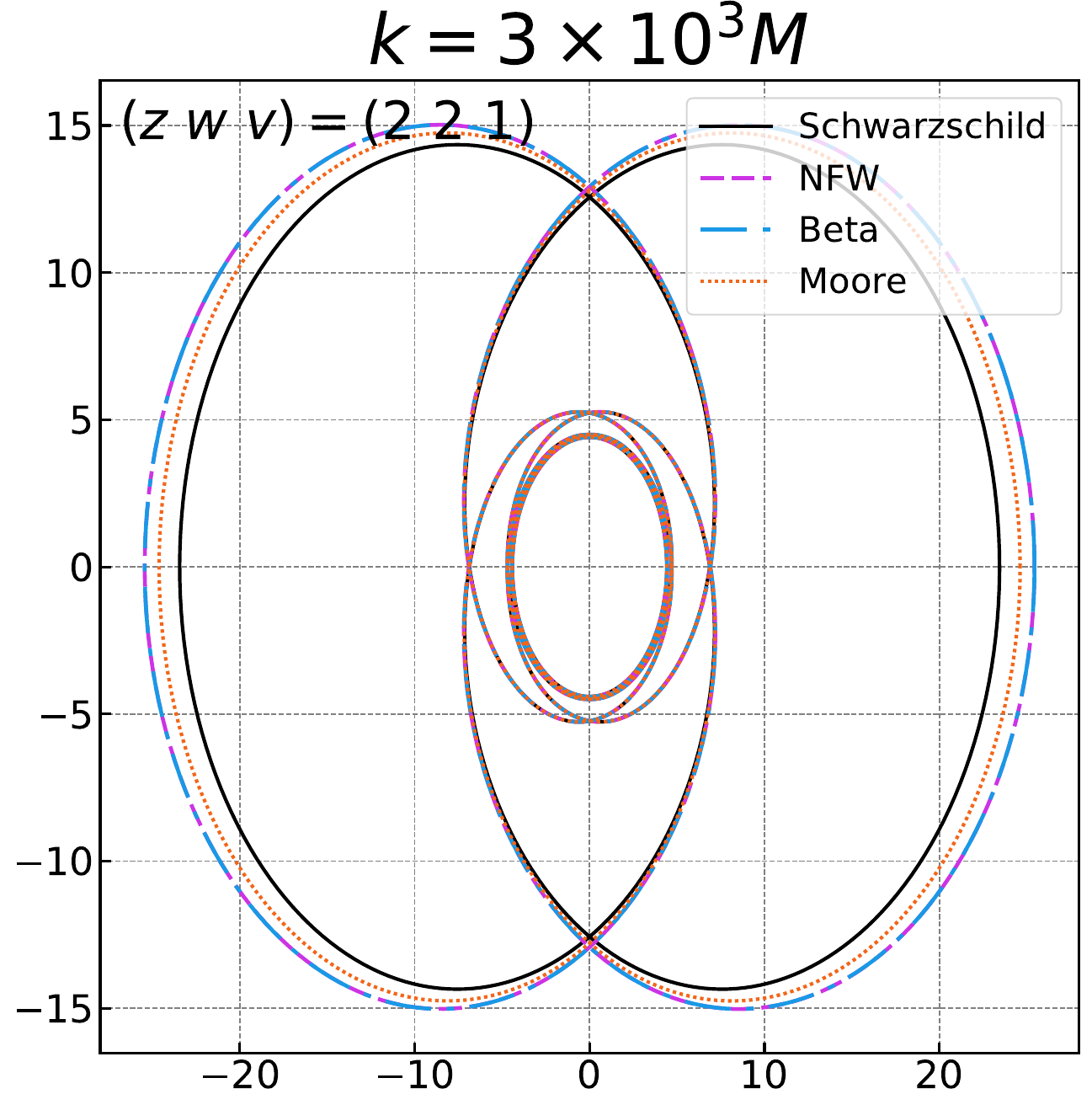}
		\includegraphics[width=0.23\linewidth]{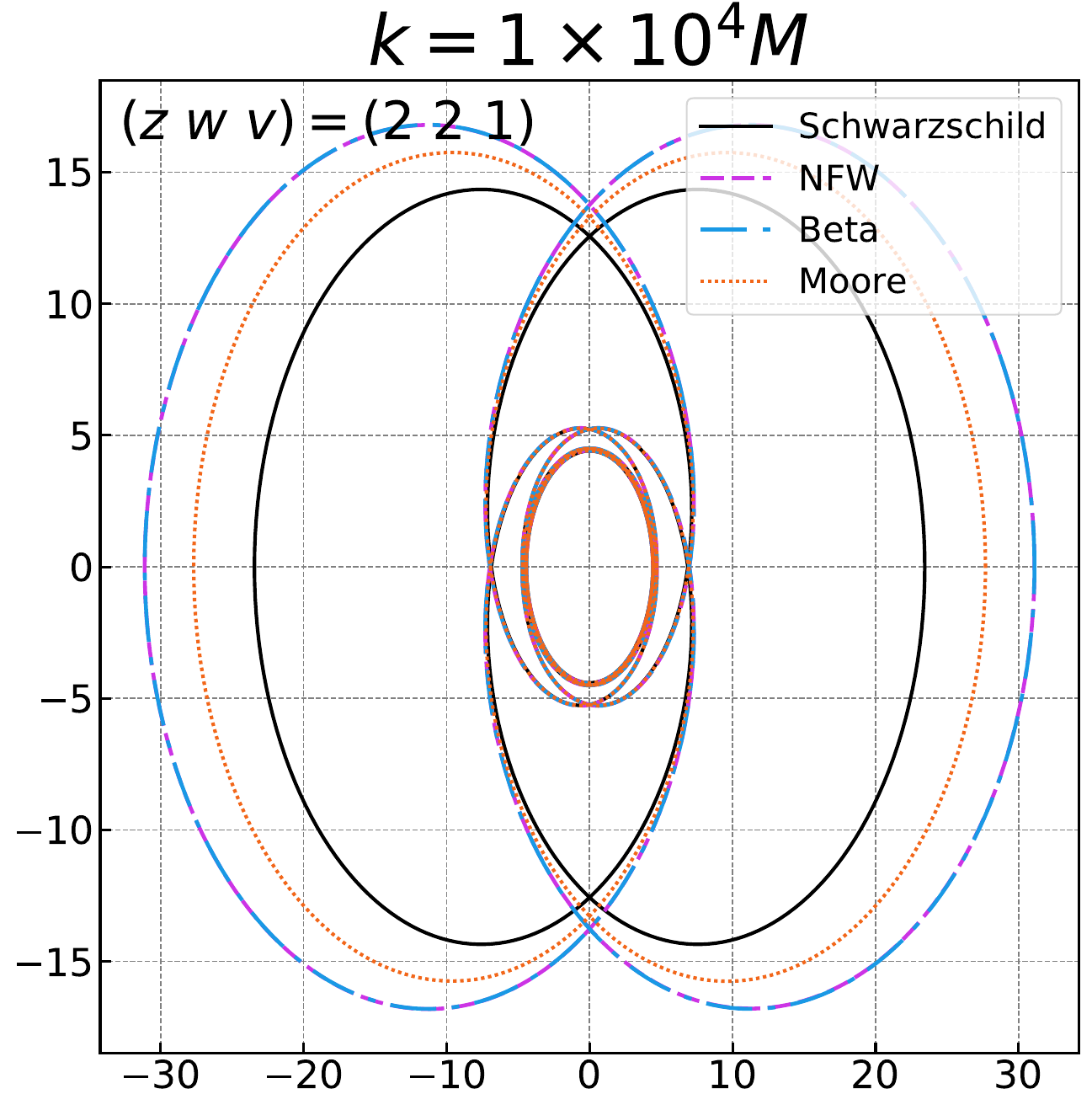}
		\includegraphics[width=0.23\linewidth]{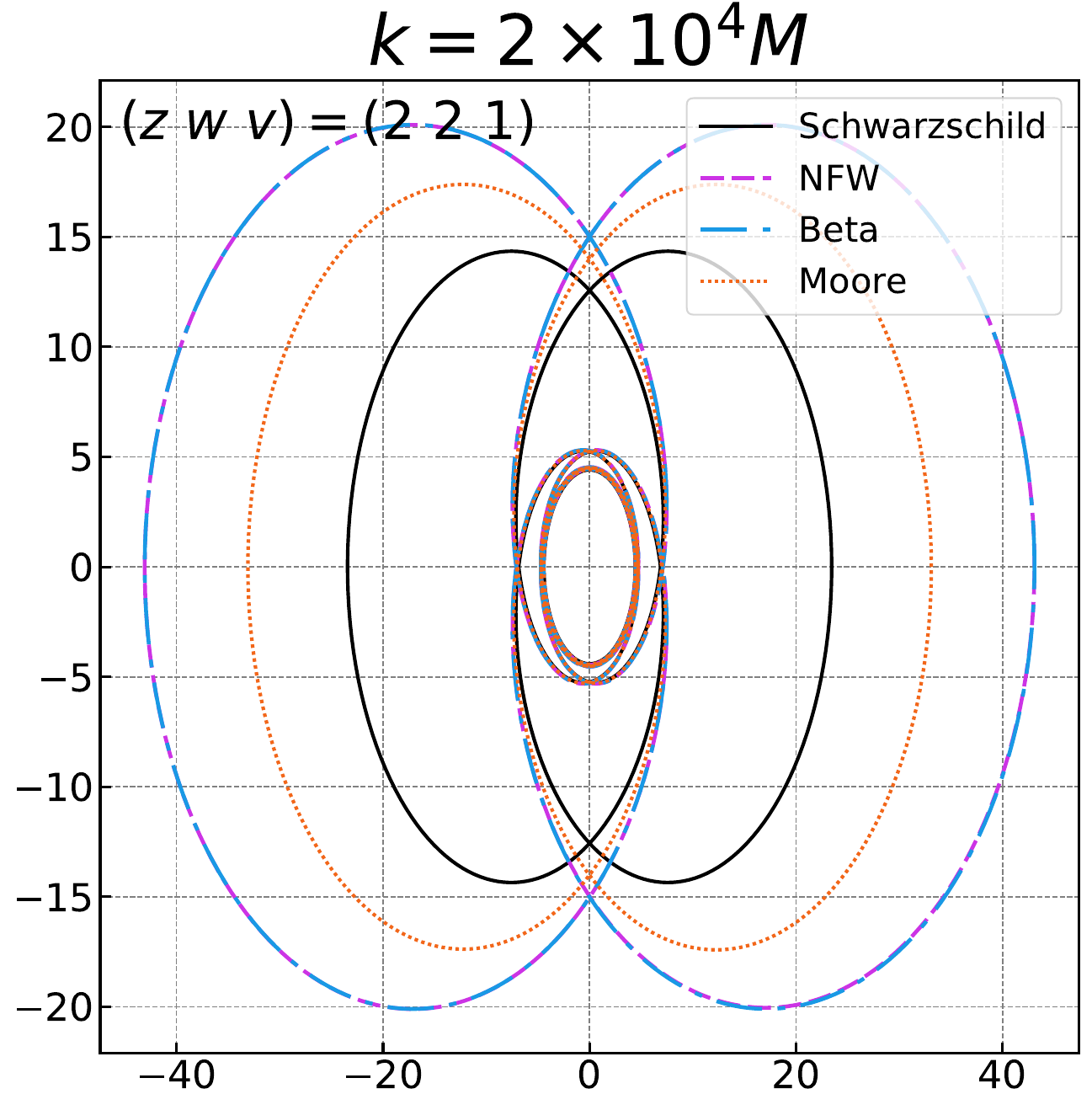}
	\end{subfigure}
	
	\caption{Periodic orbits around black holes embedded in different dark matter halos with different \( (z~w~v) \) , where the dark matter mass \( k \) ranges from \( 1\times10^3M \) $\sim$ \( 2\times10^4M \) while the dark matter halo characteristic radius remains selected at \( h = 10^7M \). The parameter for angular momentum is selected as $\varepsilon=0.5$.}
	\label{dif_k}
\end{figure}

\begin{figure}
	\centering
	\begin{subfigure}{0.3\textwidth}
		\includegraphics[width=\linewidth]{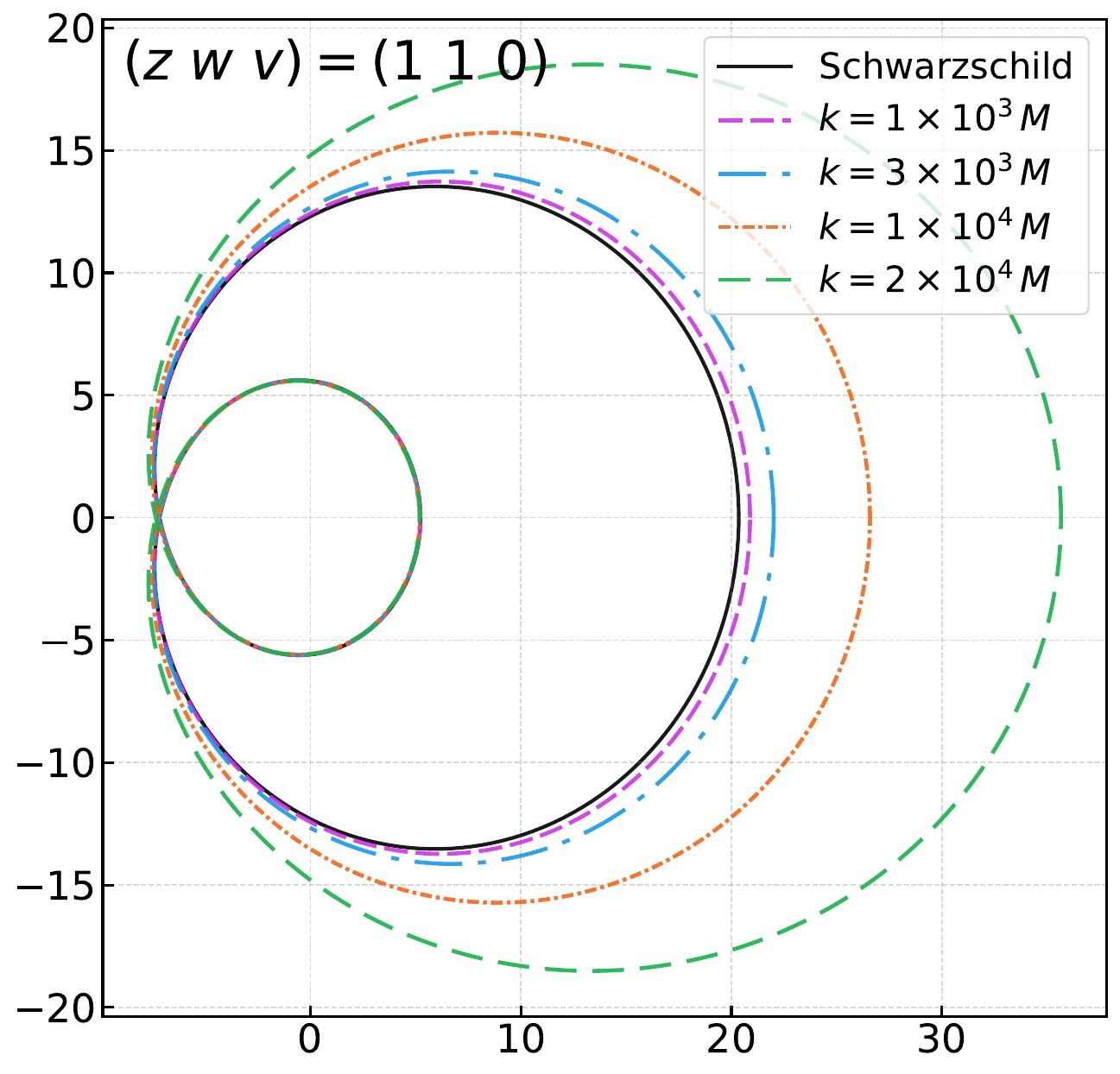}
		\caption{\( (z~w~v) \)=(1 1 0), $h= 10^7 M$}
	\end{subfigure}
	\hfill
	\begin{subfigure}{0.3\textwidth}
		\includegraphics[width=\linewidth]{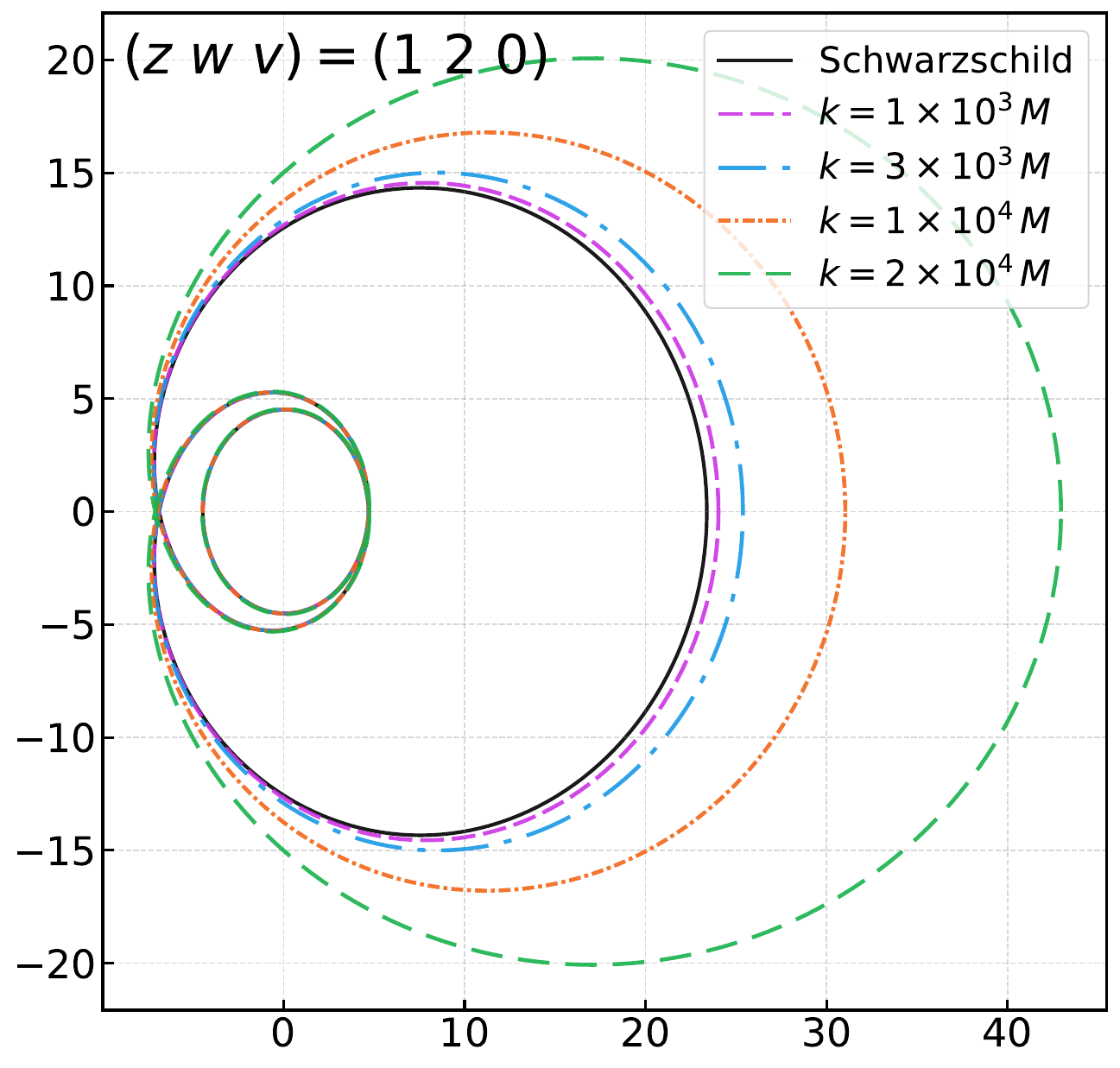}
		\caption{\( (z~w~v) \)=(1 2 0), $h= 10^7 M$}
	\end{subfigure}
	\hfill
	\begin{subfigure}{0.3\textwidth}
		\includegraphics[width=\linewidth]{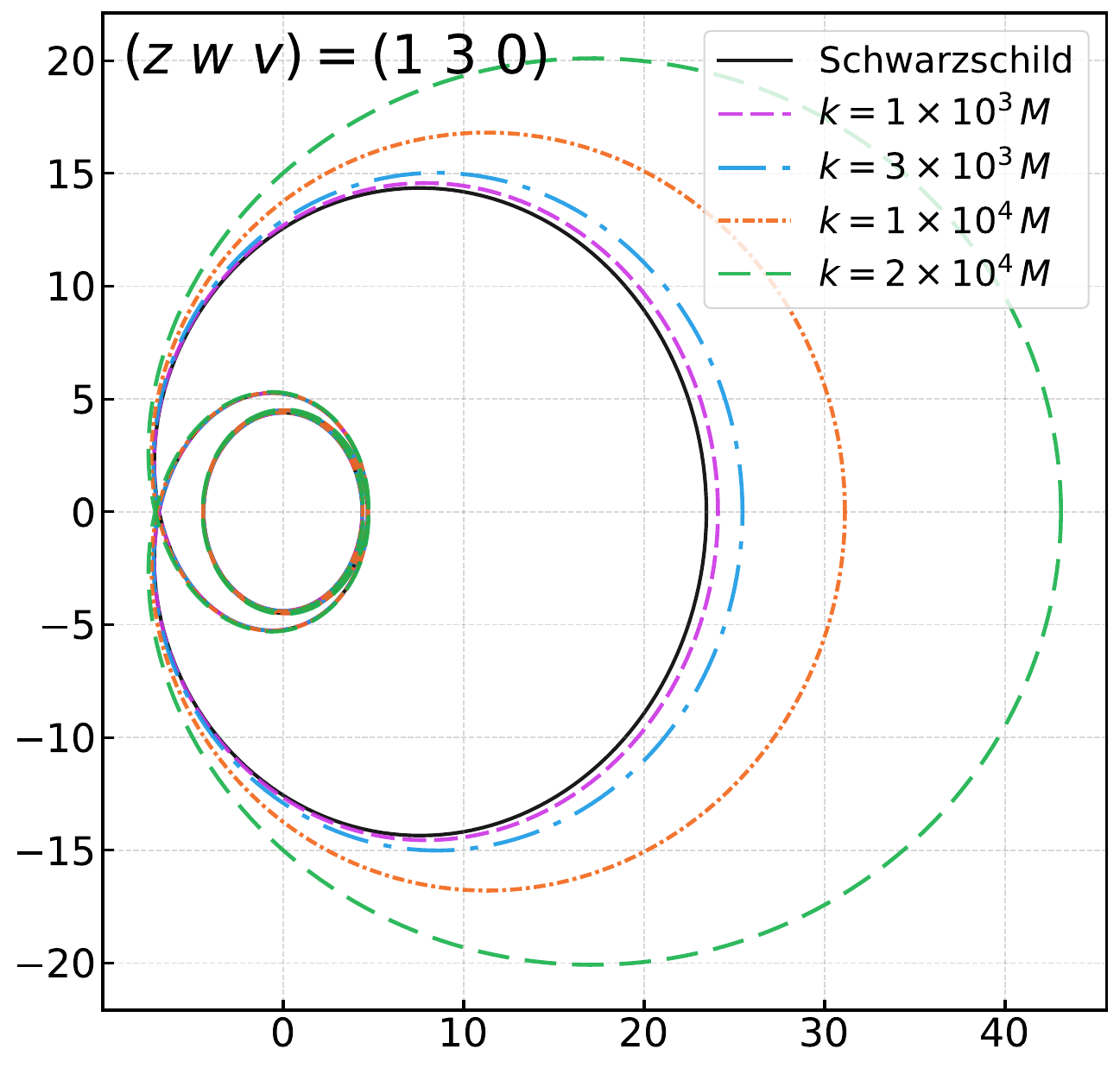}
		\caption{\( (z w~v) \)=(1 3 0), $h= 10^7 M$}
	\end{subfigure}
	\vspace{0.4cm}
	\centering
	\hspace{\fill} 
	\begin{subfigure}{0.3\textwidth}
		\includegraphics[width=\linewidth]{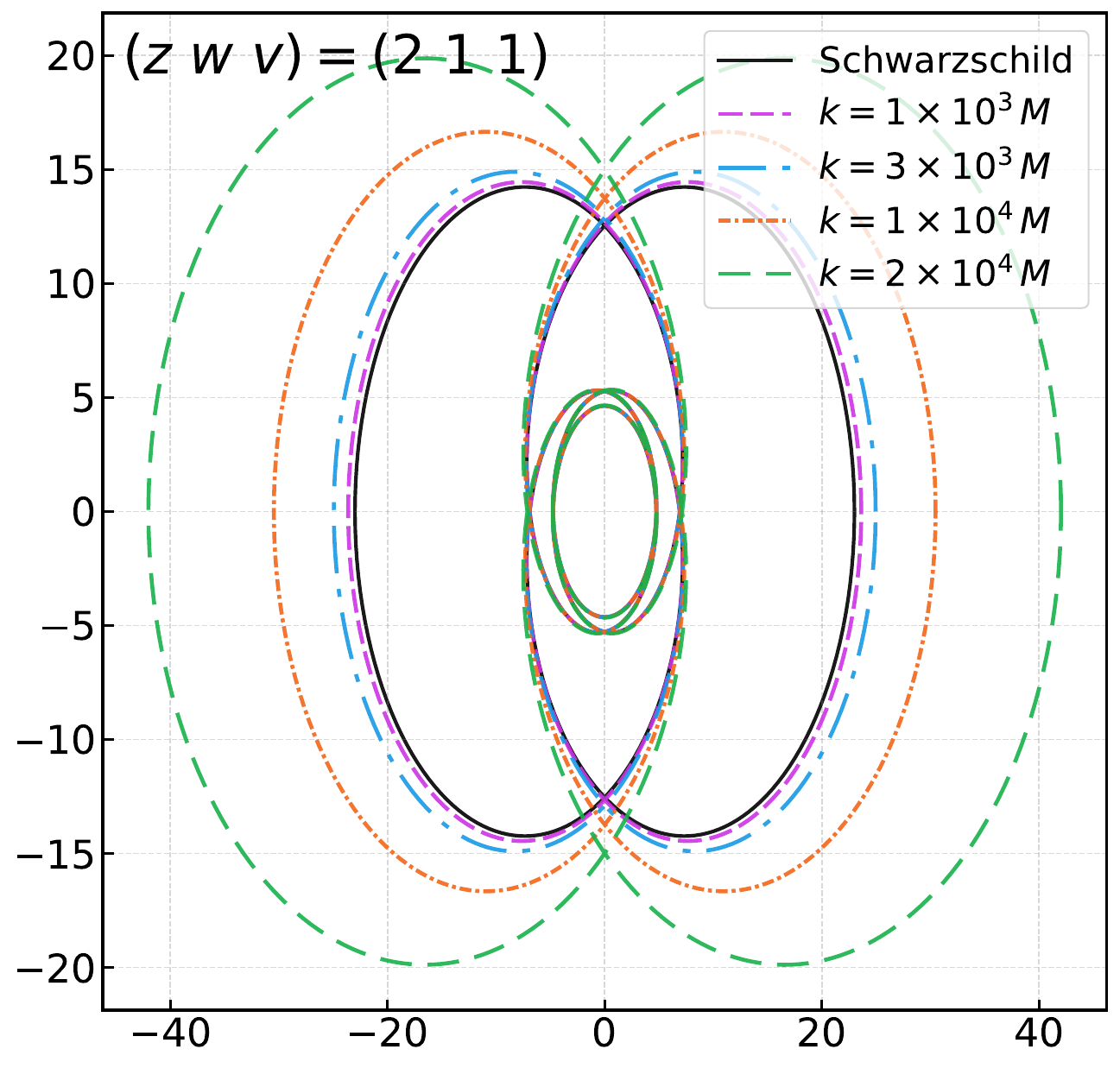}
		\caption{\( (z~w~v) \)=(2 1 1), $h= 10^7 M$}
	\end{subfigure}
	\hfill
	\begin{subfigure}{0.3\textwidth}
		\includegraphics[width=\linewidth]{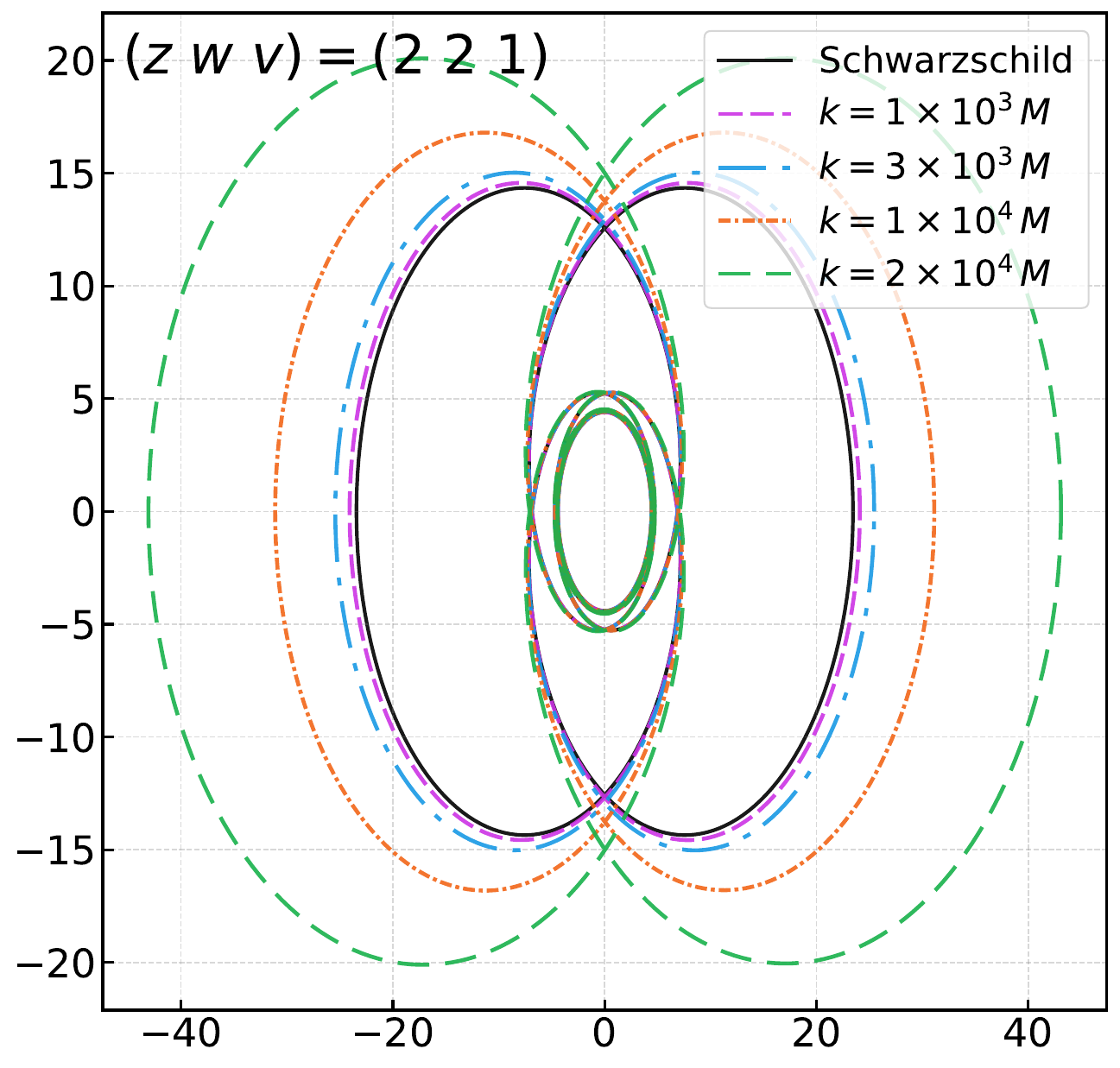}
		\caption{\( (z~w~v) \)=(2 2 1), $h= 10^7 M$}
	\end{subfigure}
	\hspace*{\fill} 
	
	\caption{The schematic diagrams of periodic orbits around black holes corresponding to different  \( k \) and different orbital configurations \( (z~w~v) \) calculated in the NFW dark matter halo model. The parameter of angular momentum is selected at $\varepsilon=0.5$.
	}
	\label{dif_k_NFW}
\end{figure}

After obtaining the corresponding orbital energy parameters $E_{(z~w~v)}$ under different dark matter masses $k$, we can solve for the trajectories of periodic orbits and systematically investigate the influence of dark matter on the geometric shapes of these periodic orbits. Fig.~\ref{dif_k} shows the periodic orbits around a Schwarzschild black hole and periodic orbits around black holes embedded in three different dark matter halos. The layout of Fig.~\ref{dif_k} is as follows: each row corresponds to the same set of orbital configurations  \((z~w~v)\), presenting the periodic orbits near the Schwarzschild black hole and those in three dark matter halo environments; each column fixes the dark matter mass \(k\), exhibiting the variation pattern of orbital shapes by varying the orbital configurations  \((z~w~v)\). From horizontal comparative analysis, it is shown that the presence of dark matter halos significantly stretches the apoapsis of periodic orbits. As the dark matter mass $k$ increases, the deviation between orbits in dark matter halo environments and Schwarzschild case gradually intensifies. The physical reason behind this phenomenon stems from the fact that, for the same precession angle $q$, the larger dark matter mass, the differences between orbital energies (dark matter halo environments vs Schwarzschild case) are more pronounced (as shown in Fig.~\ref{dif_k_q_models}). Notably, the NFW and Beta models exhibit nearly identical orbital trajectories for all dark matter masses $k$ considered in this study, which is consistent with their overlapping in precession angle curves shown in Fig.~\ref{dif_k_q}. This demonstrates that despite the different functional forms of NFW and Beta density profiles, they produce nearly indistinguishable effects on both the energetics and shapes of periodic orbits. Furthermore, Fig.~\ref{dif_k} demonstrates that periodic orbits calculated in NFW and Beta models are relatively more extended than those in Moore model and Schwarzschild black hole cases. This is caused by the fact that the orbital energies calculated in the NFW and Beta models are substantially lower than those in the Moore model for the same precession angle q (as illustrated in Figs.~\ref{dif_k_q_models} and~\ref{dif_k_q}), since orbital energy is inversely related to the apoapsis distance. This finding suggests a stronger gravitational influence from the NFW/Beta dark matter halos, which is consistent with the results for the ISCO and MBO in~\ref{sub2}. 

Furthermore, to isolate the influences of dark matter halo models, we construct periodic orbits within the NFW model using different values of $k$, each combined with five distinct orbital configurations $(z~w~v)$. The results, presented in Fig.~\ref{dif_k_NFW}, clearly demonstrate the effect of dark matter mass k on periodic orbital shapes. For example, the relationship between apoapsis distance and dark matter mass exhibits a consistent monotonic trend: an increase in dark matter mass leads to further apoapsis distance and more pronounced differences of orbital shape, compared with Schwarzschild case. Additionally, from the comparative analysis in Figs.~\ref{dif_k} and \ref{dif_k_NFW}, we observe that the orbital configurations $(z~w~v)$ directly determine the geometric shape of periodic orbits. Specifically, $(z~w~v)$ represent the number of leaves, the count of additional whirls, and the behavior of the subsequent vertex reached by the particle after leaving the initial vertex (apoapsis). As a result, larger parameter values correspond to more complex orbital structures, which is consistent with the zoom-whirl-vertex classification of periodic orbits in pioneering studies~\cite{Levin:2008mq,Levin:2008ci}.

\subsection{The Effect of Dark Matter Halo Scale on Periodic Orbits}\label{s3_2}

\begin{figure}
	\centering 
	\begin{subfigure}{0.3\textwidth}
		\includegraphics[width=\linewidth]{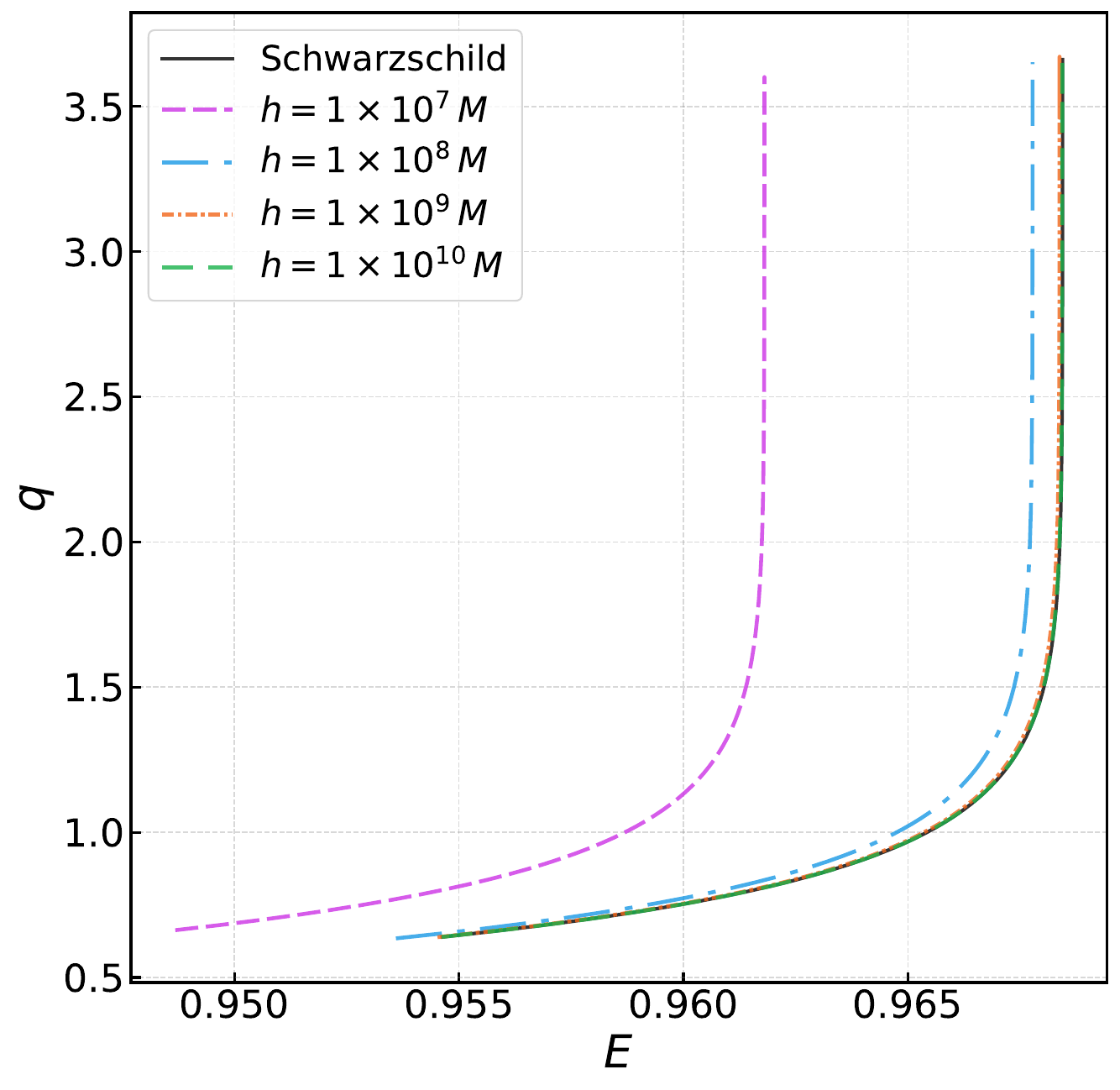}
		\caption{NFW}
	\end{subfigure}
	\hfill 
	\begin{subfigure}{0.3\textwidth}
		\includegraphics[width=\linewidth]{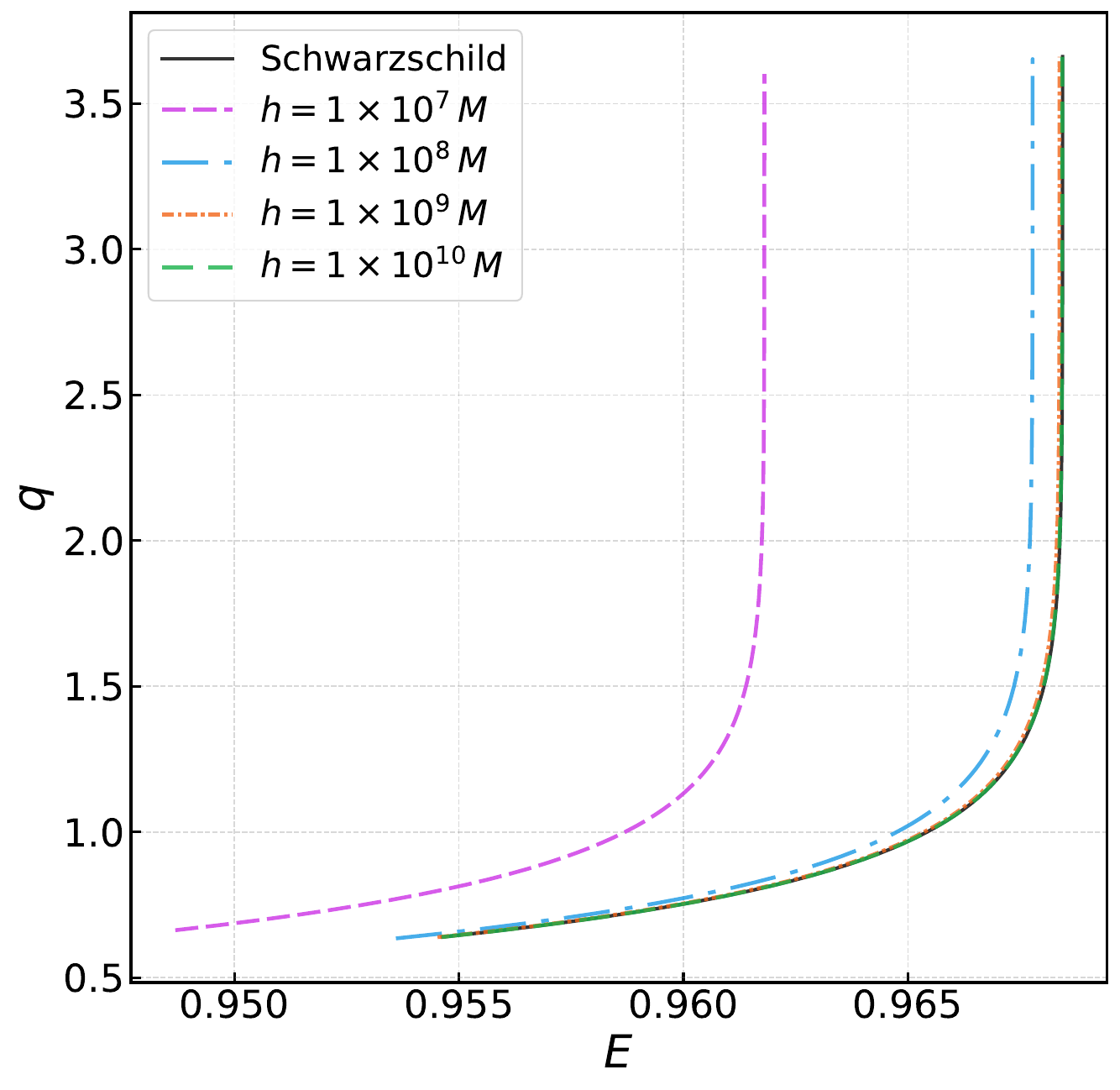}
		\caption{Beta}
	\end{subfigure}
	\hfill
	\begin{subfigure}{0.3\textwidth}
		\includegraphics[width=\linewidth]{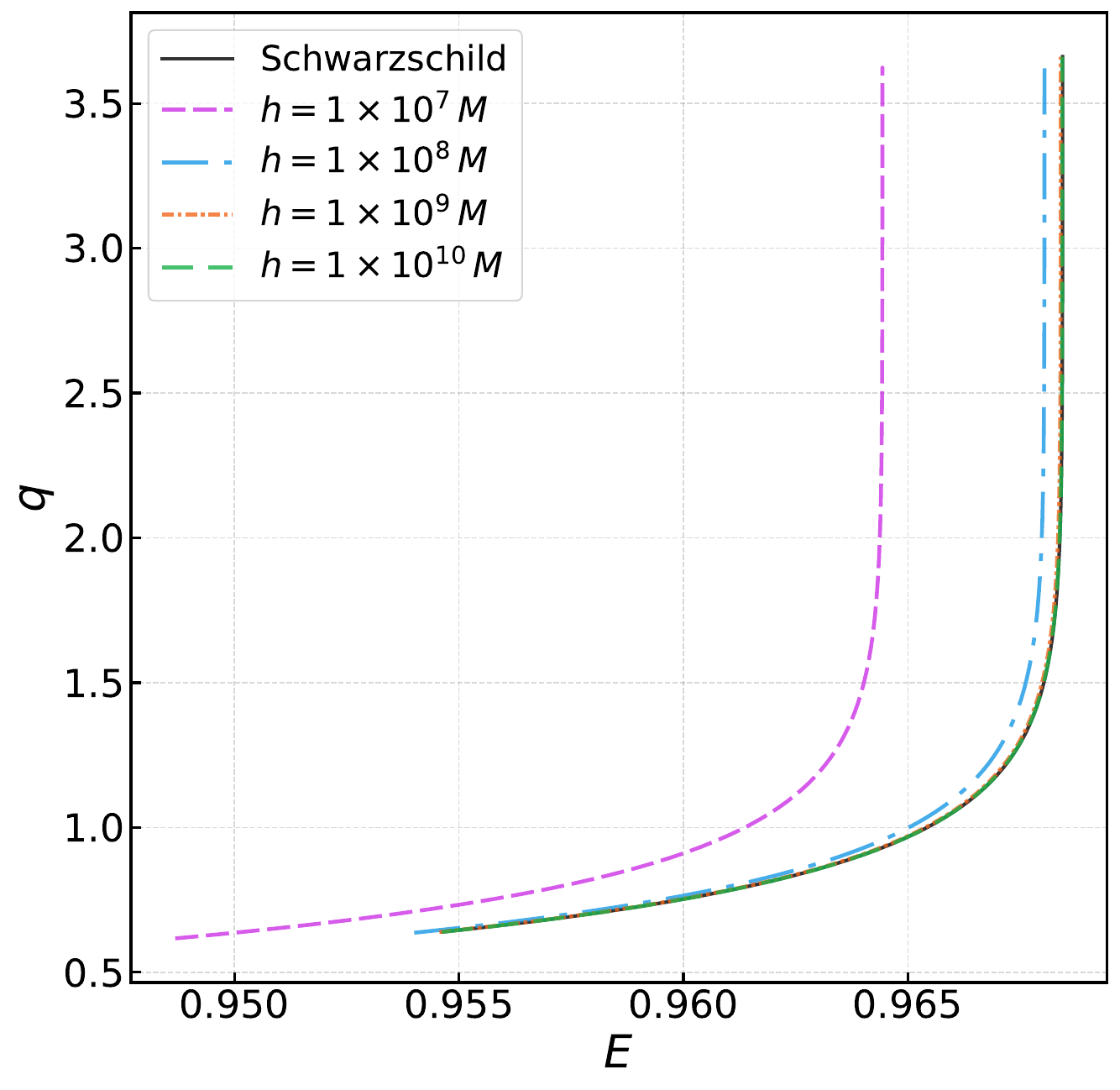}
		\caption{Moore}
	\end{subfigure}
	
	\caption{Precession angles $q$ changes with orbital energy $E$ for three dark matter halo models with fixed dark matter mass $k = 10^4 M$. Each panel shows results for the Schwarzschild metric and four different halo scales ($h = 10^7M$, $h=10^8 M$, $h=10^9 M$, $h=10^{10} M$) under (a) NFW, (b) Beta, and (c) Moore models }
	\label{dif_h_q_models}
\end{figure}

\begin{figure}[t] 
	\centering 
	\begin{subfigure}{0.22\textwidth}
		\includegraphics[width=\linewidth]{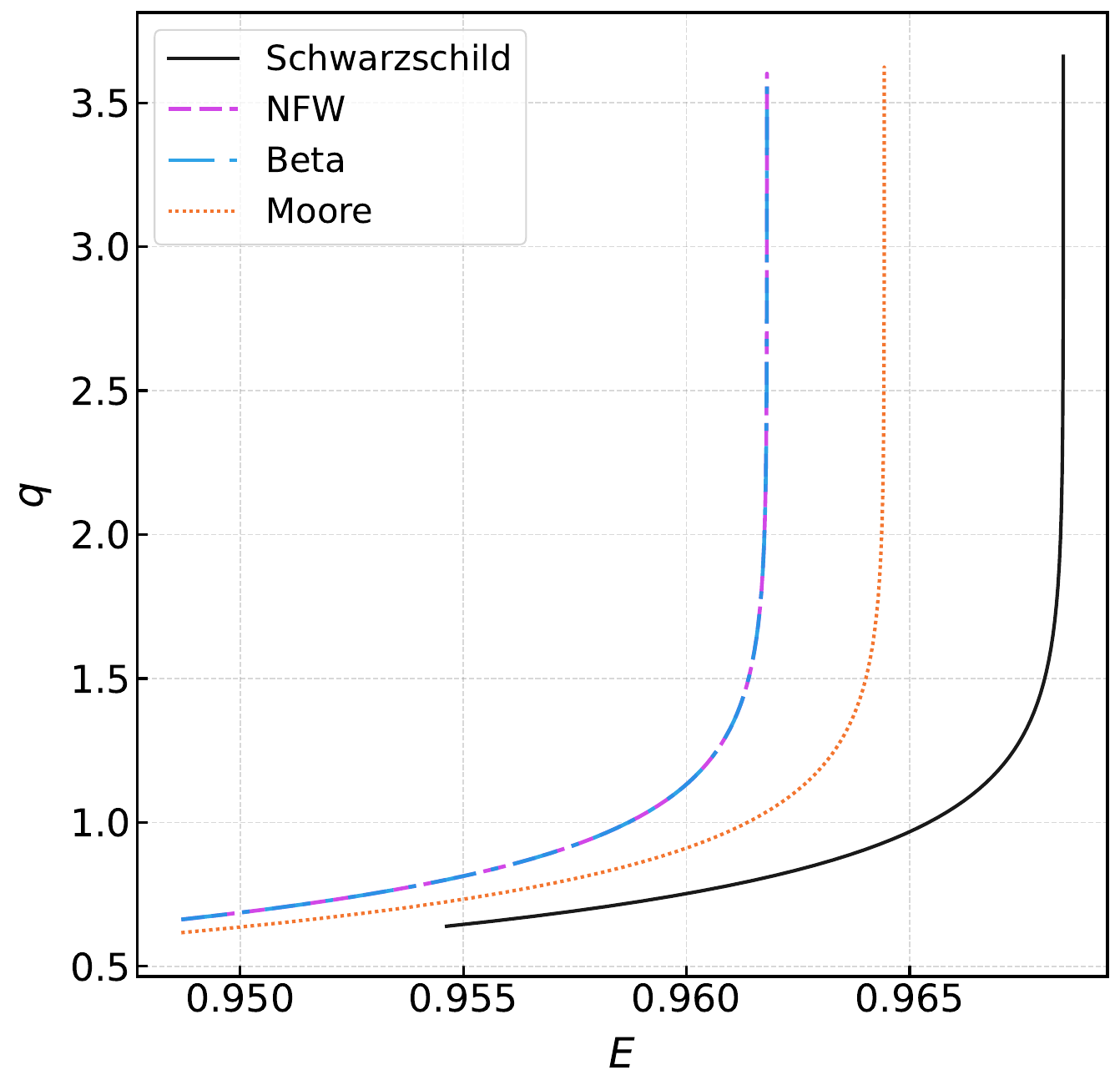}
		\caption{$h=10^7\ M$}
		\label{subfig:h1e7}
	\end{subfigure}
	\hfill 
	\begin{subfigure}{0.22\textwidth}
		\includegraphics[width=\linewidth]{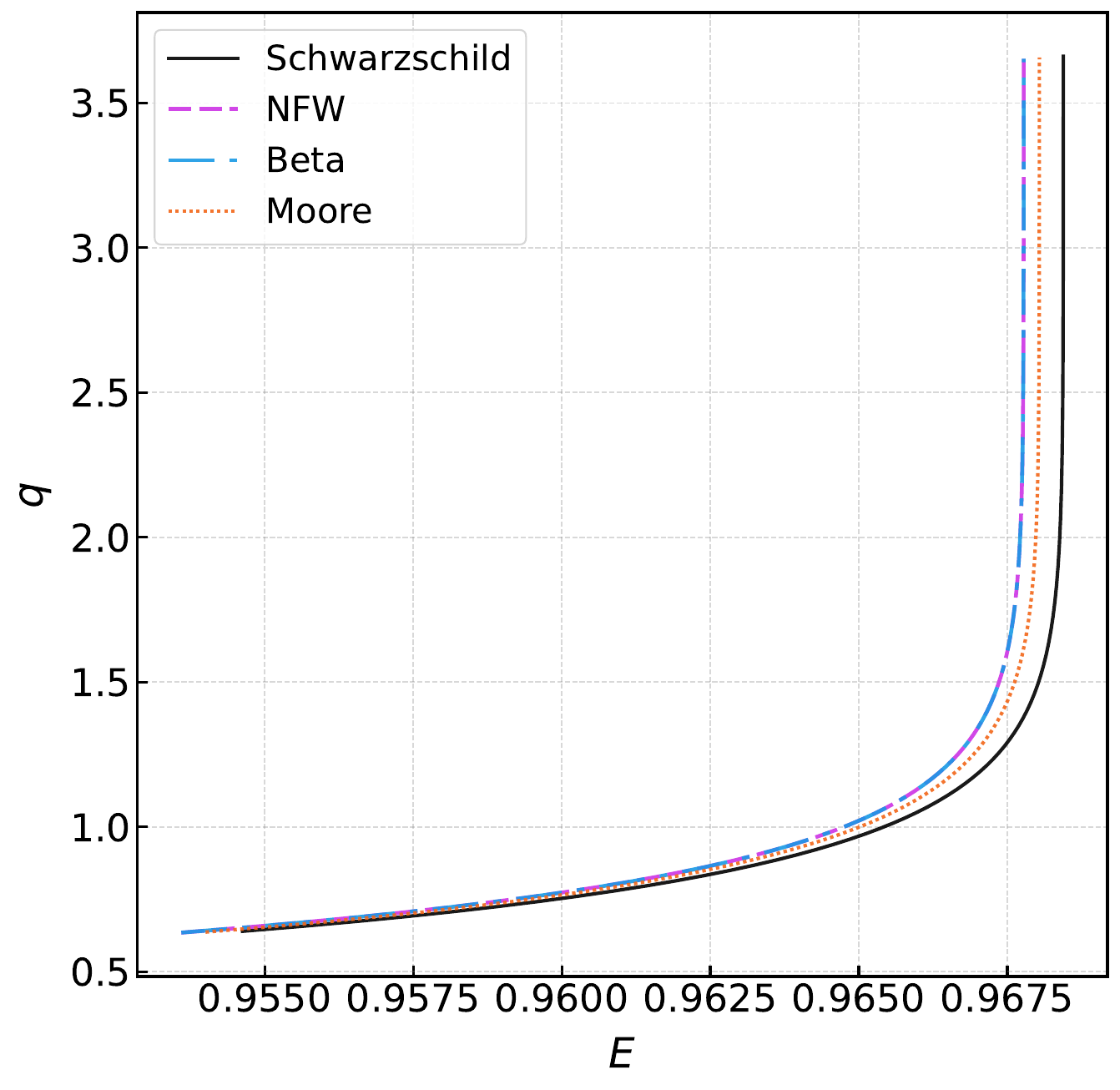}
		\caption{$h=10^8\ M$}
		\label{subfig:h3e8}
	\end{subfigure} 
	\hfill
	\begin{subfigure}{0.22\textwidth}
		\includegraphics[width=\linewidth]{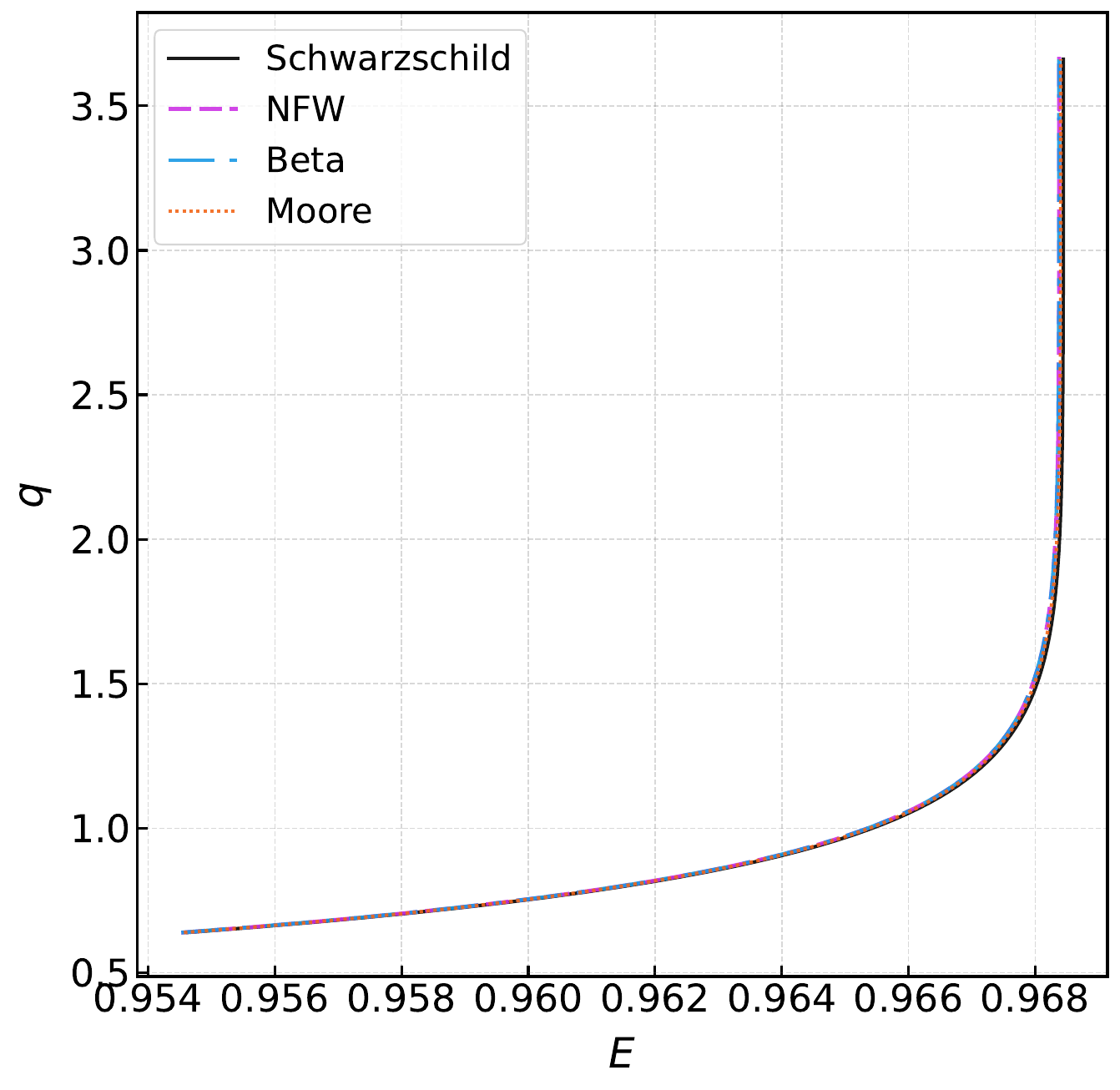}
		\caption{$h=10^9\ M$}
		\label{subfig:h1e9}
	\end{subfigure}
	\hfill 
	\begin{subfigure}{0.22\textwidth}
		\includegraphics[width=\linewidth]{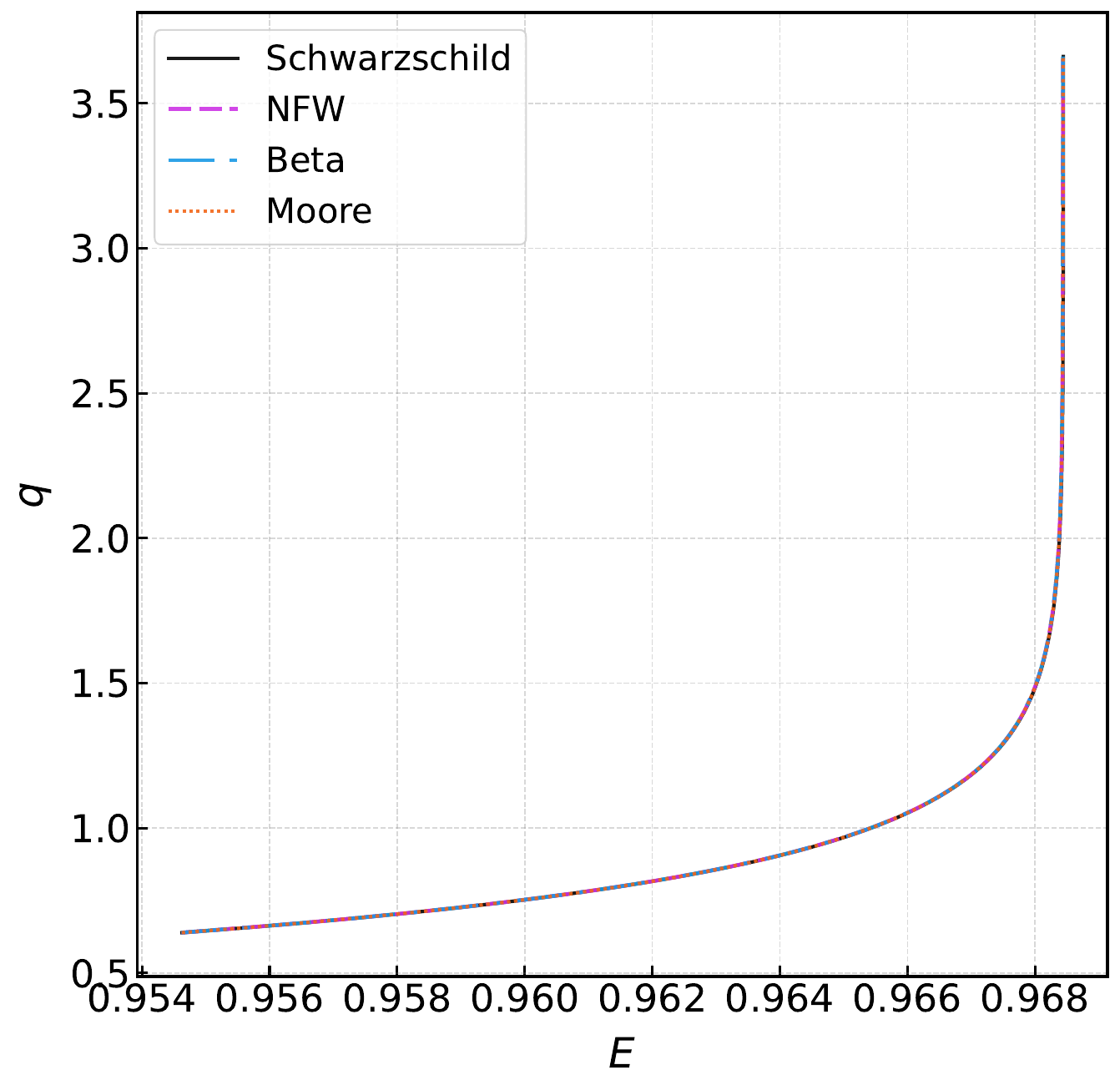}
		\caption{$h=10^{10}\ M$}
		\label{subfig:h1e10}
	\end{subfigure}
	
	\caption{Precession angle comparison across dark matter halo models for different halo scales. The figure presents precession angles calculated for three halo density profiles (NFW, Beta, and Moore), compared with the Schwarzschild metric across four different halo scales: (a) $h=10^7 M$, (b) $h=10^8 M$, (c) $h=10^9 M$, and (d) $h=10^{10} M$. }
	\label{dif_h_q}
\end{figure}

Having understood the effects of dark matter mass on periodic orbit characteristics, we then focus our attention on investigating how the scale of dark matter halo influences these periodic orbits. Fig.~\ref{dif_h_q_models} systematically illustrates the effects of dark matter halo scale $h$ on precession angle, with subfigures exhibiting results calculated using three dark matter halo models: NFW, Beta, and Moore. The analysis is carried out with the dark matter mass fixed at $k = 10^4M$ while varying the halo scale $h$ from $10^7M \sim 10^{10}M$ (with each scale represented by distinct colors). The Schwarzschild black hole results serve as the reference for comparative analysis. The results reveal a universal feature among all three dark matter halo models: an increase in the dark matter halo characteristic radius $h$ shifts the corresponding precession angle curves toward higher energy regions. This demonstrates that for any fixed precession angle $q$, achieving the same orbital configuration requires higher energy $E$ as the dark matter halo scale increases. The physical reason can be understood through the restructuring of the gravitational potential. When the total dark matter mass $k$ is fixed, increasing the scale $h$ produces a more diffuse dark matter distribution. In our normalization convention where the potential at infinity equals unity $\lim_{r \to \infty} V_{\rm eff} = 1$, a weaker gravitational field yields an effective potential $V_{\rm eff}$ closer to unity, corresponding to a shallower potential well. As shown in Fig.~\ref{Veff_2}, the diluted mass distribution results in a flatter effective potential curve that approaches unity, reflecting weaker gravitational binding in the regions where periodic orbits are located. As a result, the orbital energy $E$ needed to maintain the same precession angle also moves closer to unity. This interpretation is supported by the numerical results in Fig.~\ref{dif_h_q_models}, where diluted dark matter distributions consistently show elevated orbital energies. The combined analysis with the previous subsection reveals opposite impacts between the two halo parameters. The increasing dark matter mass $k$ systematically reduces the required energy for achieving a given precession angle $q$ (as demonstrated in Fig.~\ref{dif_k_q_models}), while increasing the halo characteristic radius $h$ produces the opposite effect on the energy thresholds for achieving the same precession angle.

\begin{figure}
	\centering 
	
	\begin{subfigure}{\textwidth}
	\includegraphics[width=0.23\linewidth]{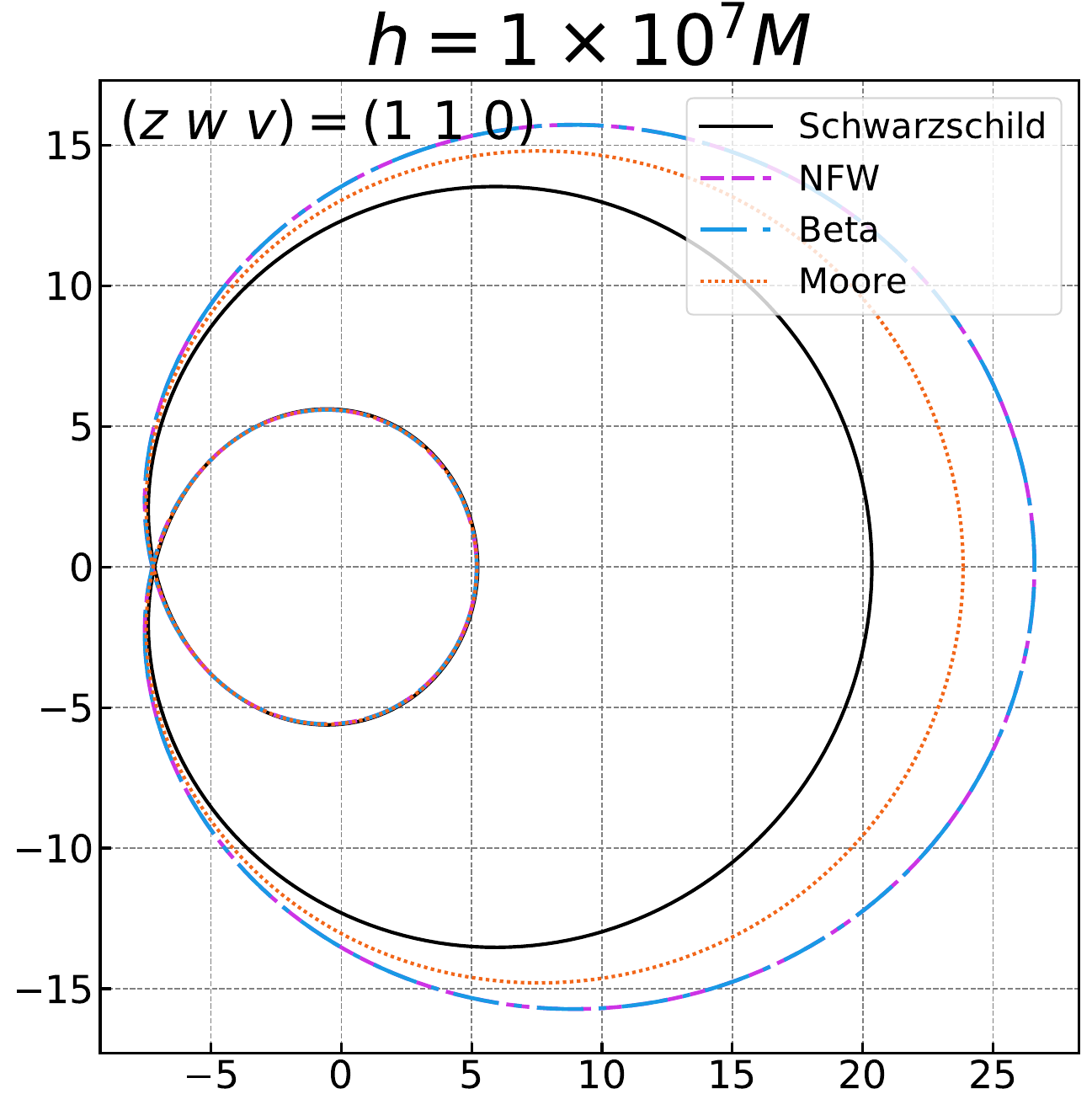}
	\includegraphics[width=0.23\linewidth]{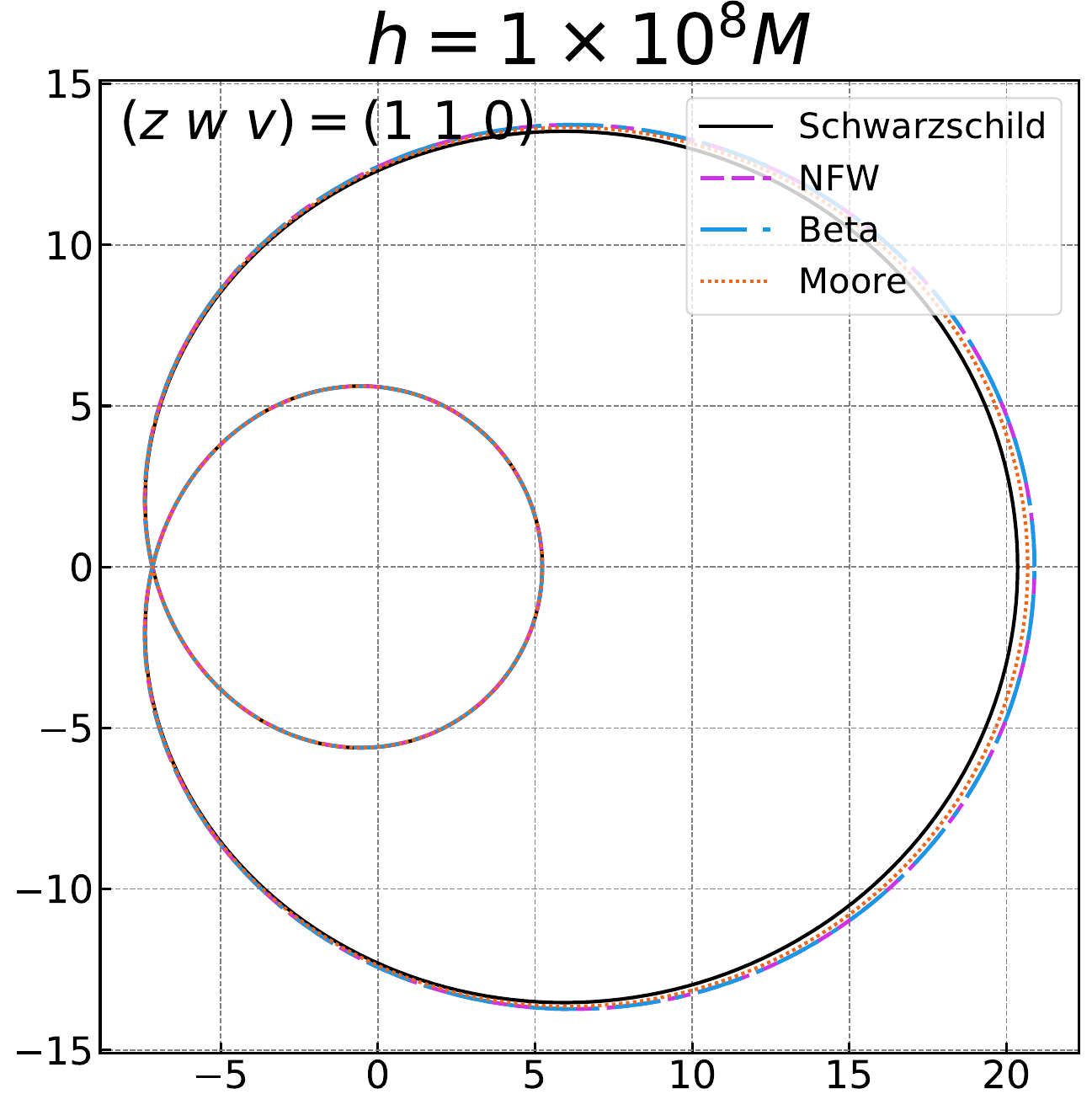}
	\includegraphics[width=0.23\linewidth]{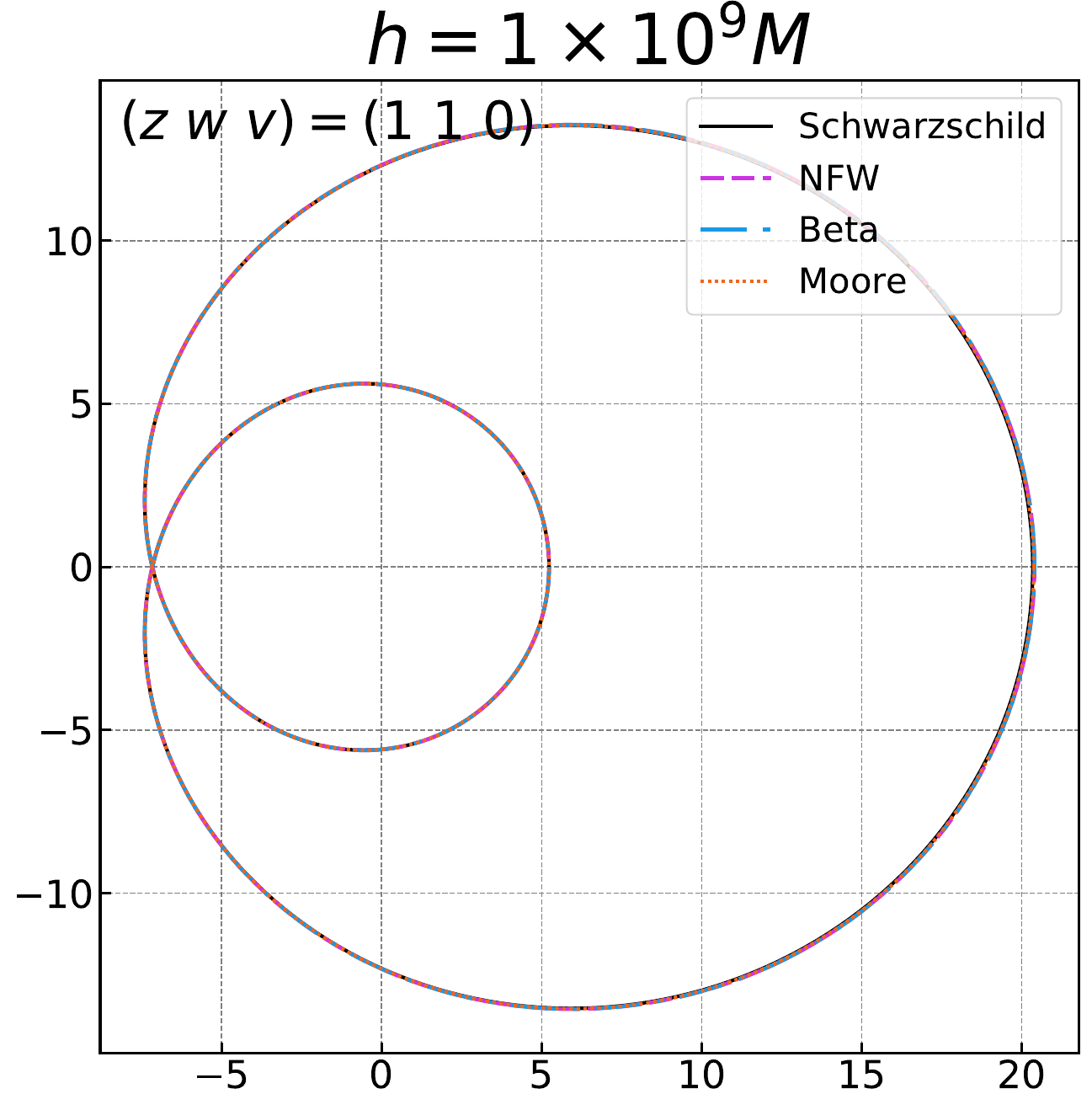}
	\includegraphics[width=0.23\linewidth]{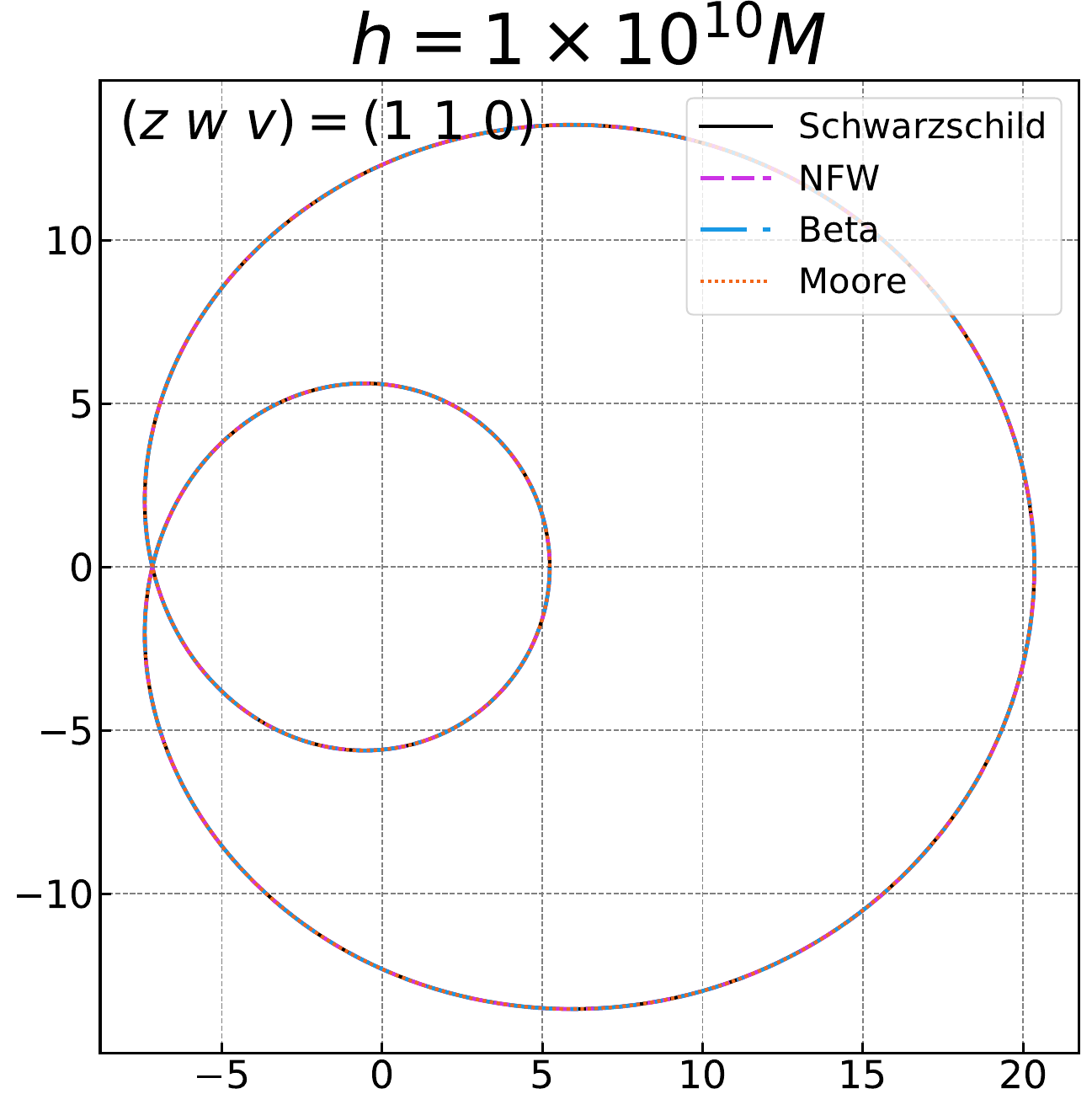}
	\end{subfigure}
	
	\vspace{0.4cm} 
	
	\begin{subfigure}{\textwidth}
	\includegraphics[width=0.23\linewidth]{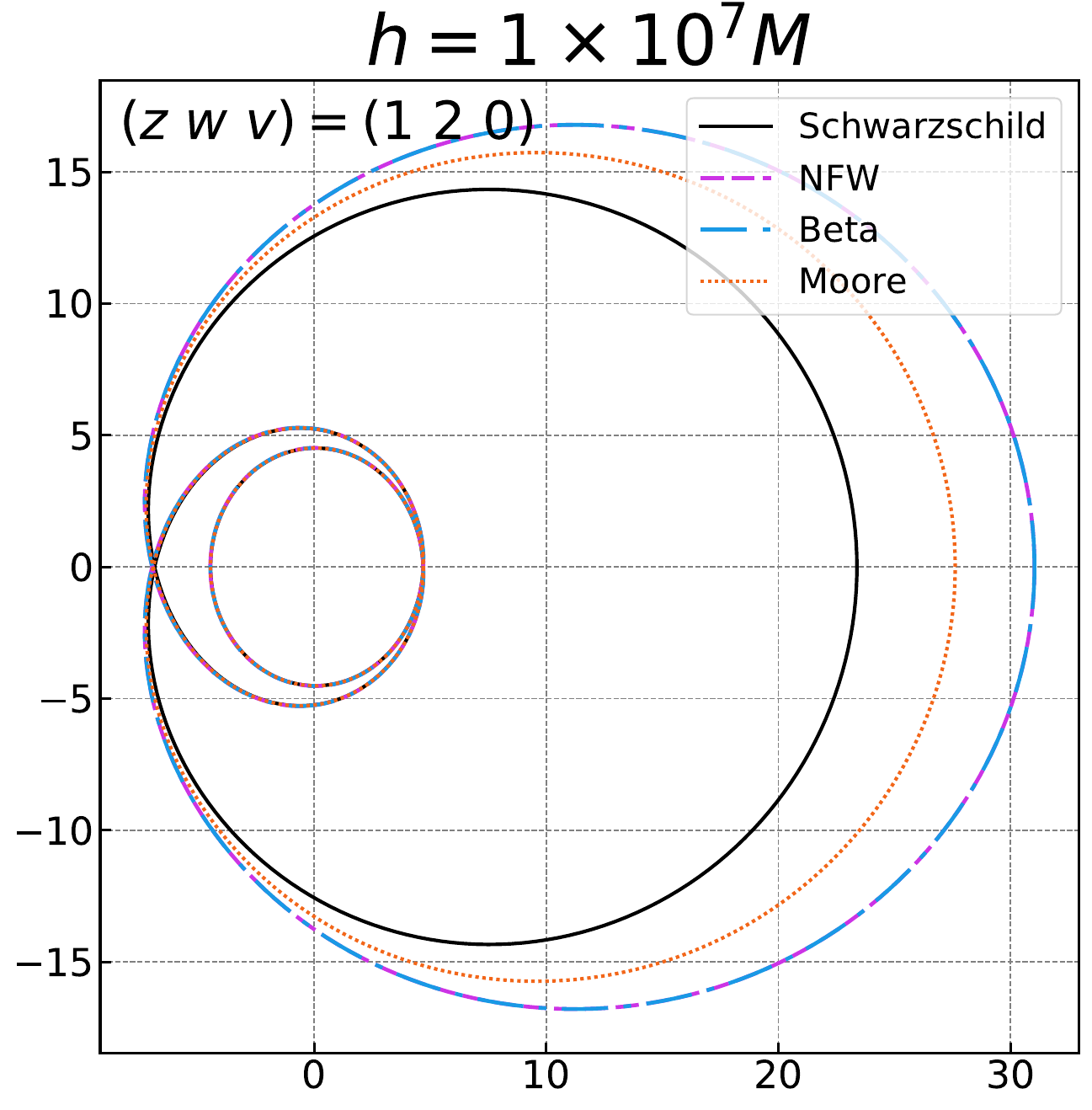}
	\includegraphics[width=0.23\linewidth]{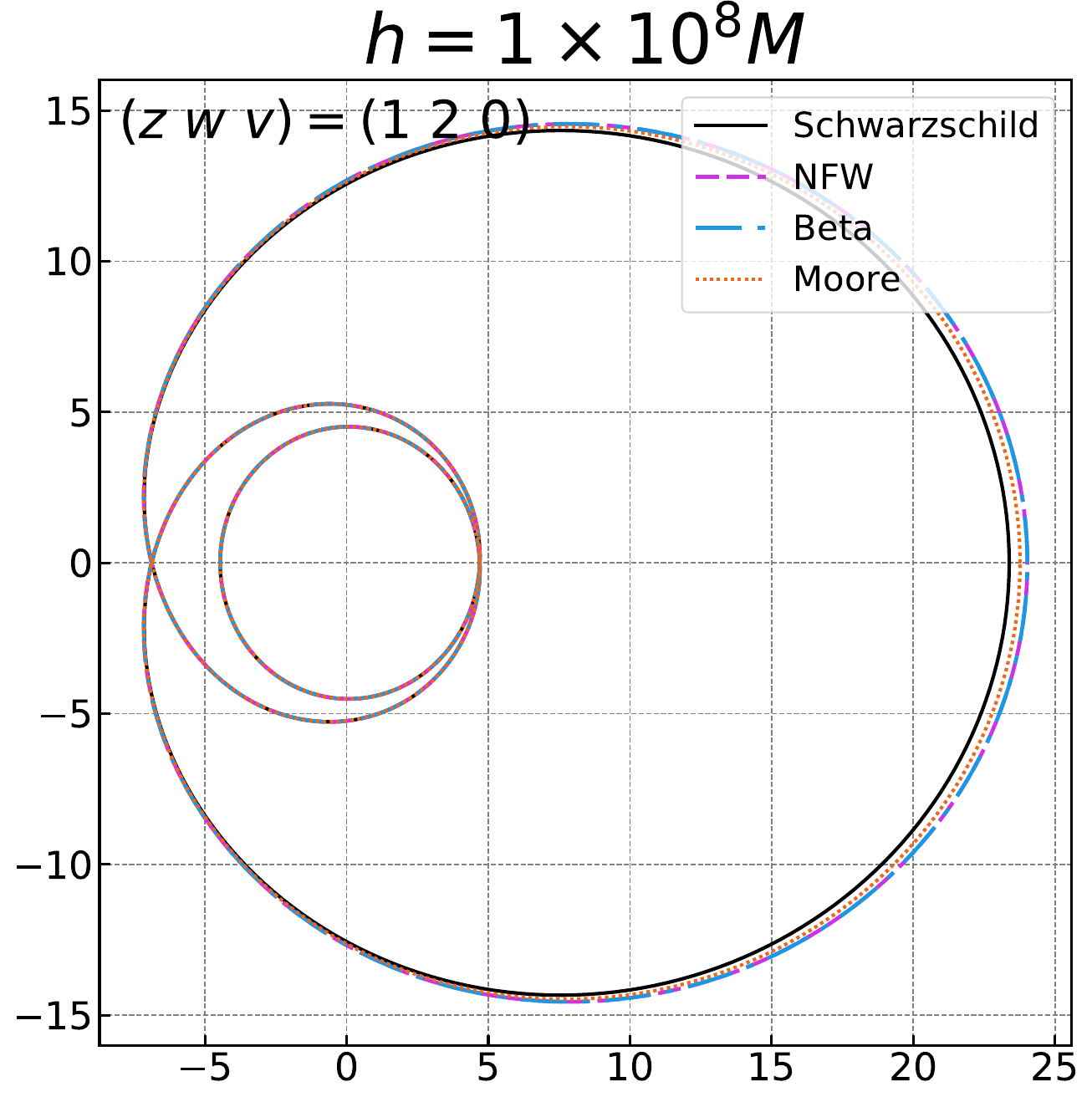}
	\includegraphics[width=0.23\linewidth]{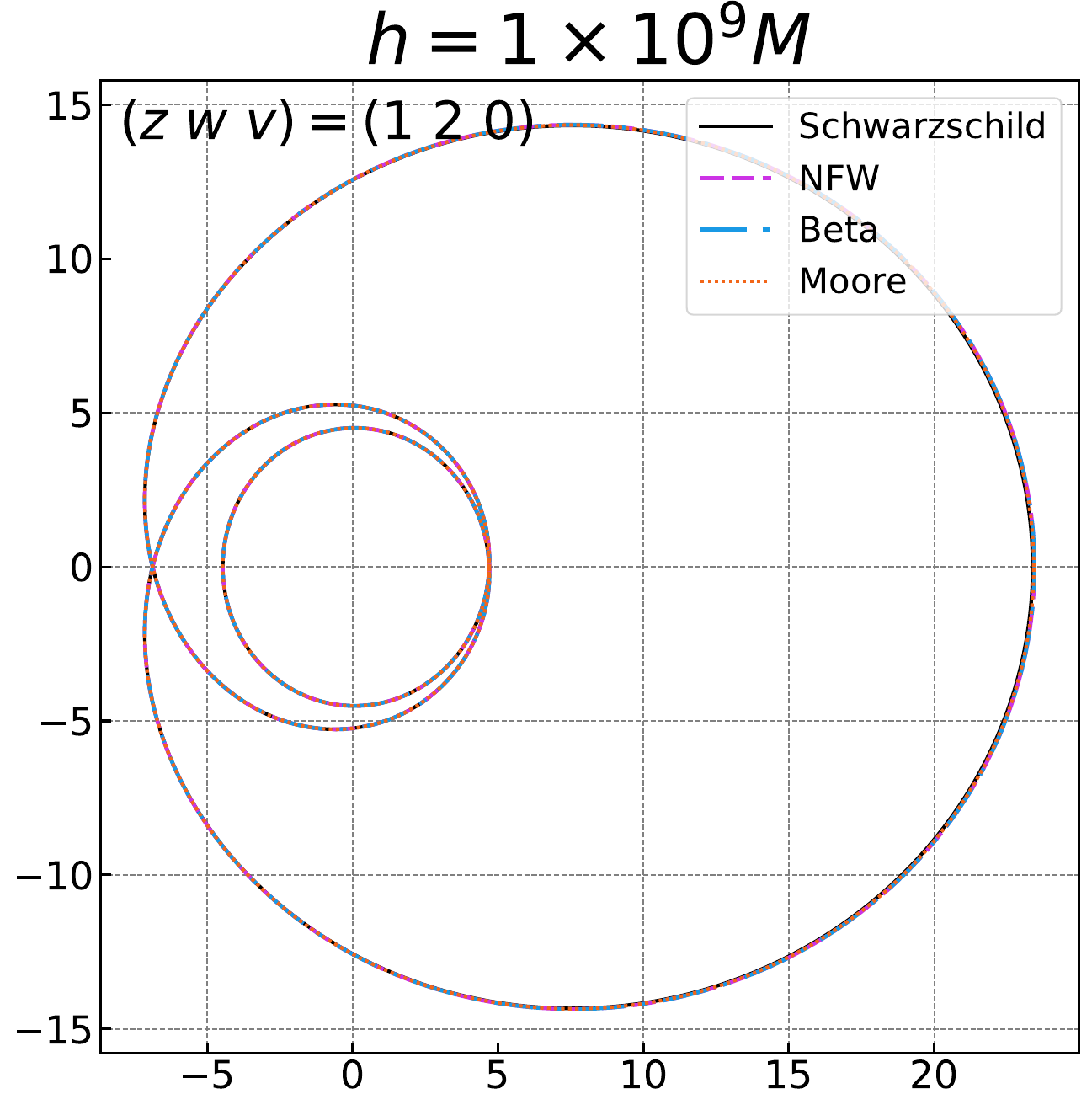}
	\includegraphics[width=0.23\linewidth]{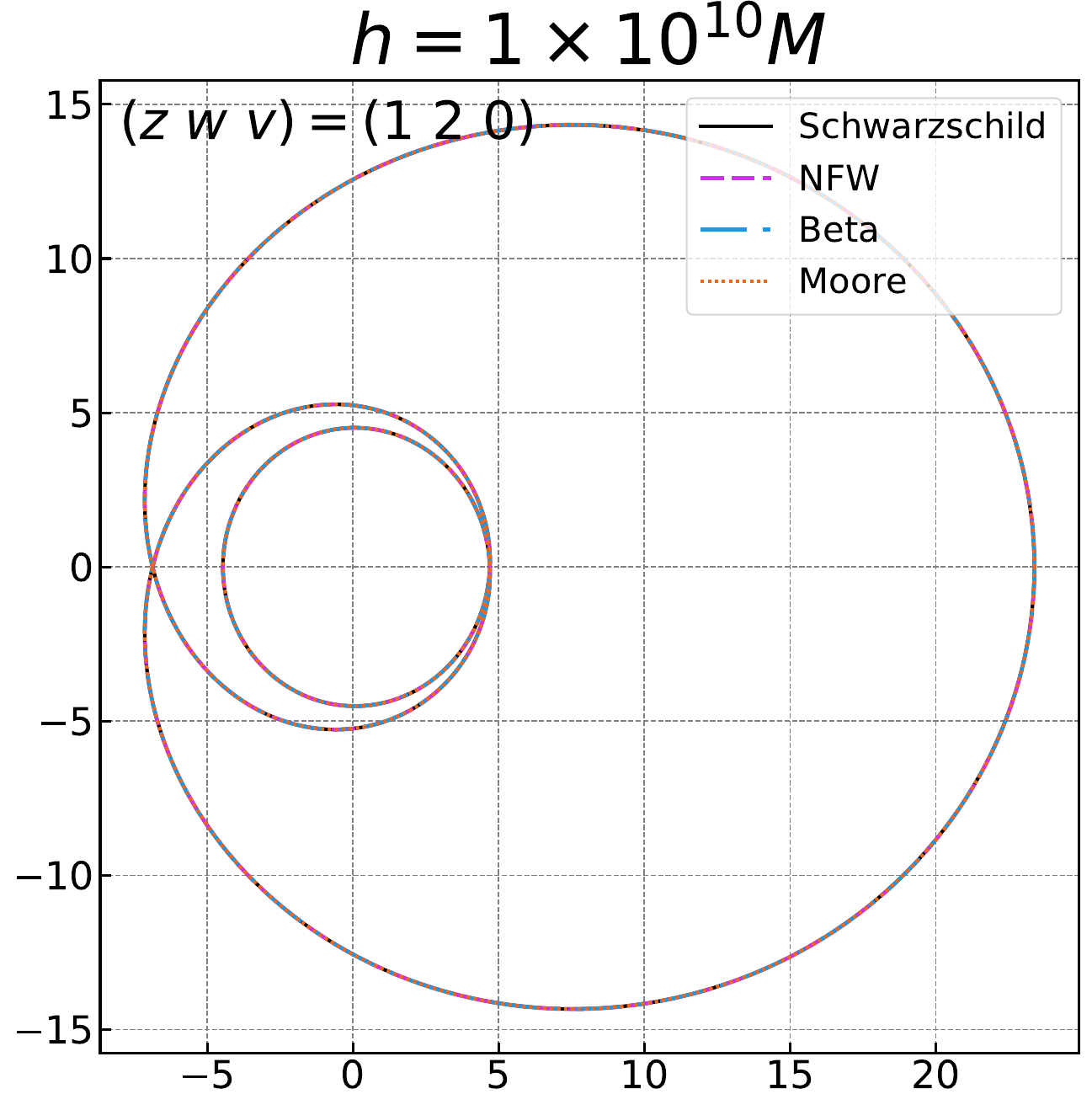}
	\end{subfigure}
	
	\vspace{0.4cm}
	
	\begin{subfigure}{\textwidth}
	\includegraphics[width=0.23\linewidth]{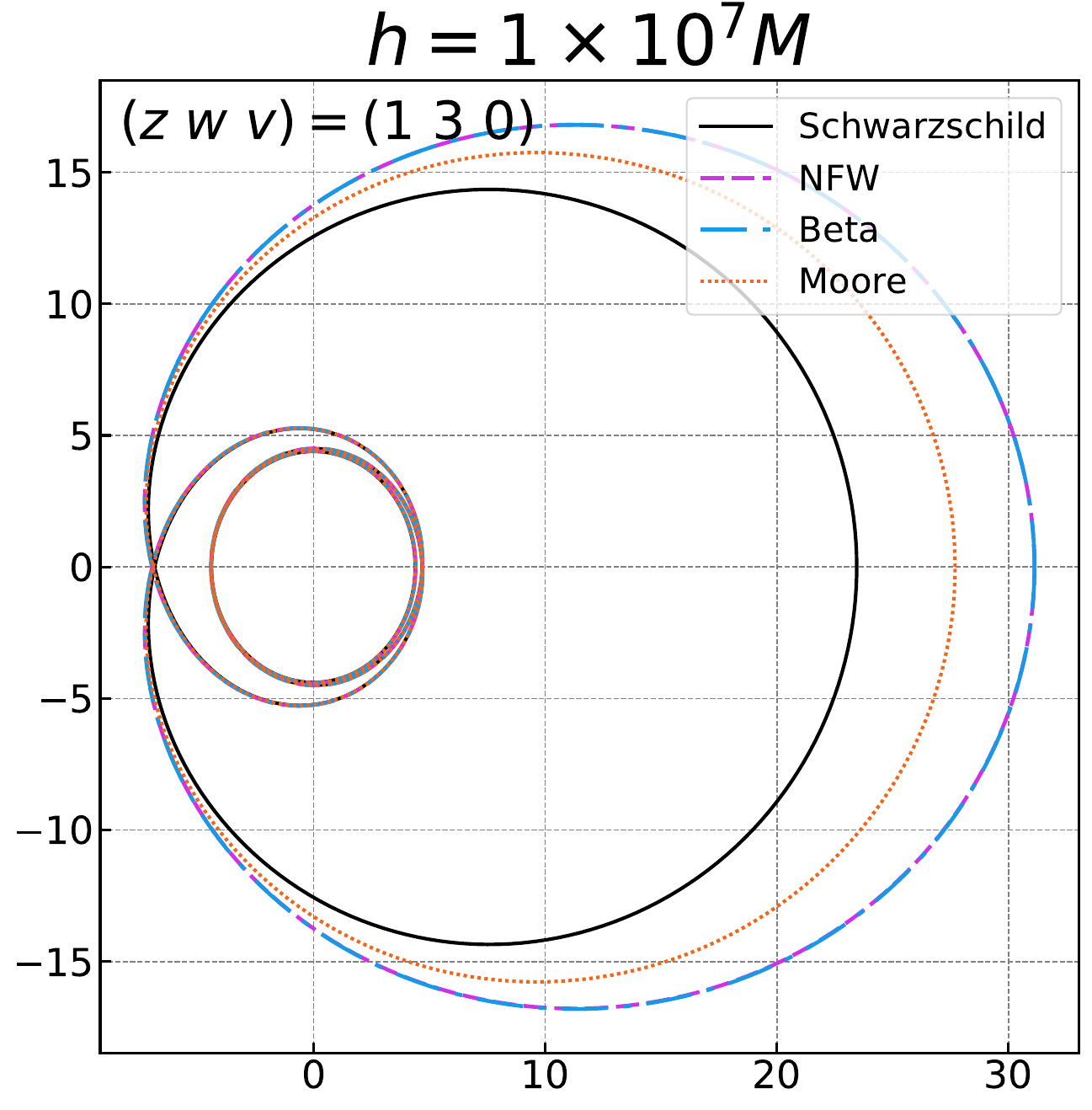}
	\includegraphics[width=0.23\linewidth]{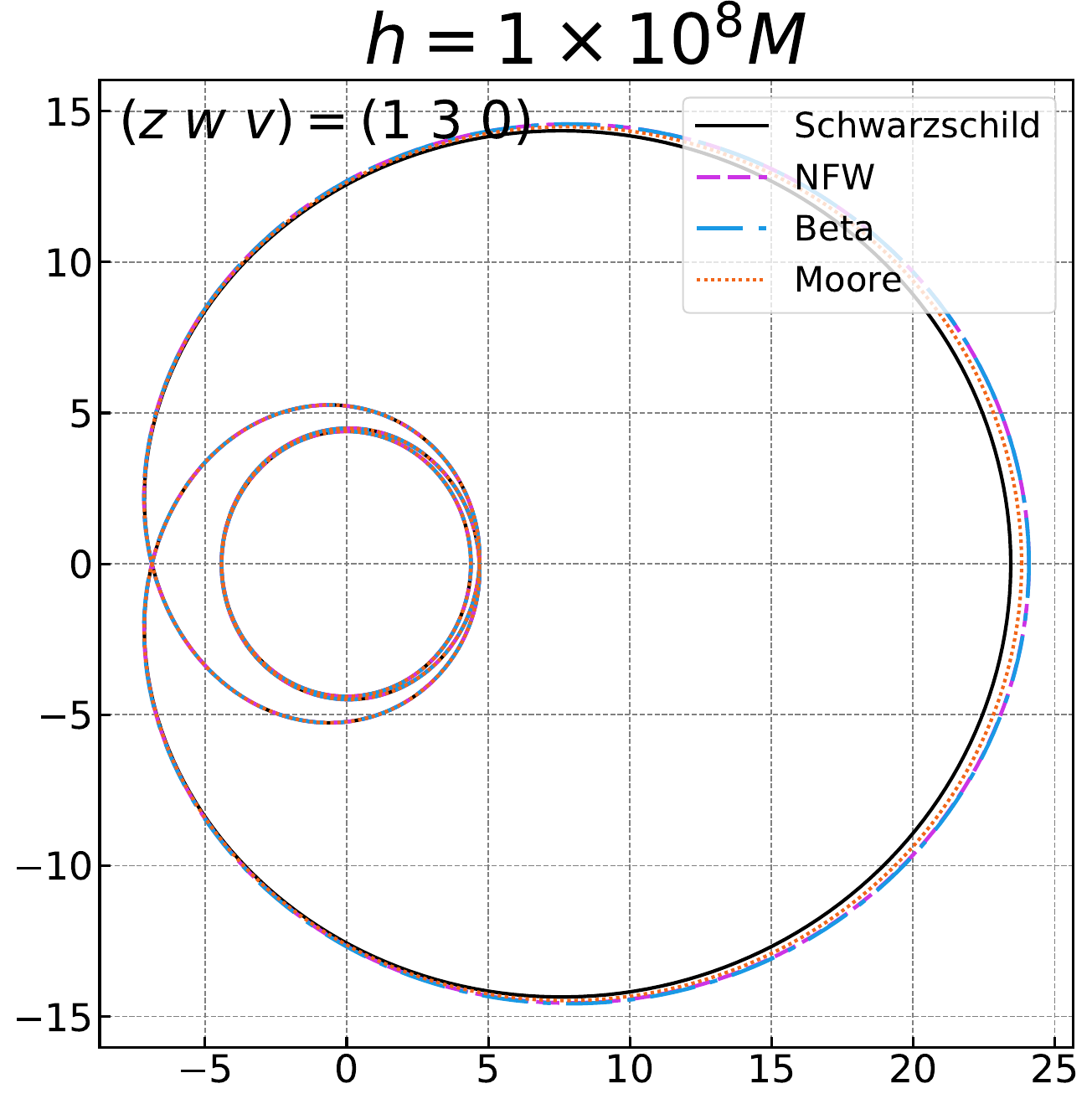}
	\includegraphics[width=0.23\linewidth]{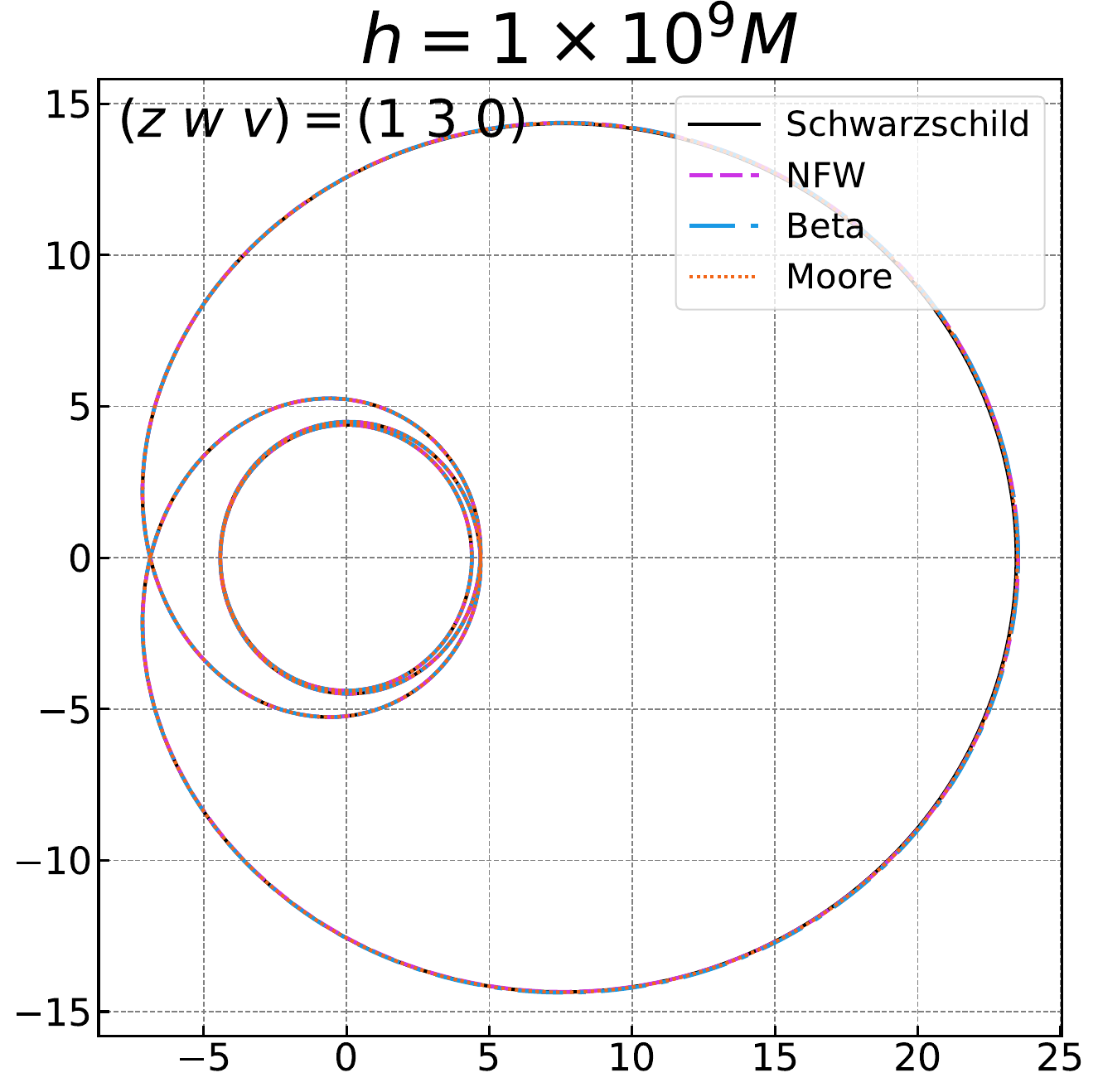}
	\includegraphics[width=0.23\linewidth]{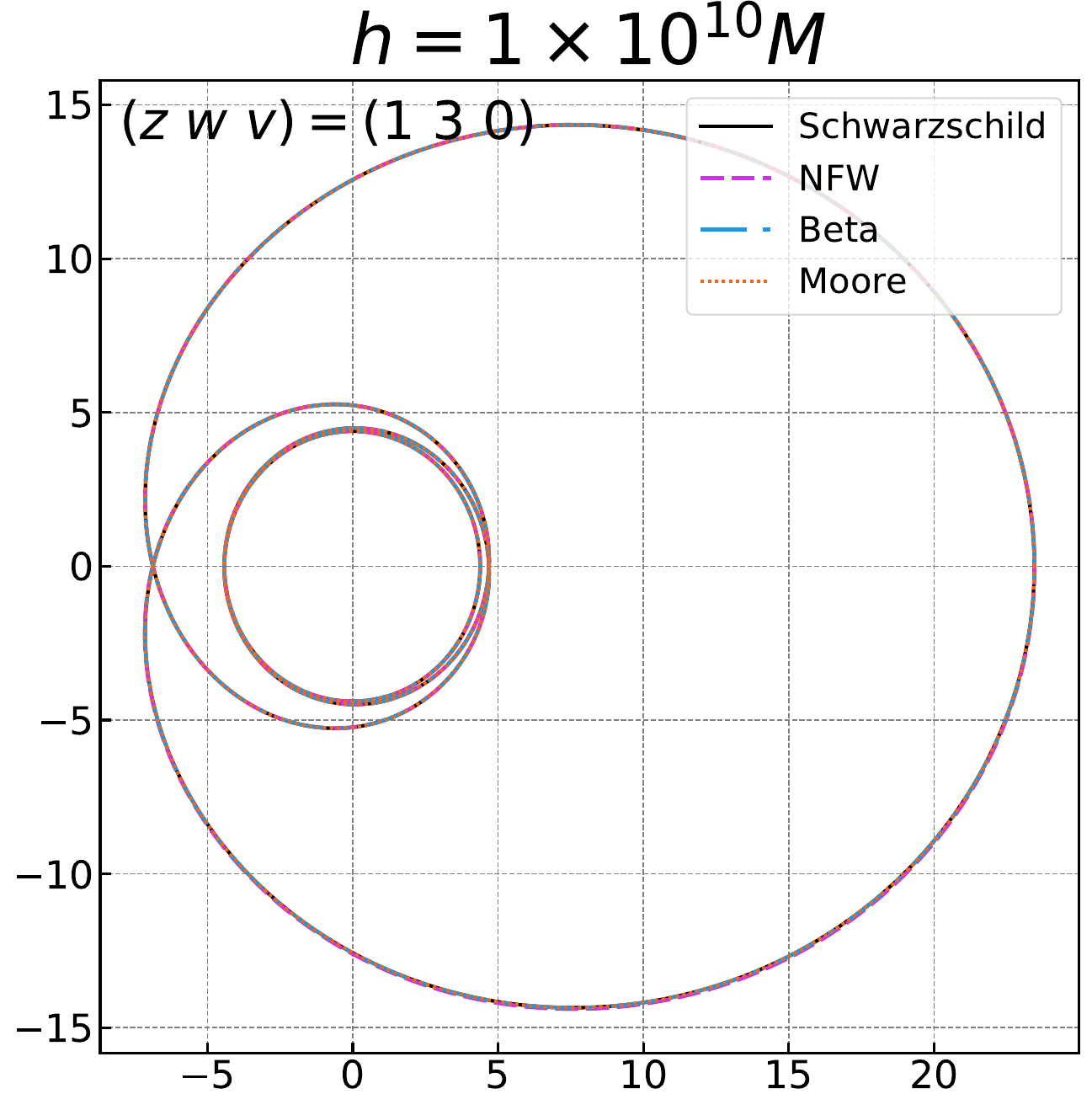}
	\end{subfigure}
	
	\vspace{0.4cm}
	
	\begin{subfigure}{\textwidth}
	\includegraphics[width=0.23\linewidth]{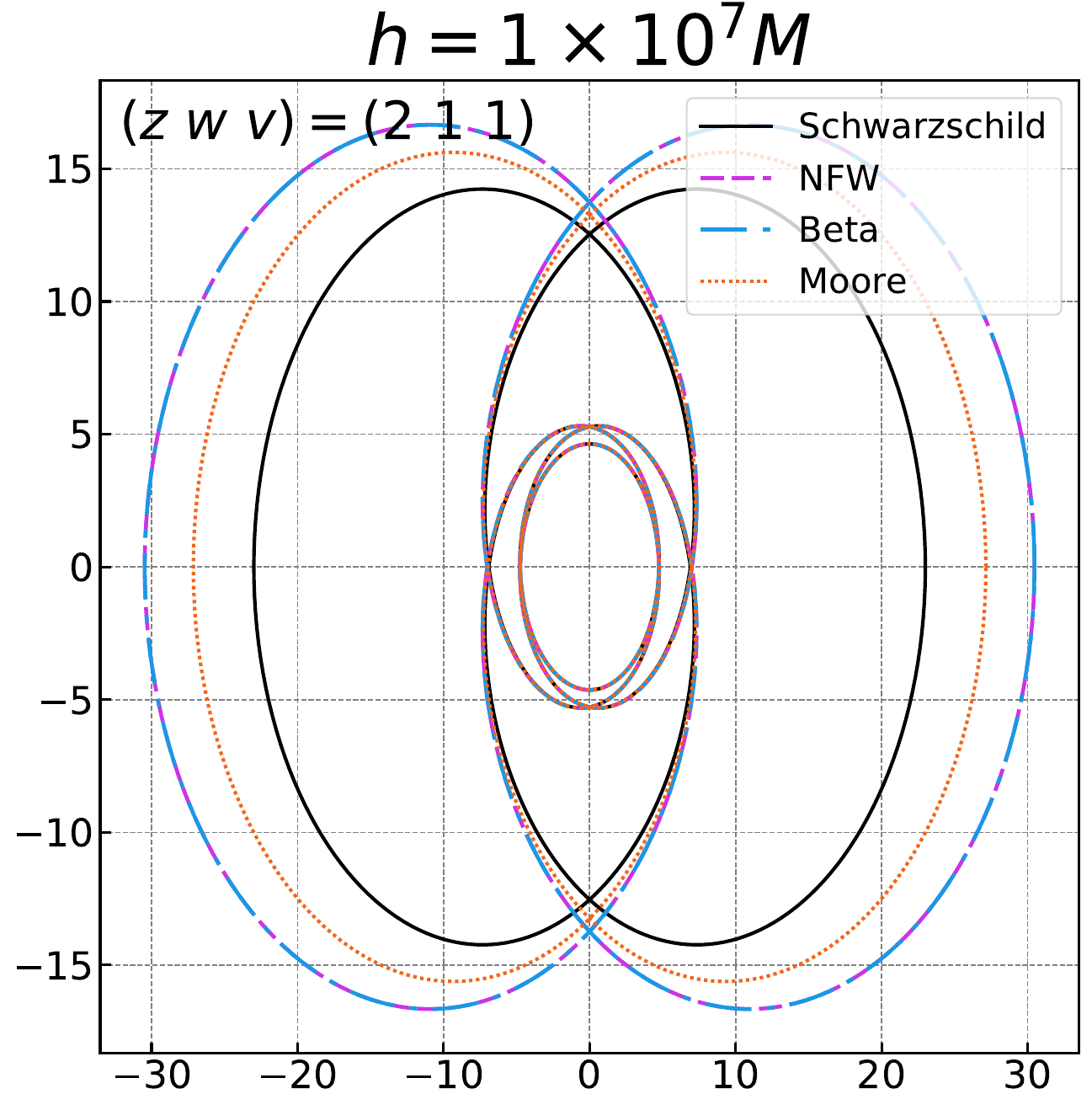}
	\includegraphics[width=0.23\linewidth]{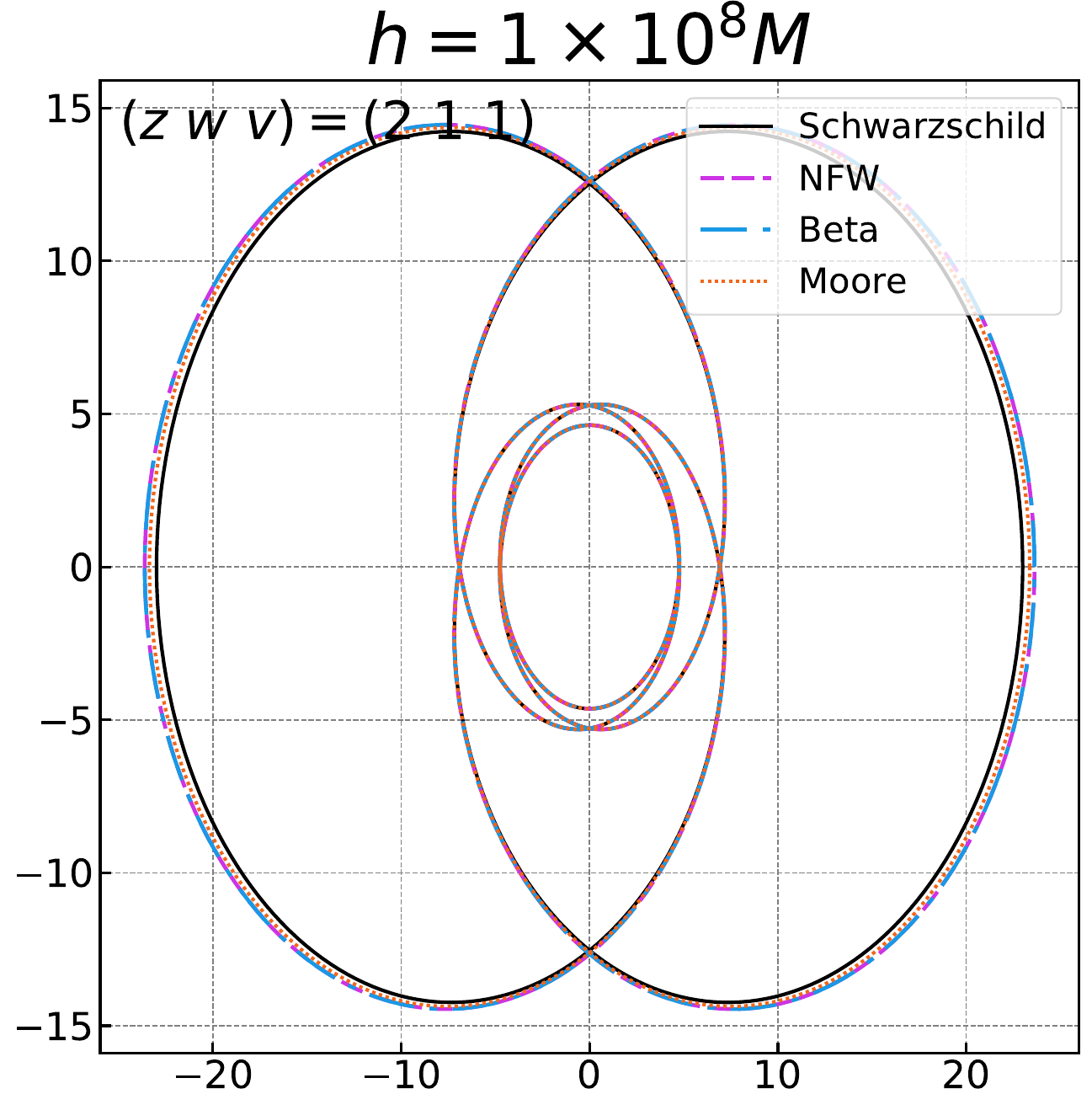}
	\includegraphics[width=0.23\linewidth]{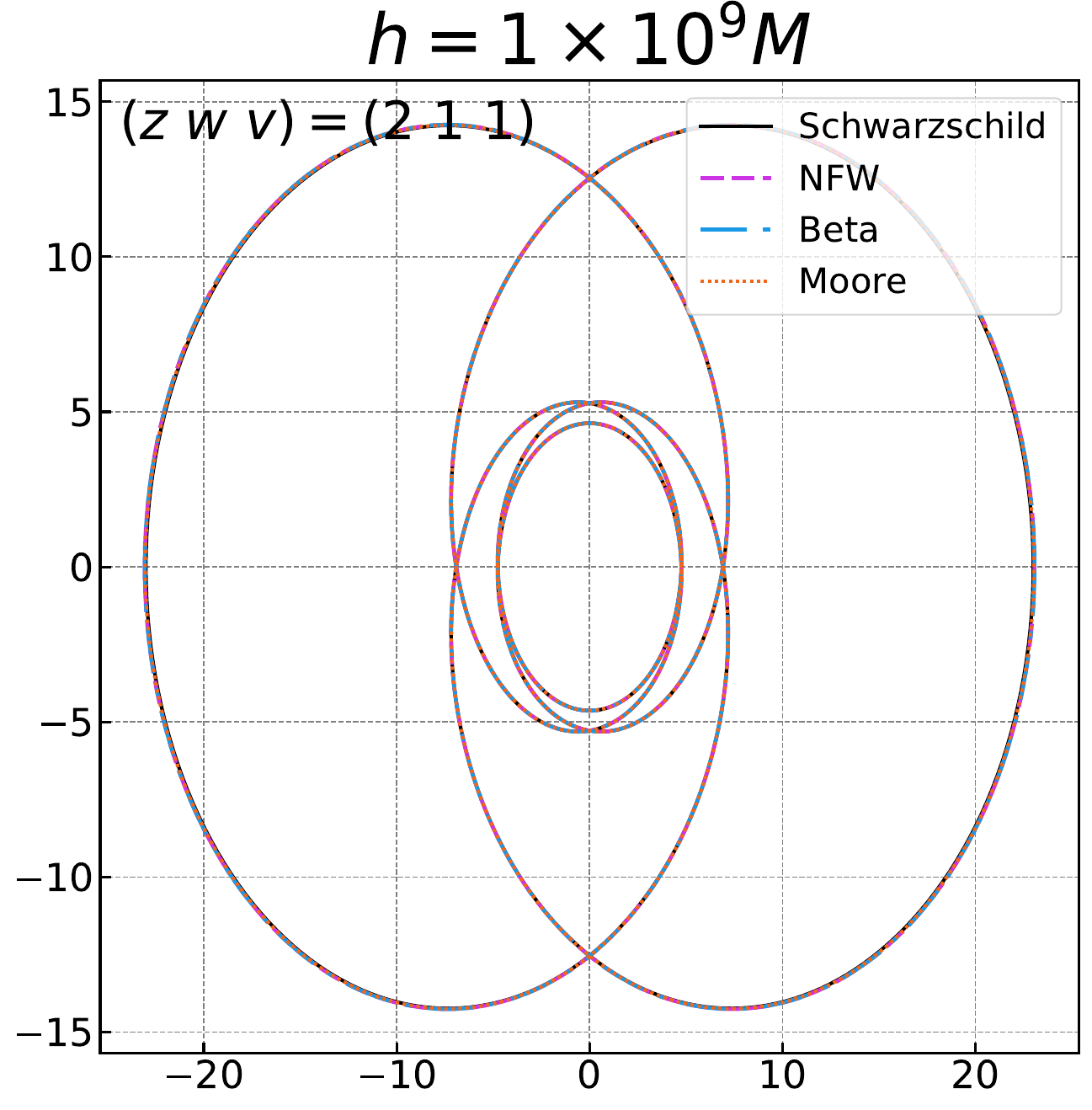}
	\includegraphics[width=0.23\linewidth]{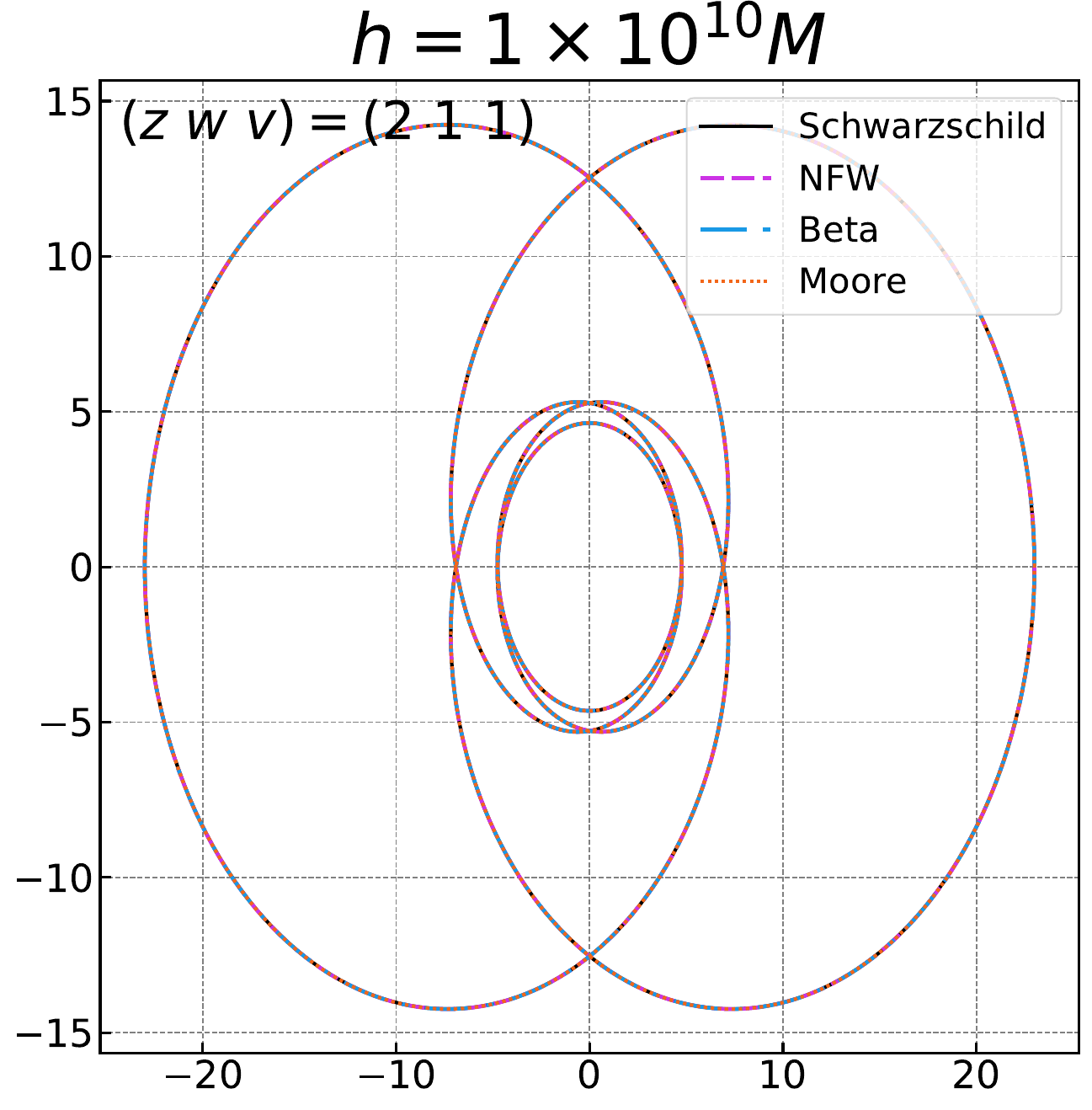}
	\end{subfigure}
	
	\vspace{0.4cm}
	
	\begin{subfigure}{\textwidth}
	\includegraphics[width=0.23\linewidth]{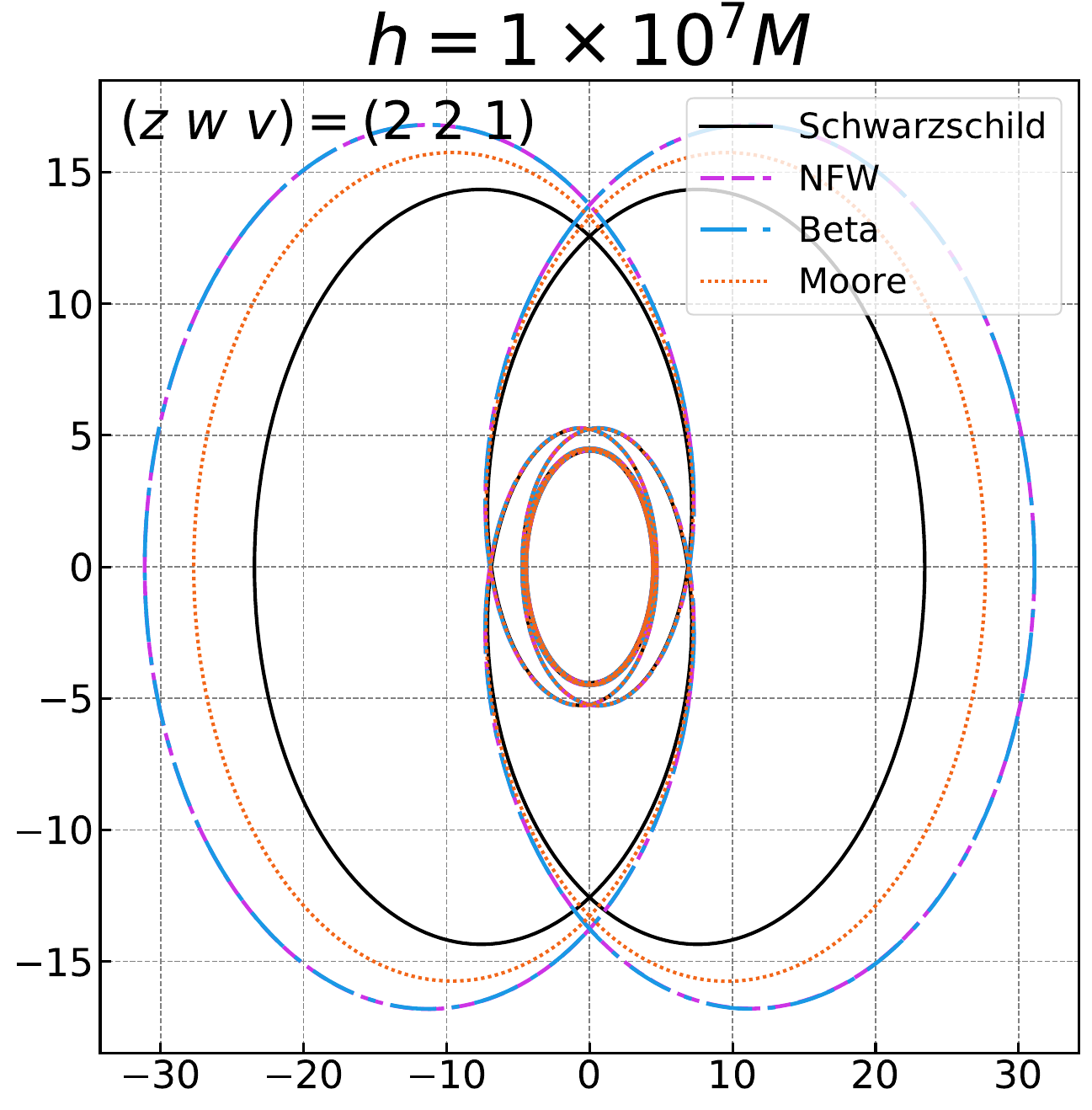}
	\includegraphics[width=0.23\linewidth]{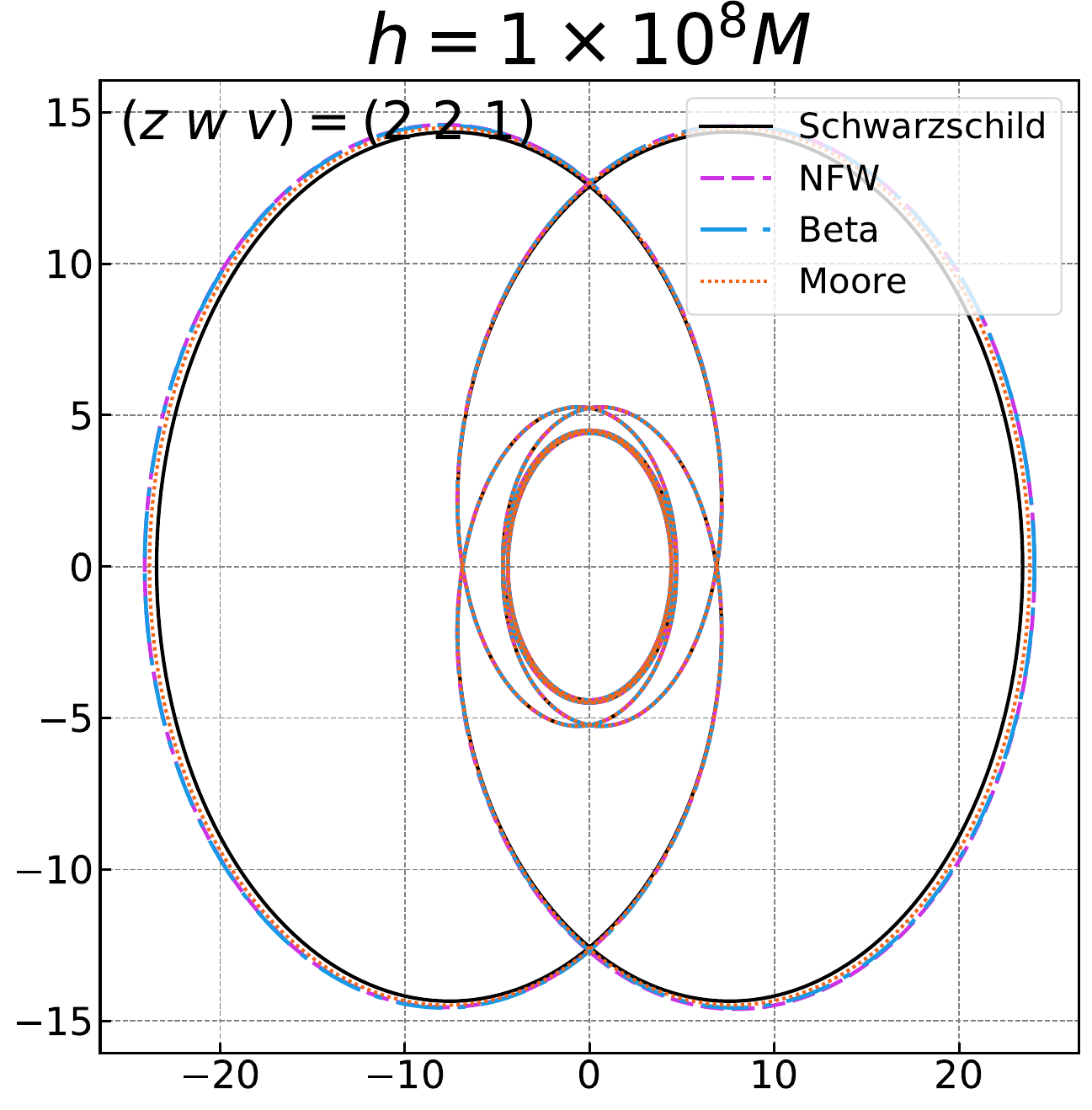}
	\includegraphics[width=0.23\linewidth]{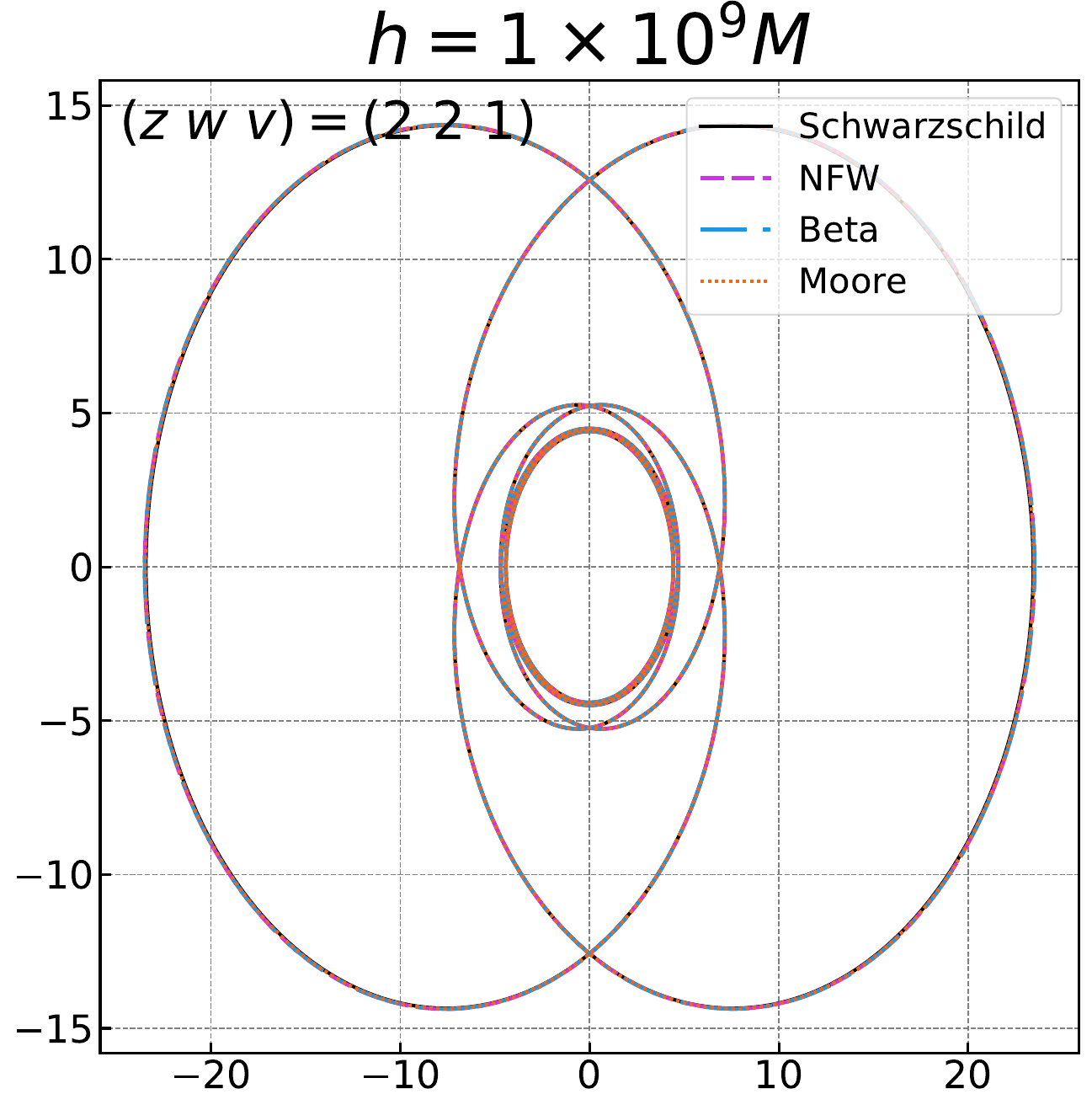}
	\includegraphics[width=0.23\linewidth]{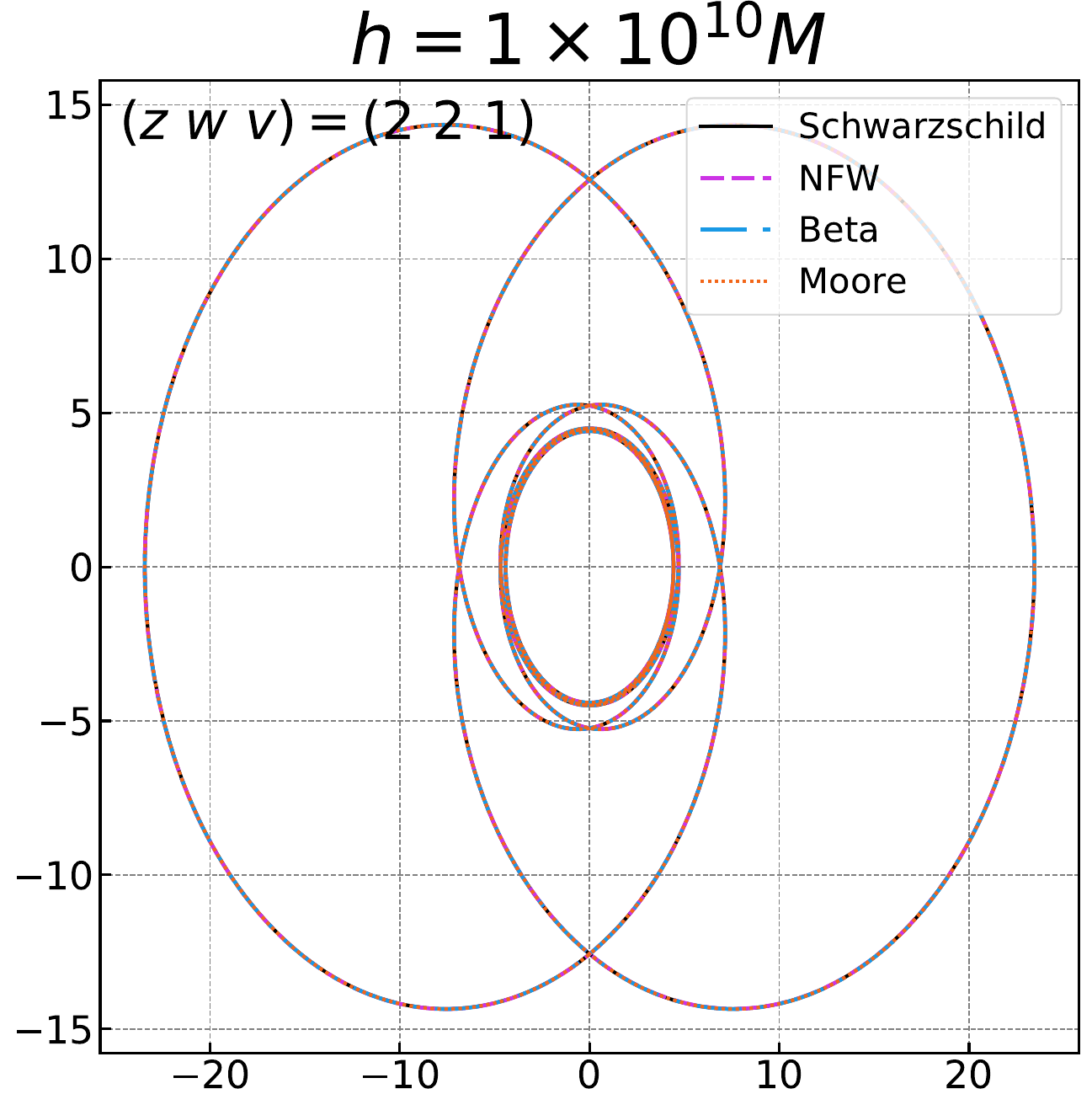}
	\end{subfigure}
	\caption{Effect of dark matter halo scale on periodic orbits around black holes. Periodic orbits with different $(z~w~v)$ configurations are shown for black holes embedded in various dark matter halos, where the halo scale $h$ varies from $10^7M \sim 10^{10}M$ and the dark matter mass parameter is fixed at $k = 10^4M$ .The parameter for angular momentum is selected as $\varepsilon=0.5$.}
	\label{dif_h}
\end{figure}

\begin{figure}[htbp]
	\centering
	\begin{subfigure}{0.3\textwidth}
		\includegraphics[width=\linewidth]{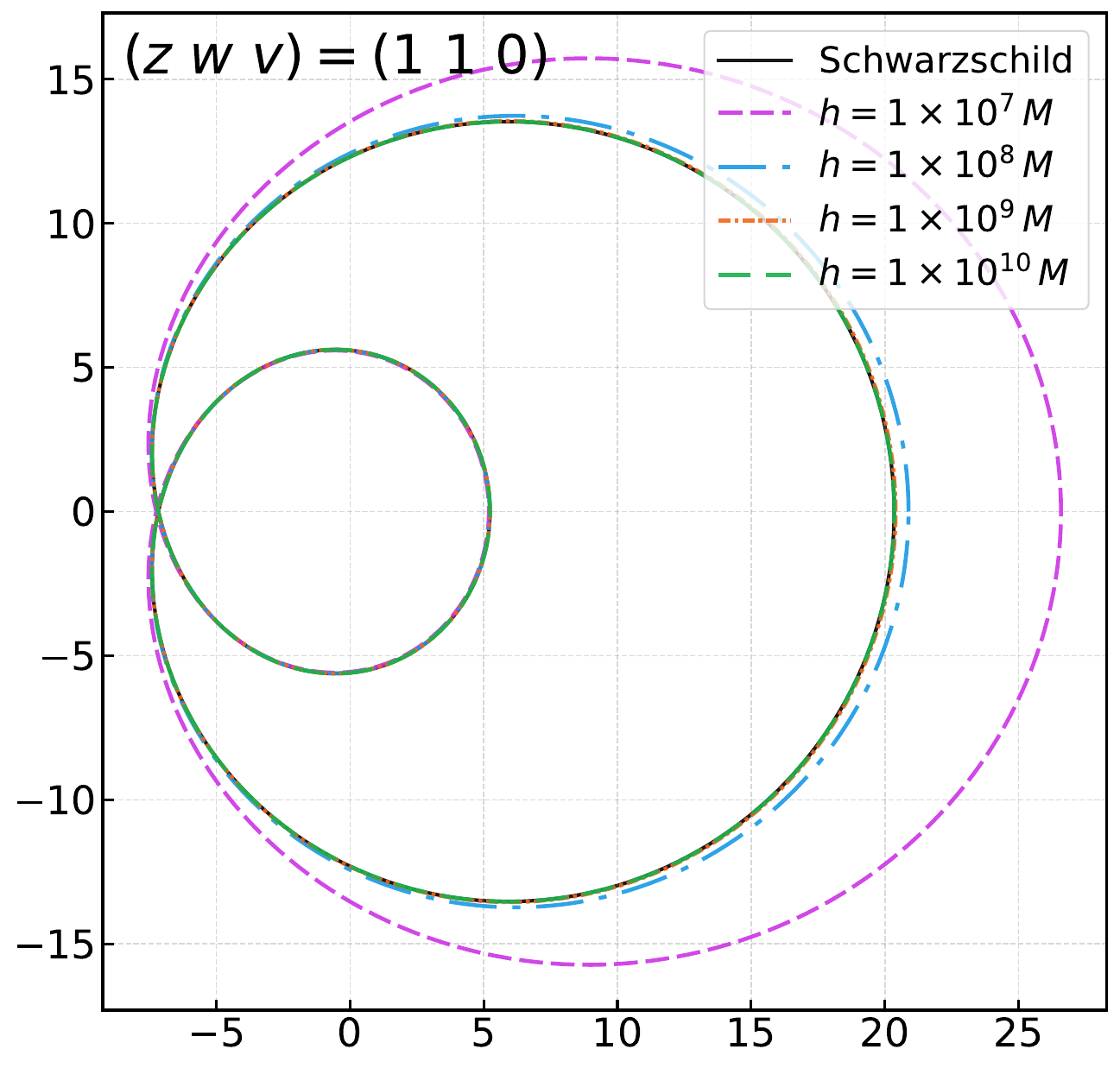}
		\caption{\( (z~w~v) \)=(1 1 0), $k= 10^4 M$}
	\end{subfigure}
	\hfill
	\begin{subfigure}{0.3\textwidth}
		\includegraphics[width=\linewidth]{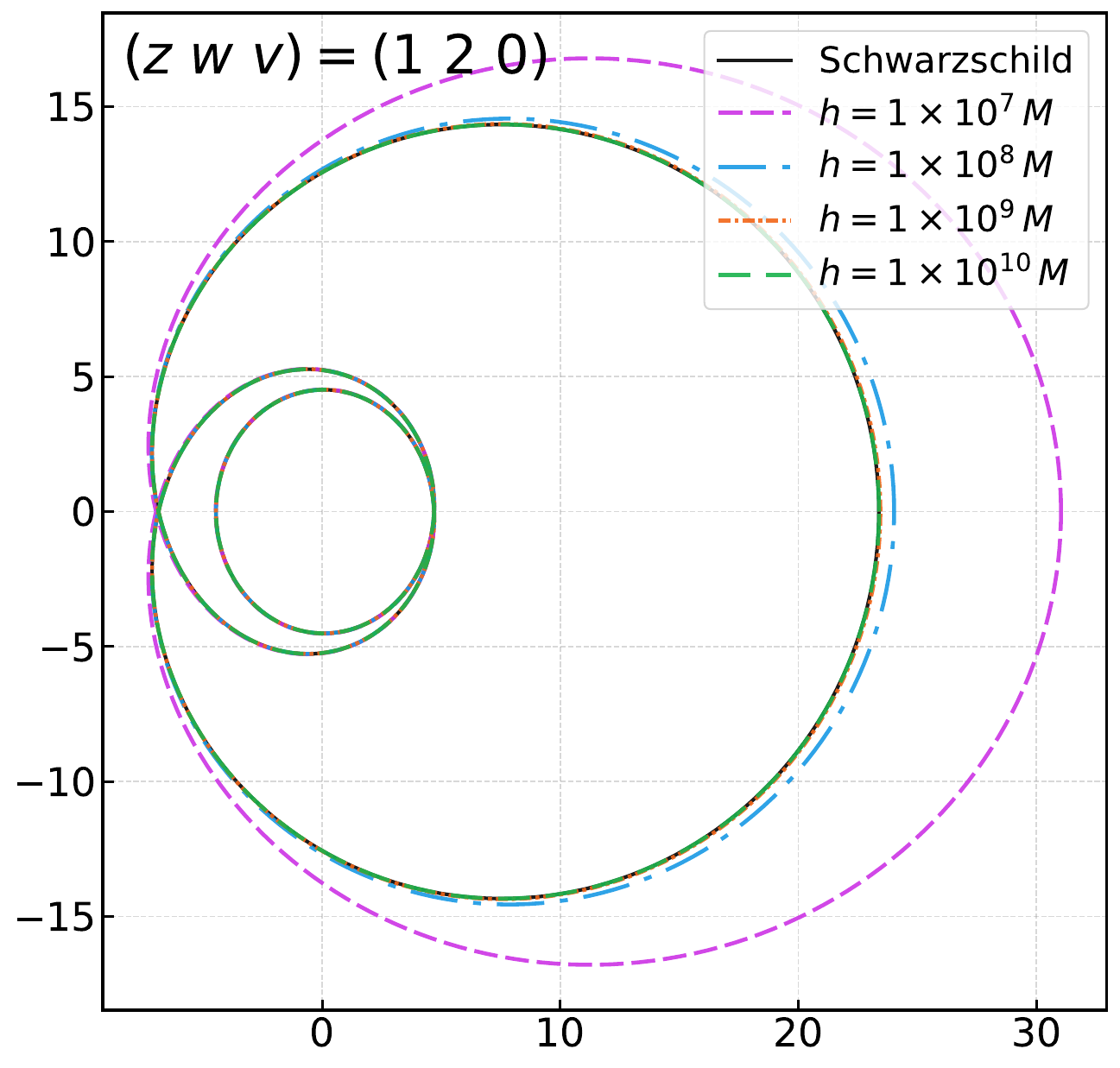}
		\caption{\( (z~w~v) \)=(1 2 0), $k= 10^4 M$}
	\end{subfigure}
	\hfill
	\begin{subfigure}{0.3\textwidth}
		\includegraphics[width=\linewidth]{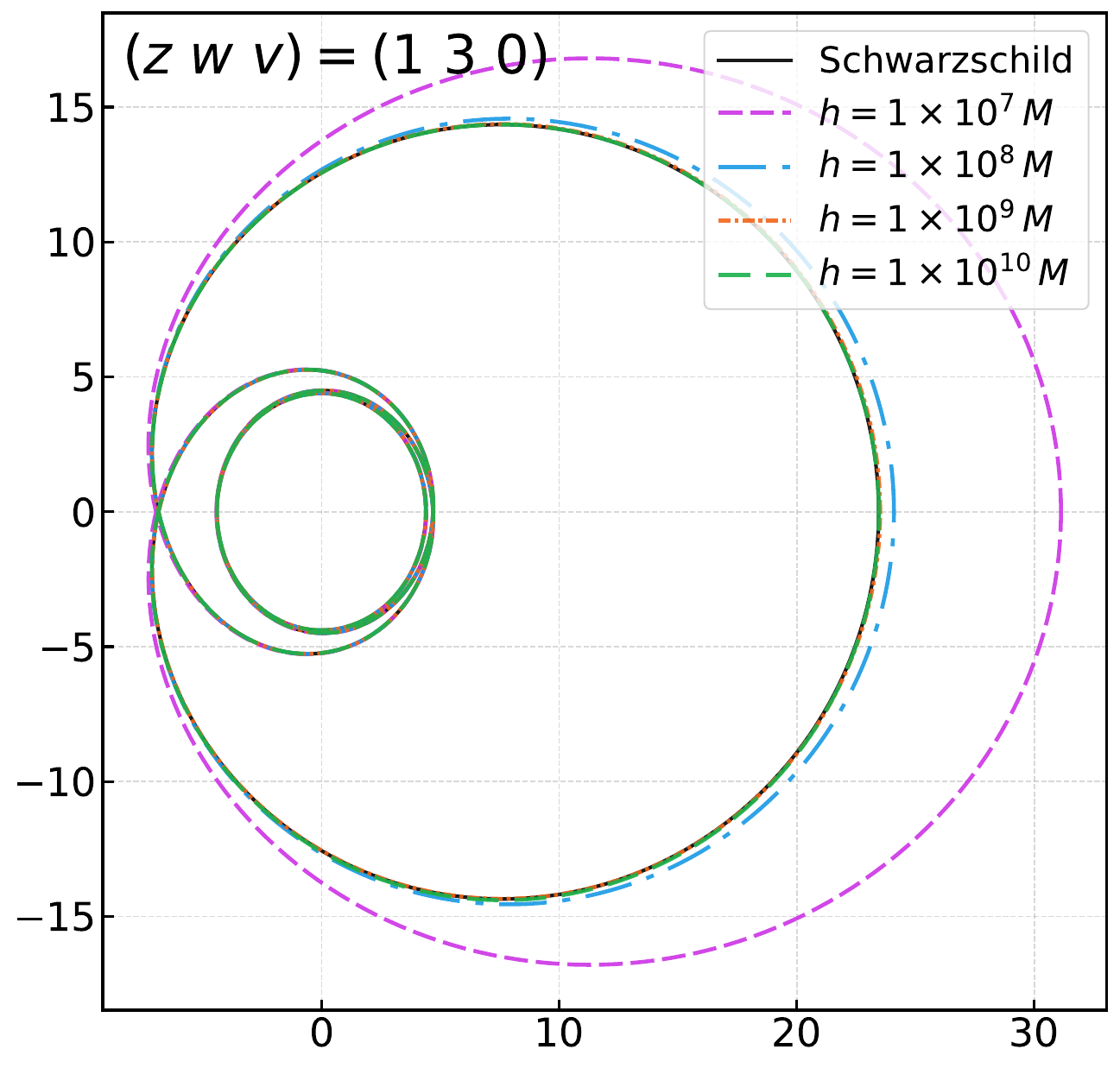}
		\caption{\( (z~w~v) \)=(1 3 0), $k= 10^4 M$}
	\end{subfigure}
	\vspace{0.4cm}
	\centering
	\hspace*{\fill} 
	\begin{subfigure}{0.3\textwidth}
		\includegraphics[width=\linewidth]{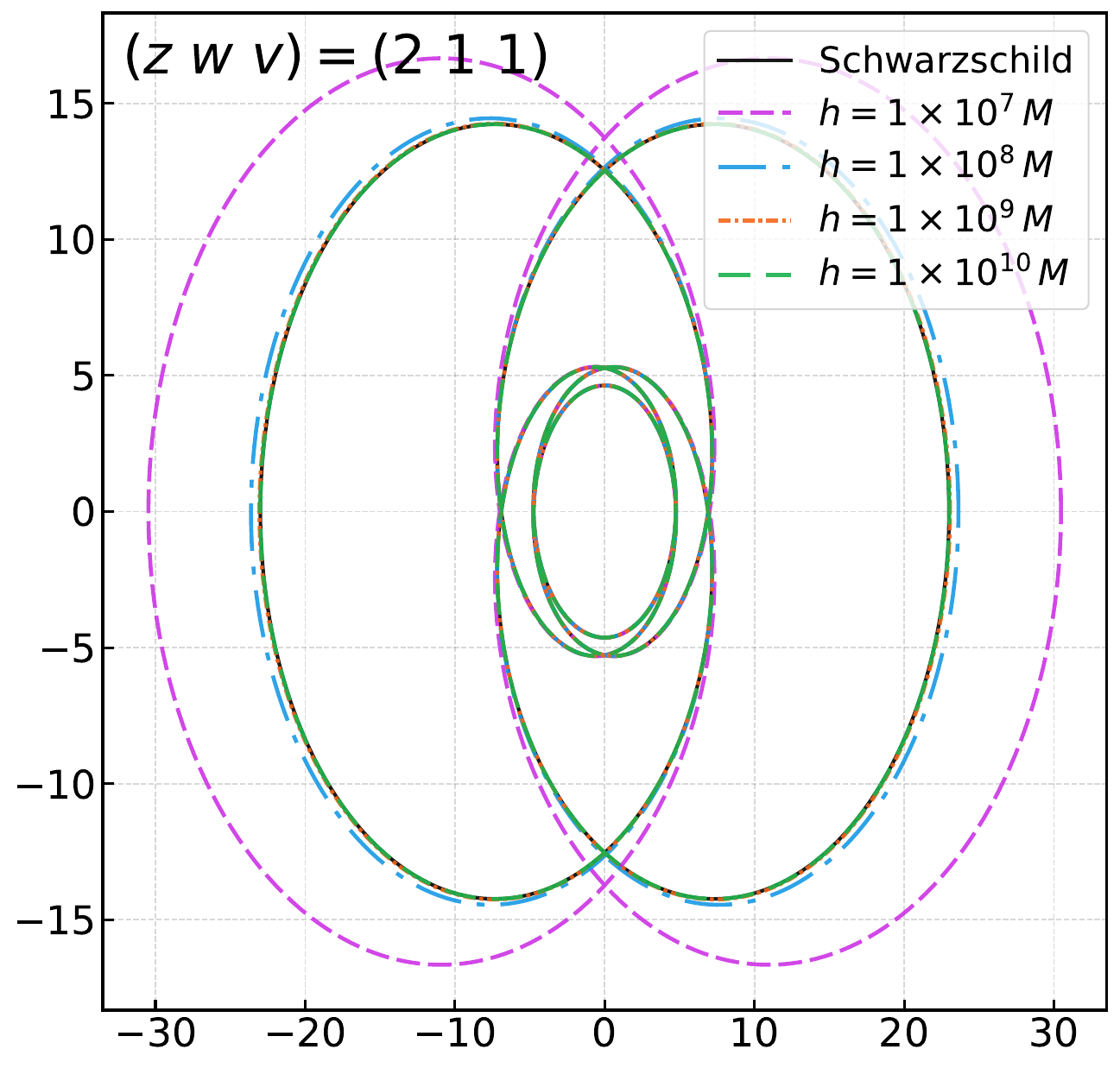}
		\caption{\( (z~w~v) \)=(2 1 1), $k= 10^4 M$}
	\end{subfigure}
	\hfill
	\begin{subfigure}{0.3\textwidth}
		\includegraphics[width=\linewidth]{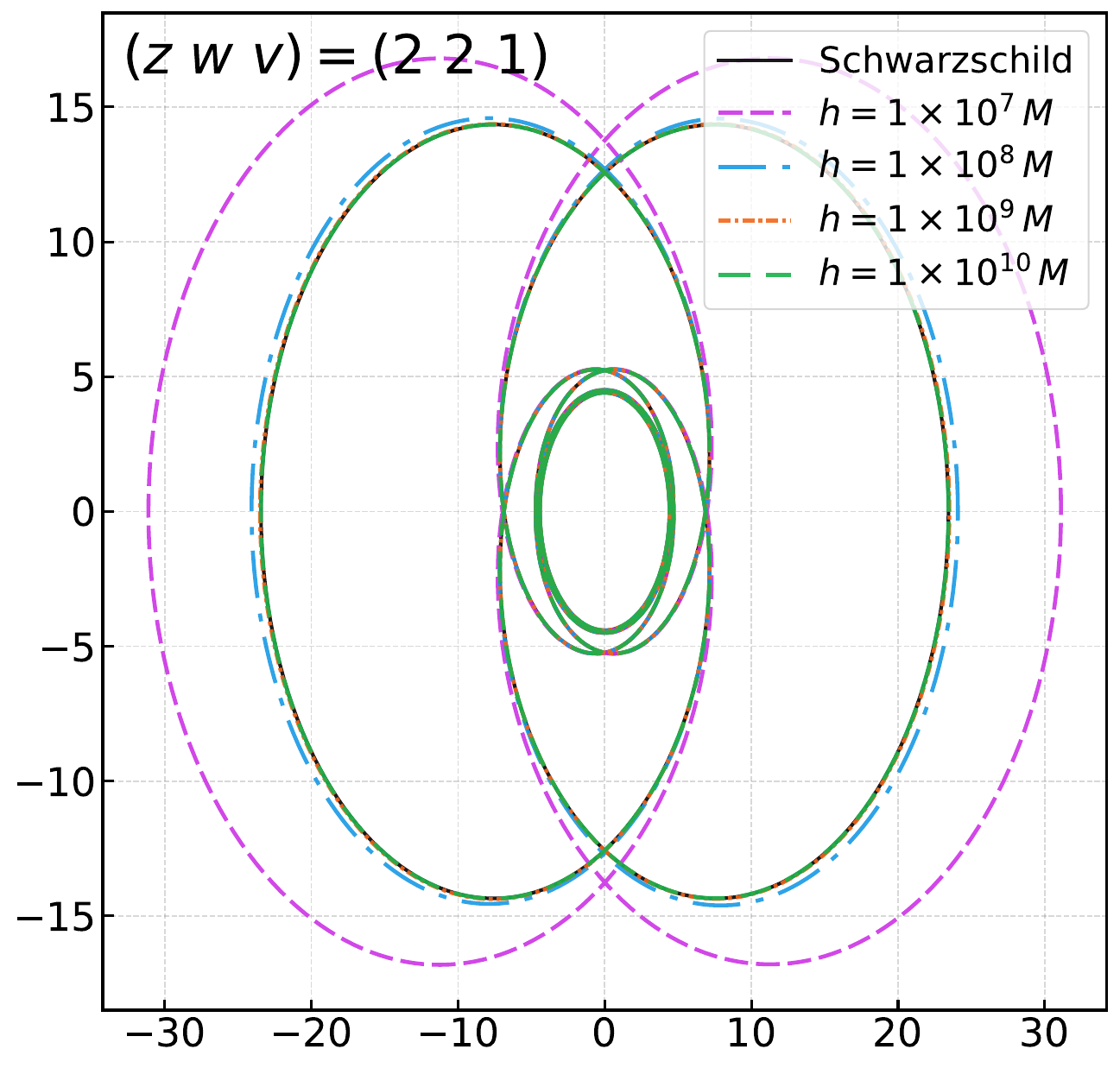}
		\caption{\( (z~w~v) \)=(2 2 1), $k= 10^4 M$}
	\end{subfigure}
	\hspace*{\fill} 
	\caption{Periodic orbits in NFW halo model with varying \( h \) and selected angular momentum parameter $\varepsilon=0.5$. In the figure, the periodic orbits in dark matter environment for a halo scale \( h = 10^{10} M \) (green dashed line) converges to the Schwarzschild black hole case (black solid line).}
	\label{dif_h_NFW}
\end{figure}

To further illustrate the effects of the dark matter halo scale, Fig.~\ref{dif_h_q} provides a detailed examination of the precession angle evaluated with four specific halo characteristic radii while maintaining a fixed dark matter mass of $k = 10^4 M$.  For all halo characteristic radii, the precession angle curves of the NFW and Beta models exhibit nearly identical behavior, with their curves overlapping almost completely. Most significantly, the results demonstrate a progressive weakening of dark matter halo influence as the characteristic radius increases. When the dark matter halo scale is $h = 10^7 M$, a significant difference is observed between the results in the dark matter halo environment and those for the Schwarzschild black hole. When $h = 10^9 M$, the effect becomes substantially diminished, with all dark matter models showing tiny deviations from the Schwarzschild black hole results (which serves as a reference baseline). When the halo characteristic radius expands to $h=10^{10}M$, the precession angle curves of all three dark matter
models precisely coincide with the Schwarzschild black hole curve. This convergence behavior demonstrates the dilution limit of dark matter halo effects: when the halo characteristic radius becomes sufficiently large, the local gravitational field closely resembles that of an isolated Schwarzschild black hole, significantly reducing the distinguishable signatures of different dark matter density profiles.

Following the analysis of precession angle, we now examine the direct impact of the dark matter halo's characteristic radius on the geometric shapes of periodic orbits. Fig.~\ref{dif_h} presents the periodic orbit trajectories for the Schwarzschild results, NFW, Beta, and Moore models results under different dark matter halo radii \(h\), with the corresponding orbital energy parameters detailed in Table~\ref{tab:h_L=L0_E} of Appendix~\ref{a2}. The layout of Fig.~\ref{dif_h} follows the same organization as the previous analysis in Fig.~\ref{dif_k}: each row corresponds to the same orbital configurations $(z~w~v)$, displaying periodic orbits within different halo models; each column fixes the dark matter halo scale \(h\) while changing the orbital configurations to illustrate the variation of orbital shapes. The dark matter mass is held constant at $k = 10^4 M$ throughout this analysis. From the row-wise comparison, a relatively small halo scale is needed to effectively separate the orbits from different halo models. It is also noteworthy that a convergence pattern emerges as the dark matter halo characteristic radius \(h\) increases. The periodic orbits of the NFW, Beta, and Moore models all gradually approach those of the Schwarzschild black hole for larger halo scales, with the orbital deviations diminishing steadily. This convergence behavior directly reflects the dilution of gravitational effects discussed in the precession angle analysis: as \(h\) increases while the total mass is kept constant, the local gravitational field strength decreases, reducing the distinguishable signatures of different density profiles. The convergence trend reaches its peak when \(h = 10^{10}M\), at which point the periodic orbits of all three dark matter models become indistinguishable from the Schwarzschild reference.

To illustrate how the scale of a dark matter halo affects periodic orbits within the same halo model, Fig.~\ref{dif_h_NFW} shows orbital trajectories in the NFW model for five different \((z~w~v)\) configurations across a range of \(h\) values. The results indicate that at a halo scale of \(h = 10^7 M\), the orbits significantly differ from those in a Schwarzschild black hole spacetime. As \(h\) increases, the trajectories under the NFW halo gradually move closer to the Schwarzschild case --- a trend also seen in other dark matter models which are not shown in this figure. When \(h\) reaches  a value \(h = 10^{10} M\), the orbits closely match those in the Schwarzschild limit. This behavior agrees with the precession angle results in Fig.~\ref{dif_h_q}, and the consistency between the two types of observables (precession angle and orbit trajectory) supports the reliability of our conclusions regarding the influence of the halo scale. The analyses of both observables converge to the same conclusion: increasing halo characteristic radius systematically weakens the gravitational influence of dark matter on periodic orbital characteristics, eventually leading to complete convergence with pure Schwarzschild behavior in the dilution limit.

\section{Gravitational waveforms from periodic orbits}\label{s4}

Extreme Mass Ratio Inspiral (EMRI) systems are key target sources for future space-based gravitational wave detectors. These systems are formed by stellar-mass compact objects orbiting around supermassive black holes, and their gravitational wave signals contain rich information about the system's dynamics and the spacetime geometry of the central black hole. When a small object orbits around a supermassive black hole enveloped by dark matter (DM) in a periodic motion, the gravitational waves it emits provide a unique avenue for studying the system's properties. This section provides an analysis of the gravitational waveforms of EMRI systems generated by periodic motions, and the theoretical framework for calculating gravitational waves from such periodic orbits is also briefly reviewed. In the calculation of gravitational waves in EMRI systems, the adiabatic approximation and Numerical Kludge waveform model, which is applicable when the small object moves in a nearly static gravitational background with circular, precessional or periodic motion, is adopted~\cite{Hughes:1999bq,Hughes:2001jr,Glampedakis:2002ya,Hughes:2005qb,Drasco:2005is,Gair:2005ih,Drasco:2005kz,Sundararajan:2008zm,Miller:2020bft,Isoyama:2021jjd}. Under this approximation, the energy and angular momentum of the smaller object can be considered constant over several orbital periods, and its trajectory can be viewed as a geodesic in the static background spacetime, with the gravitational radiation's reaction on the object's motion temporarily ignored.

The Numerical Kludge waveform model provides a practical scheme for calculating gravitational waves from periodic orbits in a DM halo environment~\cite{Babak:2006uv}. This method consists of two steps: first, the motion equations Eqs.~(\ref{e10a}) , (\ref{e10b}) and (\ref{e17}) are numerically solved to determine the orbit of the small body in the gravitational spacetime containing the DM distribution; then, the quadrupole formula for gravitational radiation is applied to this orbit to generate the corresponding waveform~\cite{Thorne:1980ru}. This method can conveniently reveal the gravitational wave signals of a slowly evolving EMRI system, providing possibilities for exploring the characteristics of the periodic orbits (or precessional orbits), central black hole, and surrounding DM distribution. For the spacetime metric perturbation \( h_{ij} \), the gravitational radiation quadrupole moment formula calculated to second order can be expressed as~\cite{Maselli:2021men,Liang:2022gdk,Zhao:2024exh},
\begin{equation}
	h_{ij} = \frac{4\mu M}{D_L} \left( v_i v_j - \frac{m}{r} n_i n_j \right),
\end{equation}
where $m$ and $M$ represent the masses of the small body and the central SMBH, respectively. Since we are studying an EMRI system, it is reasonable to set $m \ll M$. In the expression, $D_L$ represents the luminosity distance of the system to observer; $\mu = \frac{Mm}{(M+m)^2}$ is the symmetric mass ratio; $n_i$ is the unit direction vector, and $v_i$ is the velocity component of the small body. To analyze the gravitational wave signal as measured by the detector, we construct a detector-adapted coordinate system \((X, Y, Z)\). Its origin coincides with that of the original coordinate system \((r, \theta, \phi)\), with both centered on the supermassive black hole. The orientation of this new frame is determined by two angles: \(\iota\), the inclination angle of the orbital plane relative to the $X-Y$ plane, and \(\zeta\), the longitude of the pericenter measured within the orbital plane~\cite{Yang:2024lmj, gravity-book}. The basis vectors of the detector-adapted frame are expressed in the original coordinates as:
\begin{subequations}
	\begin{align}
		\mathbf{e}_X &= (\cos\zeta, -\sin\zeta, 0),\label{eX} \\
		\mathbf{e}_Y &= (\sin\iota\sin\zeta, \cos\iota\cos\zeta, -\sin\iota),\label{eY} \\
		\mathbf{e}_Z &= (\sin\iota\sin\zeta, -\sin\iota\cos\zeta, \cos\iota).\label{eZ}
	\end{align}
\end{subequations}
In general relativity, the polarization state of gravitational waves is typically decomposed into two independent modes: $+$ (orthogonal) polarization and $\times$ (cross) polarization. In the introduced coordinate system, these two polarization components \( h_+ \) and \( h_{\times} \) detected by observers can be expressed as~\cite{Zhao:2024exh,Alloqulov:2025ucf},
\begin{subequations}
	\begin{align}
		h_{+} &= \frac{1}{2} (e^i_X e^j_X - e^i_Y e^j_Y) \times h_{ij}= -\frac{2\mu M^2}{D_L r} \left(1 + \cos^2 \iota\right) \cos \left(2\phi + 2\zeta\right) \\
		h_{\times} &= \frac{1}{2} (e^i_X e^j_Y + e^i_Y e^j_X) \times h_{ij}= -\frac{4\mu M^2}{D_L r} \cos \iota \sin \left(2\phi + 2\zeta\right),
	\end{align}
\end{subequations}
Here, \(\phi\) represents the azimuthal phase angle in the orbits, and $r$ represents the radial coordinate of periodic orbits to the center. In order to demonstrate the effect of the DM halo on gravitational waves produced by different orbital configurations, we choose an EMRI system where a small body with mass \(m = 1M_\odot\) orbits a supermassive black hole with mass \(M_{\rm BH} = 10^7M_\odot\). To simplify calculations, the inclination angle \(\iota\) and latitude angle \(\zeta\) are both set to \(\pi/4\), and the luminosity distance \(D_L\) is set to 2 Gpc. Following the analysis of periodic orbits in previous sections, we now focus on how the two primary halo parameters—mass and scale—influence the characteristics of the gravitational radiation emitted by periodic orbital motion in these environments.

\subsection{The effect of dark matter mass on gravitational waves }

\begin{figure}[t]
	\centering  
	\begin{subfigure}[b]{\textwidth}
		\centering
		\begin{subfigure}[b]{0.4\textwidth}
			\centering
			\includegraphics[width=\textwidth]{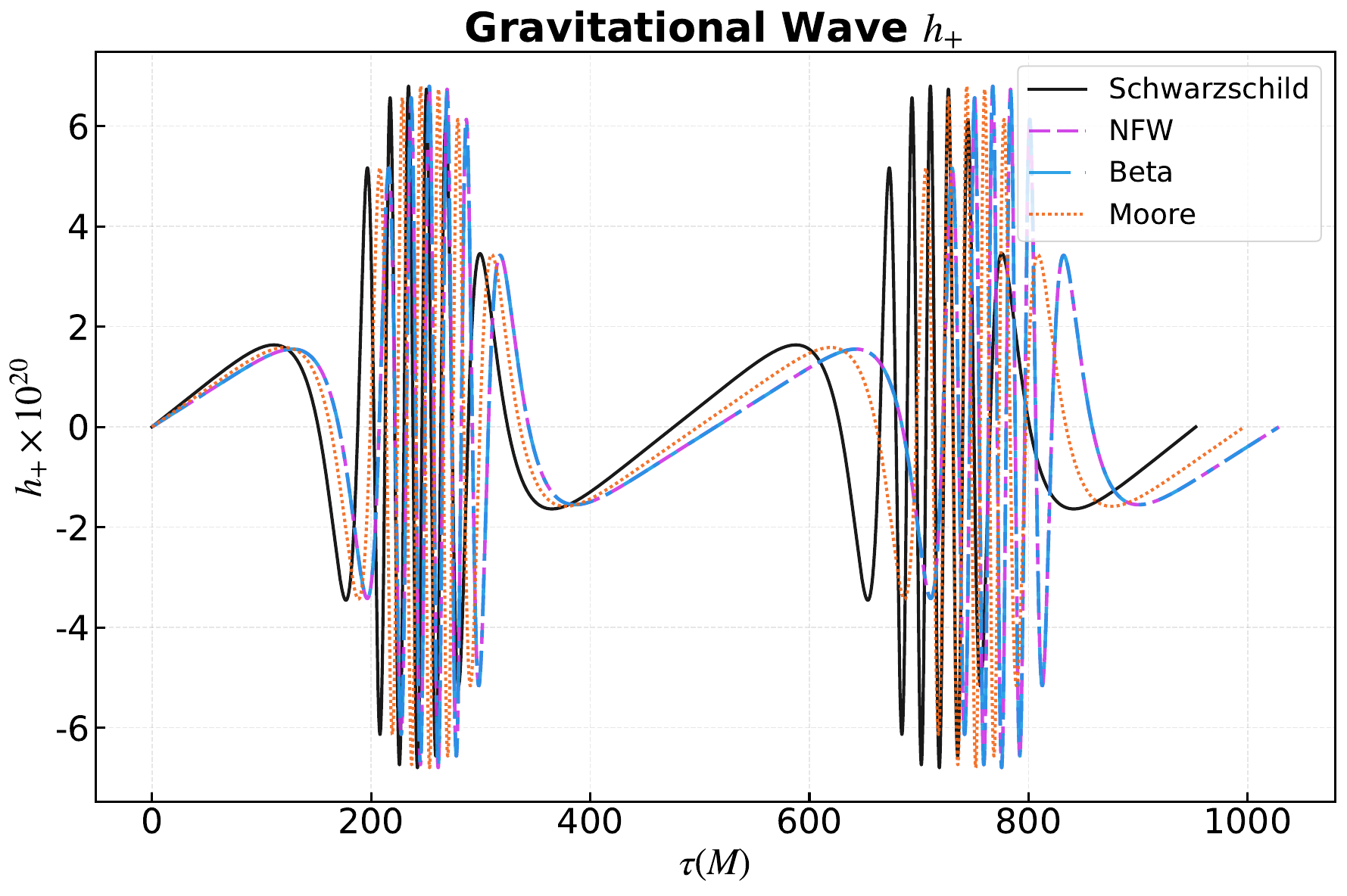}
		\end{subfigure}
		\begin{subfigure}[b]{0.4\textwidth}
			\centering
			\includegraphics[width=\textwidth]{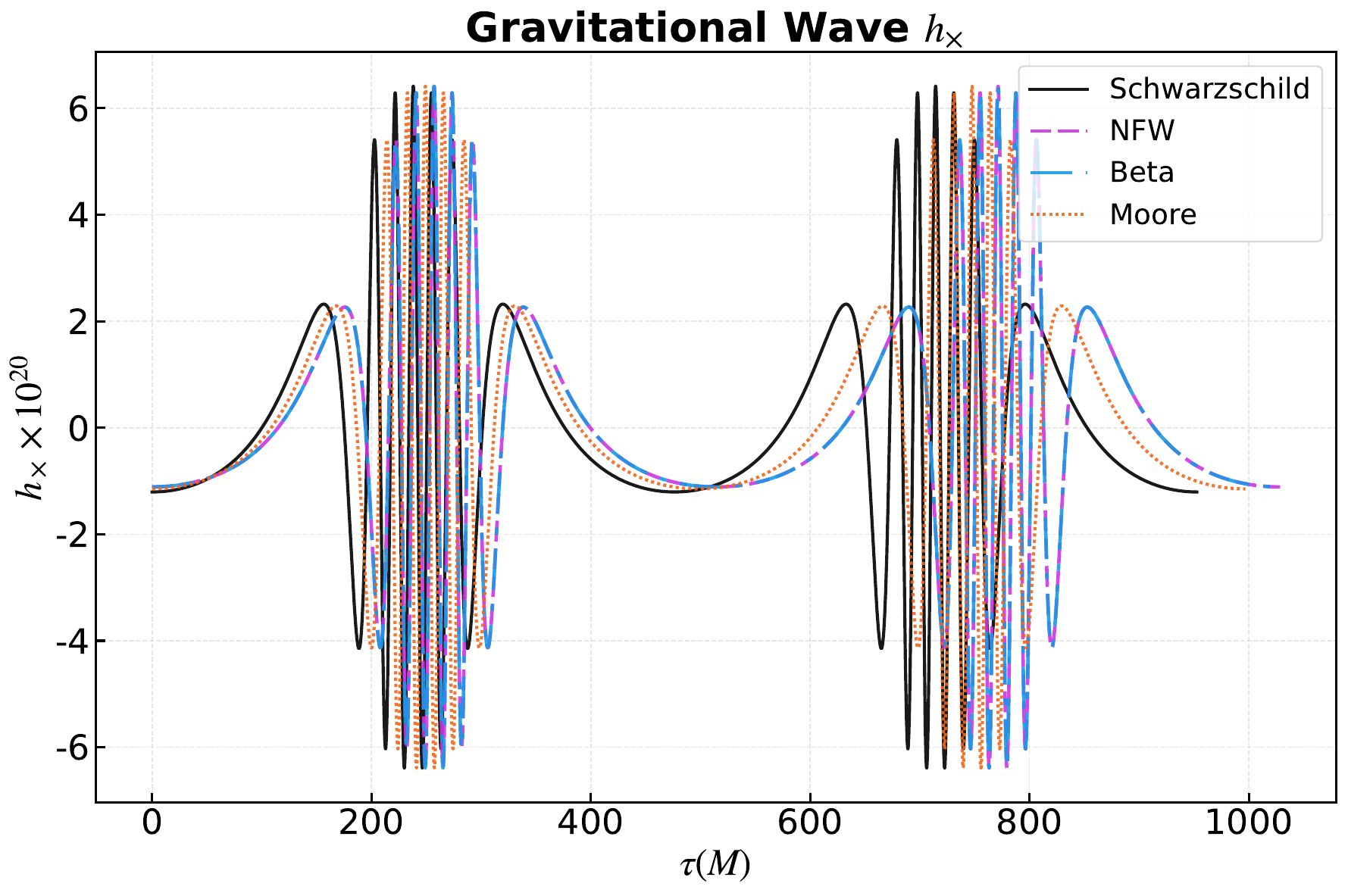}
		\end{subfigure}
		\caption{$k=3 \times 10^3M$}
	\end{subfigure}
	
	\vspace{0.2cm}
	
	\begin{subfigure}[b]{\textwidth}
		\centering
		\begin{subfigure}[b]{0.4\textwidth}
			\centering
			\includegraphics[width=\textwidth]{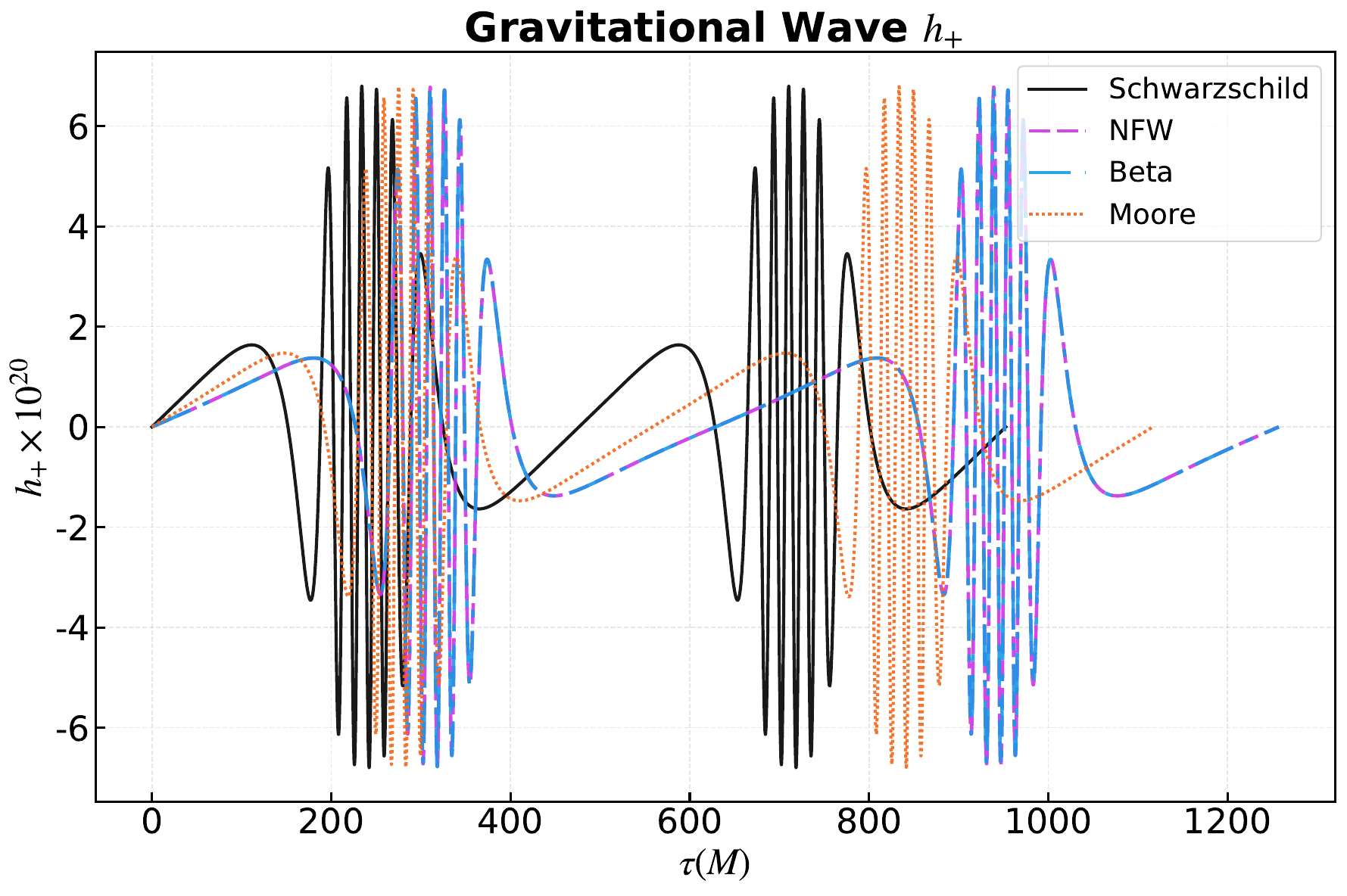}
		\end{subfigure}
		\begin{subfigure}[b]{0.4\textwidth}
			\centering
			\includegraphics[width=\textwidth]{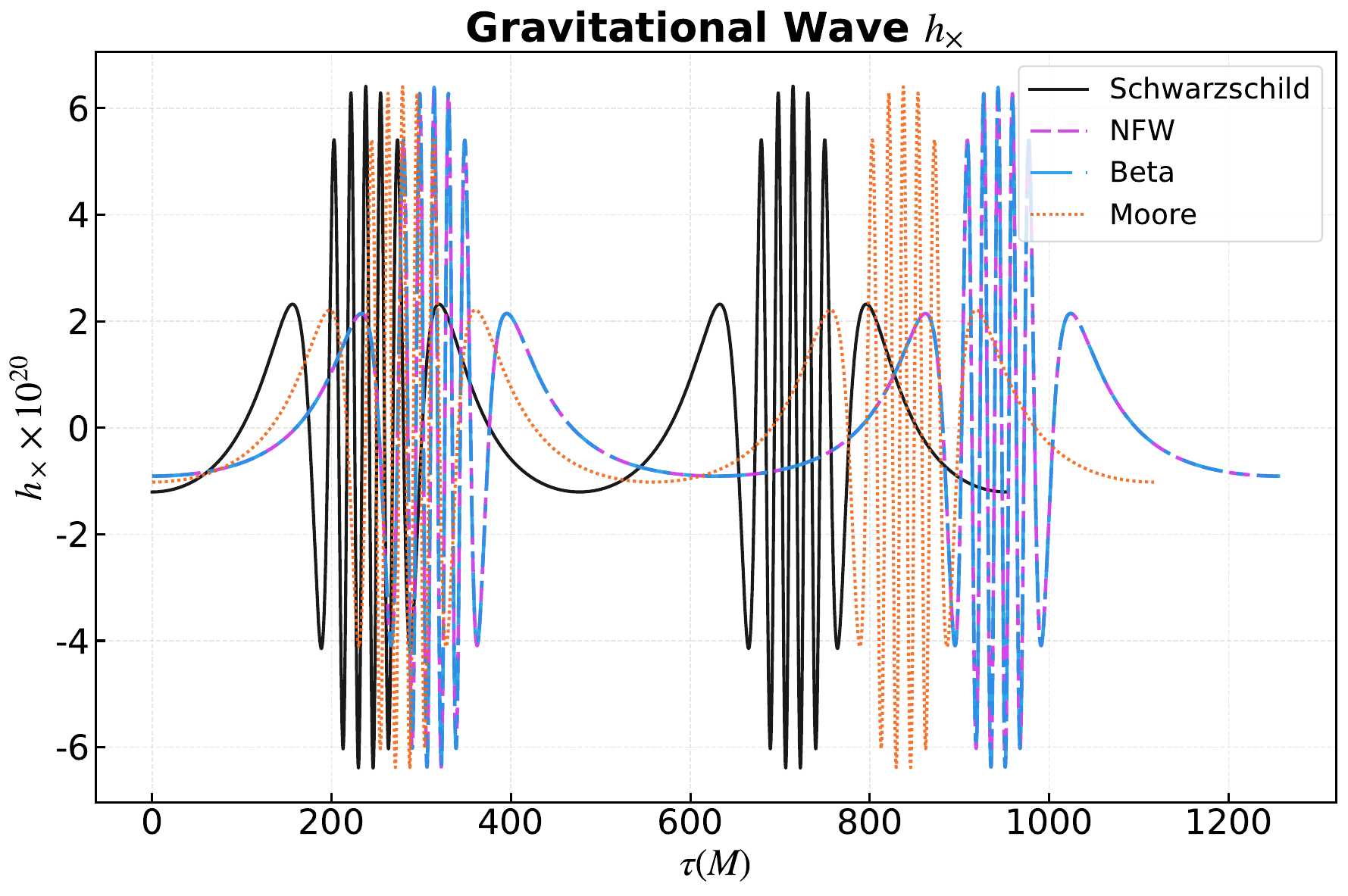}
		\end{subfigure}
		\caption{$k=1 \times 10^4M$}
	\end{subfigure}
	
	\vspace{0.2cm}
	
	\begin{subfigure}[b]{\textwidth}
		\centering
		\begin{subfigure}[b]{0.4\textwidth}
			\centering
			\includegraphics[width=\textwidth]{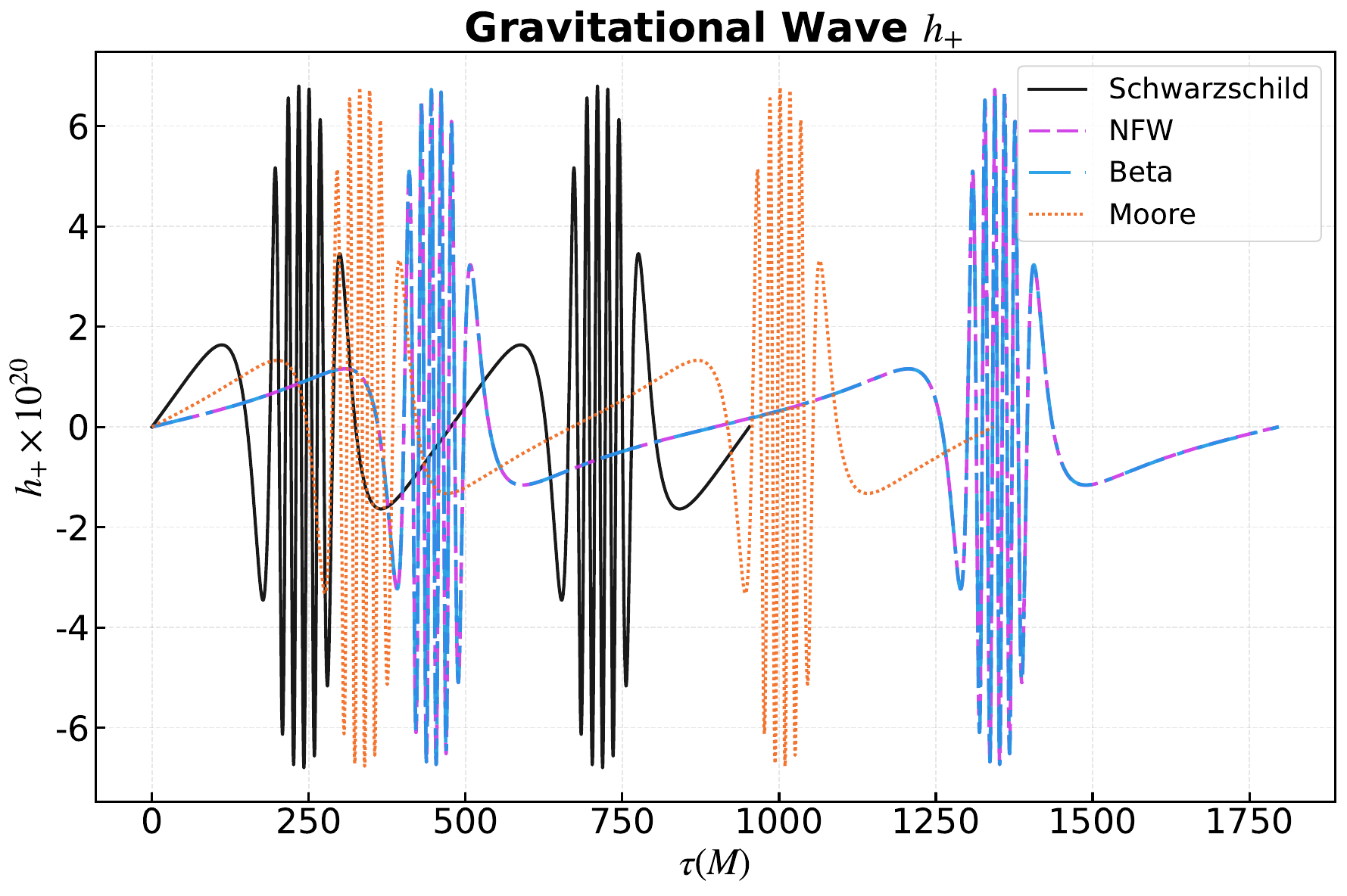}
		\end{subfigure}
		\begin{subfigure}[b]{0.4\textwidth}
			\centering
			\includegraphics[width=\textwidth]{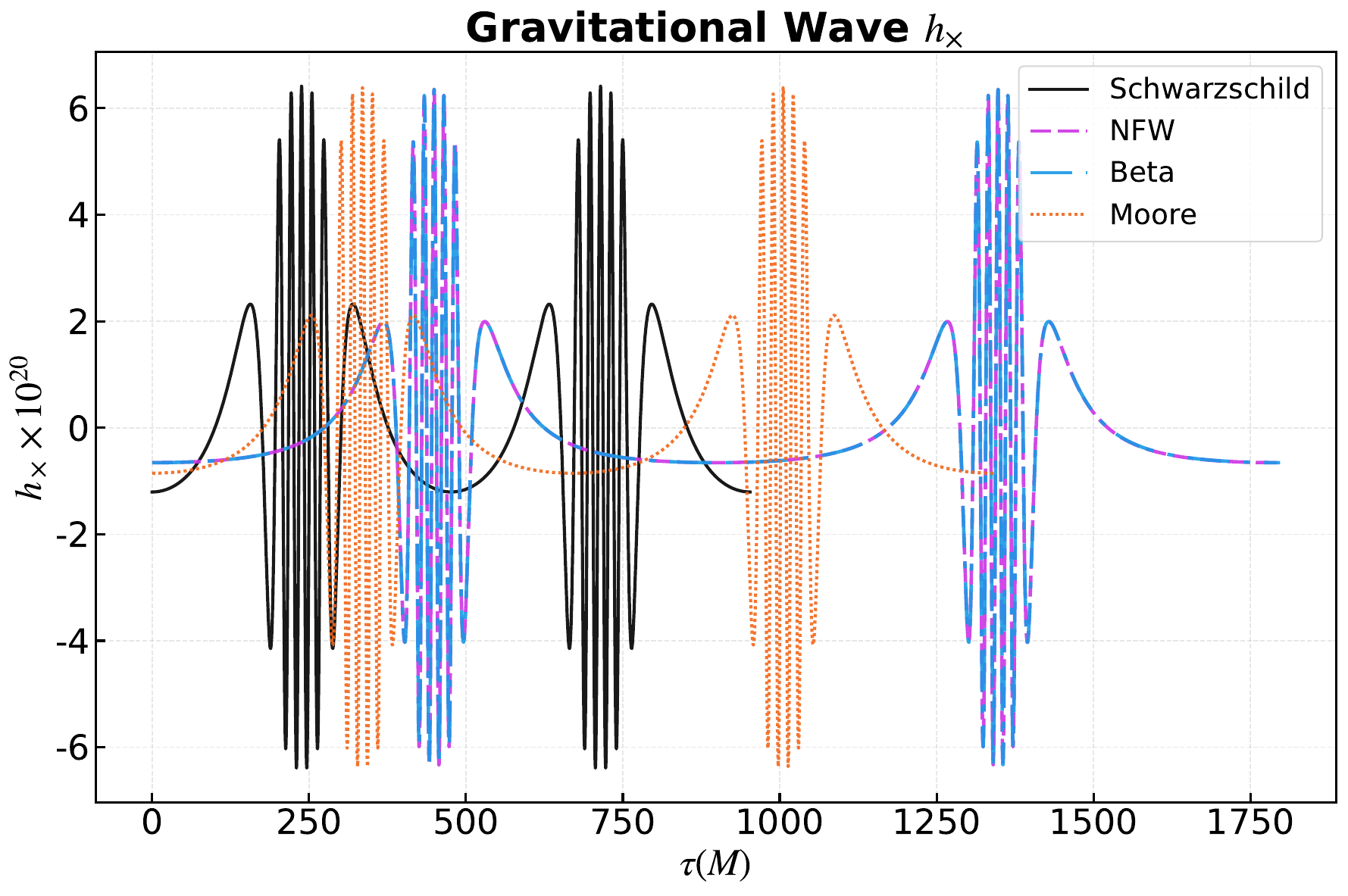}
		\end{subfigure}
		\caption{$k=2 \times 10^4M$}
	\end{subfigure}
	\caption{Gravitational waveforms generated by the $(2~2~1)$ orbital configuration under different dark matter masses for the NFW, Beta, and Moore models. The dark matter halo characteristic radius is fixed at $h = 10^7 M$. Each panel shows results for a specific dark matter mass: (a) $k = 3 \times 10^3 M$; (b) $k = 1 \times 10^4 M$; (c) $k = 2 \times 10^4 M$. The left column displays the $h_+$ polarization component, while the right column shows the $h_\times$ component. The Schwarzschild black hole results (black solid curves) serve as reference baselines. This figure displays the gravitational wave signals over one orbital period.}
	\label{GW_k}
\end{figure}

We begin with an analysis of the dark matter mass effects on the gravitational waveform of periodic orbits for a given orbital configuration $(z~w~v)$. Fig.~\ref{GW_k} presents the gravitational waveforms generated by the $(2~2~1)$ orbital configuration under different dark matter masses for the NFW, Beta, and Moore models. The dark matter halo characteristic radius is fixed at $h = 10^7M$, while the dark matter mass $k$ varies from $1 \times 10^3M \sim 2 \times 10^4M$. The waveforms display both $h+$ and $h_\times$ polarization components, which encode information about the orbital evolution and the spacetime geometry. From the figure, we observe that increasing the dark matter mass $k$ leads to progressive changes in the gravitational wave signals. Specifically, larger dark matter masses induce that waveforms deviate more substantially from the Schwarzschild reference case. The signal period duration extends significantly, with the waveform spanning longer time intervals as $k$ increases, reflecting the rapid changes in the orbital periods of particles in stronger dark matter halo environments. The physical reason for this behavior can be traced back to the orbital shapes discussed earlier. As demonstrated in Fig.~\ref{dif_k}, increasing dark matter mass $k$ stretches the apoapsis of periodic orbits, leading to larger orbital dimensions. Since gravitational wave emission depends on the time-varying quadrupole moment of the mass distribution, these geometric changes of periodic orbits directly lead to modified waveform characteristics. The extended orbital dimensions result in increased orbital periods, which directly correspond to the longer period duration observed in the gravitational waveforms. Furthermore, the slower orbital motion at larger distances leads to reduced instantaneous frequencies in the
gravitational waveform, causing the phase to accumulate more gradually over the same physical time interval. These effects combine to produce the characteristic stretching of the waveform patterns visible in Fig.~\ref{GW_k}. Therefore, the influence of the dark matter halo manifests as a broadening of the waveforms in the time scales. This effect primarily extends the orbital periods and correspondingly lengthens the characteristic timescales of the gravitational wave signals, resulting from the gravitational potential contributed by the dark matter halo. The comparative analysis between different dark matter models reveals that the NFW and Beta models produce nearly identical gravitational waveforms across all mass values of $k$ examined in this work. These findings are consistent with the periodic orbital characteristics shown in Figs.~\ref{dif_k_q} and~\ref{dif_k}. The Moore model, consistent with its distinct orbital properties, generates waveforms that differ from the NFW and Beta models, though the qualitative trends in the waveform with varying dark matter mass remain similar. Examining the waveform structure in detail, we observe that the gravitational wave signals exhibit distinct zoom and whirl stages within one orbital period. The relatively calm sections of the waveforms correspond to the zoom stage, where the small compact object traverses the nearly elliptical portions of its orbit far from the black hole. In contrast, the sections with rapid oscillations correspond to the whirl stage, where the small object approaches the black hole and exhibits near-circular motion during this stage. The gravitational wave frequency rises sharply, producing intense oscillations. This zoom-whirl structure in the waveforms directly reflects the zoom-whirl orbital behavior discussed in the periodic orbit classification.

\begin{figure} 
	\centering  
	\begin{subfigure}[b]{\textwidth}
		\centering
		\begin{subfigure}[b]{0.32\textwidth}
			\centering
			\includegraphics[width=\textwidth]{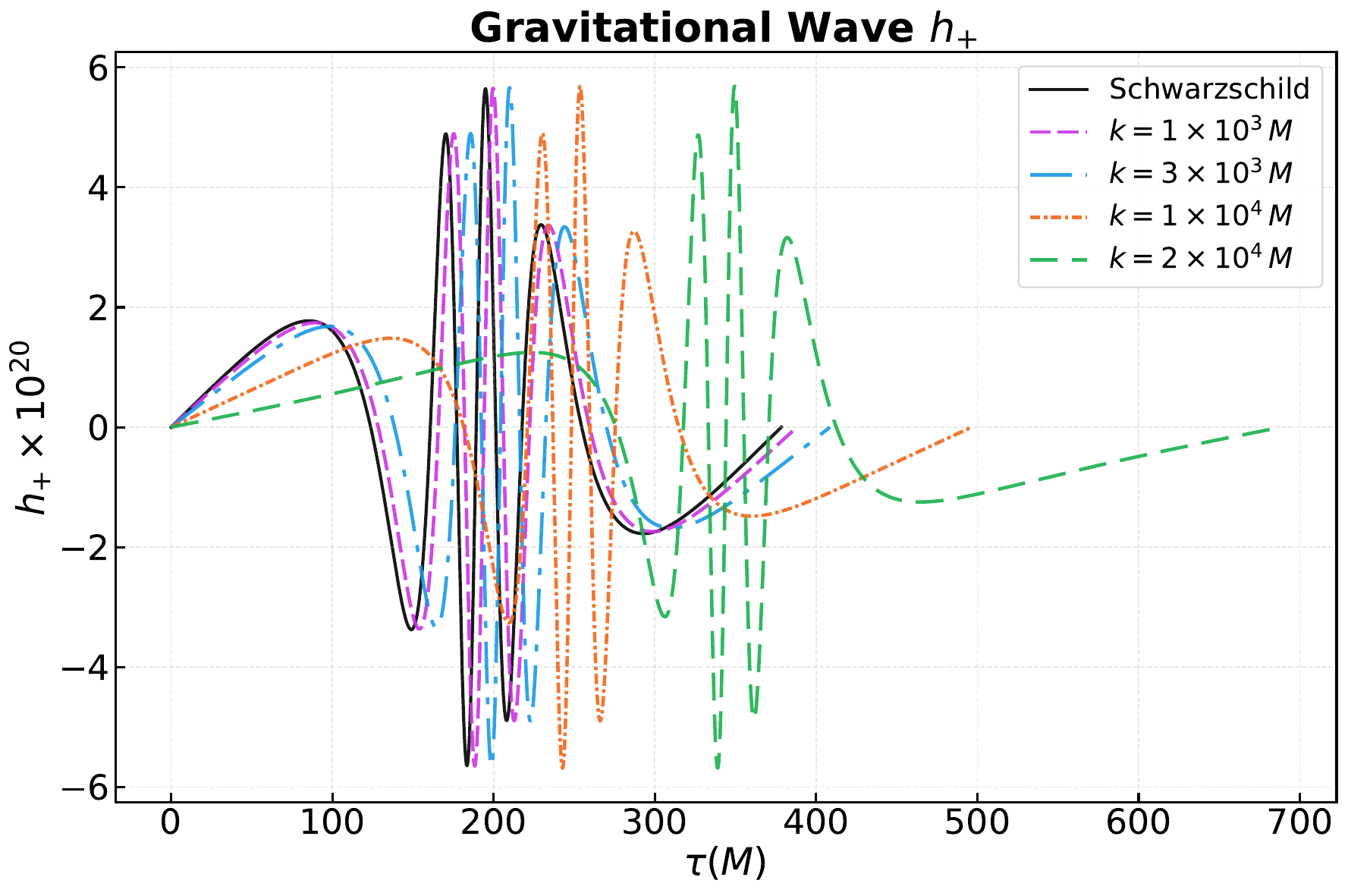}
		\end{subfigure}
		\begin{subfigure}[b]{0.32\textwidth}
			\centering
			\includegraphics[width=\textwidth]{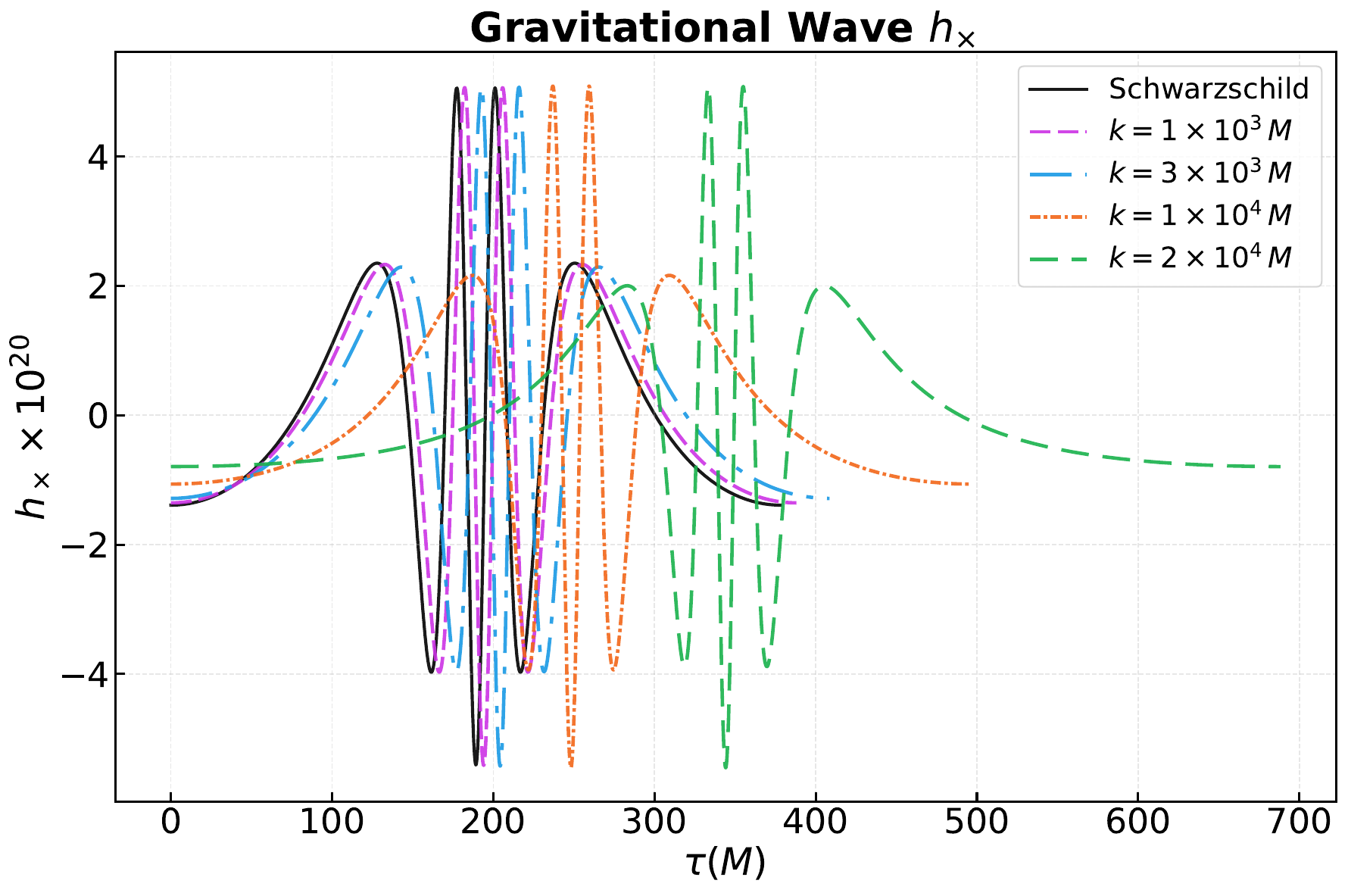}
		\end{subfigure}
		\caption{$(z\ w\ v)~= ~(1\ 1\ 0)$}
		\label{fig:gw_110}
	\end{subfigure}
	
	\vspace{0.3cm}
	
	\begin{subfigure}[b]{\textwidth}
		\centering
		\begin{subfigure}[b]{0.32\textwidth}
			\centering
			\includegraphics[width=\textwidth]{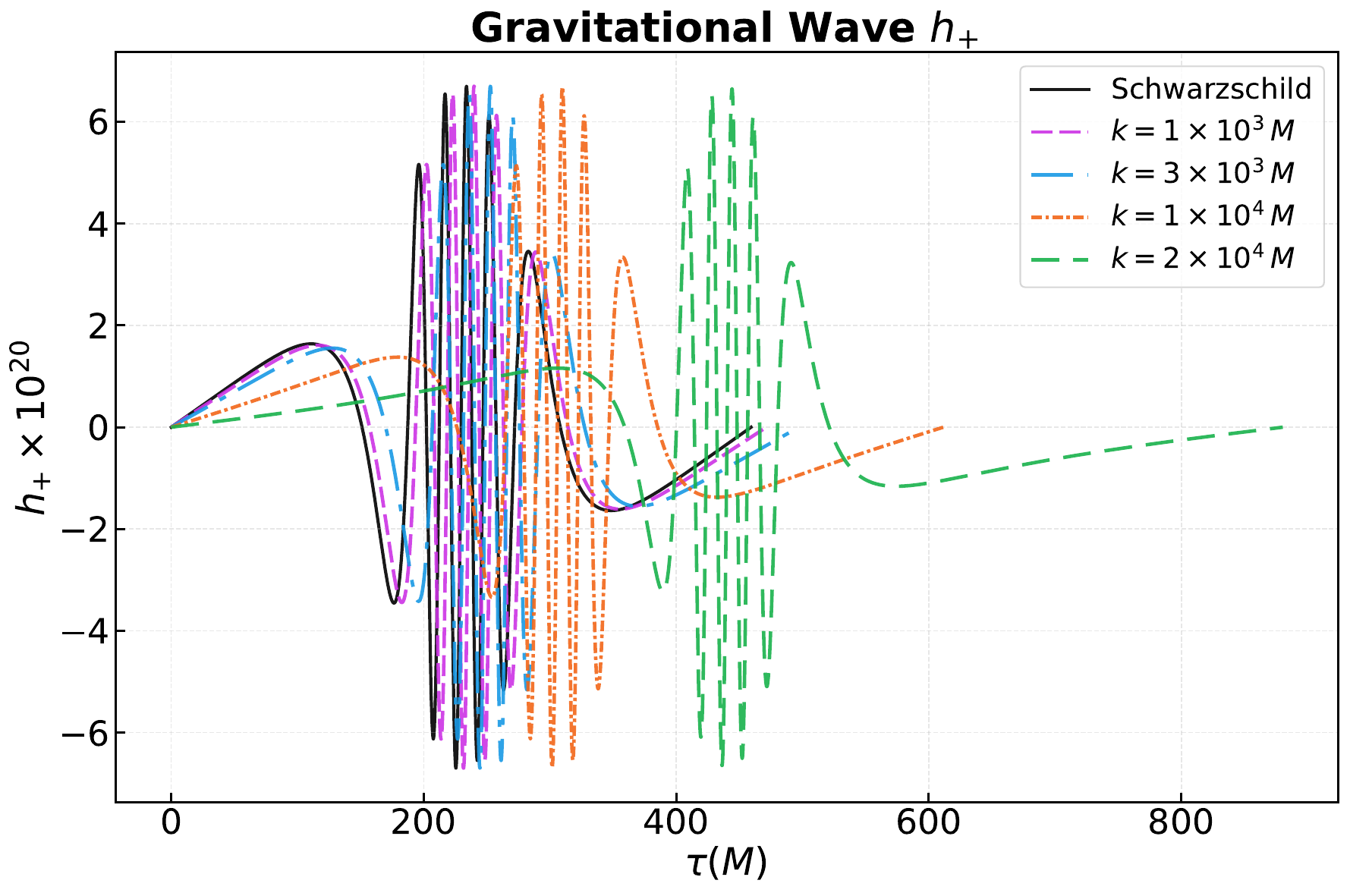}
		\end{subfigure}
		\begin{subfigure}[b]{0.32\textwidth}
			\centering
			\includegraphics[width=\textwidth]{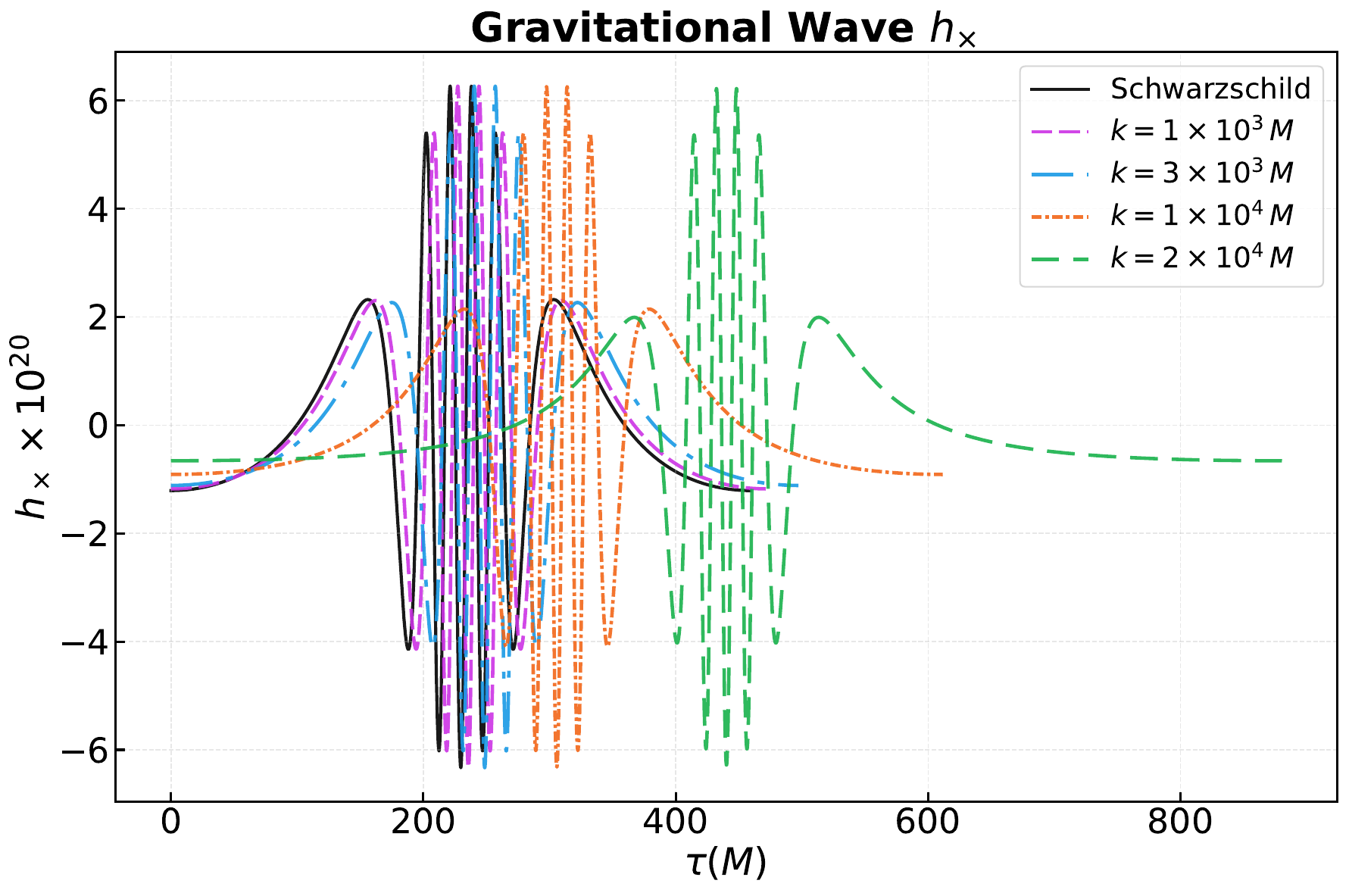}
		\end{subfigure}
		\caption{$(z\ w\ v)~= ~(1\ 2\ 0)$}
		\label{fig:gw_120}
	\end{subfigure}
	
	\vspace{0.3cm}
	
	\begin{subfigure}[b]{\textwidth}
		\centering
		\begin{subfigure}[b]{0.32\textwidth}
			\centering
			\includegraphics[width=\textwidth]{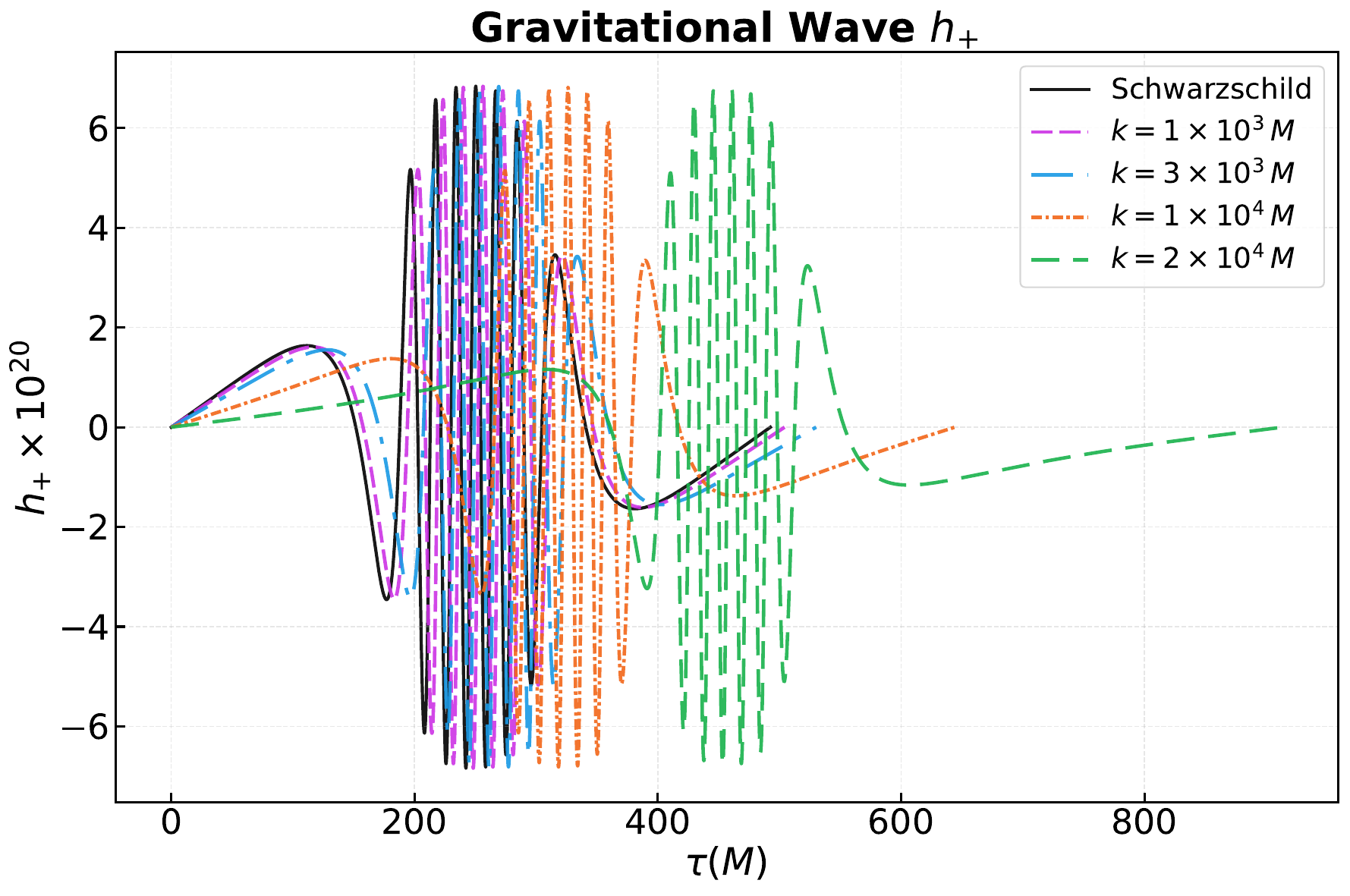}
		\end{subfigure}
		\begin{subfigure}[b]{0.32\textwidth}
			\centering
			\includegraphics[width=\textwidth]{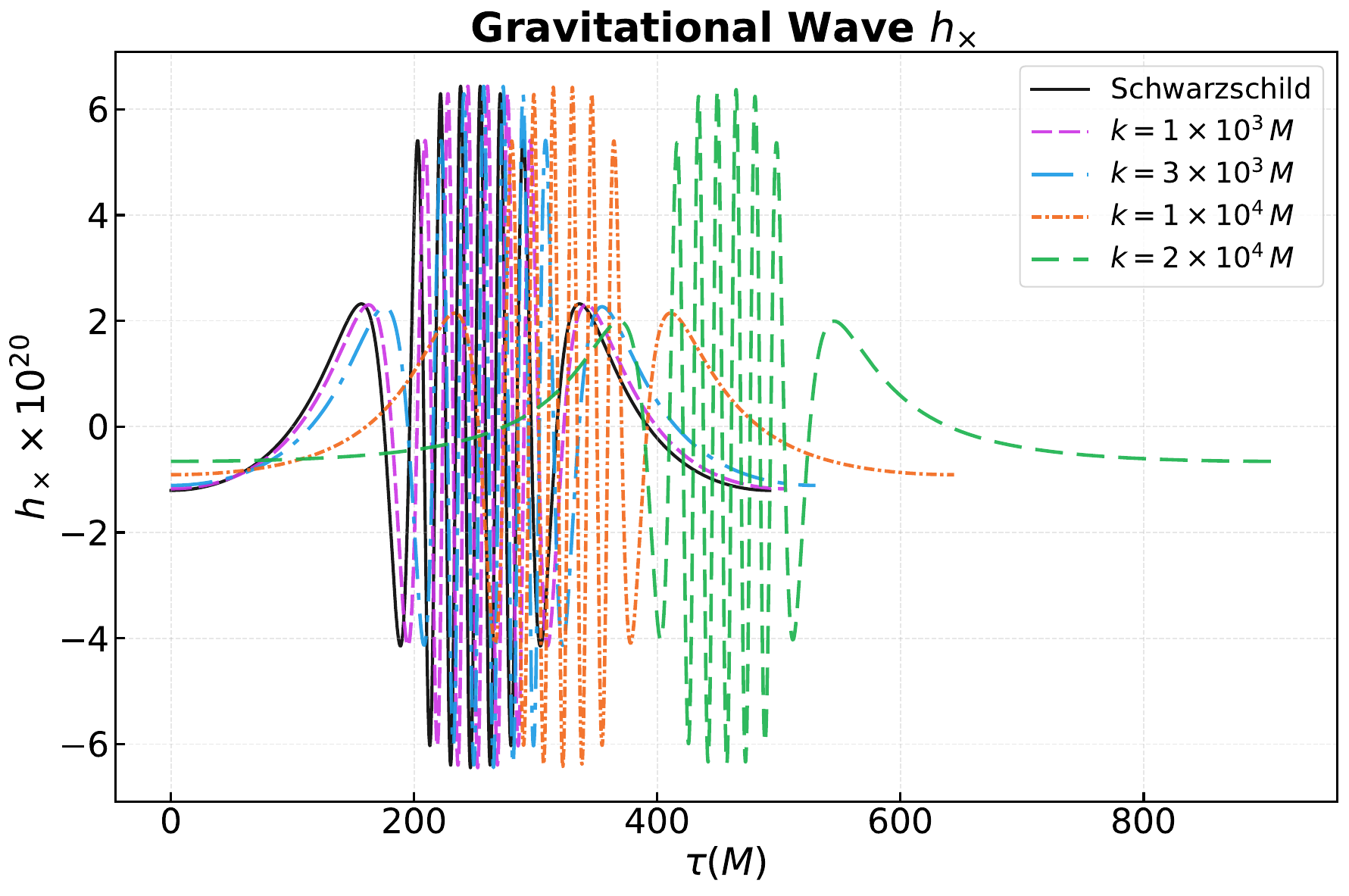}
		\end{subfigure}
		\caption{$(z\ w\ v)~= ~(1\ 3\ 0)$}
	\end{subfigure}
	
	\vspace{0.3cm}
	
	\begin{subfigure}[b]{\textwidth}
		\centering
		\begin{subfigure}[b]{0.32\textwidth}
			\centering
			\includegraphics[width=\textwidth]{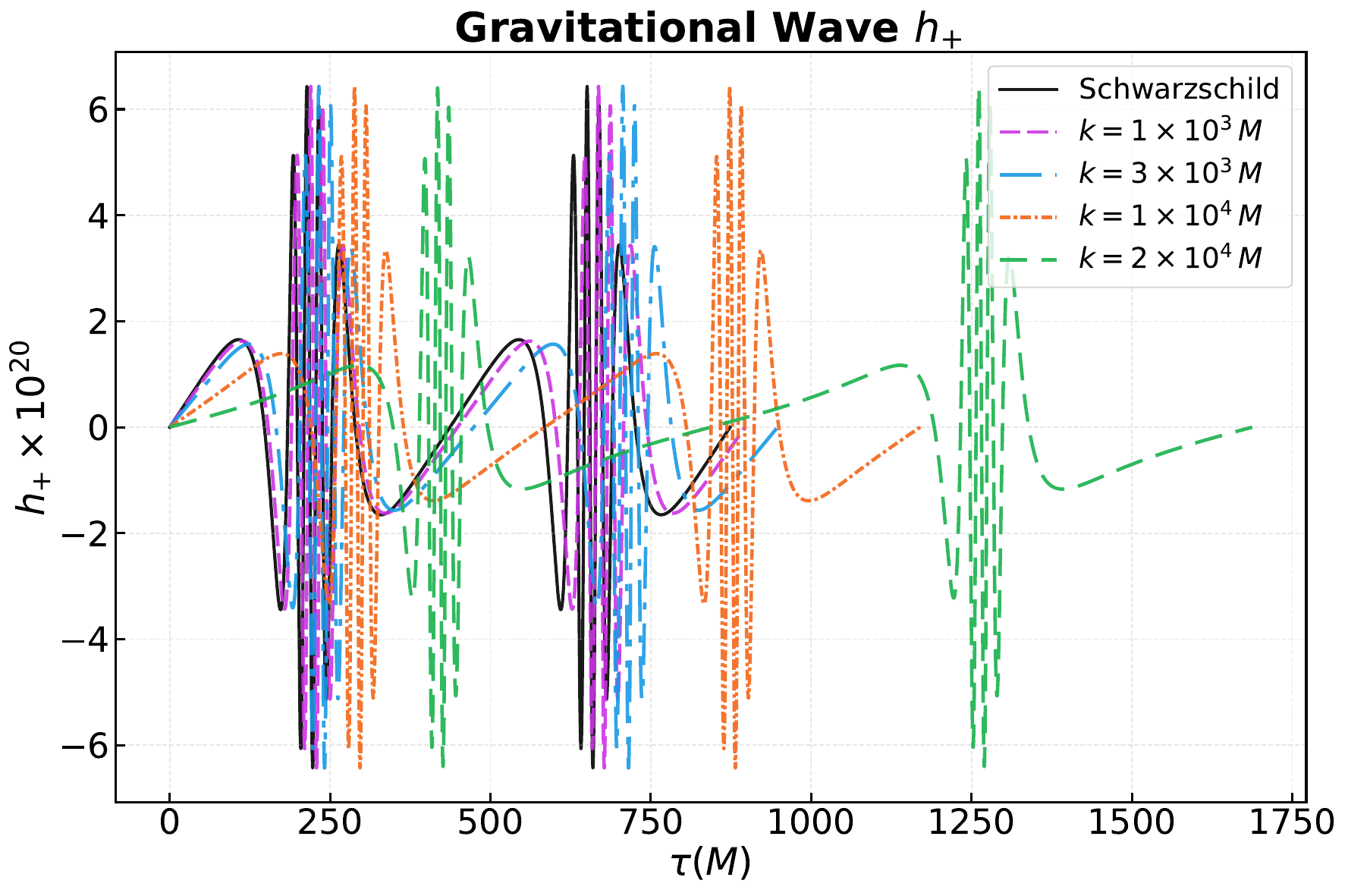}
		\end{subfigure}
		\begin{subfigure}[b]{0.32\textwidth}
			\centering
			\includegraphics[width=\textwidth]{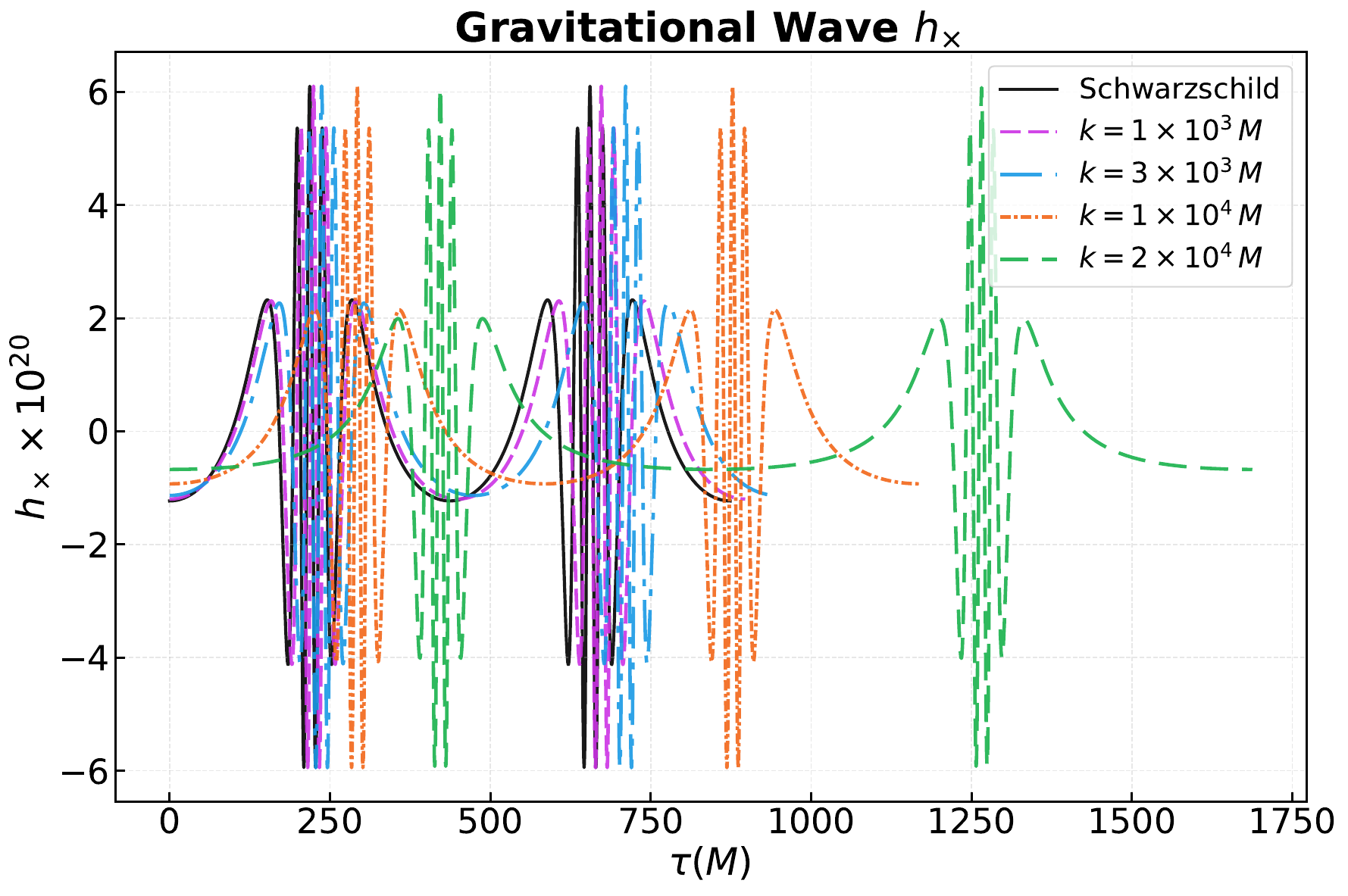}
		\end{subfigure}
		\caption{$(z\ w\ v)~= ~(2\ 1\ 1)$}
		\label{fig:gw_211}
	\end{subfigure}
	
	\vspace{0.3cm}
	
	\begin{subfigure}[b]{\textwidth}
		\centering
		\begin{subfigure}[b]{0.32\textwidth}
			\centering
			\includegraphics[width=\textwidth]{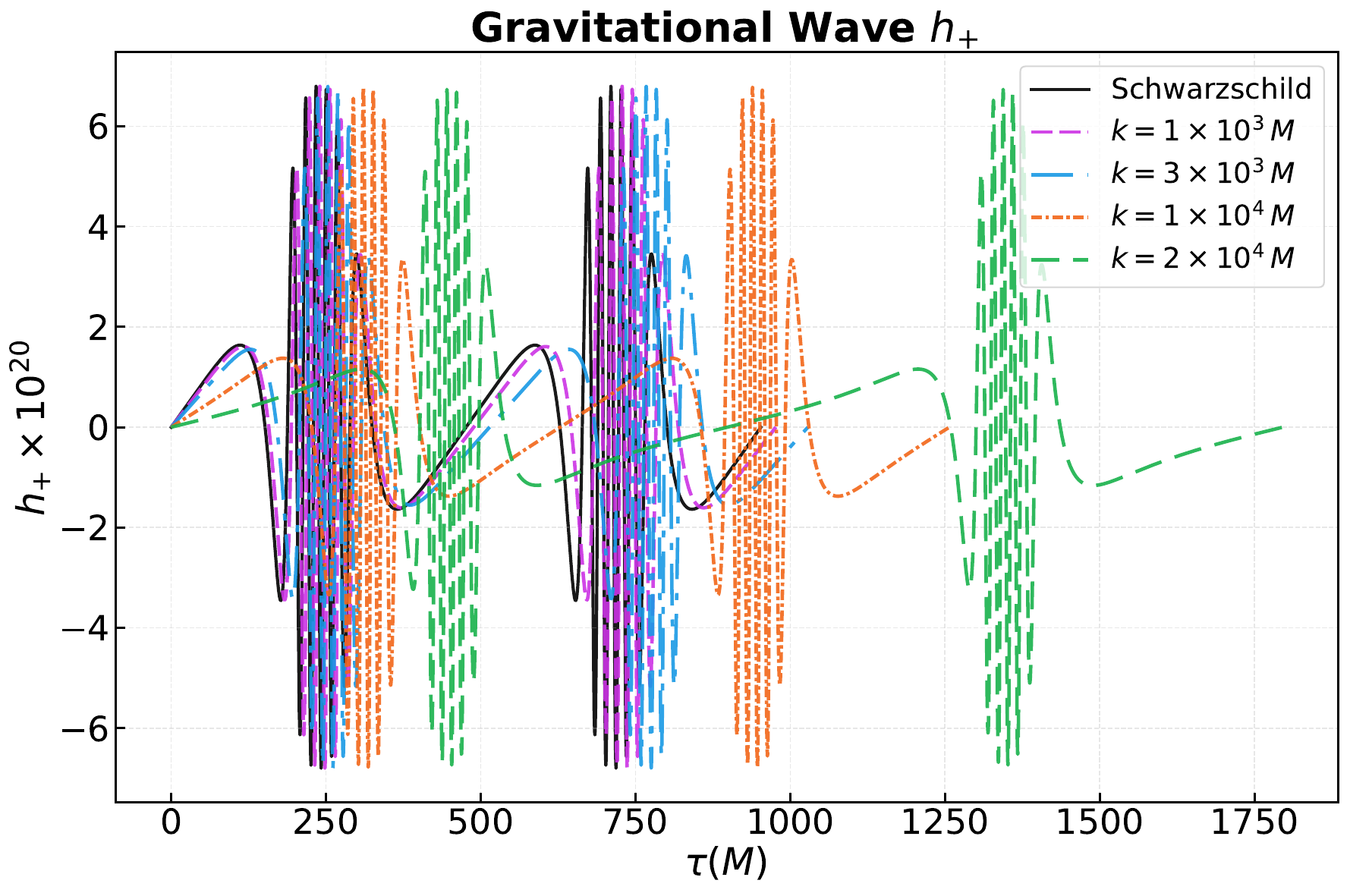}
		\end{subfigure}
		\begin{subfigure}[b]{0.32\textwidth}
			\centering
			\includegraphics[width=\textwidth]{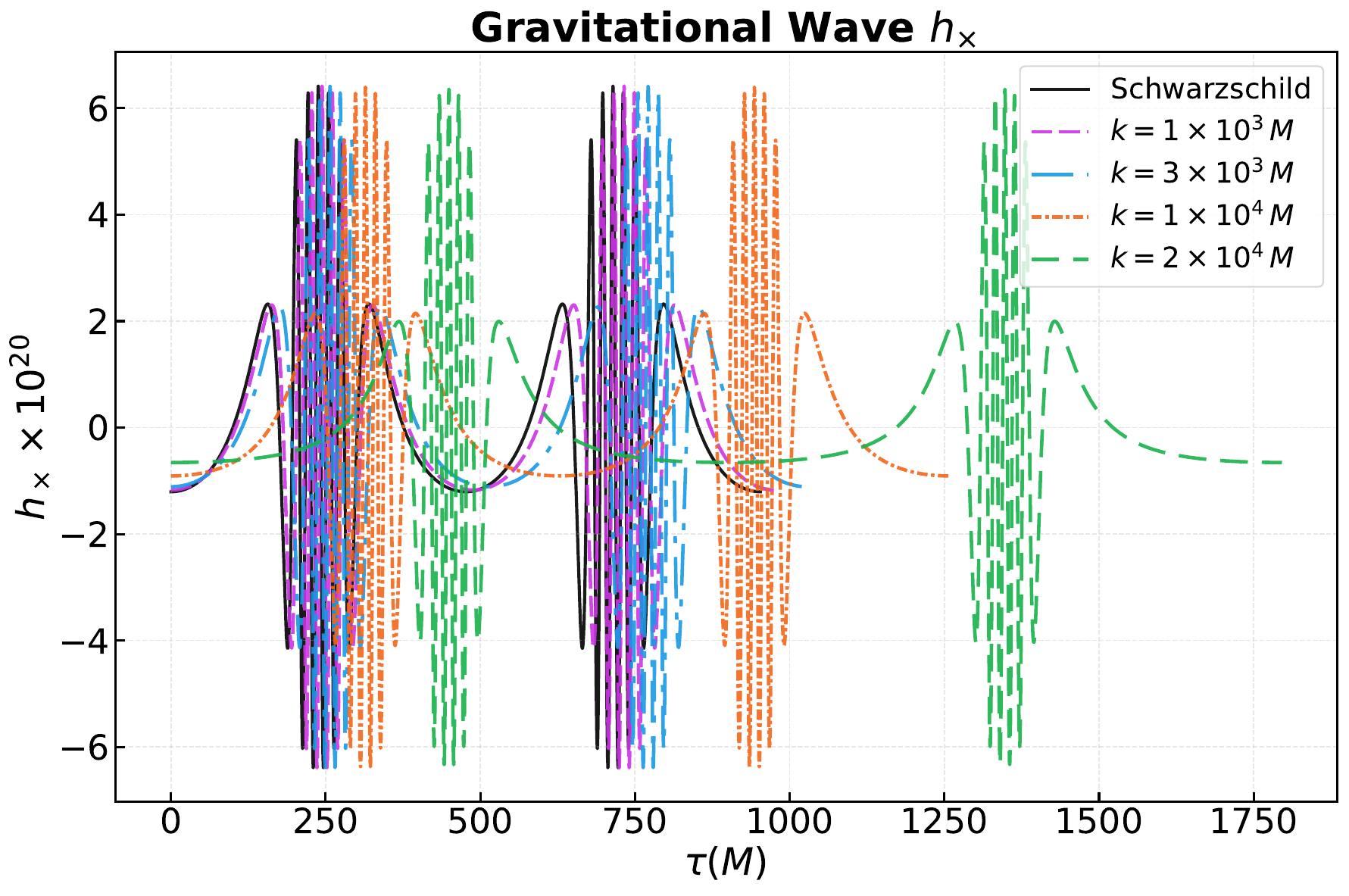}
		\end{subfigure}
		\caption{$(z\ w\ v) ~= (2\ 2\ 1)$}
		\label{fig:gw_221}
	\end{subfigure}
	
	\caption{Comparison of gravitational waveforms under different dark masses in the NFW model for five orbital configurations. The left column displays the $h_+$ polarization component, while the right column shows the $h_\times$ component. Different colors represent varying dark matter masses ranging from $1 \times 10^3 M \sim 2 \times 10^4 M$. This figure displays the gravitational wave signals over one orbital period.}
	\label{GW_k_NFW}
\end{figure}

To further illustrate the mass-dependent effects across different orbital configurations, Fig.~\ref{GW_k_NFW} provides a comprehensive comparison of gravitational waveforms under varying dark matter masses in the NFW model. 
Comparing across different orbital configurations $(z~w~v)$, it is demonstrated that the gravitational waveforms from periodic orbits with larger zoom-whirl-vertex numbers exhibit richer 
substructures. We also observe that more complex orbital configurations---those with larger $(z~w~v)$ ---generate longer-duration gravitational wave signals in a period with correspondingly more oscillations. This relationship follows directly from the orbital geometry: larger configuration values $(z~w~v)$ correspond to more intricate orbital structures and more extended orbital dimensions, as illustrated in Fig.~\ref{dif_k_NFW}. As a result, the $(1~1~0)$ configuration produces relatively simple waveforms with fewer oscillations spanning approximately $700M$ in time scale, while the $(2~2~1)$ configuration generates complex waveforms extending beyond $1750M$ with significantly more oscillations. The waveform duration and the total number of oscillations therefore serve as direct indicators of the orbital complexity. Furthermore, a closer examination reveals a quantitative correspondence between the orbital configurations and specific waveform features. The number $z$ exhibits a direct relationship with the number of low-frequency regions in the gravitational wave signal, which exactly matches the count of these calm regions in the waveform corresponding to the zoom stages. Meanwhile, the whirl number $w$ leads to more rapid oscillations during the whirl stages, manifesting in a steeper frequency evolution~\cite{Alloqulov:2025ucf,Haroon:2025rzx}. This enhancement in oscillation frequency reflects the increasingly vigorous whirl motion in the near-black hole region. The orbital configurations $(z~w~v)$, as fundamental quantities characterizing the geometric structure of periodic motion, thus directly determine the time-domain features of the gravitational wave signals, leading to a dramatic change in orbital dimensions and the time duration of an orbital period. Additionally, combining the results shown in Figs.~\ref{GW_k} and~\ref{GW_k_NFW}, the waveform duration within one orbital period is also greatly influenced by the mass of dark matter. The dark matter mass $k$ influences the waveforms indirectly by modifying the gravitational potential and effective spacetime metric. A larger 
value of $k$ leads to a stronger modification of the potential and more distinct waveform deviations from the Schwarzschild case for all orbital configurations.

\subsection{ The effect of dark matter halo scale on gravitational waves}

\begin{figure}
	\centering  
	\begin{subfigure}[b]{\textwidth}
		\centering
		\begin{subfigure}[b]{0.4\textwidth}
			\centering
			\includegraphics[width=\textwidth]{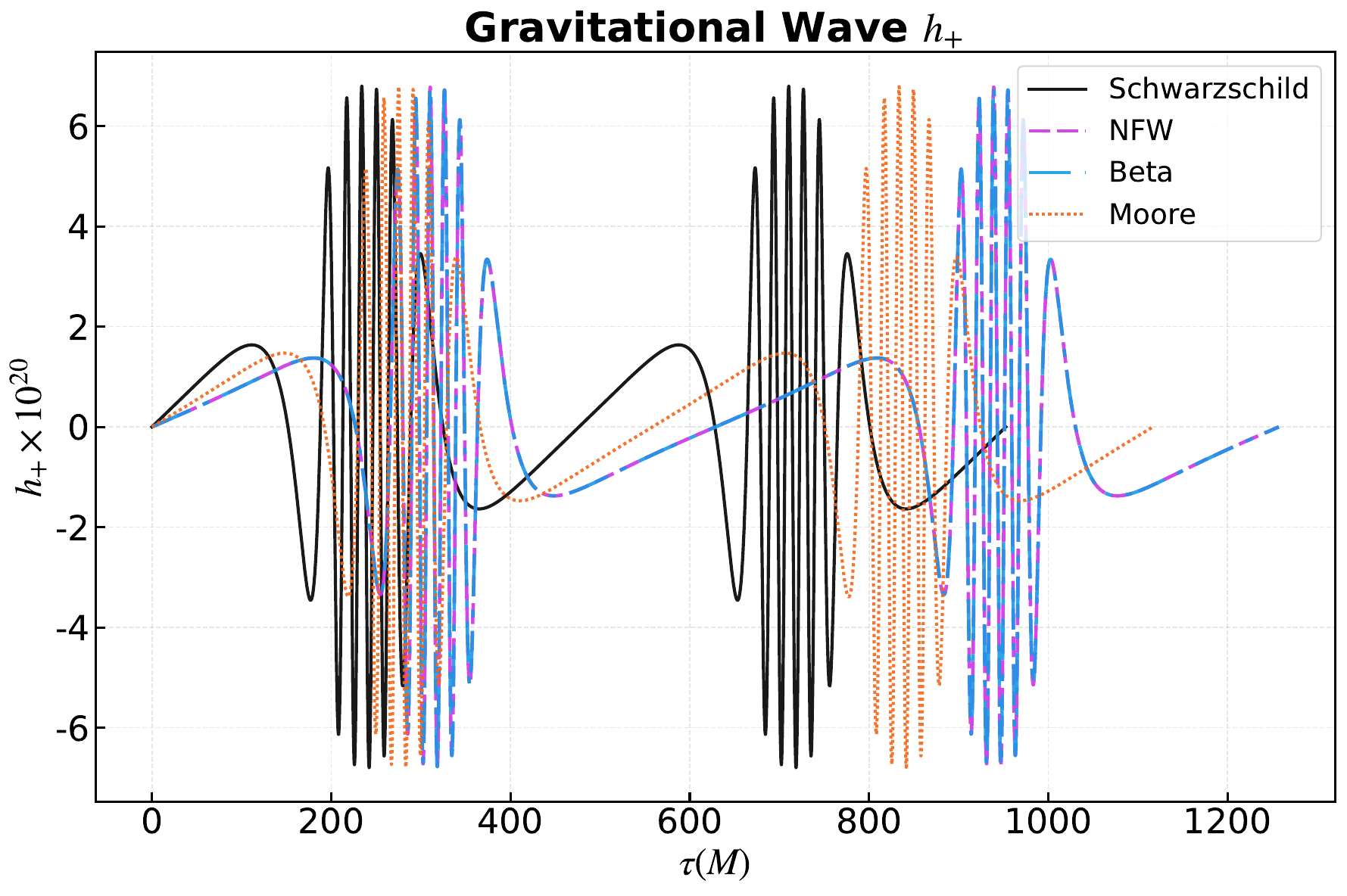}
		\end{subfigure}
		\begin{subfigure}[b]{0.4\textwidth}
			\centering
			\includegraphics[width=\textwidth]{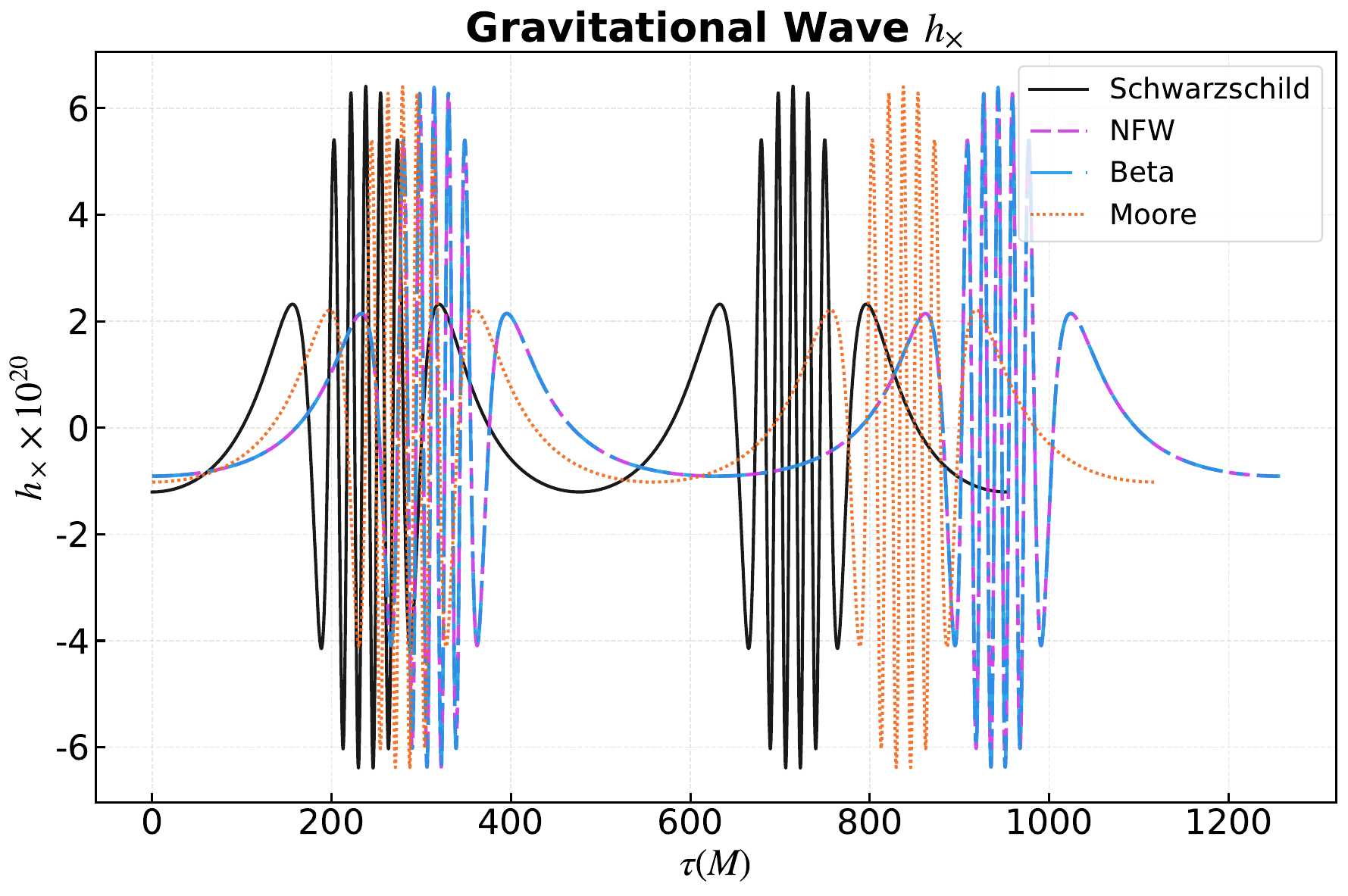}
		\end{subfigure}
		\caption{$h=10^7M$}
	\end{subfigure}
	
	\vspace{0.2cm}
	
	\begin{subfigure}[b]{\textwidth}
		\centering
		\begin{subfigure}[b]{0.4\textwidth}
			\centering
			\includegraphics[width=\textwidth]{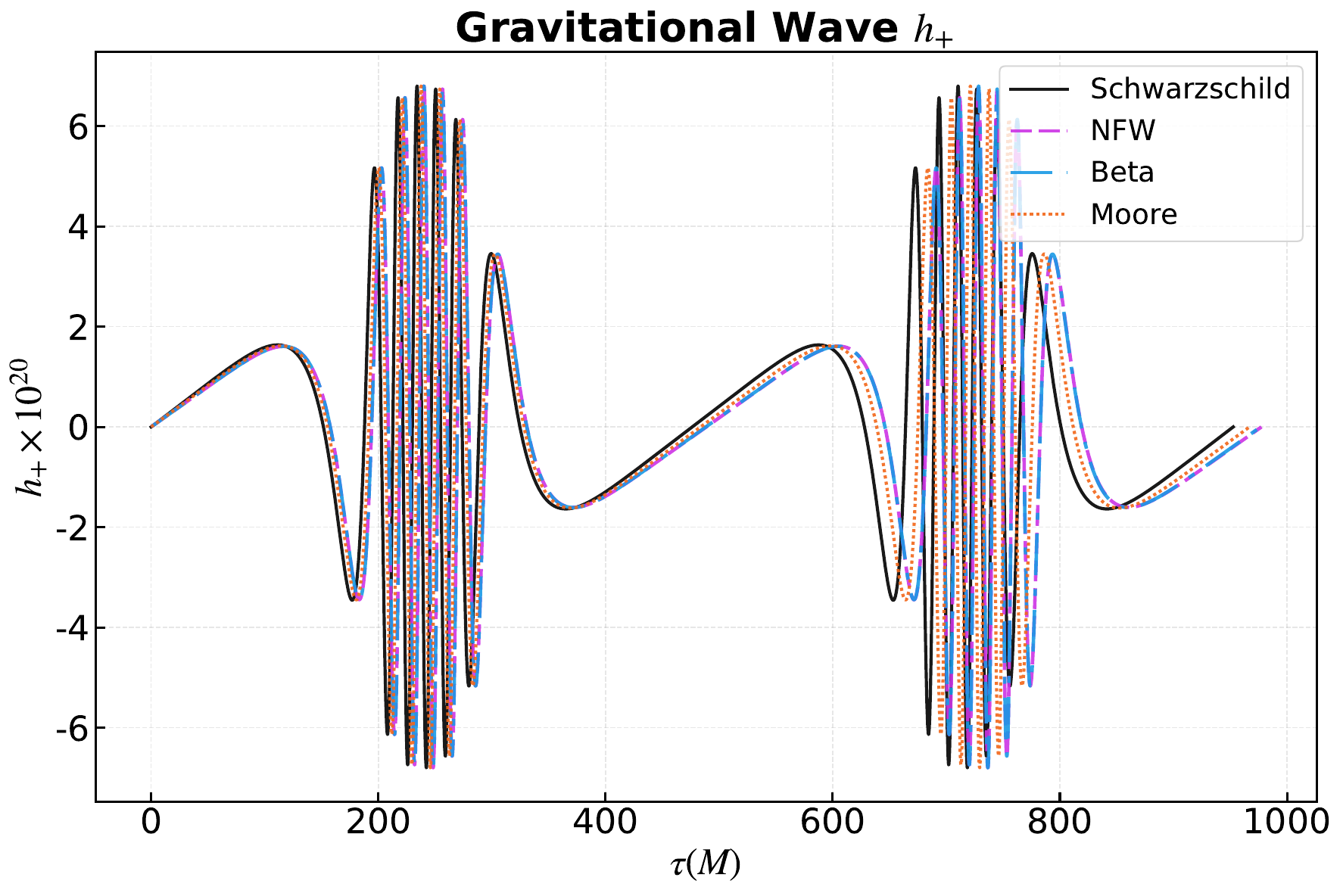}
		\end{subfigure}
		\begin{subfigure}[b]{0.4\textwidth}
			\centering
			\includegraphics[width=\textwidth]{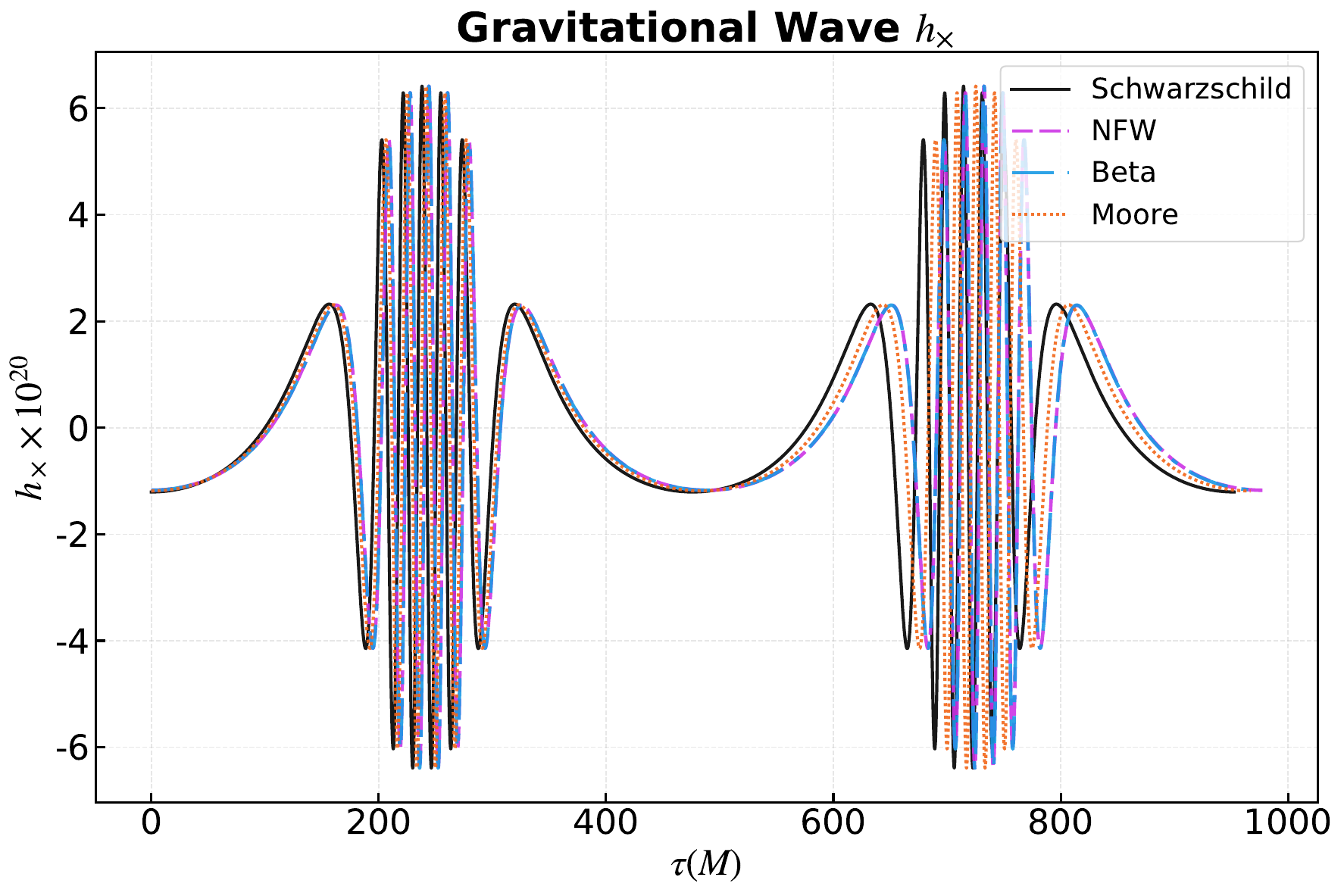}
		\end{subfigure}
		\caption{$h=10^8M$}
	\end{subfigure}
	
	\vspace{0.2cm}
	
	\begin{subfigure}[b]{\textwidth}
		\centering
		\begin{subfigure}[b]{0.4\textwidth}
			\centering
			\includegraphics[width=\textwidth]{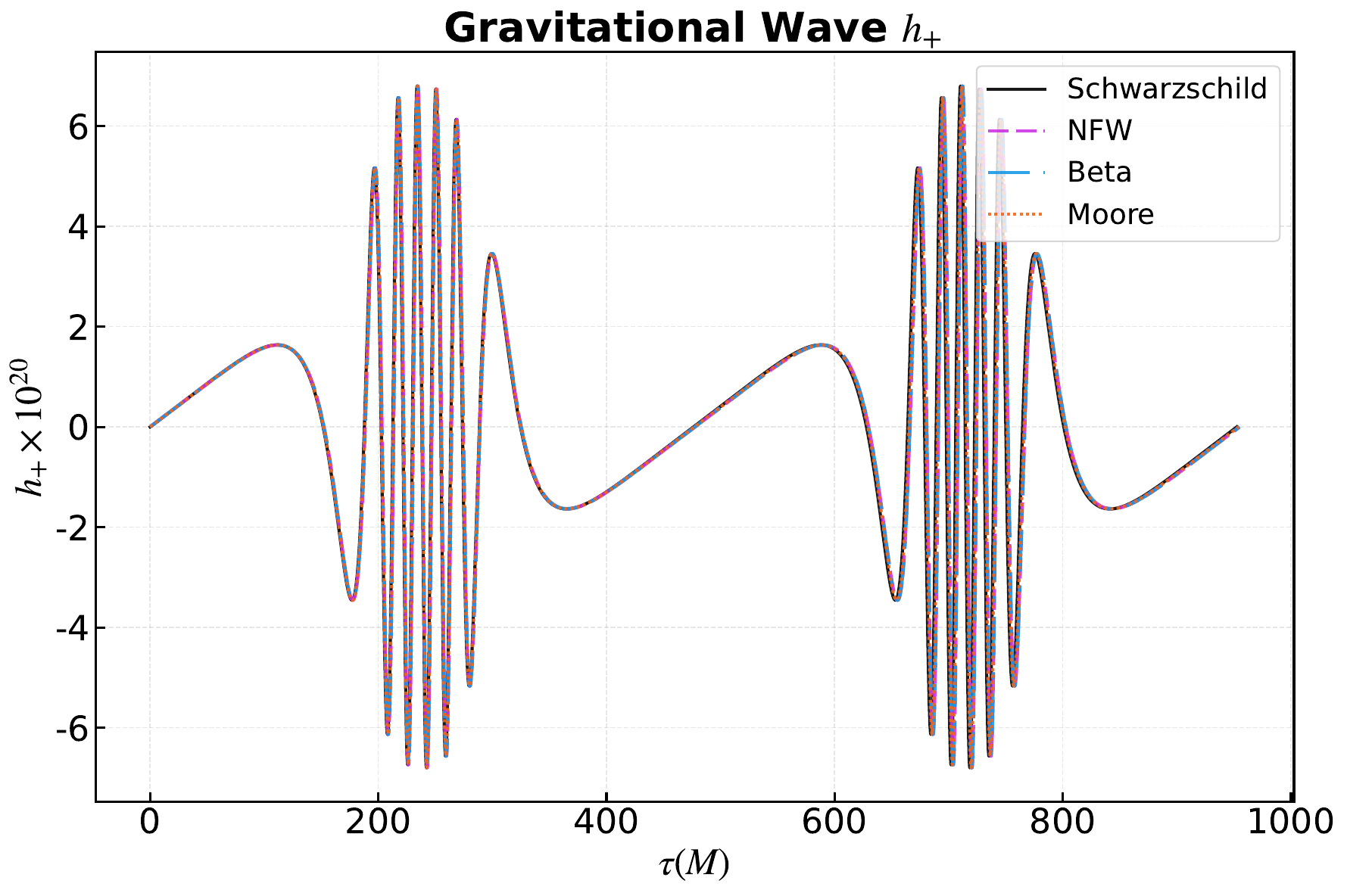}
		\end{subfigure}
		\begin{subfigure}[b]{0.4\textwidth}
			\centering
			\includegraphics[width=\textwidth]{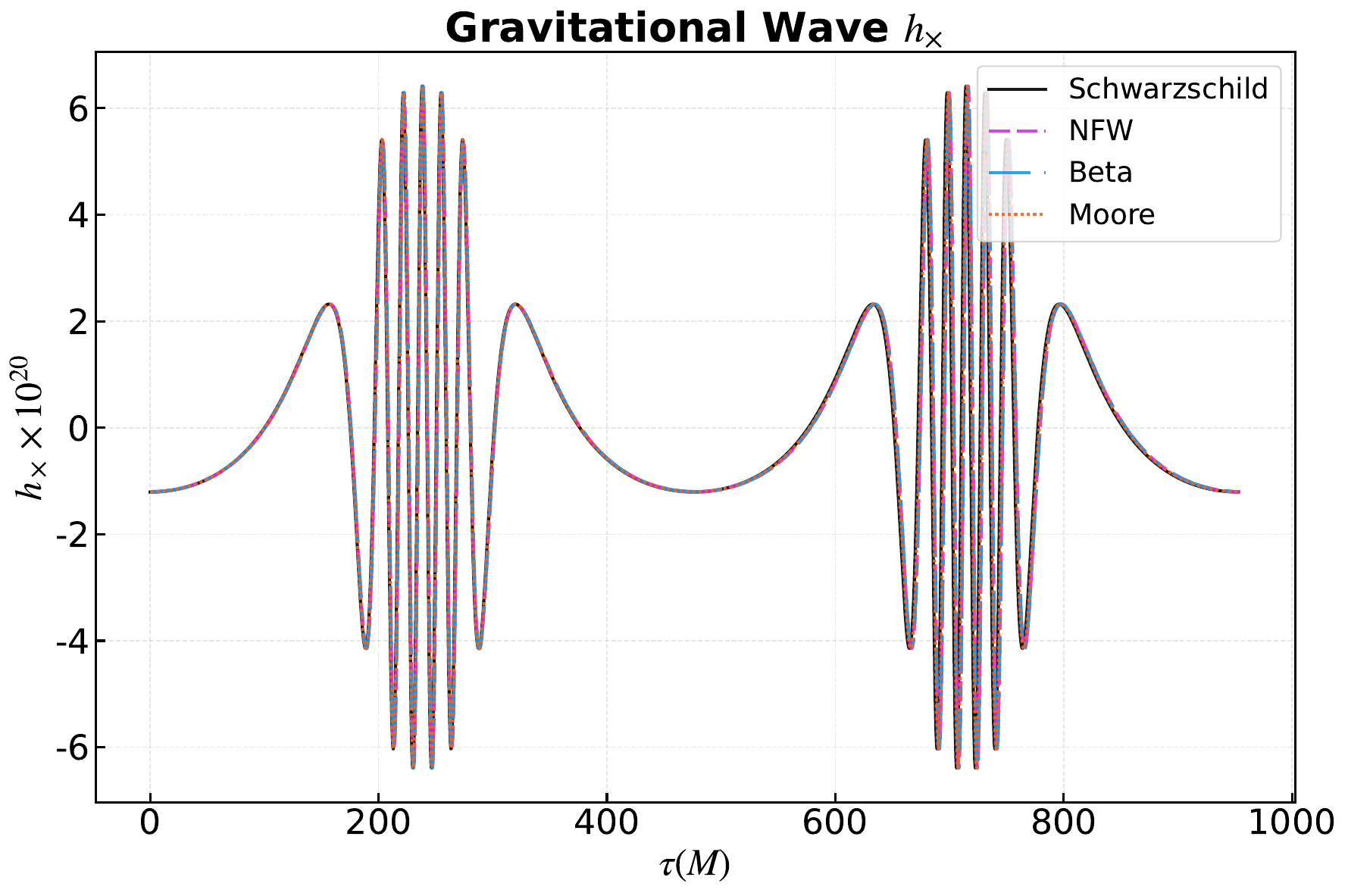}
		\end{subfigure}
		\caption{$h=10^9M$}
	\end{subfigure}
	
	\caption{Gravitational waveforms produced by the $(2~2~1)$ orbital configuration for varying dark matter halo scale in the NFW, Beta, and Moore models. The dark matter mass is maintained constant as $k = 10^4 M$. Each panel shows results for a specific halo characteristic radius: (a) $h = 10^7 M$; (b) $h = 10^8 M$; (c) $h = 10^9 M$. The left column displays the $h_+$ polarization component, while the right column shows the $h_\times$ component. This figure displays the gravitational wave signals over one orbital period.}
	
	\label{GW_h}
\end{figure}

\begin{figure}
	\centering  
	\begin{subfigure}[b]{\textwidth}
		\centering
		\begin{subfigure}[b]{0.32\textwidth}
			\centering
			\includegraphics[width=\textwidth]{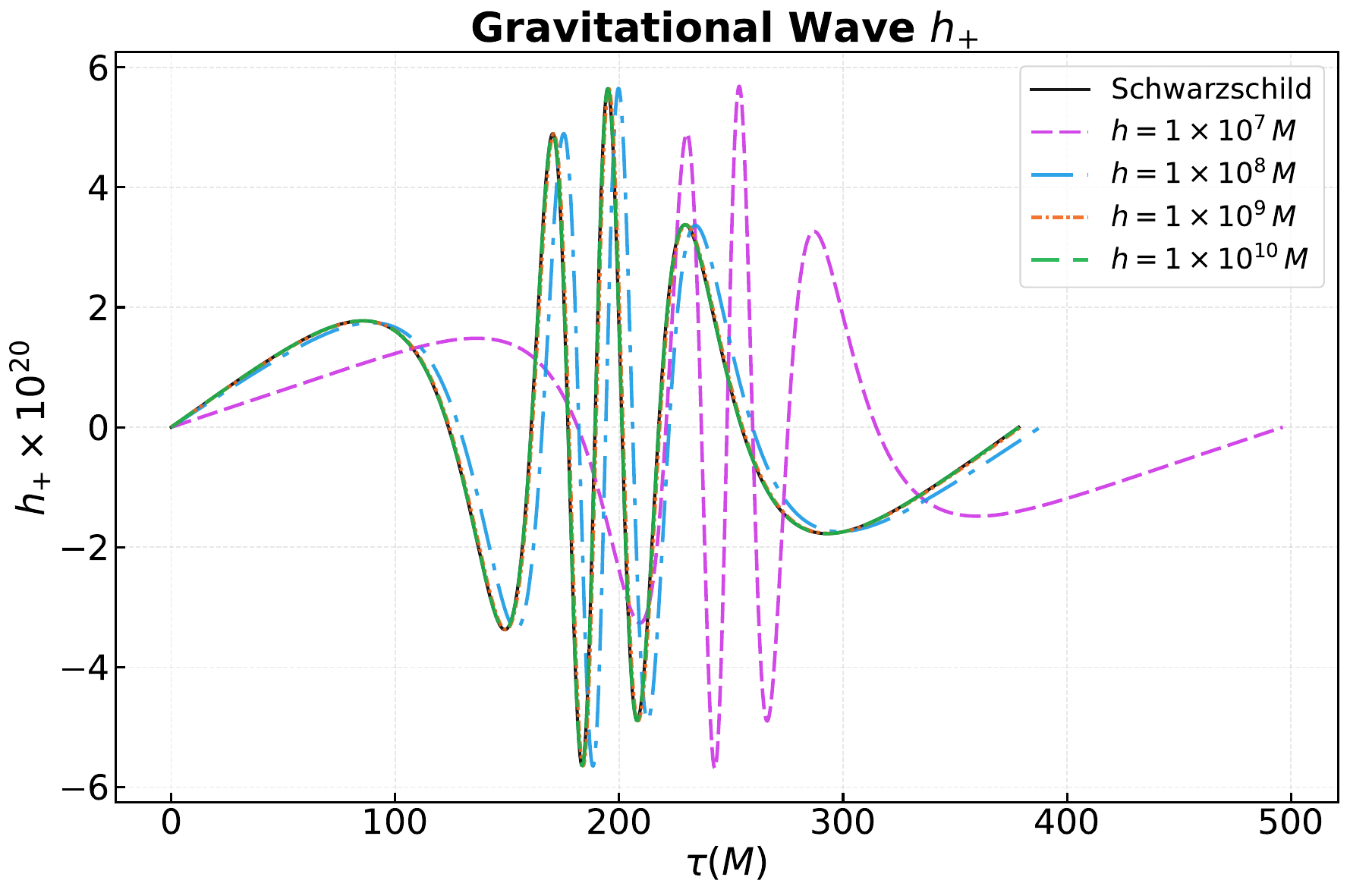}
		\end{subfigure}
		\begin{subfigure}[b]{0.32\textwidth}
			\centering
			\includegraphics[width=\textwidth]{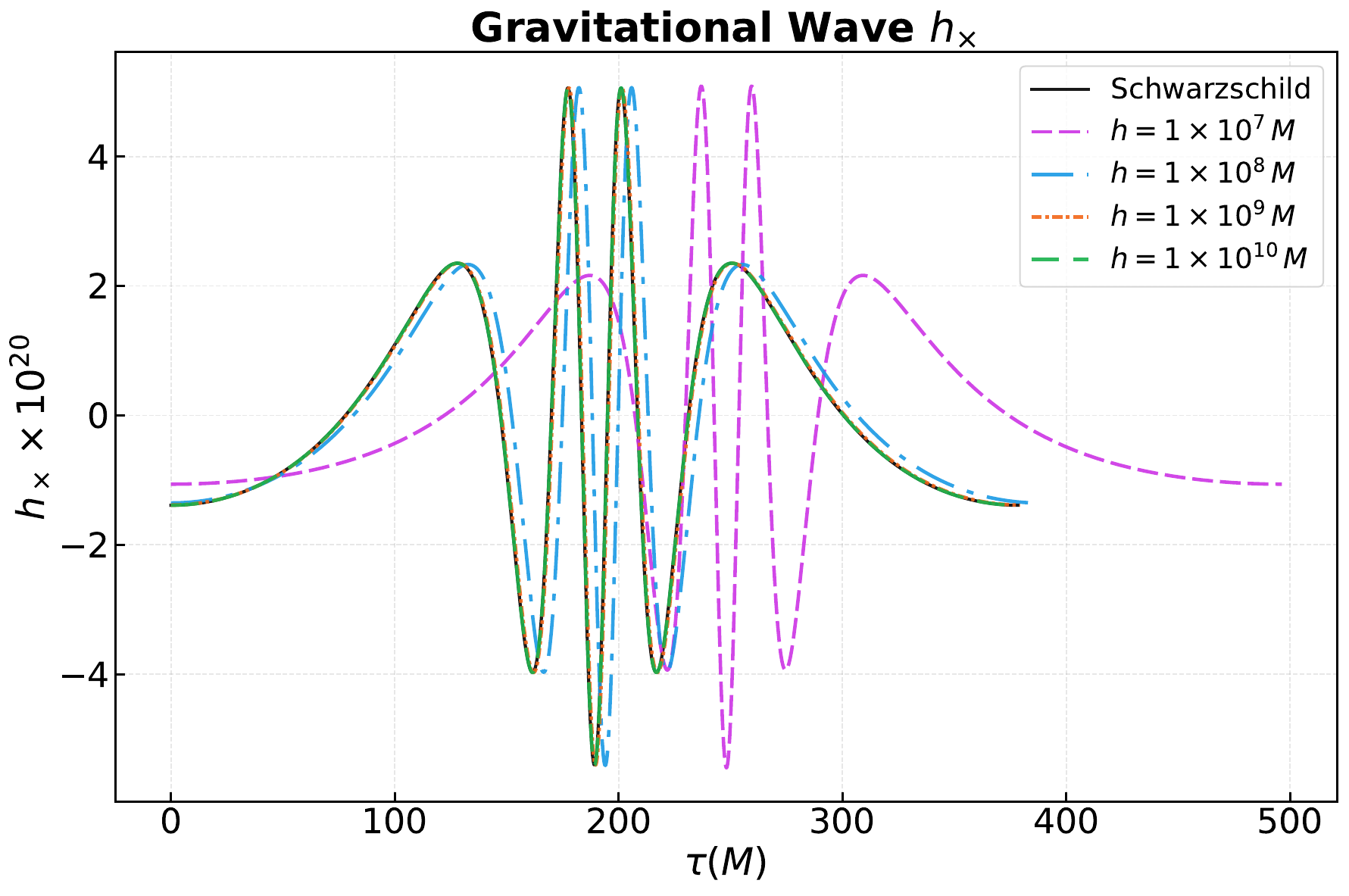}
		\end{subfigure}
		\caption{$(z\ w\ v)~=~(1\ 1\ 0)$}
	\end{subfigure}
	
	\vspace{0.3cm}
	
	\begin{subfigure}[b]{\textwidth}
		\centering
		\begin{subfigure}[b]{0.32\textwidth}
			\centering
			\includegraphics[width=\textwidth]{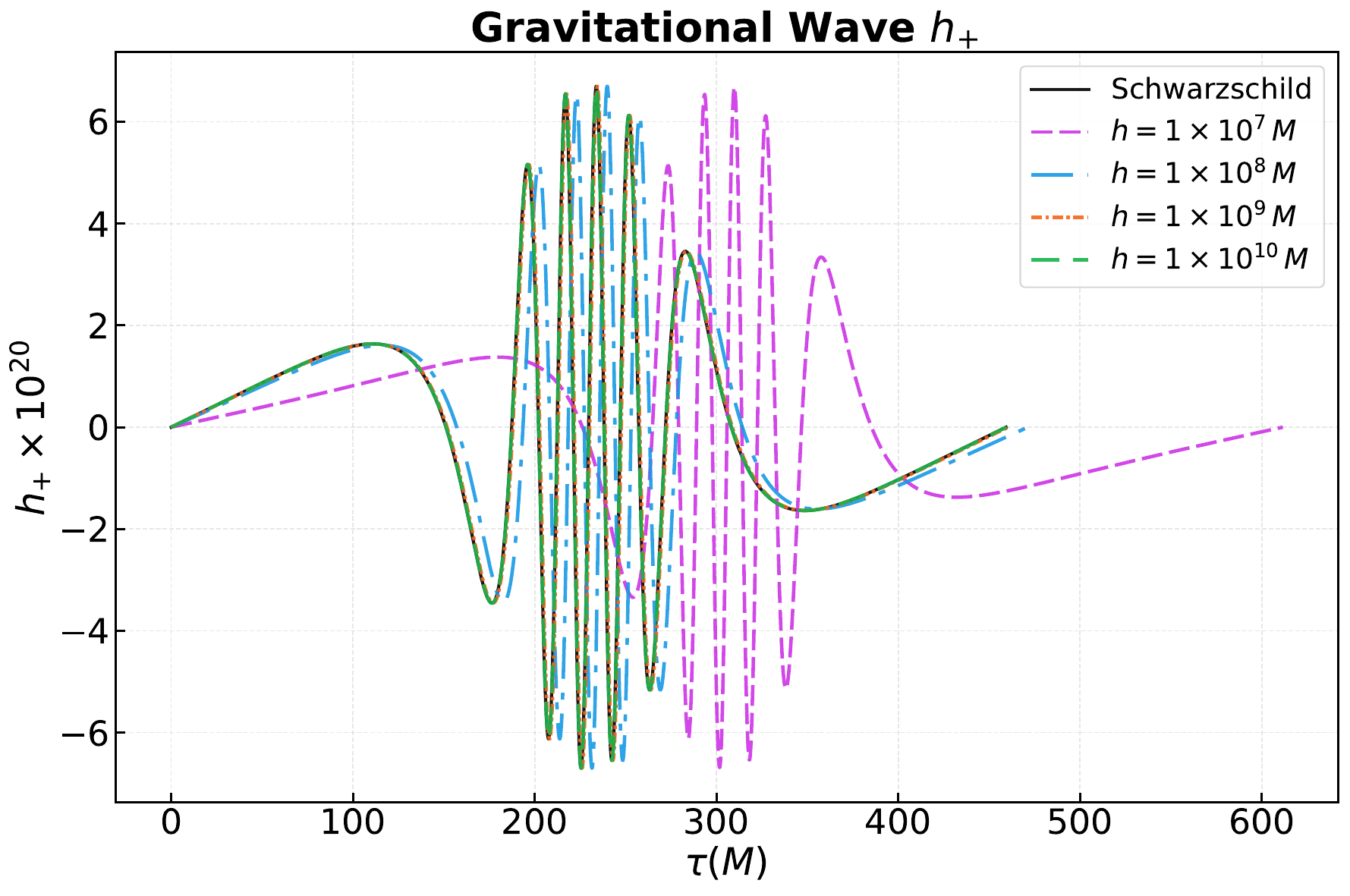}
		\end{subfigure}
		\begin{subfigure}[b]{0.32\textwidth}
			\centering
			\includegraphics[width=\textwidth]{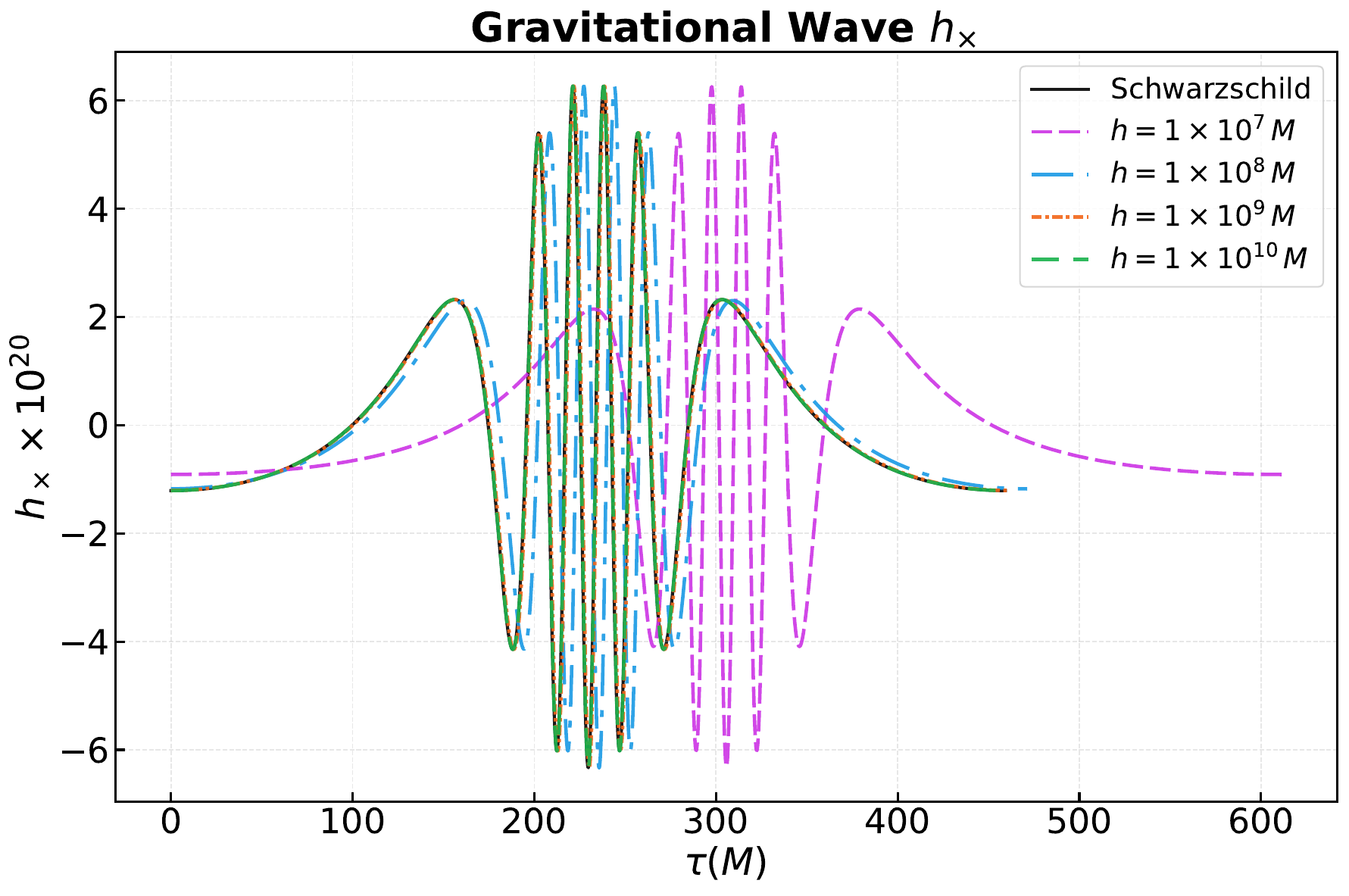}
		\end{subfigure}
		\caption{$(z\ w\ v)~=~(1\ 2\ 0)$}
	\end{subfigure}
	
	\vspace{0.3cm}
	
	\begin{subfigure}[b]{\textwidth}
		\centering
		\begin{subfigure}[b]{0.32\textwidth}
			\centering
			\includegraphics[width=\textwidth]{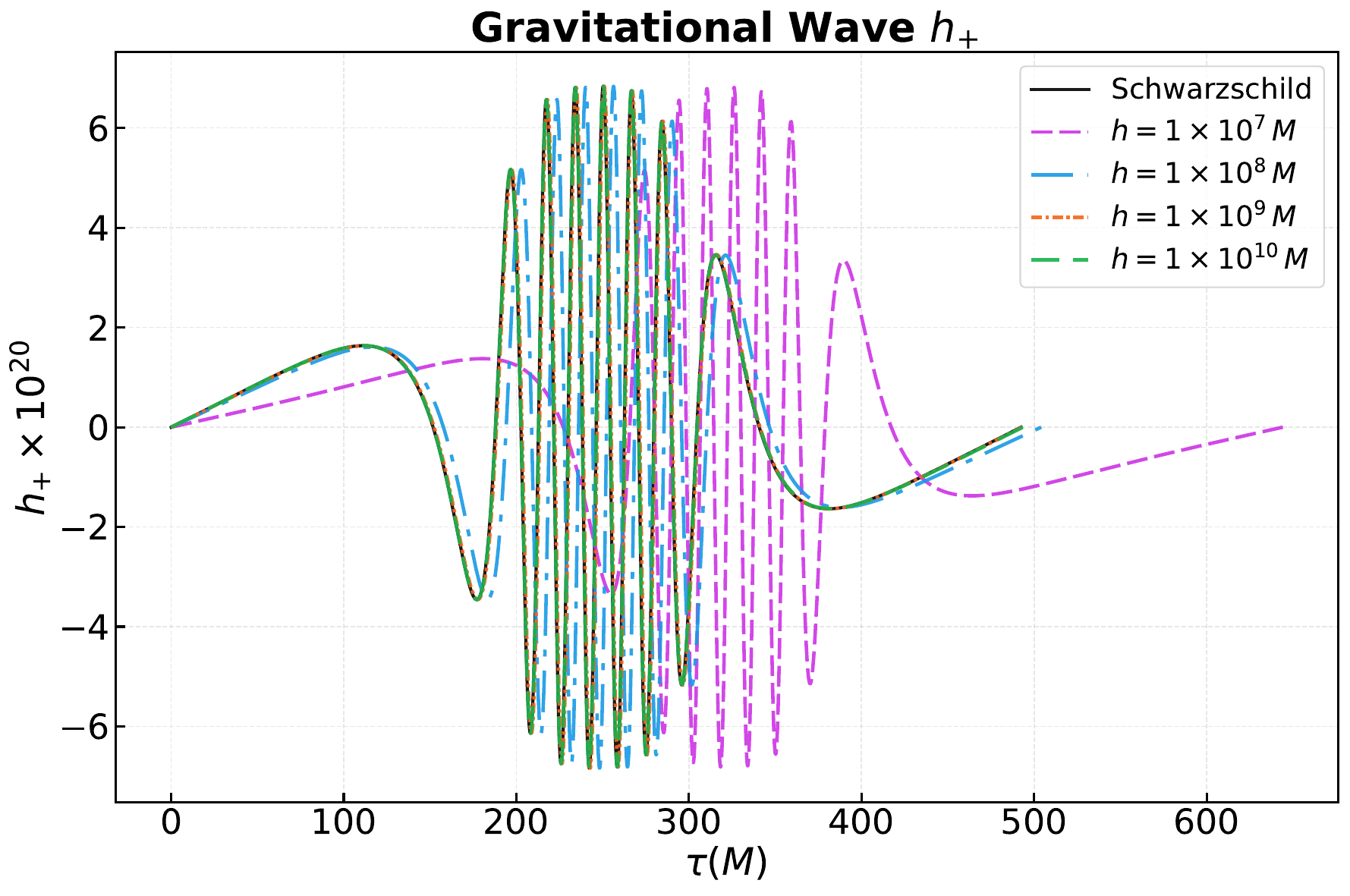}
		\end{subfigure}
		\begin{subfigure}[b]{0.32\textwidth}
			\centering
			\includegraphics[width=\textwidth]{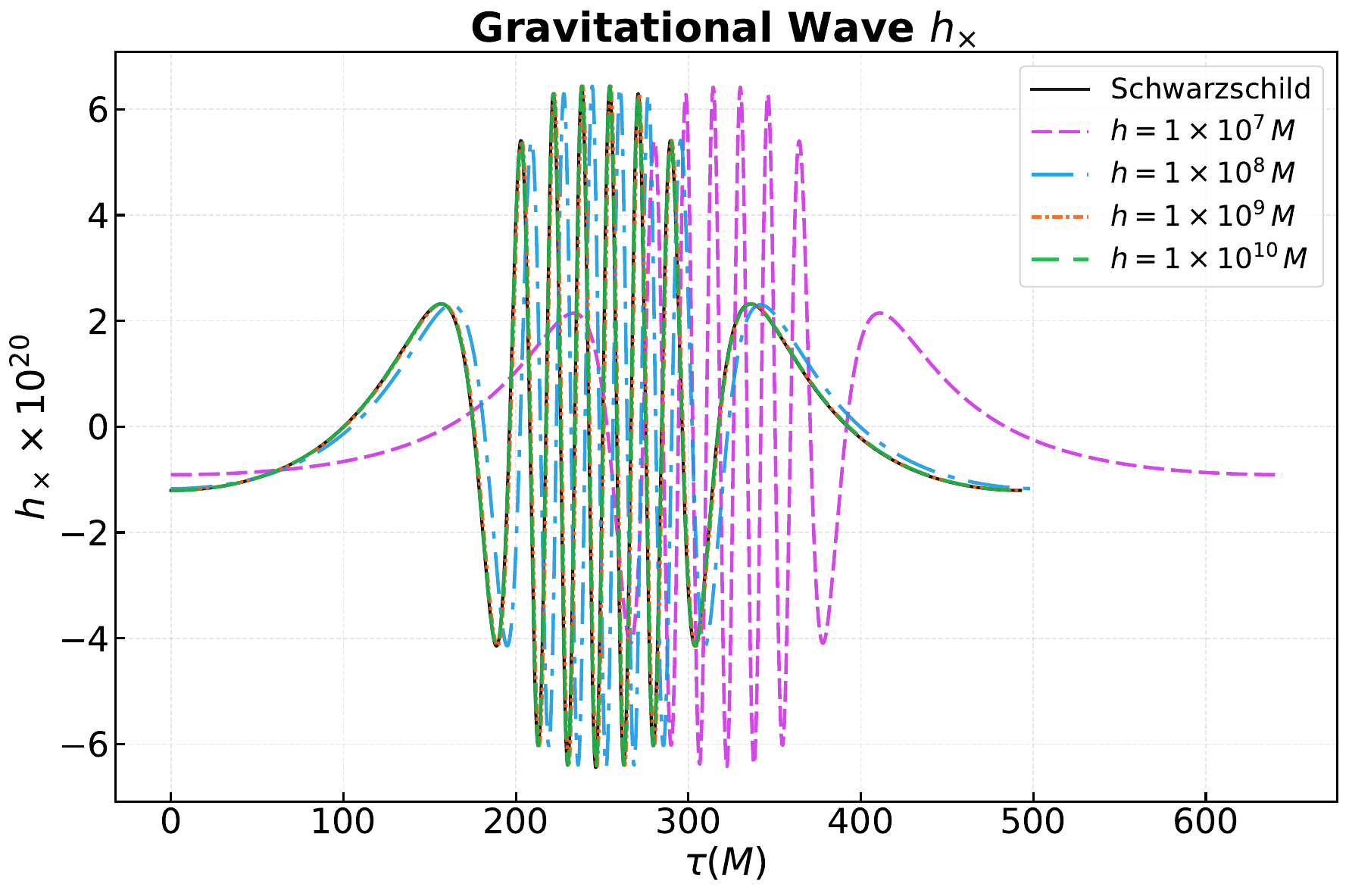}
		\end{subfigure}
		\caption{$(z\ w\ v)~=~(1\ 3\ 0)$}
	\end{subfigure}
	
	\vspace{0.3cm}
	
	\begin{subfigure}[b]{\textwidth}
		\centering
		\begin{subfigure}[b]{0.32\textwidth}
			\centering
			\includegraphics[width=\textwidth]{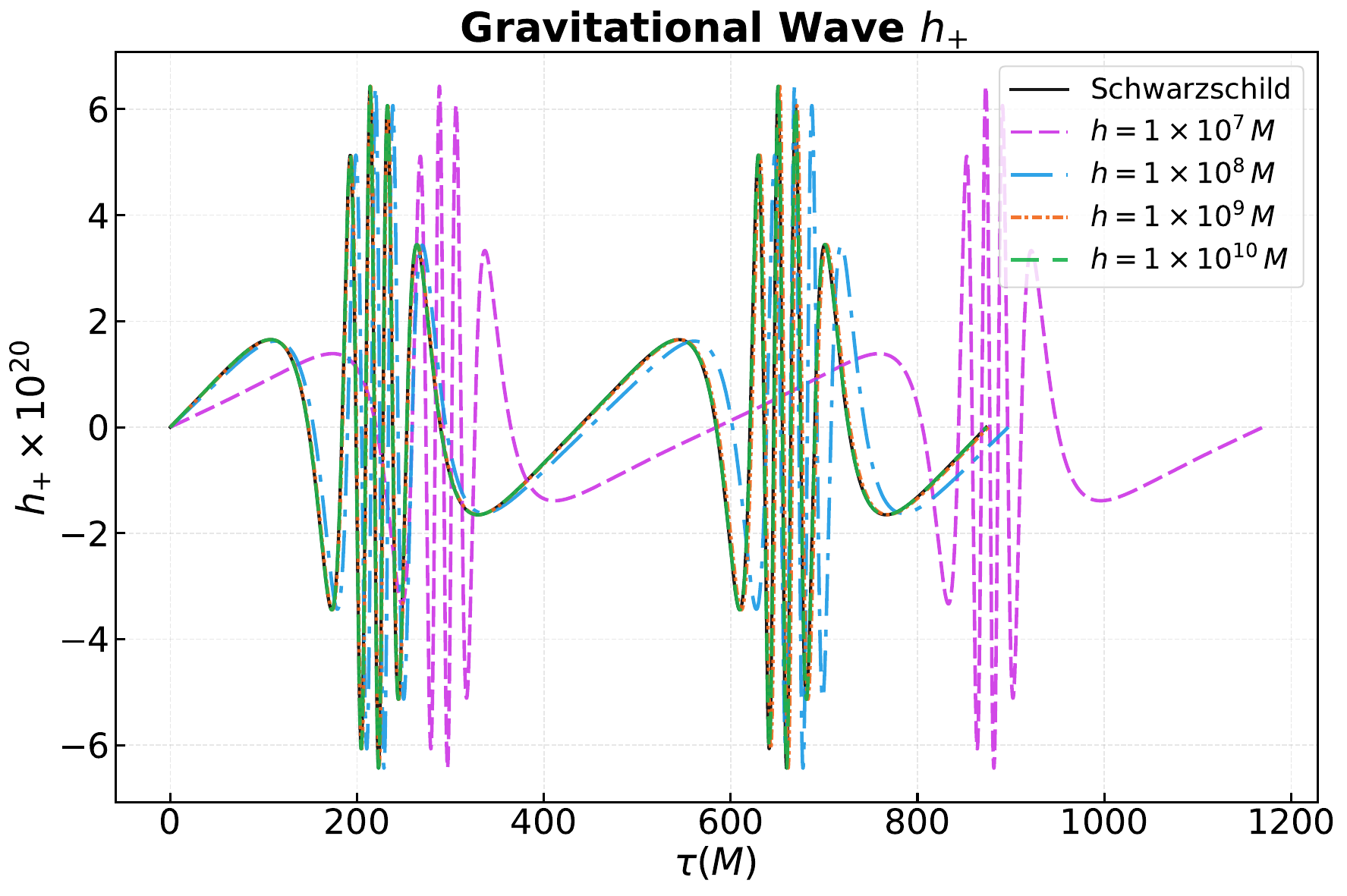}
		\end{subfigure}
		\begin{subfigure}[b]{0.32\textwidth}
			\centering
			\includegraphics[width=\textwidth]{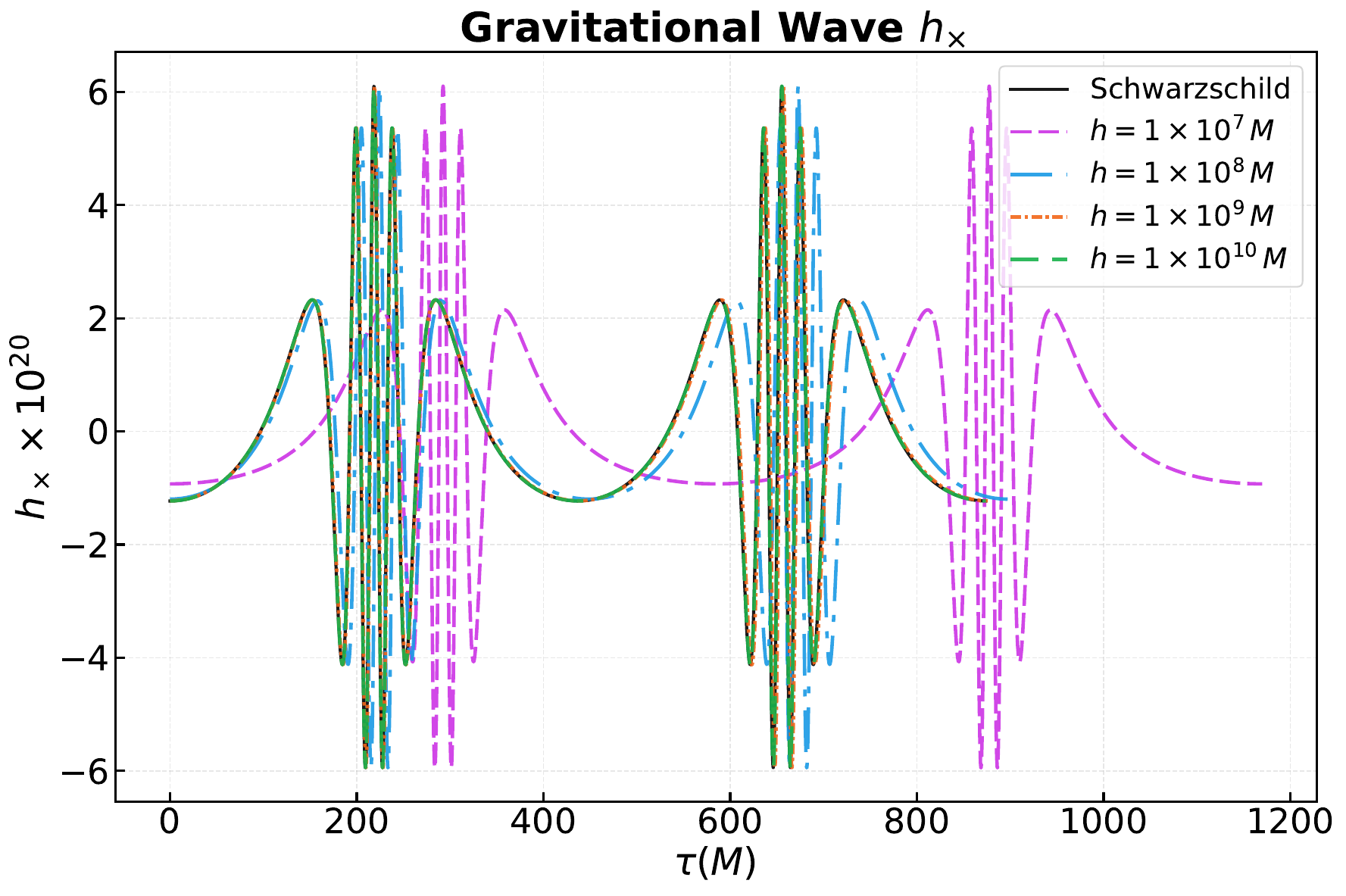}
		\end{subfigure}
		\caption{$(z\ w\ v)~=~(2\ 1\ 1)$}
	\end{subfigure}
	
	\vspace{0.3cm}
	
	\begin{subfigure}[b]{\textwidth}
		\centering
		\begin{subfigure}[b]{0.32\textwidth}
			\centering
			\includegraphics[width=\textwidth]{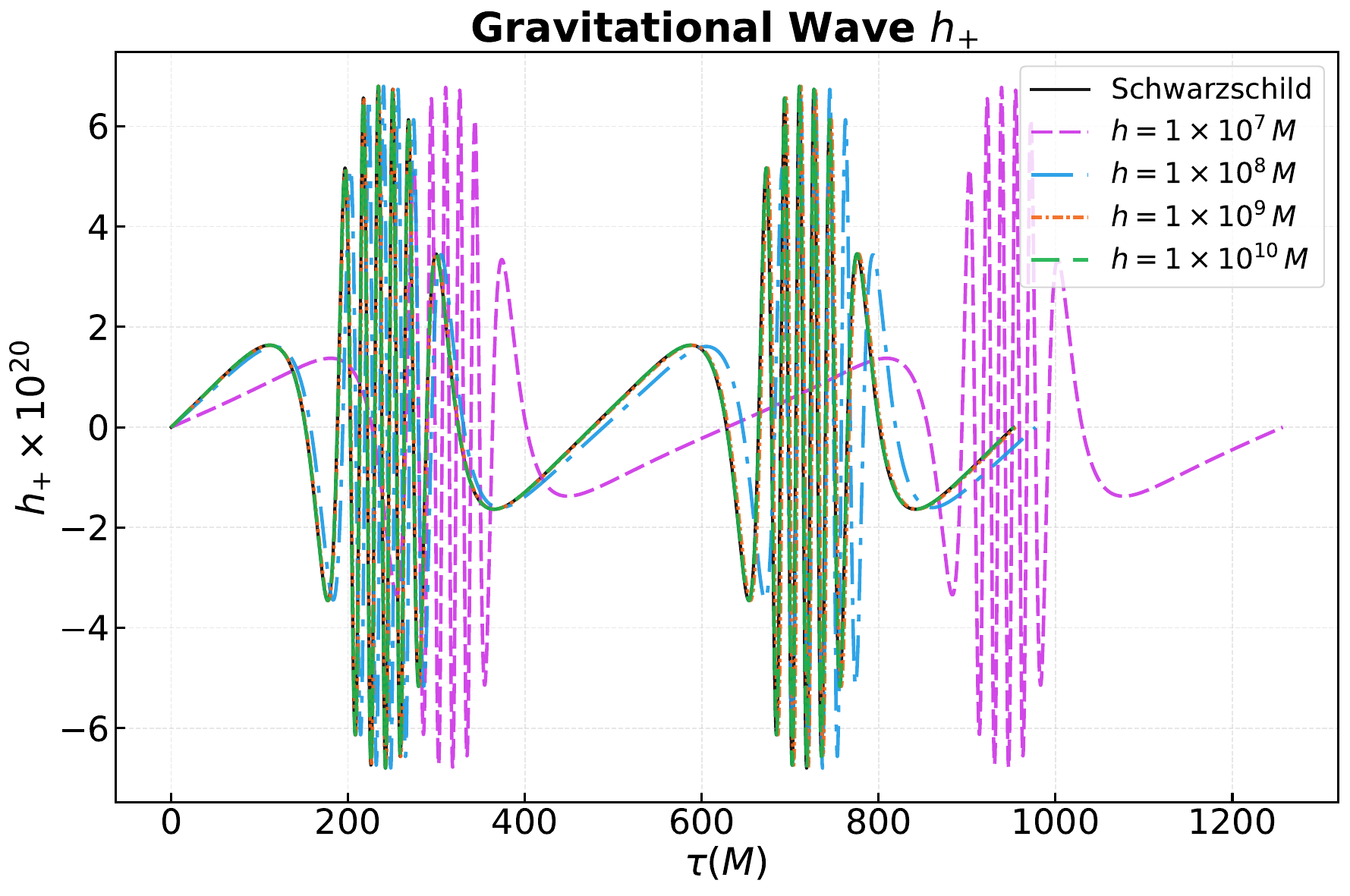}
		\end{subfigure}
		\begin{subfigure}[b]{0.32\textwidth}
			\centering
			\includegraphics[width=\textwidth]{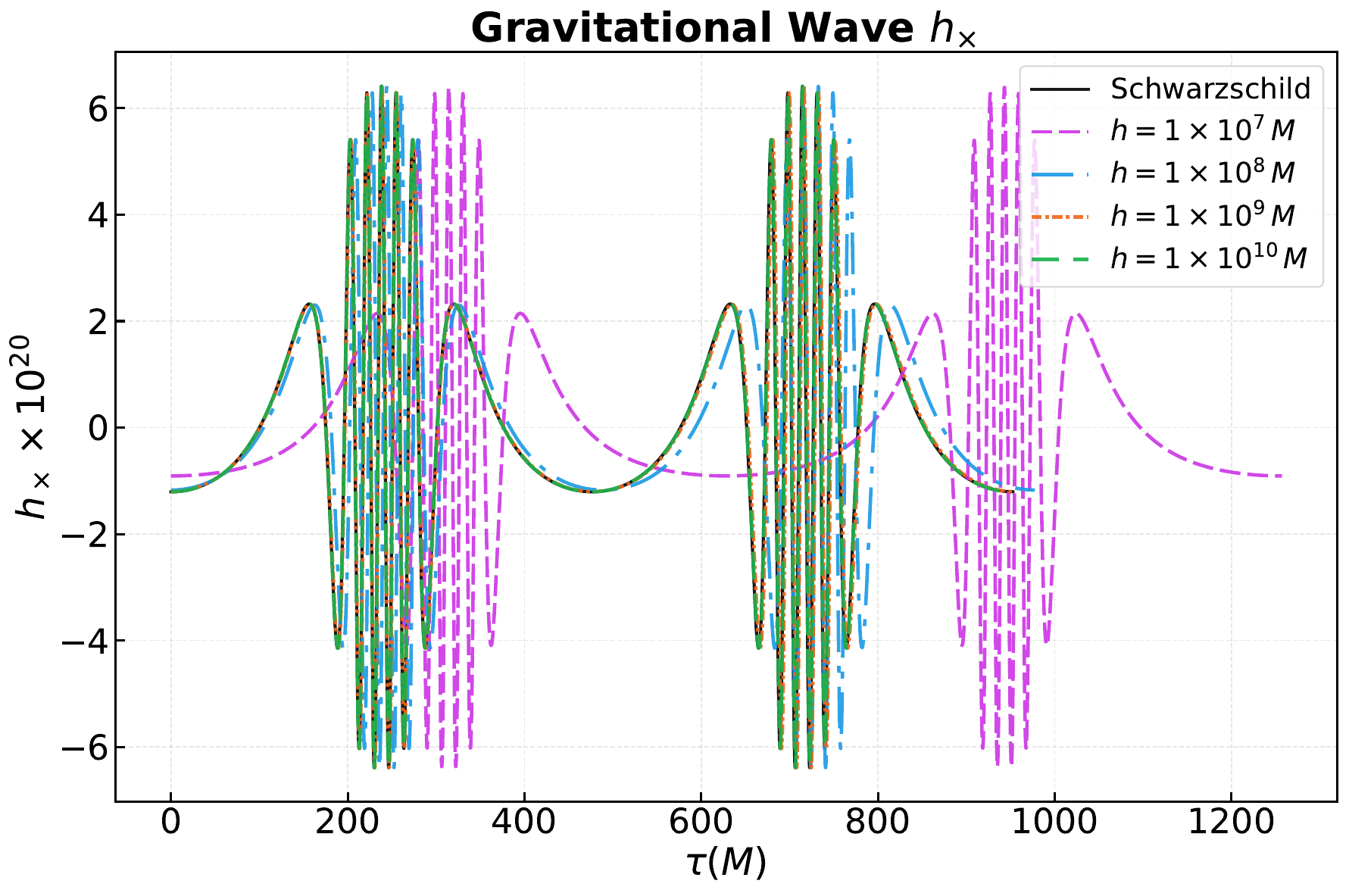}
		\end{subfigure}
		\caption{$(z\ w\ v)~=~(2\ 2\ 1)$}
	\end{subfigure}
	\caption{Gravitational waveform comparison for varying NFW halo characteristic radii for five orbital configurations. The left column displays the $h_+$ polarization component, while the right column shows the $h_\times$ component. The Schwarzschild case (black solid curves) serves as the reference baseline.  This figure displays the gravitational wave signals over one orbital period.}
	\label{GW_h_NFW}
\end{figure}

Following the analysis of dark matter mass effects, we now examine how the dark matter halo scale $h$ influences gravitational wave signals. Fig.~\ref{GW_h} presents the waveforms produced by the $(2~2~1)$ orbital configuration for varying dark matter halo scales in the three models. The dark matter mass is maintained constant as $k = 10^4M$, while the halo characteristic radius $h$ varies from $10^7M \sim 10^{10}M$. The several dark matter models produce distinguishable features in gravitational waveforms for a relatively small dark matter halo scale, and the results reveal a convergence pattern as the halo characteristic radius increases. When $h = 10^7M$, the gravitational waveforms obtained in three dark matter models exhibit noticeable deviations from the Schwarzschild reference baseline, reflecting the significant influence of the dark matter halo on the spacetime geometry and particle orbits. These deviations are evident in waveform characteristics: the signal duration extends to approximately $1200M$ in an orbital time period (compared to roughly $950M$ for Schwarzschild results), and the the oscillations exhibit noticeably distinct characteristics. As the halo characteristic radius expands to $h = 10^8M$, these deviations diminish substantially, with the waveforms moving closer to the Schwarzschild case. The signal duration contracts toward the Schwarzschild results, and the oscillation behaviors become more similar. This convergence process continues as $h$ increases to $h= 10^9M$, at which point the waveforms of all dark matter models become nearly indistinguishable from the Schwarzschild reference baseline in terms of signal duration, oscillation behaviors and amplitude structure.  When \(h > 10^{9} M\), the periodic orbits of the dark matter halo models completely coincide with those of the Schwarzschild black hole, and the corresponding gravitational waveforms are identical as well. For this reason, the results for such cases are not presented separately. This convergence behavior directly parallels the halo effects observed in the precession angle analysis (Figs.~\ref{dif_h_q_models} and~\ref{dif_h_q}) and orbital trajectories (Fig.~\ref{dif_h}). The physical interpretation remains consistent: when the dark matter halo characteristic radius becomes sufficiently large while maintaining dark matter total mass as constant, the dark matter halo is diluted and the local gravitational field gradually approaches that of an isolated Schwarzschild black hole. Thereby reducing the distinguishable signatures of different dark matter density profiles in the gravitational waves.

To provide a more comprehensive view of dark matter halo's scale effects across different orbital configurations, Fig.~\ref{GW_h_NFW} presents gravitational waveform comparisons for varying NFW halo scales using five orbital configurations: $(1~1~0)$, $(1~2~0)$, $(1~3~0)$, $(2~1~1)$, and $(2~2~1)$. This figure examines gravitational waveforms for four dark matter scale values: $h = 10^7M$, $h = 10^8M$, $h =10^9M$ and $h=10^{10} M$, with the Schwarzschild case serving as the reference baseline. 
These results not only reveal how different orbital configurations \((z~w~v)\) modulate the gravitational waveforms (consistent with findings in previous sections), but also indicate the clear trend: larger dark matter halo scales lead to weaker effects on the gravitational potential, with waveforms approaching the Schwarzschild black hole case.

\section{Detectability of Periodic Orbits in Space-Based Gravitational Wave Detections}\label{s5}

To further evaluate the observational prospects of periodic orbits in extreme mass ratio inspirals (EMRIs) in various dark matter environments, we investigate the gravitational wave frequency spectra that emitted from these periodic orbits. Recently, a number of studies have suggested that frequency spectra play indispensable roles in the observational searching of periodic orbits \cite{Yang:2024lmj,Ahmed:2025azu,Xamidov:2026kqs,Agrawal:2026rwu}. These frequency spectra, when compared with the detector sensitivity thresholds, enable an assessment of whether these orbits can be effectively detected by future space-based gravitational wave observatories.

Unlike typical inspiraling trajectories that produce continuous chirping signals, the strict time periodicity of these periodic orbits leads to a discrete gravitational wave spectrum, with emission solely at isolated frequencies. Now we briefly introduce how to numerically obtain gravitational wave spectra at discrete frequencies. In the numerical calculation, gravitational wave strains emitted from periodic orbits are calculated at discrete time series, i.e. $h_{+,\times}(t_n)$ with $n=0,1,2, ..., N-1$ (where $N$ is the total number of sample points used in the numerical simulations). Accordingly, the frequency-domain spectrum can be computed via the discrete Fourier transform ~\cite{Agrawal:2026rwu}:
\begin{eqnarray}
	&                 &
	\tilde{h}(f) = \int_{-\infty}^{\infty} h(t) e^{-2\pi i f t} dt \nonumber 
	\\
	& \Rightarrow & 
	\tilde{h}_{+,\times}(f_l) 
	= \sum_{n=0}^{N-1} h_{+,\times}(t_n) e^{-2\pi i f_l t_n} 
	\cdot \Delta t
\end{eqnarray}
where $f_l = \frac{l}{N \Delta t}$ denotes the discrete frequency bins, and $\Delta t = t_{n+1} - t_n$ is the time interval between two sample points. By analyzing the relative amplitude distribution and structures in the frequency spectra obtained using different halo models, we can identify the spectral features that encode the properties of the surrounding dark matter halo. Notably, the gravitational signal caused by a periodic orbit is a real-valued time series. Therefore, its Fourier transform satisfies the Hermitian symmetry $\tilde{h}(-f)=\tilde{h}(f)^{*}$, and we only need to analyze the positive frequency components $f_l$ with $l=0, 1, 2, ..., N/2$ (assuming $N$ is even).

\begin{figure}
	\centering
	\begin{subfigure}{0.8\textwidth}
		\centering
		\includegraphics[width=\textwidth]{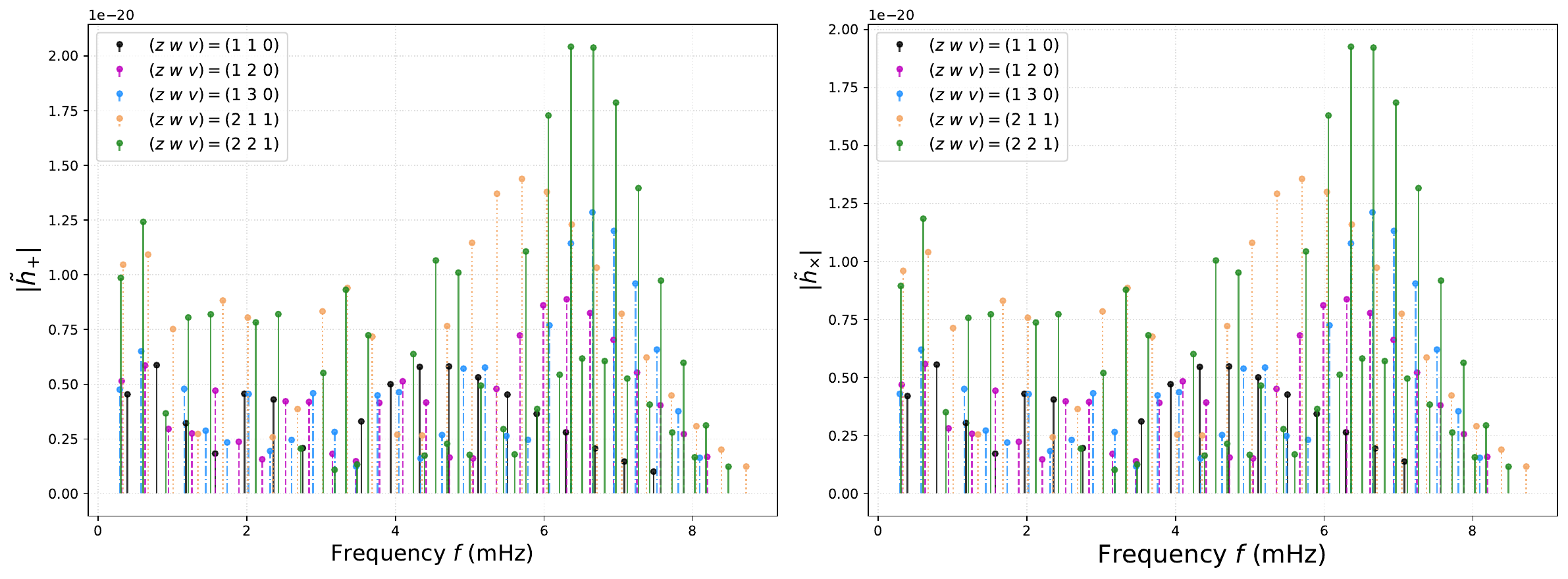}
		\caption{$k = 10^4 M$}
		\label{fig:moore_spectra_a}
	\end{subfigure}
	
	\vspace{0.5cm} 
	
	\begin{subfigure}{0.8\textwidth}
		\centering
		\includegraphics[width=\textwidth]{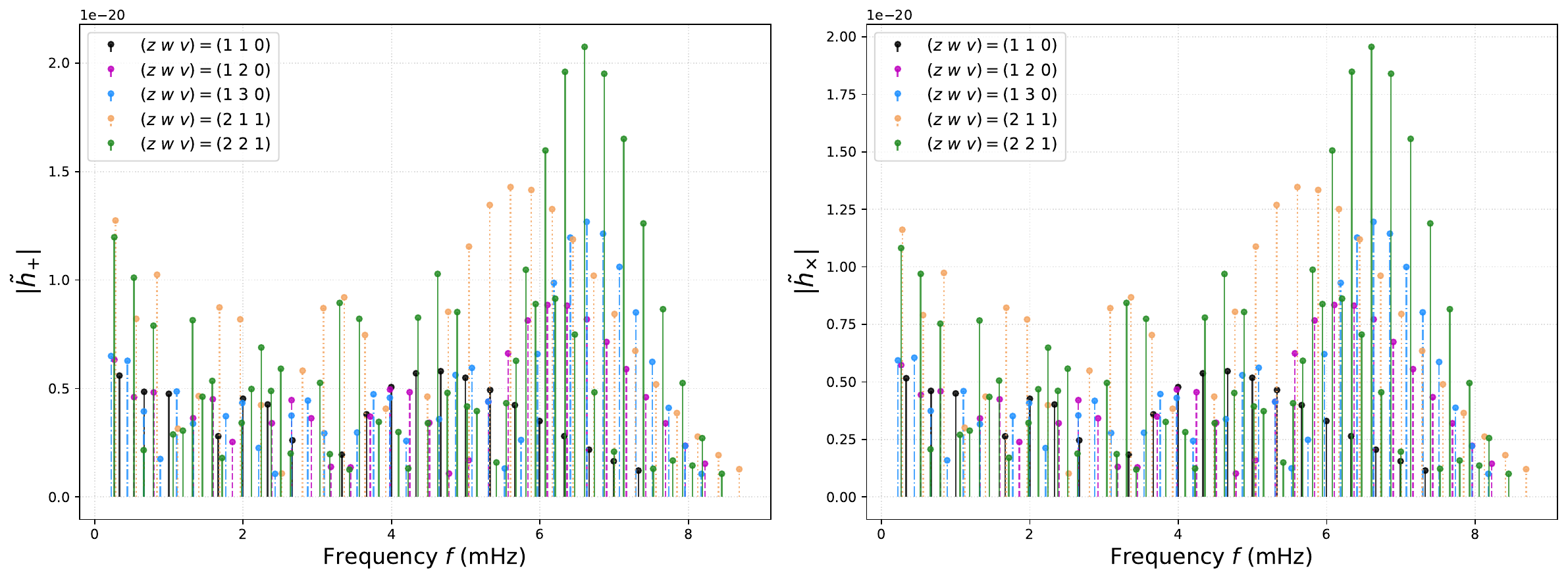}
		\caption{$k = 2 \times 10^4 M$}
		\label{fig:moore_spectra_b}
	\end{subfigure}
	\caption{The frequency spectra of five distinct periodic orbital configurations (1 1 0), (1 2 0), (1 3 0), (2 1 1), (2 2 1) calculated using the Moore dark matter profile. The upper panel  corresponds to a halo parameter $k = 10^4 M$, while the lower panel corresponds to $k = 2 \times 10^4 M$. In this figure, we have assumed a central black hole mass $M = 10^6 M_\odot$, a small compact object's mass $m = 10 M_\odot$, a halo scale parameter $h = 10^7 M$, and a luminosity distance $D_L = 200\,\mathrm{Mpc}$.}
	\label{fig:moore_spectra}
\end{figure}

\begin{figure}
	\centering
	\begin{subfigure}{0.8\textwidth}
		\centering
		\includegraphics[width=\textwidth]{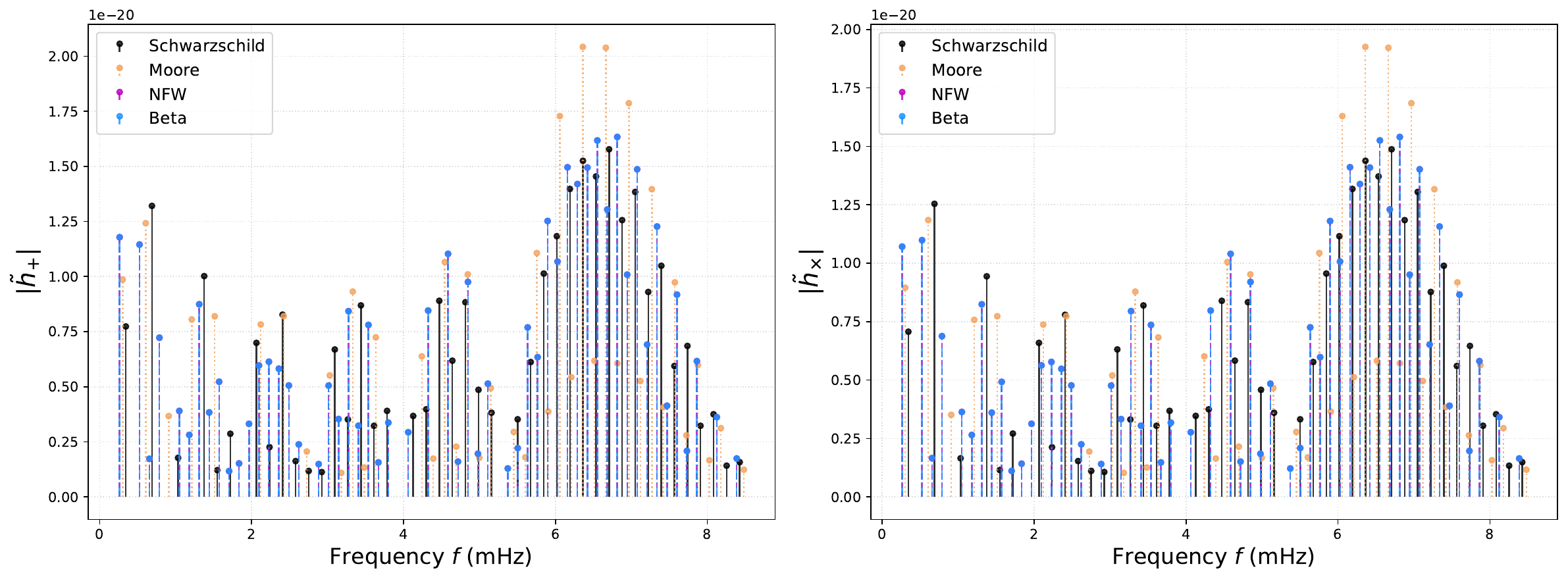}
		\caption{$k = 10^4 M$}
		\label{fig:orbit_comparison_a}
	\end{subfigure}
	
	\vspace{0.5cm} 
	
	\begin{subfigure}{0.8\textwidth}
		\centering
		\includegraphics[width=\textwidth]{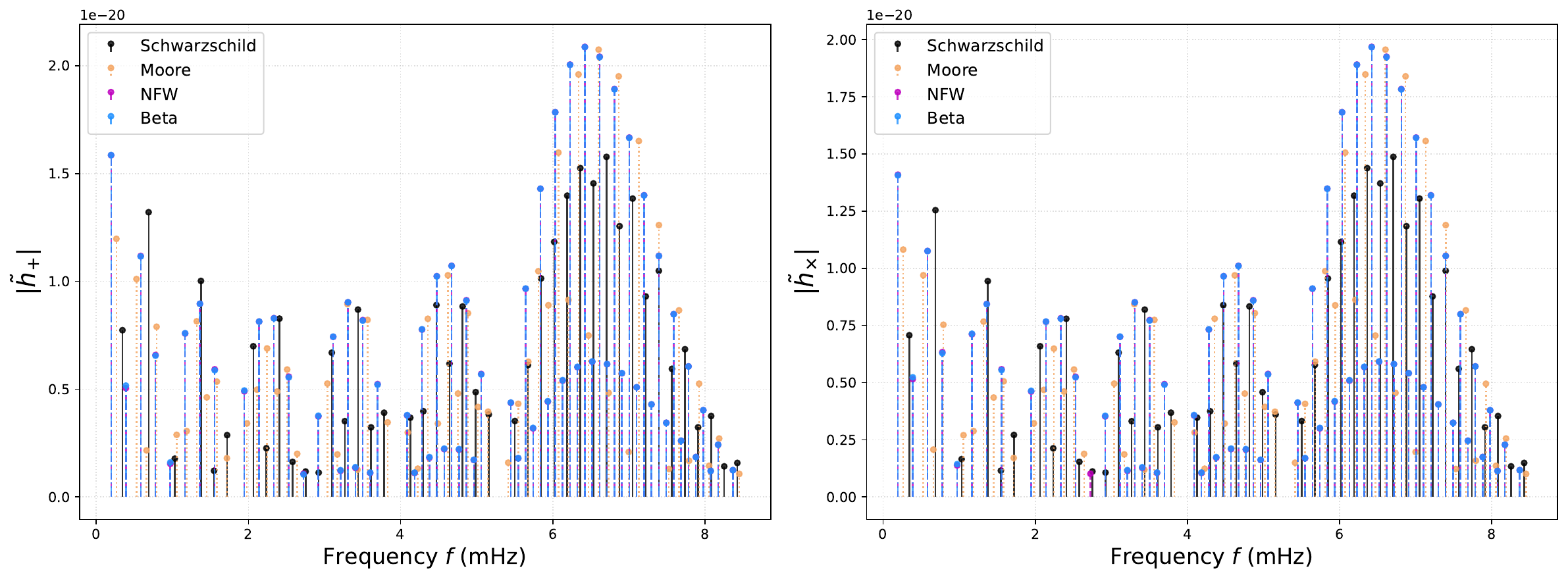}
		\caption{$k = 2 \times 10^4 M$}
		\label{fig:orbit_comparison_b}
	\end{subfigure}
	\caption{Comparison of the gravitational wave frequency spectra for the periodic orbital configuration $(2~2~1)$ obtained using different halo models: NFW, Beta, and Moore. The upper panel shows the results for a dark matter mass parameter $k = 10^4 M$, and the lower panel shows the results for $k = 2 \times 10^4 M$. The Schwarzschild case is also included in this figure as a reference baseline. In this figure, we set a central black hole mass $M = 10^6 M_\odot$, a small compact object's mass $m = 10 M_\odot$, a characteristic radius $h = 10^7 M$ and a luminosity distance $D_L = 200\,\mathrm{Mpc}$.}
	\label{fig:orbit_comparison}
\end{figure}

Firstly, we focus on the gravitational frequency spectra arising from different orbital configurations corresponding to a specific dark matter density distribution. Fig.~\ref{fig:moore_spectra} presents the discrete frequency spectra for five distinct periodic orbital configurations (1 1 0), (1 2 0), (1 3 0), (2 1 1), (2 2 1) under the Moore profile, considering dark matter mass parameters $k = 10^4 M$ and $k = 2 \times 10^4 M$. It is clearly illustrated that each orbital configuration $(z~w~v)$ exhibits a distinct spectral pattern. As the dark matter mass $k$ increases, the dark matter distribution becomes denser and deepens the effective gravitational potential (see Fig.~\ref{Veff}). This deepened potential slightly shifts the frequency components of periodic orbits, changing the structure of the frequency spectrum, especially in the low frequency region. Complementary to varying the orbital configurations, we then analyze the distinct spectral signatures induced by different dark matter environments for a fixed orbital configuration. Fig.~\ref{fig:orbit_comparison} compares the frequency spectra of the periodic orbital configuration $(2~2~1)$ predicted by different halo models. Using the Schwarzschild vacuum case as a baseline, we observe that dark matter environments induce noticeable deviations in the relative amplitude distribution and the structure of frequency spectra (see the discrete peaks). The spectra produced by the NFW and Beta models are nearly indistinguishable in the frequency domain, which is consistent with the findings for the orbital trajectories and gravitational waveforms. However, the Moore model is clearly distinct from the others, because the structure of its discrete peaks exhibits rather distinctive signatures. For instance, the upper panel of Fig.~\ref{fig:orbit_comparison} shows that the Moore model results in more pronounced amplitudes in frequency spectrum when $f \approx 6.5$ mHz for the $k=10^4 M$ case.

\begin{figure}
	\centering
	\begin{subfigure}{0.495\textwidth}
		\centering
		\includegraphics[width=\textwidth]{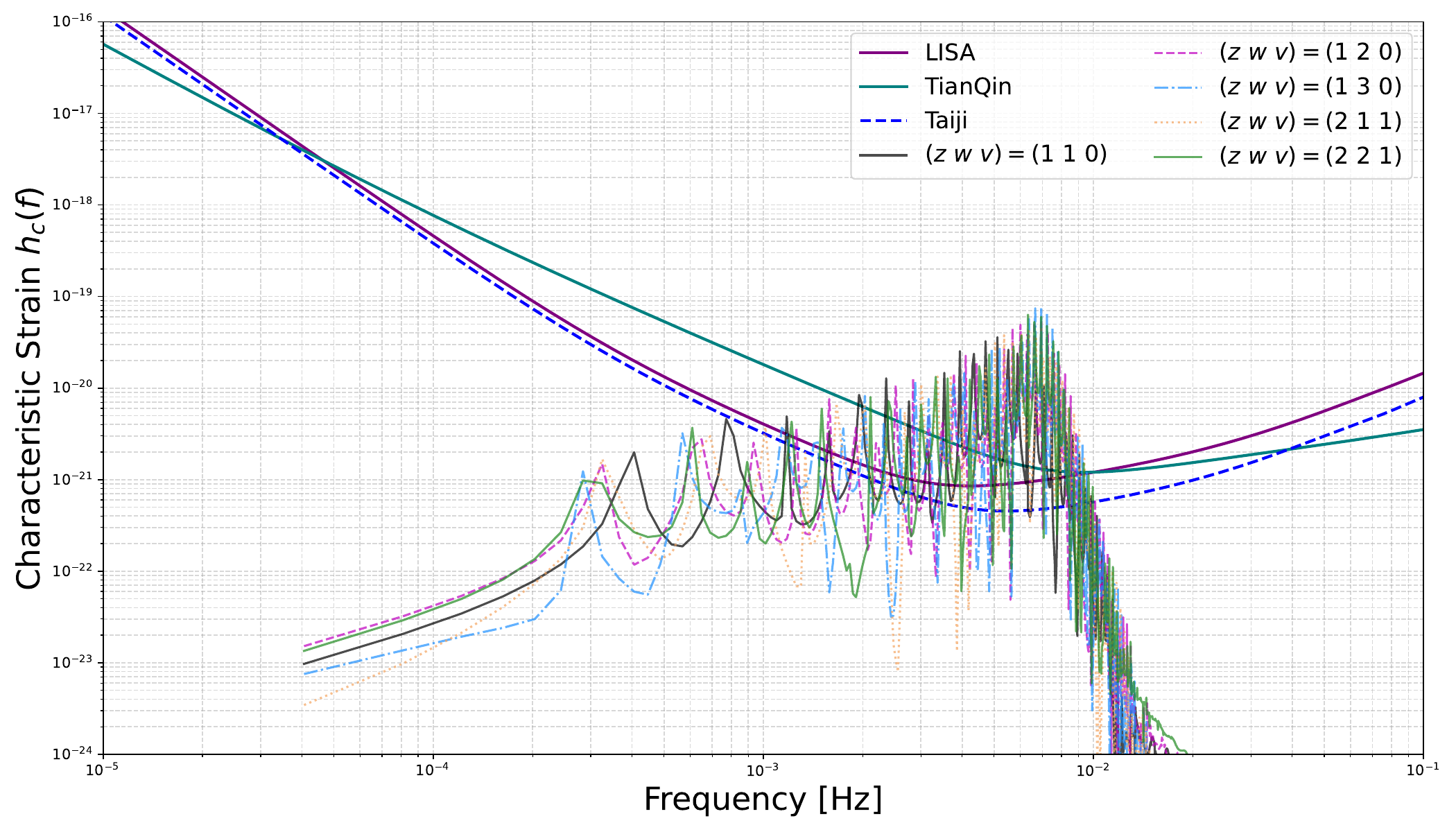}
		\caption{}
		\label{fig:hc_moore_orbits}
	\end{subfigure}
	\hfill
	\begin{subfigure}{0.495\textwidth}
		\centering
		\includegraphics[width=\textwidth]{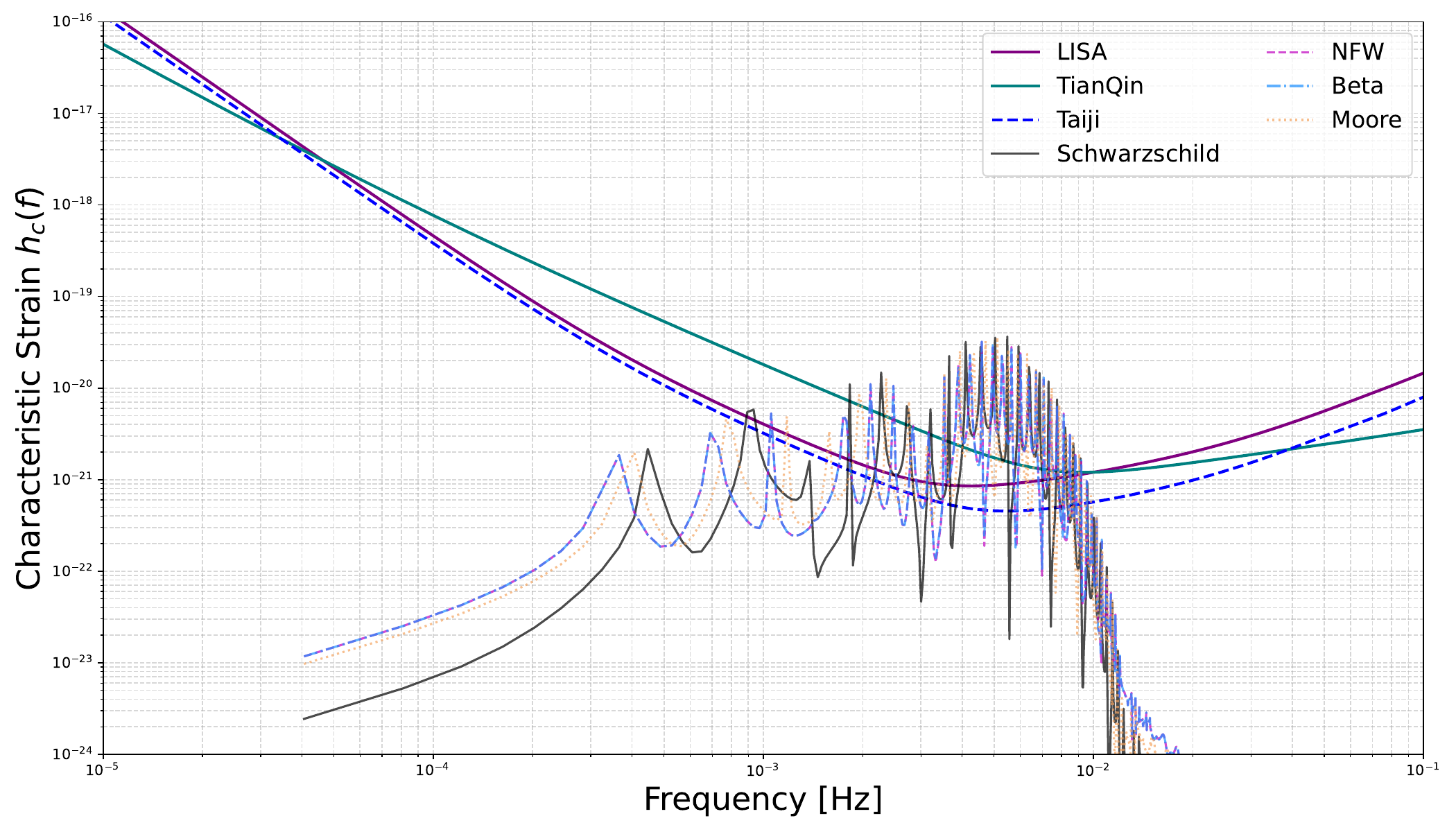}
		\caption{}
		\label{fig:hc_orbit_221}
	\end{subfigure}
	\caption{Characteristic strain of gravitational waves emitted from periodic orbits, compared with the sensitivity curves of LISA, Taiji and TianQin, for a dark matter mass parameter $k = 10^4 M$. The left panel displays the characteristic strain for five distinct orbital configurations under the Moore dark matter profile. The right panel compares the characteristic strain of the periodic orbital configuration $(1~1~0)$ across the NFW, Beta, and Moore halo models. The Schwarzschild case is also included in the right panel as a reference baseline. The system is characterized by a central black hole mass $M = 10^6 M_\odot$, a small compact object's mass $m = 10 M_\odot$, a characteristic halo radius $h = 10^7 M$, and a luminosity distance $D_L = 200\,\mathrm{Mpc}$. The observation time window is $T_{\text{obs}} = 5000 M$, and a Tukey window function is applied to the signal during the Fourier analysis.}
	\label{fig:hc_combined}
\end{figure}

To directly assess whether these discrete spectral features can be resolved by future space-based detectors, it is necessary to compare the characteristic strain of gravitational wave signals in the frequency domain with the instrumental sensitivity thresholds. The characteristic strain of gravitational waves is given by
\begin{equation}
	h_c(f) = 2f \left( \left| \tilde{h}_+(f) \right|^2 + \left| \tilde{h}_\times(f) \right|^2 \right)^{1/2}.
\end{equation}
Fig. \ref{fig:hc_combined} provides a comparison between the gravitational wave strain of periodic orbits with detector sensitivity thresholds. In our numerical evaluation, the total observation time window is set to $T_{\text{obs}} = 5000 M$ (which is approximately 6.8 hours for $M = 10^6 M_\odot$). To mitigate spectral leakage arising from the finite duration of observed signals, a Tukey window function is applied during the Fourier analysis to obtain the characteristic strain of gravitational waves. The resulting characteristic strain is then compared with the projected sensitivity curves of three space-based gravitational wave interferometers: LISA, Taiji, and TianQin. Fig.~\ref{fig:hc_moore_orbits} presents the characteristic strain for five distinct periodic orbital configurations obtained under the Moore profile, assuming a dark matter mass $k = 10^4 M$ and a characteristic halo radius $h=10^7 M$. For each orbital configuration, the corresponding characteristic strains are lie prominently above the instrumental sensitivity curves. Given that these characteristic strains exceed the sensitivity thresholds of LISA, Taiji, and TianQin, the periodic orbits of EMRIs embedded in the Moore halo are anticipated to yield substantial signals, indicating a high probability of being detected. To complement the analysis of varied orbital configurations, Fig.~\ref{fig:hc_orbit_221} gives the characteristic strain of the periodic orbital configuration $(1~1~0)$ across the NFW, Beta, and Moore halo models. The Schwarzschild case is also included in this figure as a reference baseline to show the influence of the central black hole in the absence of dark matter. Relative to the Schwarzschild baseline, the presence of dark matter environments slightly changes the structure of the characteristic strain, thereby improving the overall distinguishability of the signals emitted from periodic orbits with and without dark matter. Notably, this distinguishability suggests that dark matter halos are potentially detectable by future space-based interferometers (LISA, Taiji, and TianQin). Consistent with the preceding spectral analysis, the characteristic strains associated with the NFW and Beta models remain observationally degenerate. Conversely, the Moore model induces a significantly different structure in the characteristic strain, compared with the NFW and Beta models.

Furthermore, it should be mentioned that the analysis presented in this section focuses only on the short-term gravitational wave observations of periodic orbits, in which neither dissipative mechanisms nor dynamical evolution of orbits are taken into consideration. If one aims to study the long-time dynamical evolution of these periodic orbits and their gravitational wave signatures, the dissipative mechanisms cannot be omitted, and the orbital energy $E$ and angular momentum $L$ can no longer be treated as conserved quantities. This would break the periodicity of these orbits. In appendix \ref{a3} , we examine how long these orbits can maintain their periodicity over a long-time dynamical evolution, where the gravitational wave reaction is incorporated as a dissipative effect. The numerical results suggest that periodic orbits can retain their periodicity signatures on short observation timescales, confirming the validity of the methods for analyzing instantaneous waveform (namely the methods presented in section \ref{s4} and this section). A simpler orbital configuration (such as (1~1~0)) and a relatively small mass ratio can significantly extend the timescale over which periodicity is preserved. In particular, for the mass ratio $m/M=10^{-6}$, the periodicity of the simplest orbital configuration (1~1~0) can persist for more than one month.

\section{Conclusions and Perspectives}\label{s6}

This study systematically investigates the influence of dark matter halos on periodic orbits around a black hole (with zoom-whirl-vertex behavior) and the produced gravitational wave radiation. This work is carried out within the framework of spherically symmetric dark matter halo models, which have been widely used in astrophysical studies. Through detailed analysis of three dark matter halo models—NFW, Beta, and Moore—we examine how dark matter halo parameters affect spacetime metrics, effective potentials, and properties of periodic orbits around black holes, as well as the resulting gravitational wave signals.

The effects of dark matter halos on spacetime geometry, periodic orbits and gravitational waves can be incorporated into two parameters: the dark matter mass $k$ and halo characteristic radius $h$. Our analysis of the effective potential reveals that the existence of dark matter halos reduces the extrema of the effective potential. This reduction becomes more obvious as the dark matter mass $k$ increases, directly affecting the MBO and ISCO of test particles. The presence of dark matter halos increases the angular momentum for both MBO and ISCO. Particularly, the NFW and Beta models showing nearly identical behavior on ISCO (or MBO), while the Moore model exhibits lower angular momentum values of ISCO (or MBO). 

To comprehensively explore periodic orbits, we examine the effects of both dark matter mass $k$ and halo characteristic radius $h$ under the condition of selected angular momentum $L = L_{\text{ISCO}} + \varepsilon(L_{\text{MBO}} - L_{\text{ISCO}})$ and the selected energy $E$ (in Appendix~\ref{a2}). For the selected angular momentum condition, it is demonstrated that for a given precession angle $q$, larger dark matter masses require lower orbital energies $E$. The NFW and Beta models require substantially lower energies $E_{(z~w~v)}$ than the Moore model for the same precession angle and orbital configurations $(z~w~v)$. This energy difference directly affects the orbital shapes, producing the stretching effects on periodic orbital shapes. An increasing of dark matter mass $k$ greatly stretches the apoapsis of periodic orbits, which leads to larger orbital dimensions. On the other hand, larger dark matter halo scales require higher orbital energy for a same precession angle $q$. At smaller halo scales ($h \sim 10^{7}M$), periodic orbital shapes in dark matter environments significantly differ from the Schwarzschild case. As the halo scale increases, the orbits gradually converge towards the Schwarzschild's orbital trajectories, closely approaching  them at large halo scales. This behavior reflects the dilution of gravitational effects: when the dark matter mass remains constant while the halo scale increases, the gravitational binding becomes weakens in the spacetime. Objects must then possess higher energies (closer to the asymptotic value 
$\lim_{r \to \infty} V_{\text{eff}} = 1$ without gravitation) to maintain the same precession characteristics. This scale-dependent effect is clearly shown and explained with five different periodic orbital configurations $(z~w~v)$, where orbits gradually approach the Schwarzschild case as $h$ increases. In particular, when the dark matter halo scale is \( h \sim 10^{10} M \), the periodic orbits of all three dark matter models become indistinguishable from the Schwarzschild reference, exhibiting the dilution limit of dark matter halo effects. Furthermore, the analysis of periodic orbits with fixed orbital energy $E$ (in appendix) yields consistent conclusions with those obtained for fixed angular momentum $L$.

To explore the observational signatures of these periodic orbits, we analyze the extreme-mass-ratio inspiral (EMRI) system with a typical mass ratio and luminosity distance. The gravitational wave analysis shows strong correspondence between the gravitational waveforms and orbital properties. Larger dark matter mass leads to more pronounced differences in generated gravitational waves between periodic orbits around SMBH in dark matter halos and those of Schwarzschild black holes, enhancing the distinguishability among different dark matter halo models. The waveforms exhibit stretching trends, particularly an increase in the time duration of each complete orbital period, which stems directly from the modification of the gravitational potential by the dark matter halo. In contrast to the dark matter mass effects, a larger dark matter halo scale $h$ results in a more diffuse mass distribution, weakening its effect on the gravitational potential. The gravitational waveforms generated by periodic orbits exhibit significant differences between various dark matter halo models under a relatively small halo scale ($h \sim 10^{7}M$), which can be used to constrain dark matter models with future observations. The gravitational waveforms gradually approach those for Schwarzschild black holes when dark matter halo scales become sufficiently large. Particularly, when the characteristic radius of the dark matter halo is \( h \sim 10^{10} M \), the periodic orbits and their gravitational waveforms converge precisely with those of the Schwarzschild black hole. Examining the waveform structure reveals that the orbital configurations $(z~w~v)$, as fundamental quantities characterizing the geometric shapes of periodic motion, directly determine the spatiotemporal features (the zoom-whirl stages) of the gravitational wave signals. Furthermore, our calculations of the characteristic strain for these periodic orbits show that they fall well above the sensitivity curves of the LISA, TianQin, and Taiji detectors. This demonstrates that such distinct gravitational wave signatures are potentially detectable by future space-based gravitational wave observatories.

Throughout our analysis, we observe that the NFW and Beta models produce nearly indistinguishable results for both orbital characteristics and gravitational wave signals for all parameter ranges examined in this study. This is in accordance with the conclusions of precession angles, effective potentials, and ISCO/MBO, suggesting that these two density profiles have very similar effects on particle orbits and gravitational radiation in the parameter ranges relevant to galactic dark matter halos. The Moore model, while following similar qualitative trends, generates different quantitative results compared with Beta and NFW models, which become more pronounced for larger dark matter masses.

In summary, this study provides an systematic analysis on how dark matter halo parameters influence periodic orbits around black holes and their gravitational wave radiation, through the comparisons of three dark matter halo models. The complementary effects of the dark matter mass and its characteristic radius offer a pathway to understand the complex interplay between dark matter halo and SMBH in galactic center, as well as provide theoretical predictions for gravitational waveforms that can be compared with observations. These findings may provide useful insights and guidance for interpreting gravitational wave signals from EMRIs detected by future space-based observatories, such as LISA, TianQin, and Taiji.

The present work serves as a preliminary study of the complex environmental effects on celestial object's orbits in EMRIs. More rigorous statistical methods (such as the Fisher matrix analysis) and a more comprehensive treatment of the interactions between dark matter and EMIRIs (such as the fully relativistic method proposed in references~\cite{Vicente:2025gsg,Karydas:2025bkj}) will be incorporated into our future works to better distinguish the imprints of various dark matter environments from the gravitational waves of periodic orbits in EMRIs, and to give stricter constraints on the parameter estimation among different dark matter models.

\appendix

\section{Periodic Orbits under the Condition of Selected Energy \( E \)}\label{a1}

\begin{figure}[t]
	\centering 
	\begin{subfigure}{0.3\textwidth}
	\includegraphics[width=\linewidth]{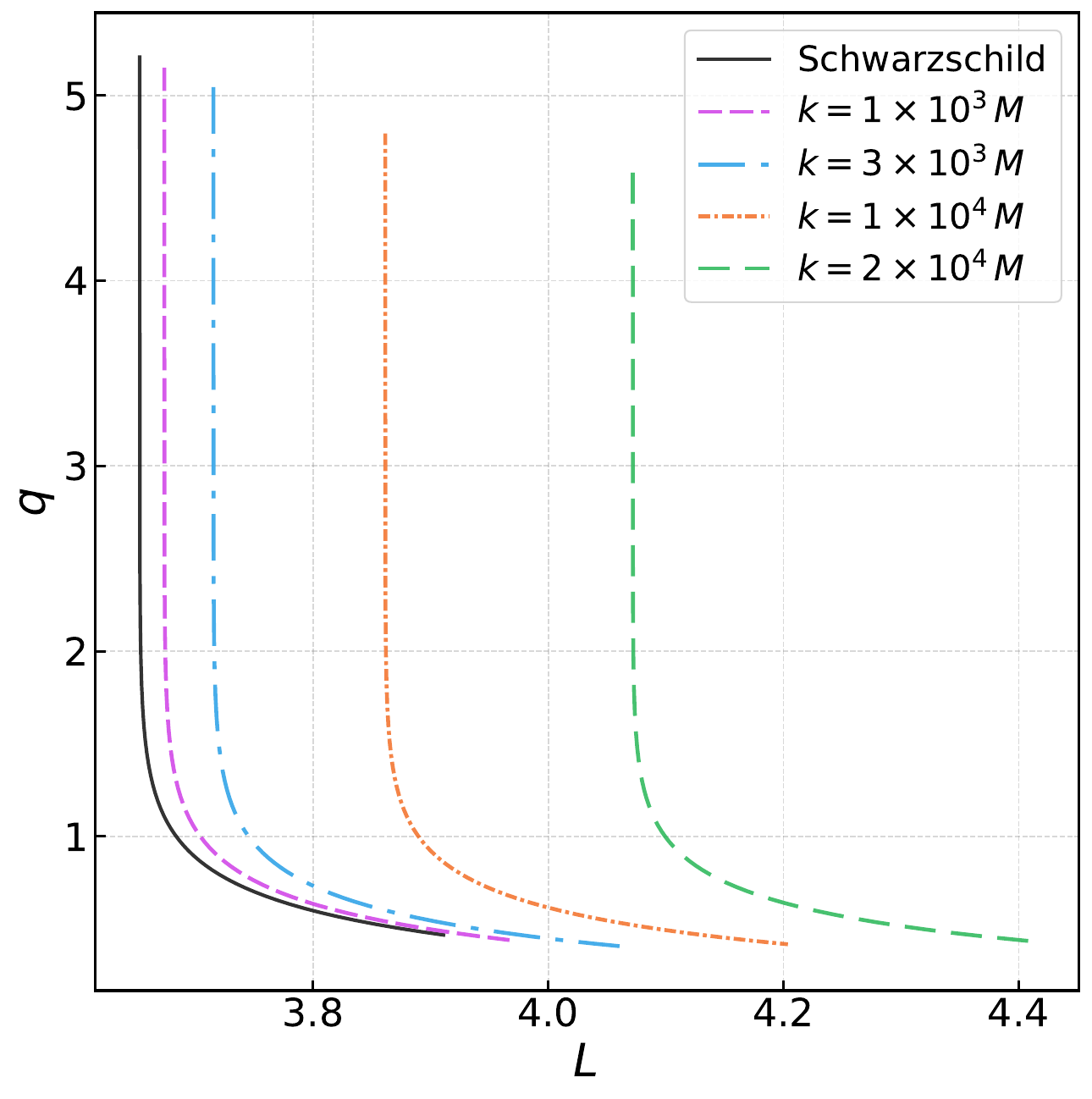}
	\caption{NFW}
	\end{subfigure}
	\hfill 
	\begin{subfigure}{0.3\textwidth}
	\includegraphics[width=\linewidth]{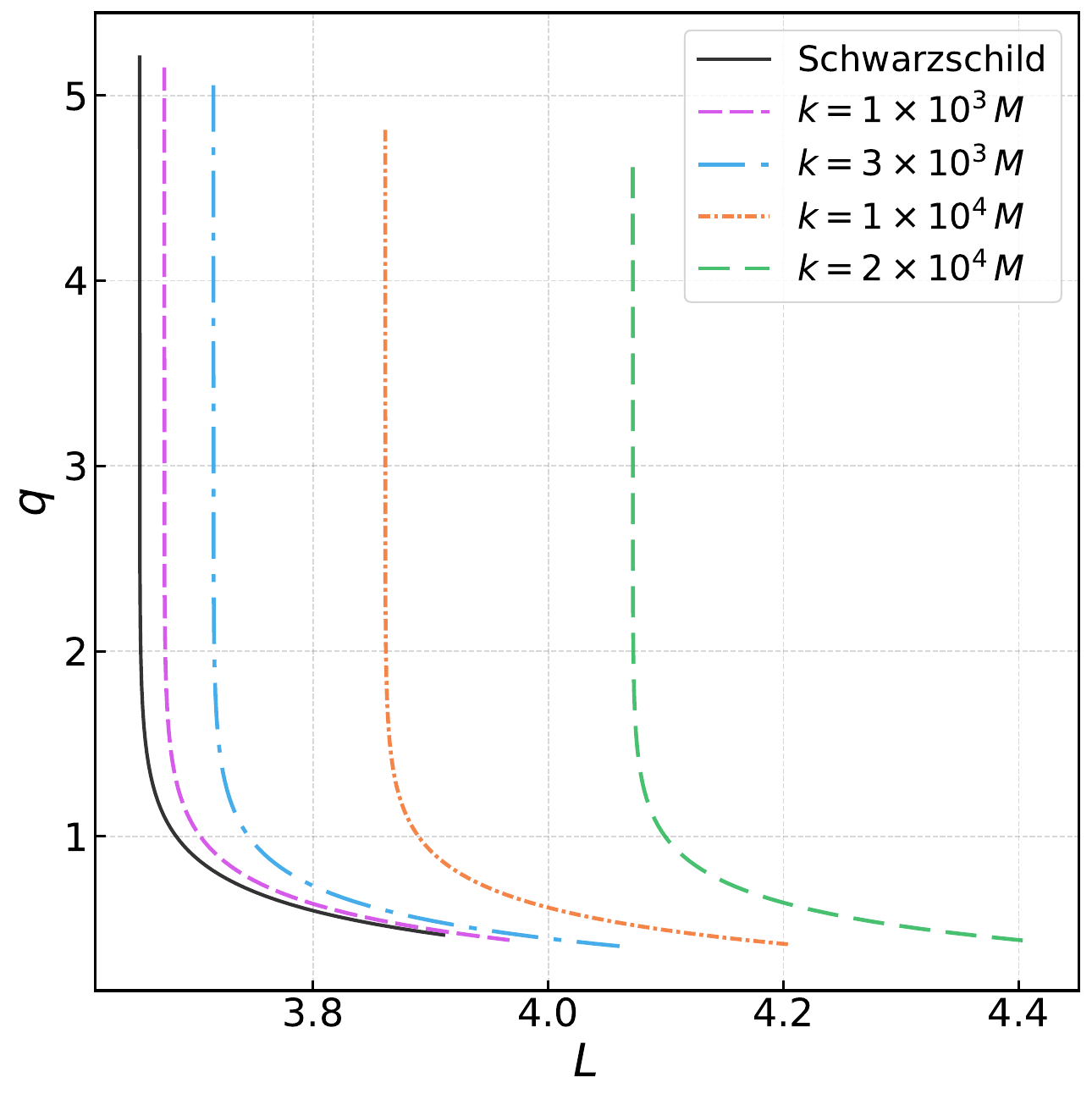}
	\caption{Beta}
	\end{subfigure}
	\hfill
	\begin{subfigure}{0.3\textwidth}
	\includegraphics[width=\linewidth]{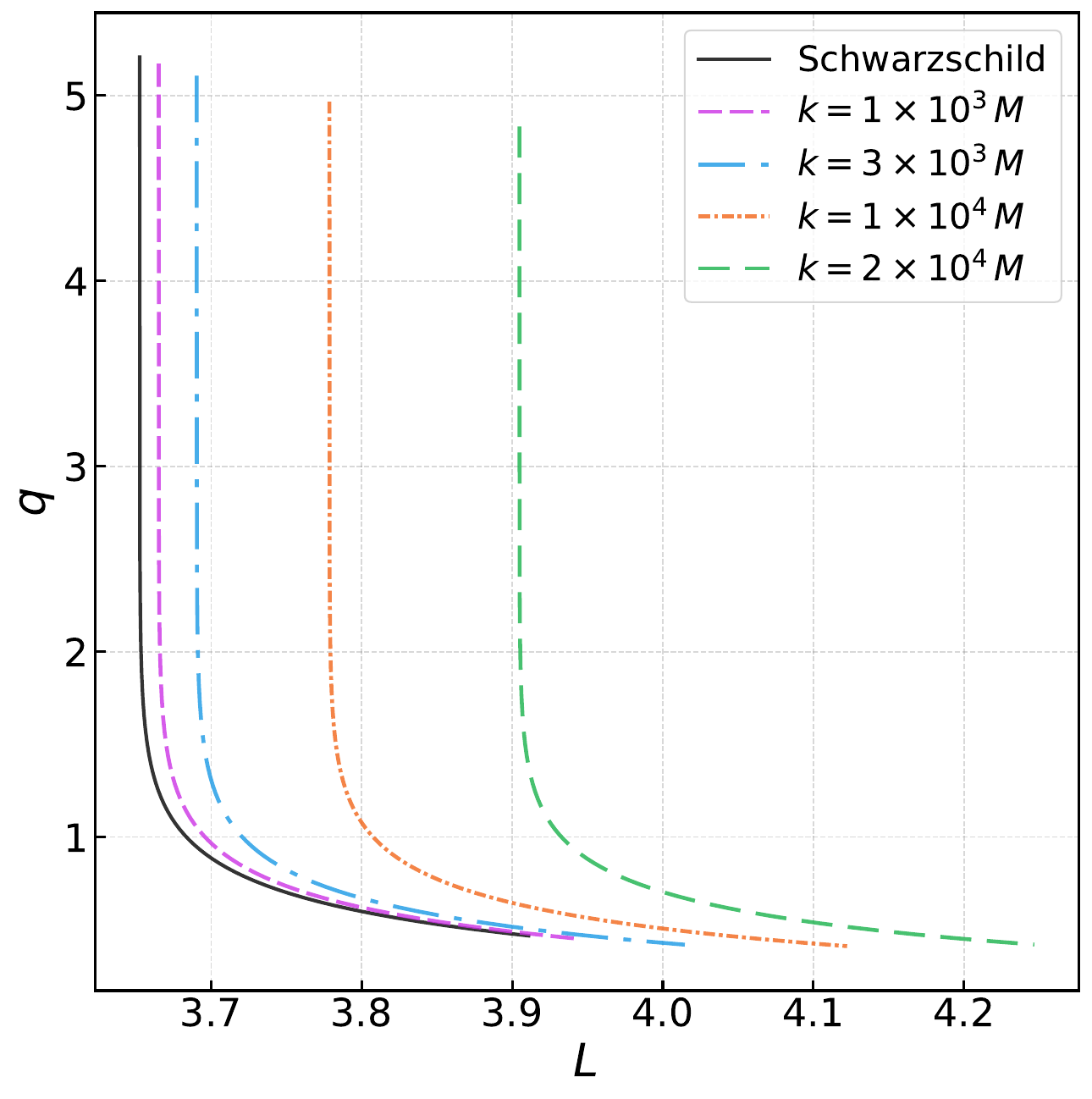}
	\caption{Moore}
	\end{subfigure}
	\caption{Precession angles changed with varying angular momentum under three dark matter halo models with $E=0.96$. The dark matter halo scale is fixed at $h = 10^7 M$, and the effects of different dark matter masses $k$ (ranging from $1 \times 10^3 M \sim 2 \times 10^4 M$) on the precession angles are compared: (a) NFW model; (b) Beta model; (c) Moore model. The Schwarzschild black hole results (dashed black lines) serve as reference baselines.}
	\label{dif_k_E=E0_q_models}
\end{figure}

\begin{figure}[t]
	\centering 
	\begin{subfigure}{0.22\textwidth}
	\includegraphics[width=\linewidth]{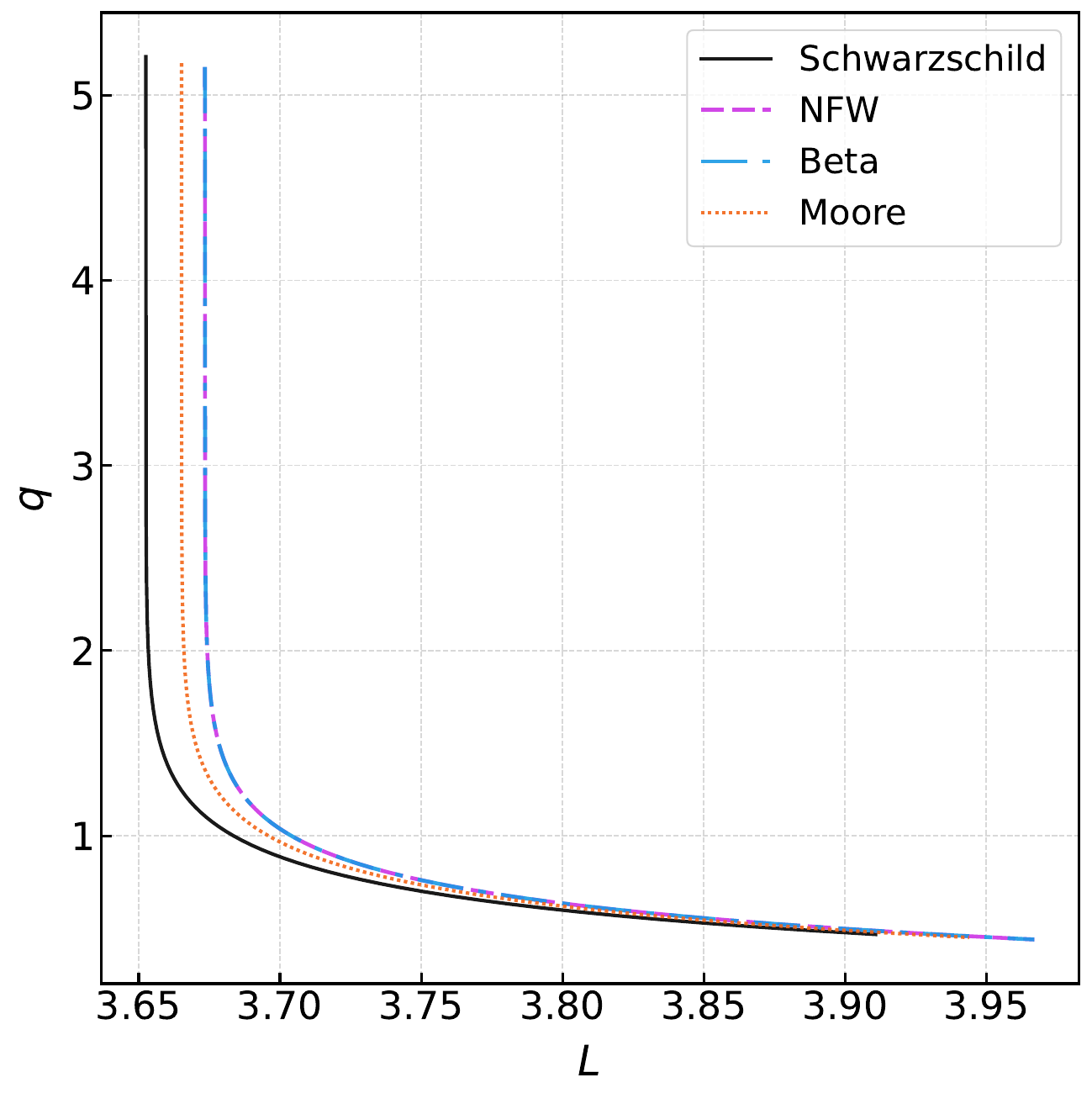}
	\caption{$k=1 \times 10^3\,M$}
	\end{subfigure}
	\hfill 
	\begin{subfigure}{0.22\textwidth}
	\includegraphics[width=\linewidth]{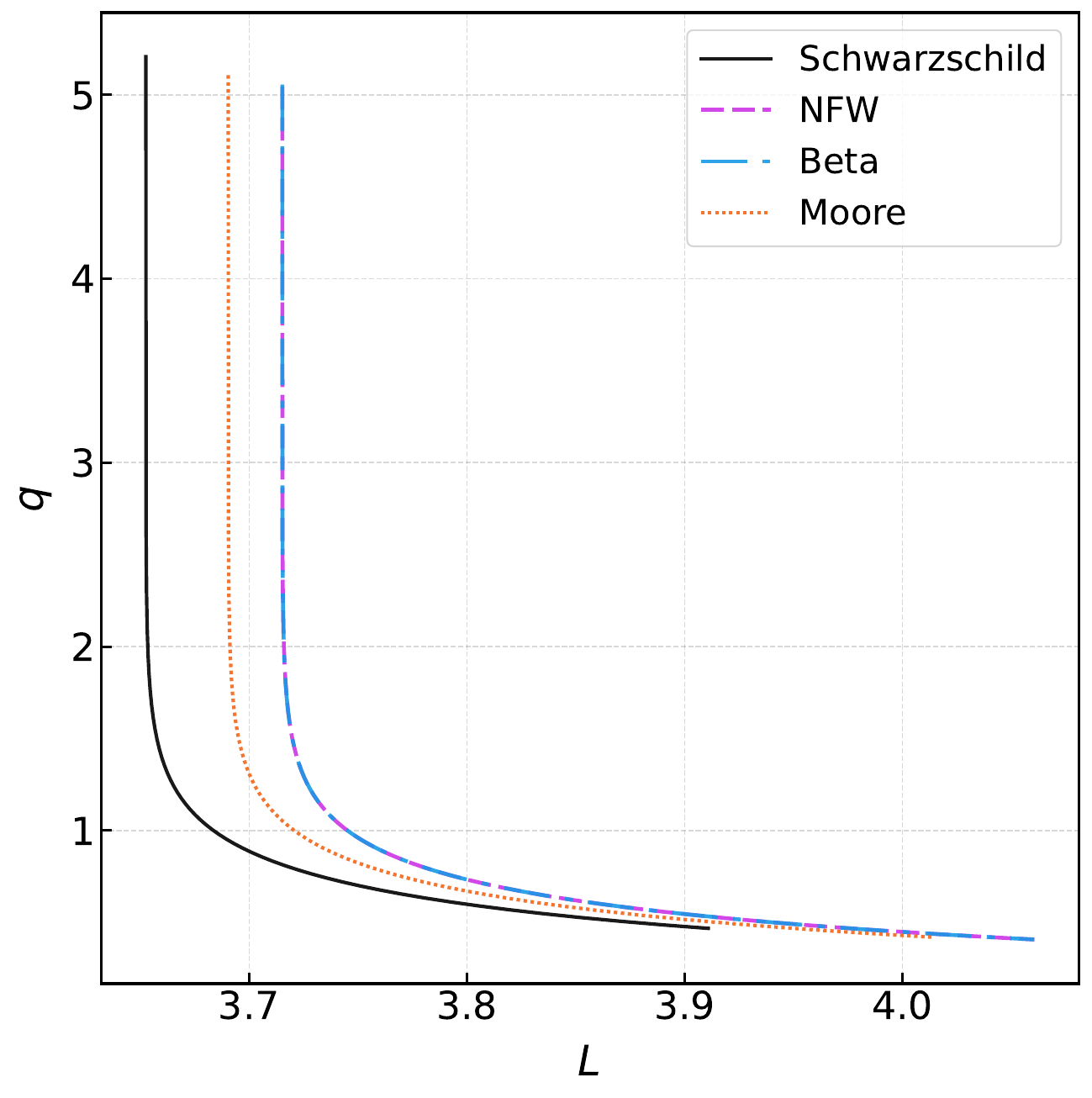}
	\caption{$k=3 \times 10^3\,M$}
	\end{subfigure} 
	\hfill 
	\begin{subfigure}{0.22\textwidth}
	\includegraphics[width=\linewidth]{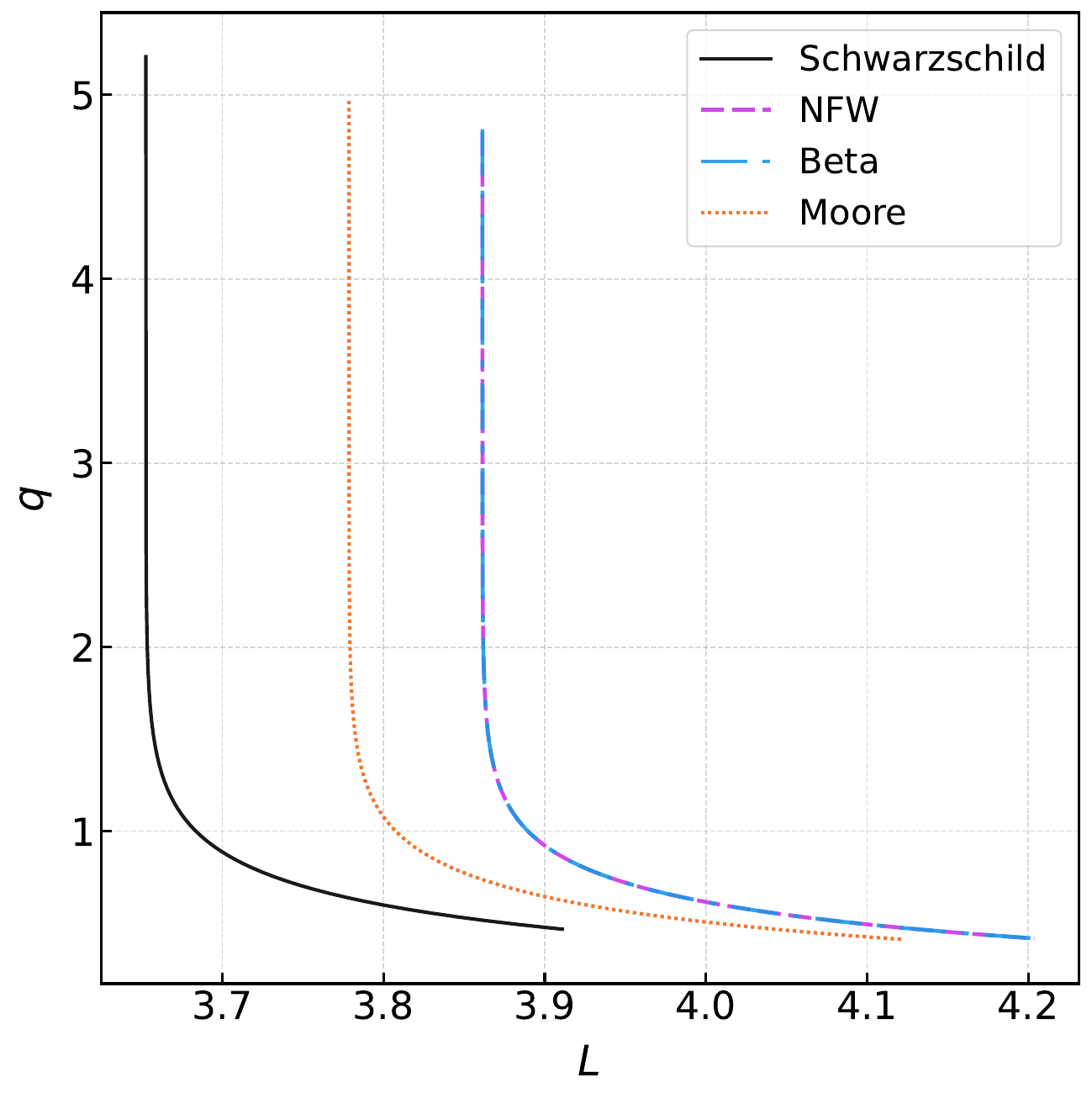}
	\caption{$k=1 \times 10^4\,M$}
	\end{subfigure}
	\hfill 
	\begin{subfigure}{0.22\textwidth}
	\includegraphics[width=\linewidth]{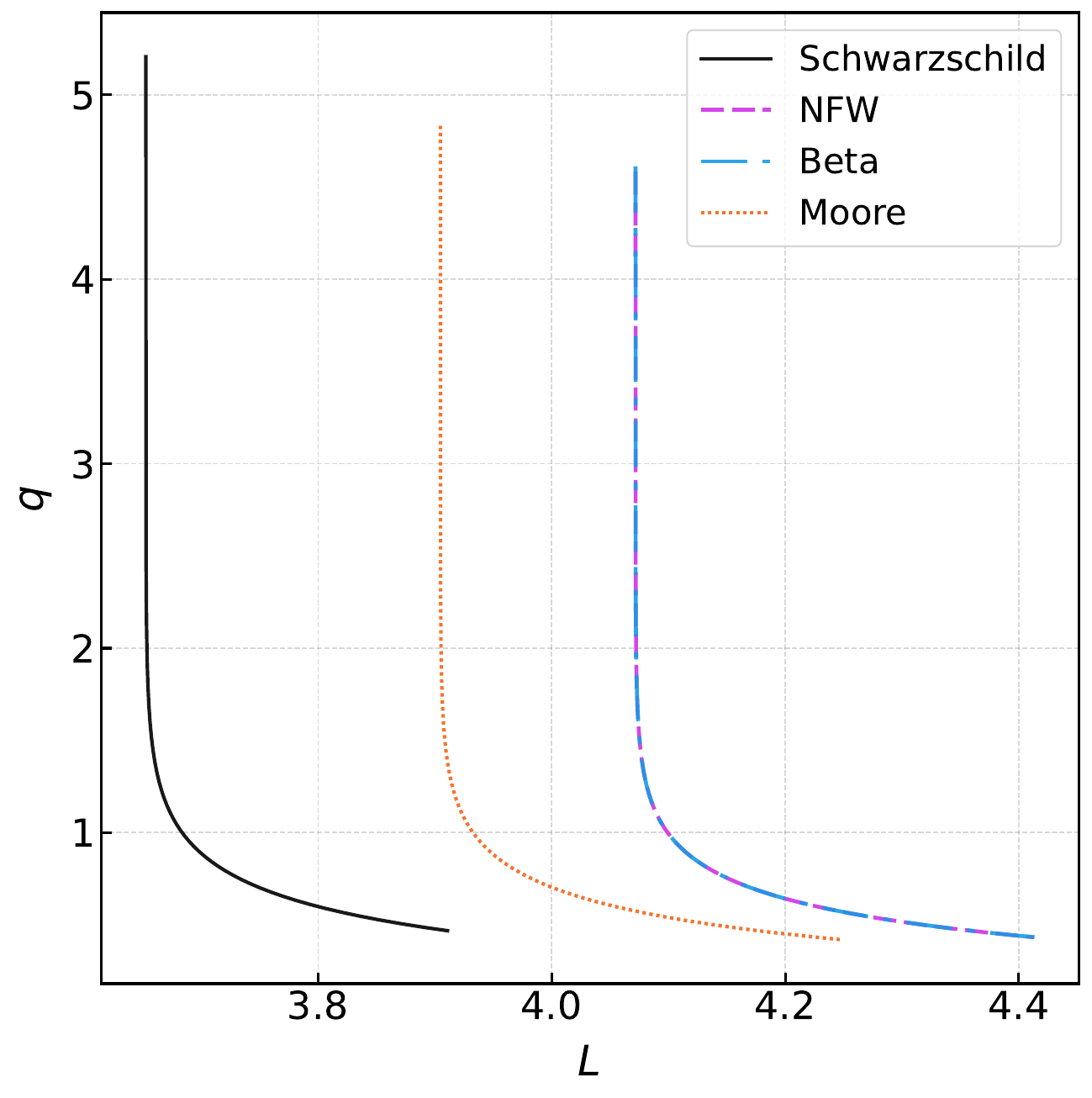}
	\caption{$k=2 \times 10^4\,M$}
	\end{subfigure}
	
	\caption{Comparison of precession angles for different models under different dark matter masses for \( E = 0.96 \). The dark matter halo scale is fixed at $h = 10^7 M$, and the figure illustrates the differences between models under four distinct dark matter masses, including (a) $k = 1 \times 10^3 M$; (b) $k = 3 \times 10^3 M$; (c) $k = 1 \times 10^4 M$; and (d) $k = 2 \times 10^4 M$.}
	\label{dif_E=E0_k_q}
\end{figure}

The periodic orbits are determined by the orbit parameters $(E, L)$. In Section~\ref{s3}, we characterize periodic orbits by fixing the angular momentum at $L = L_{\rm ISCO} + \varepsilon (L_{\rm MBO} - L_{\rm ISCO})$ and varying the energy $E_{(z~w~v)}$ across different zoom-whirl-vertex integers $(z~w~v)$. In this Appendix, we adopt an alternative but equivalent approach: fixing the energy parameter $E$ and obtaining the angular momentum $L_{(z~w~v)}$ to explore different periodic orbital configurations $(z~w~v)$.

\begin{figure} 
	\centering 
	\begin{subfigure}{\textwidth}
	\includegraphics[width=0.23\linewidth]{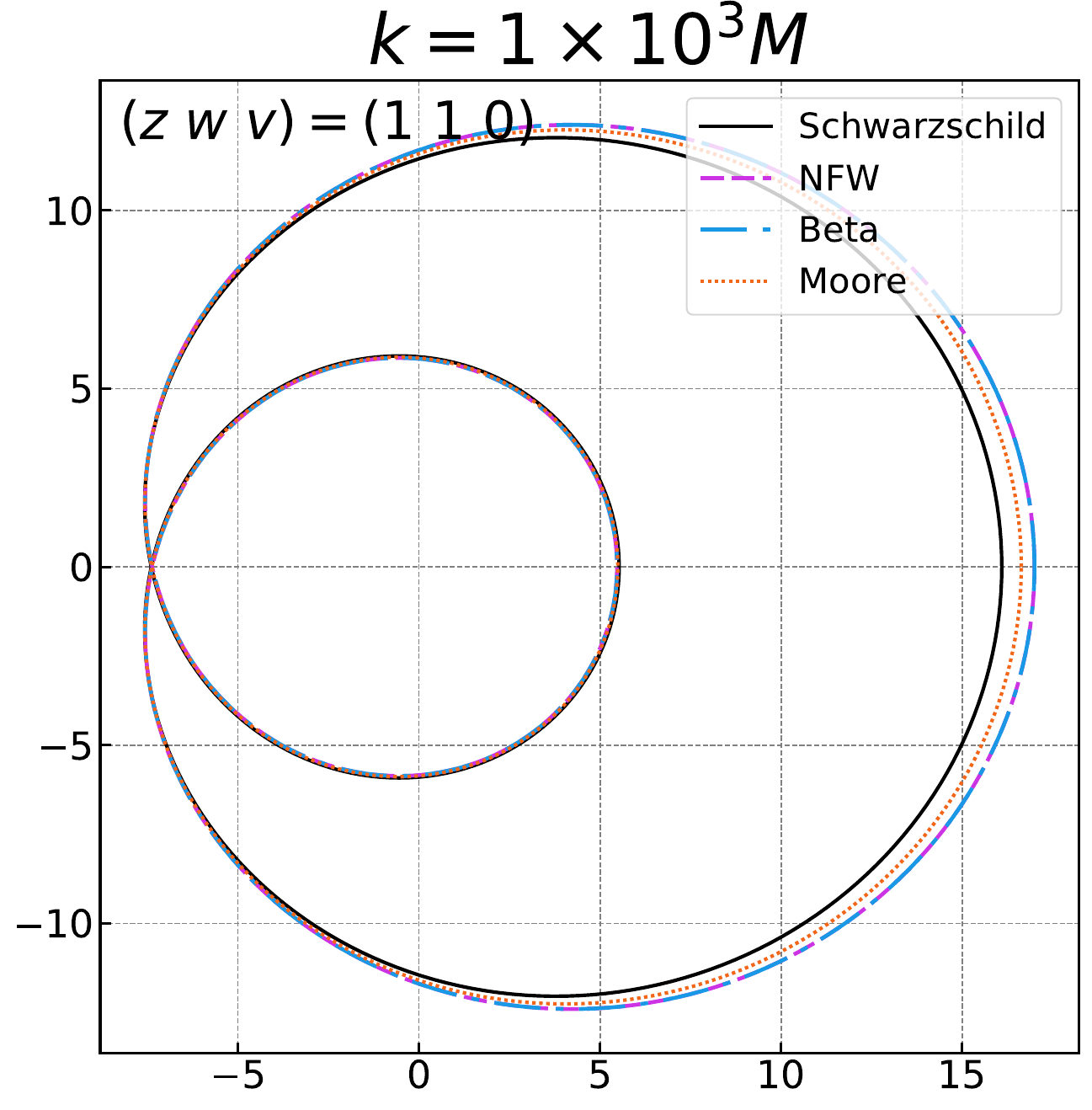}
	\includegraphics[width=0.23\linewidth]{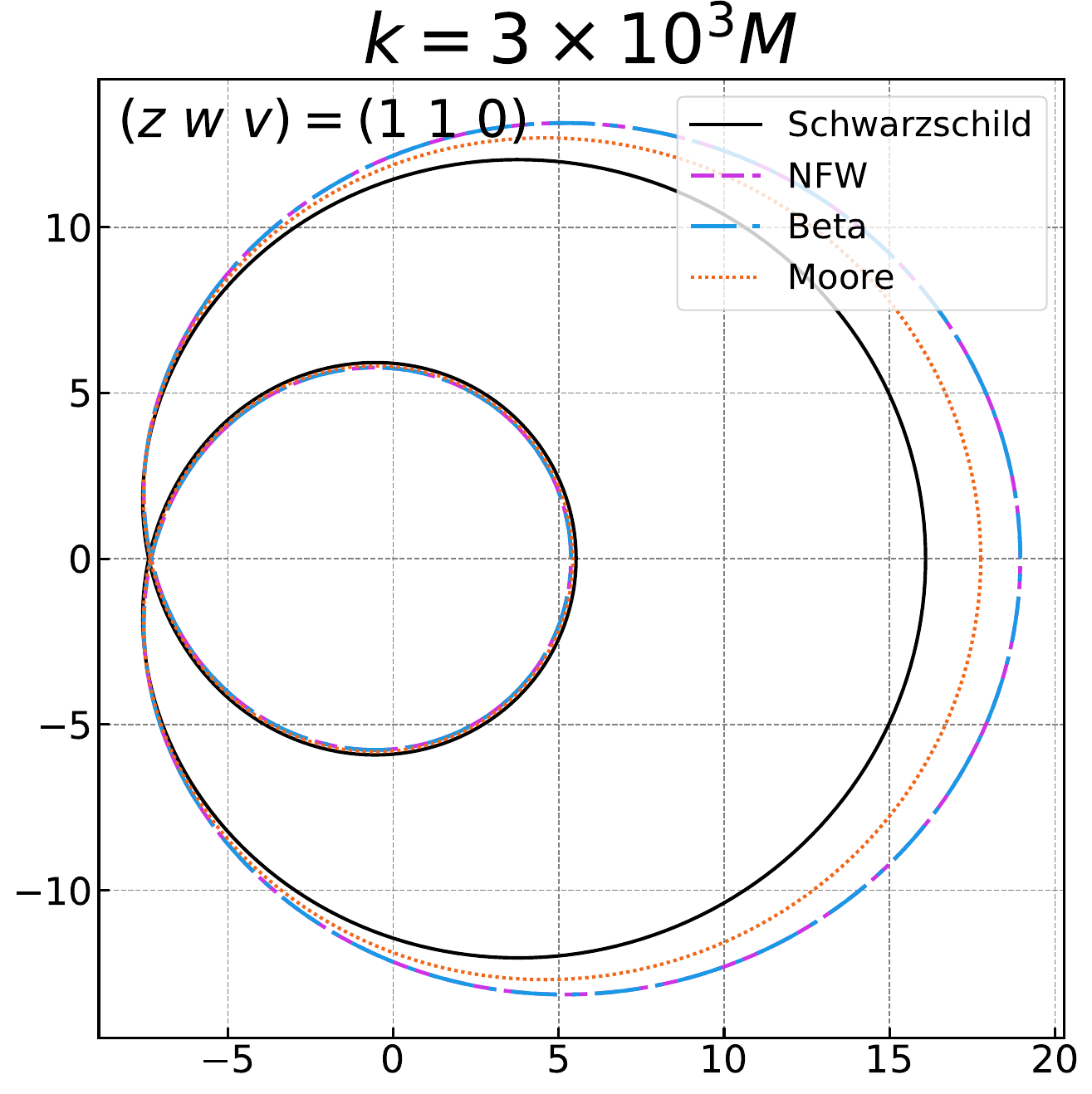}
	\includegraphics[width=0.23\linewidth]{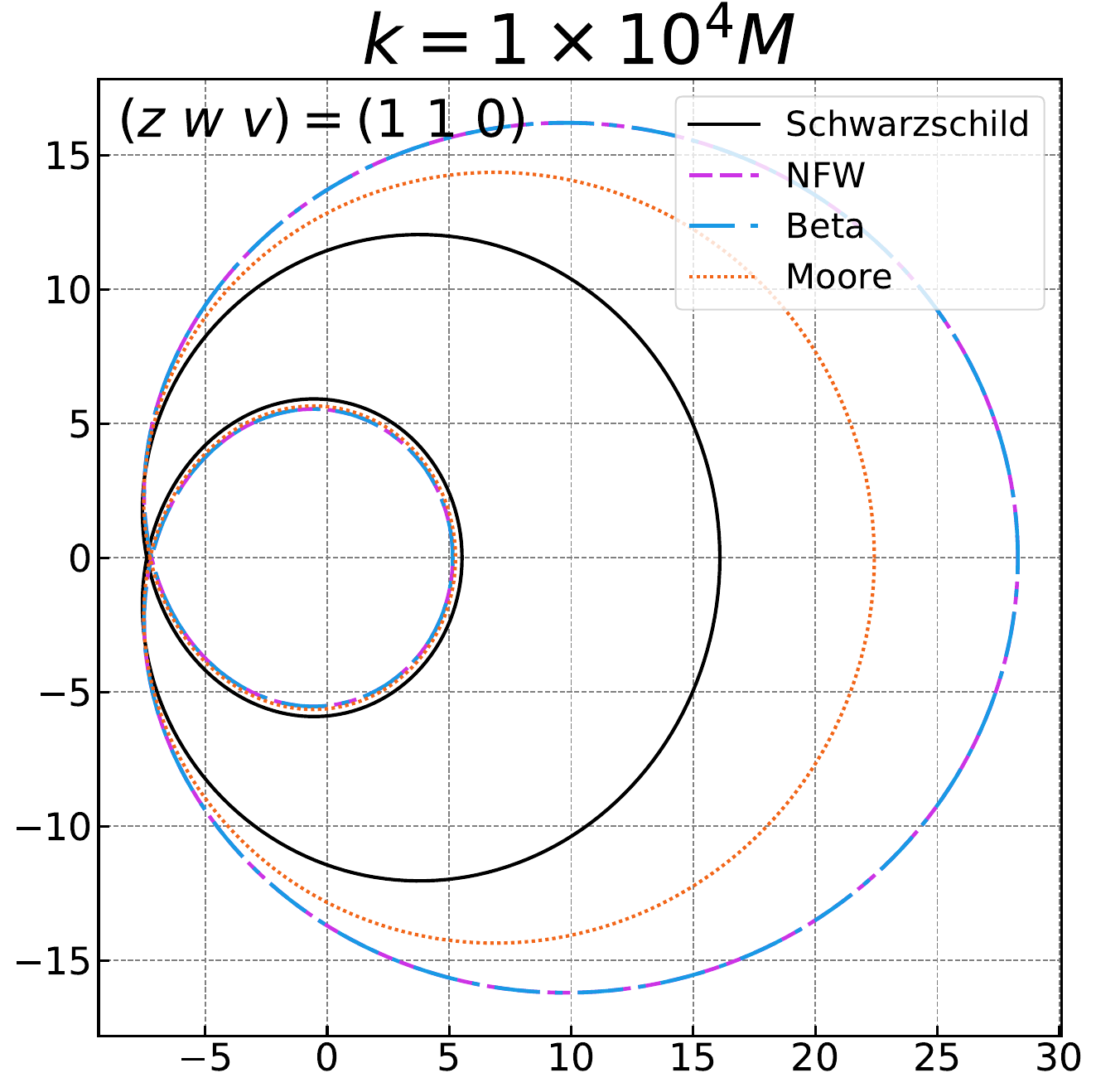}
	\includegraphics[width=0.23\linewidth]{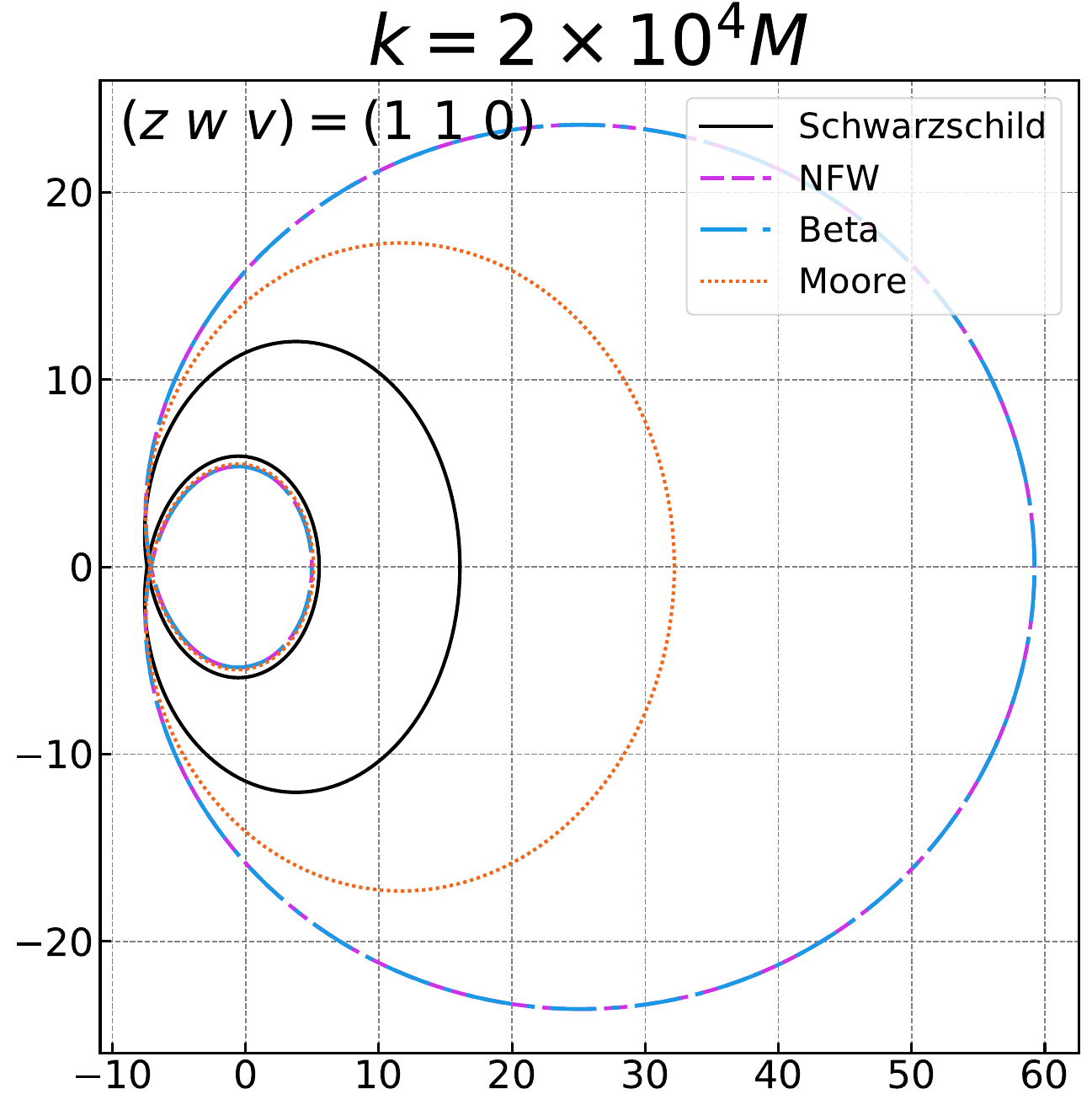}
	\end{subfigure}
	
	\vspace{0.4cm} 
	
	\begin{subfigure}{\textwidth}
	\includegraphics[width=0.23\linewidth]{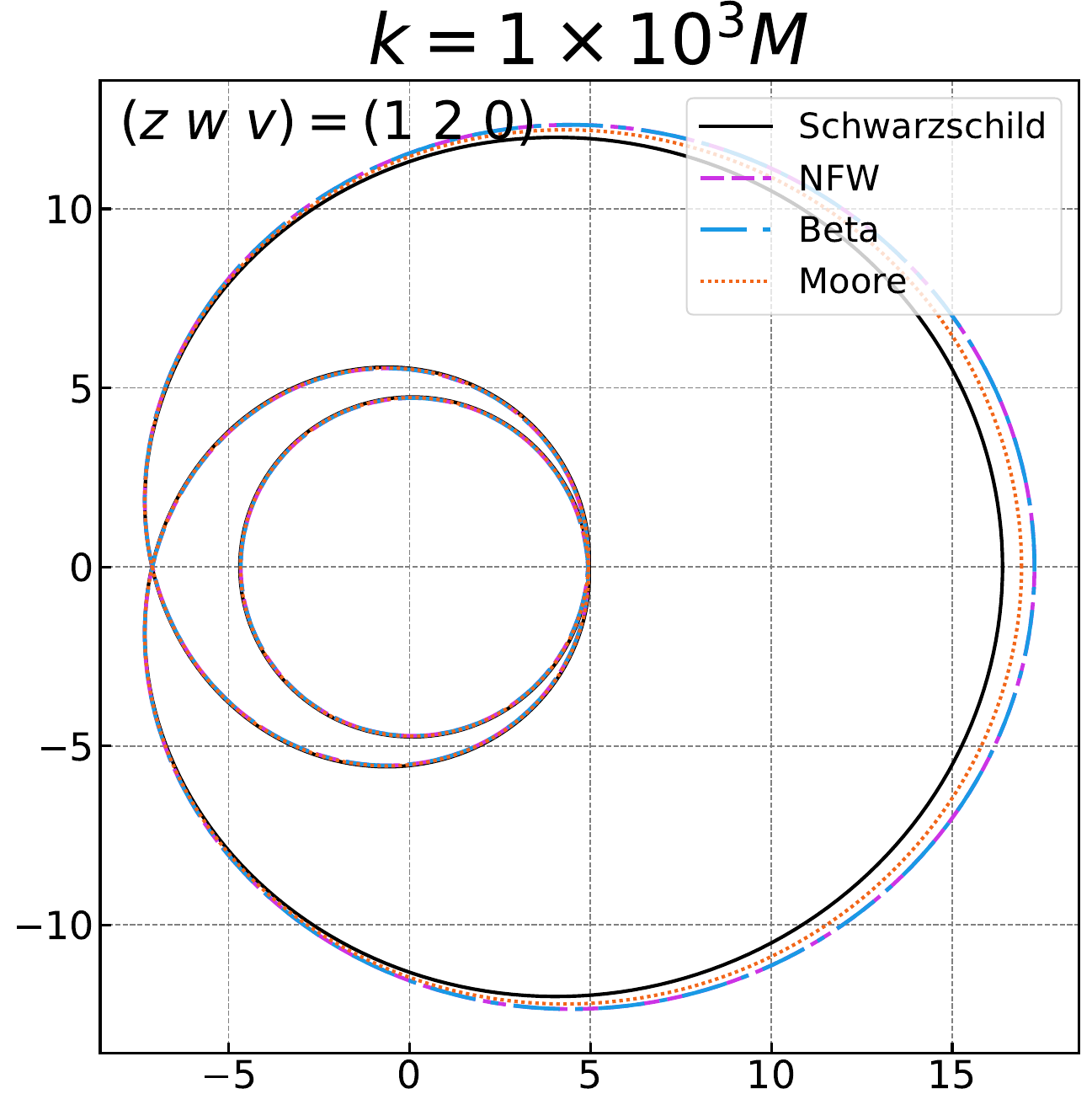}
	\includegraphics[width=0.23\linewidth]{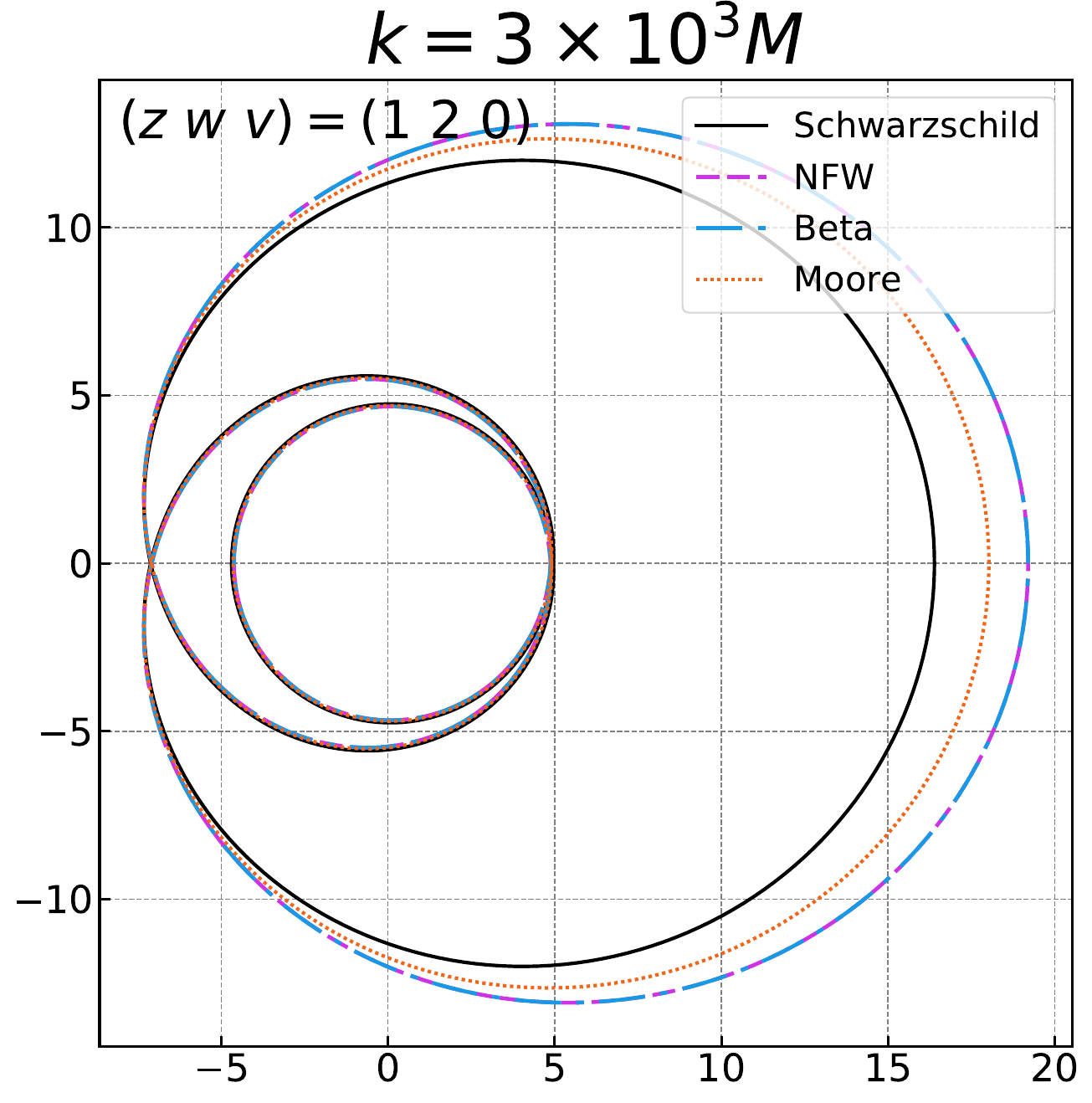}
	\includegraphics[width=0.23\linewidth]{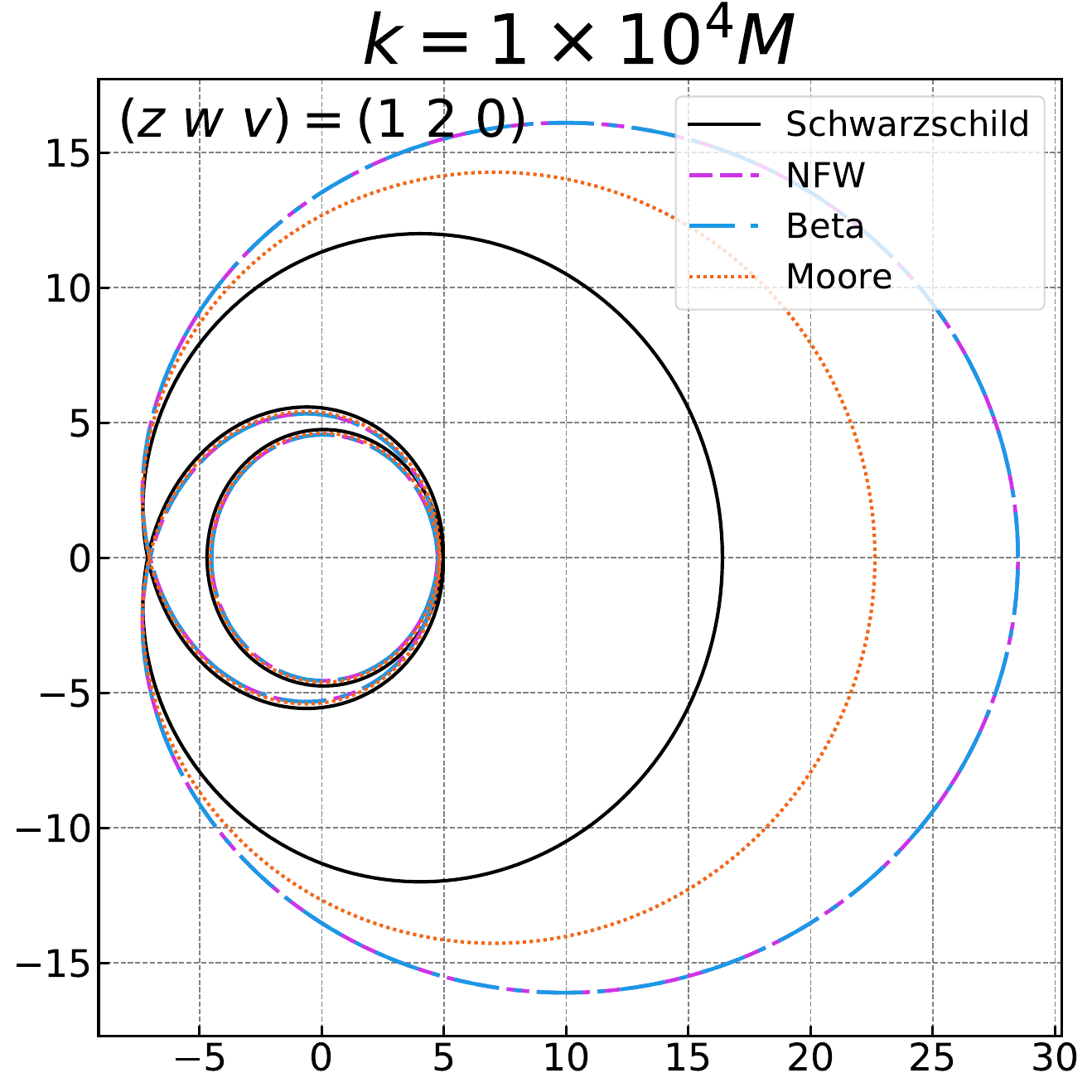}
	\includegraphics[width=0.23\linewidth]{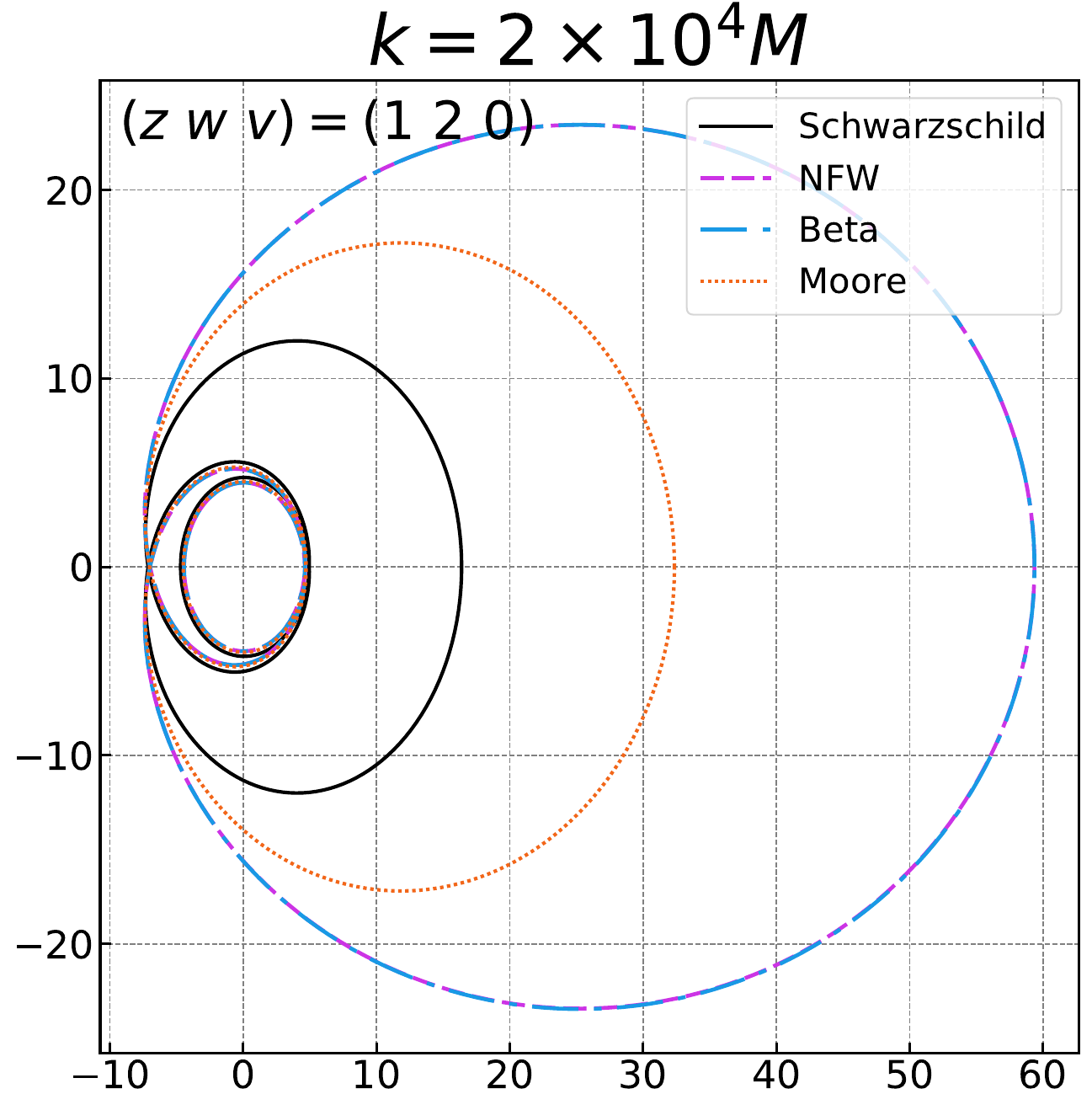}
	\end{subfigure}
	
	\vspace{0.4cm}
	
	\begin{subfigure}{\textwidth}
	\includegraphics[width=0.23\linewidth]{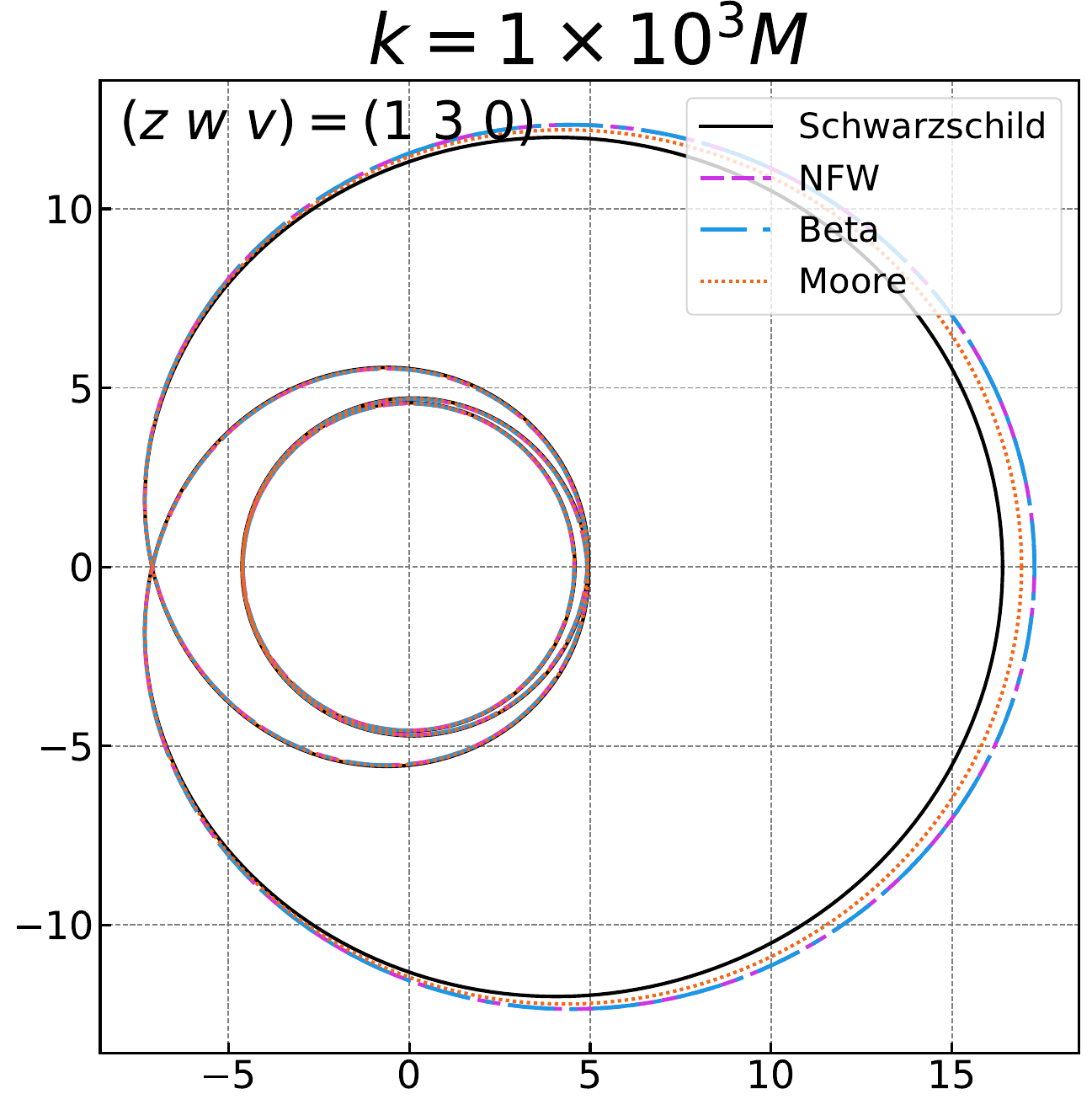}
	\includegraphics[width=0.23\linewidth]{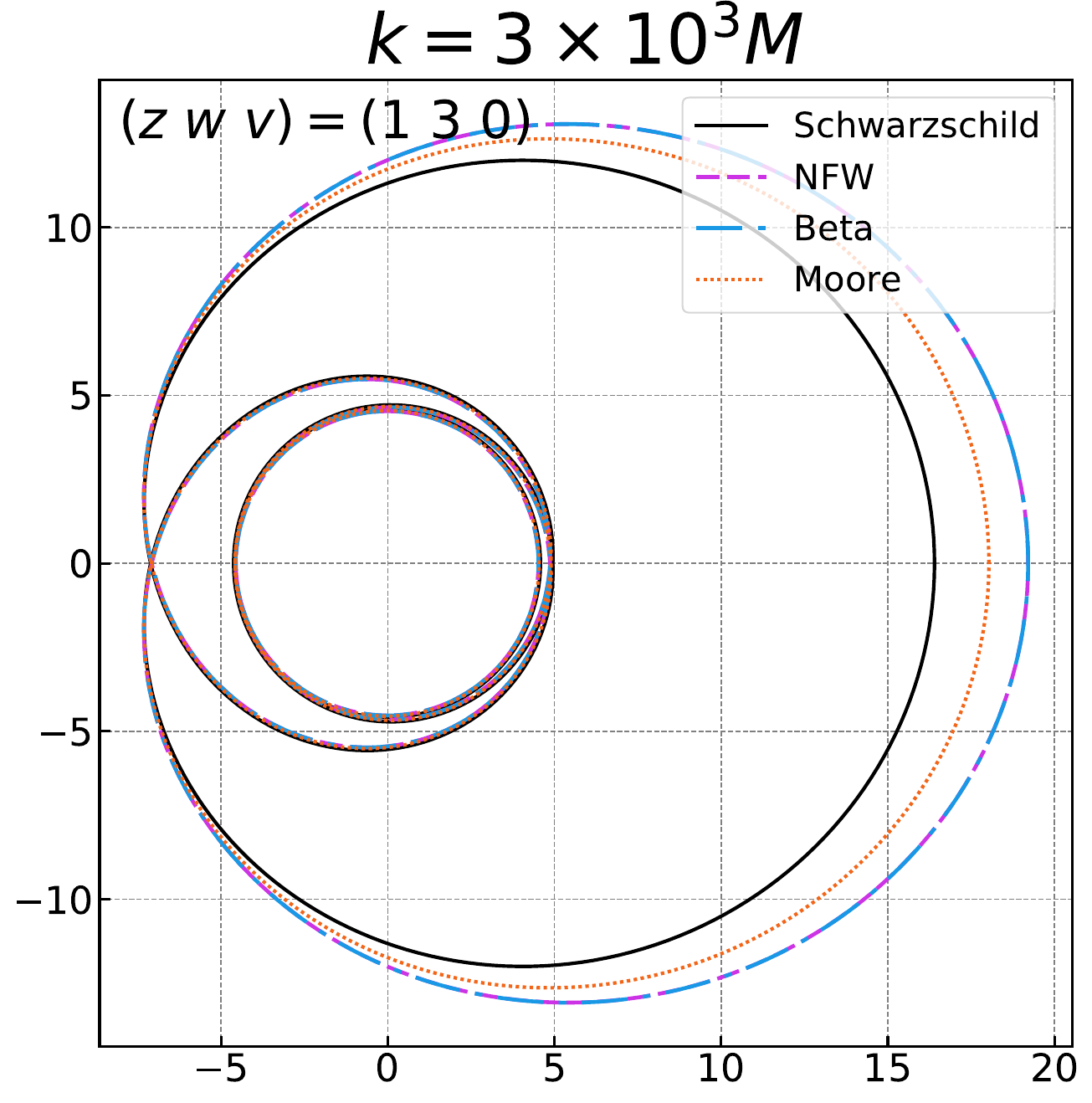}
	\includegraphics[width=0.23\linewidth]{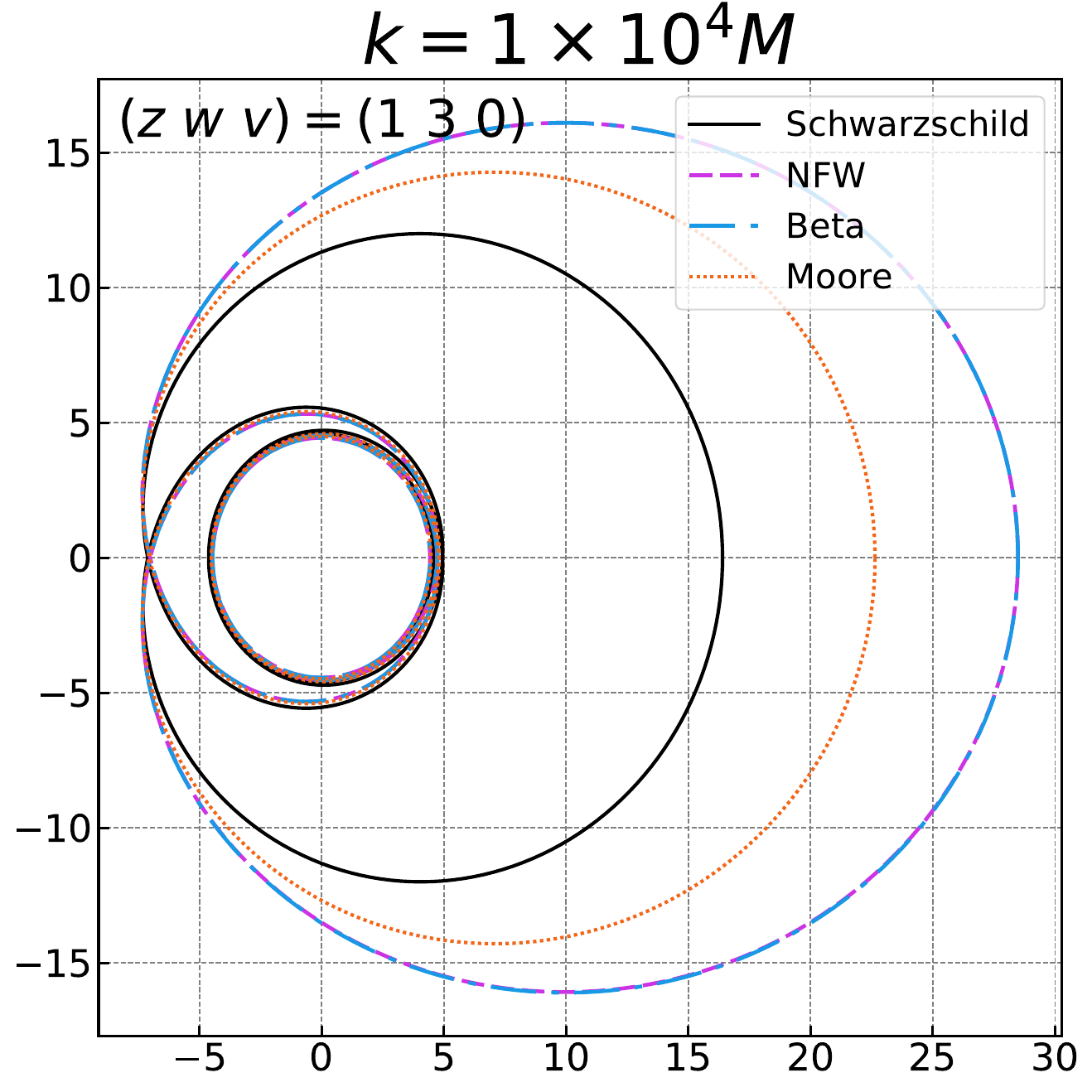}
	\includegraphics[width=0.23\linewidth]{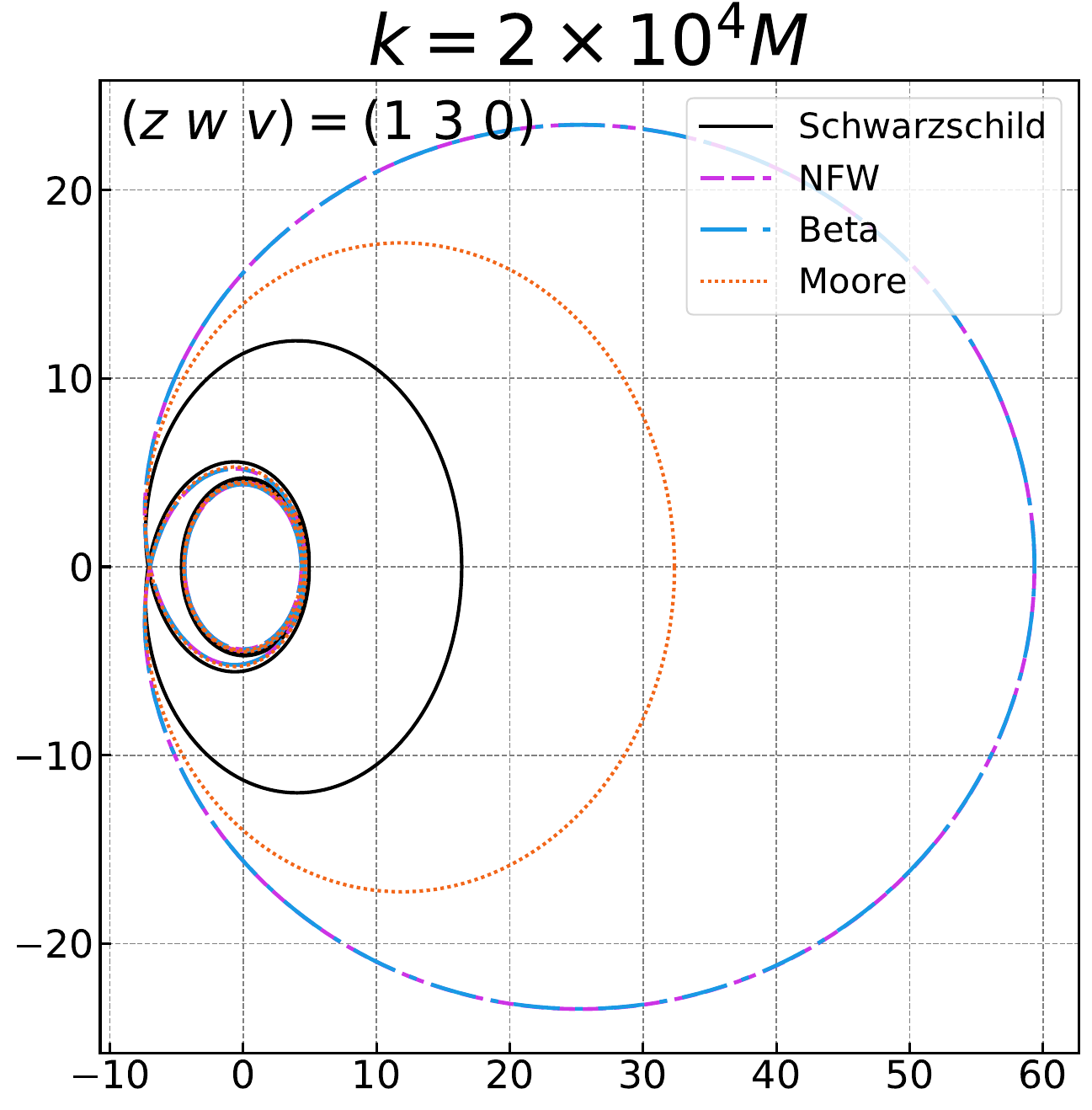}
	\end{subfigure}
	
	\vspace{0.4cm}
	
	\begin{subfigure}{\textwidth}
	\includegraphics[width=0.23\linewidth]{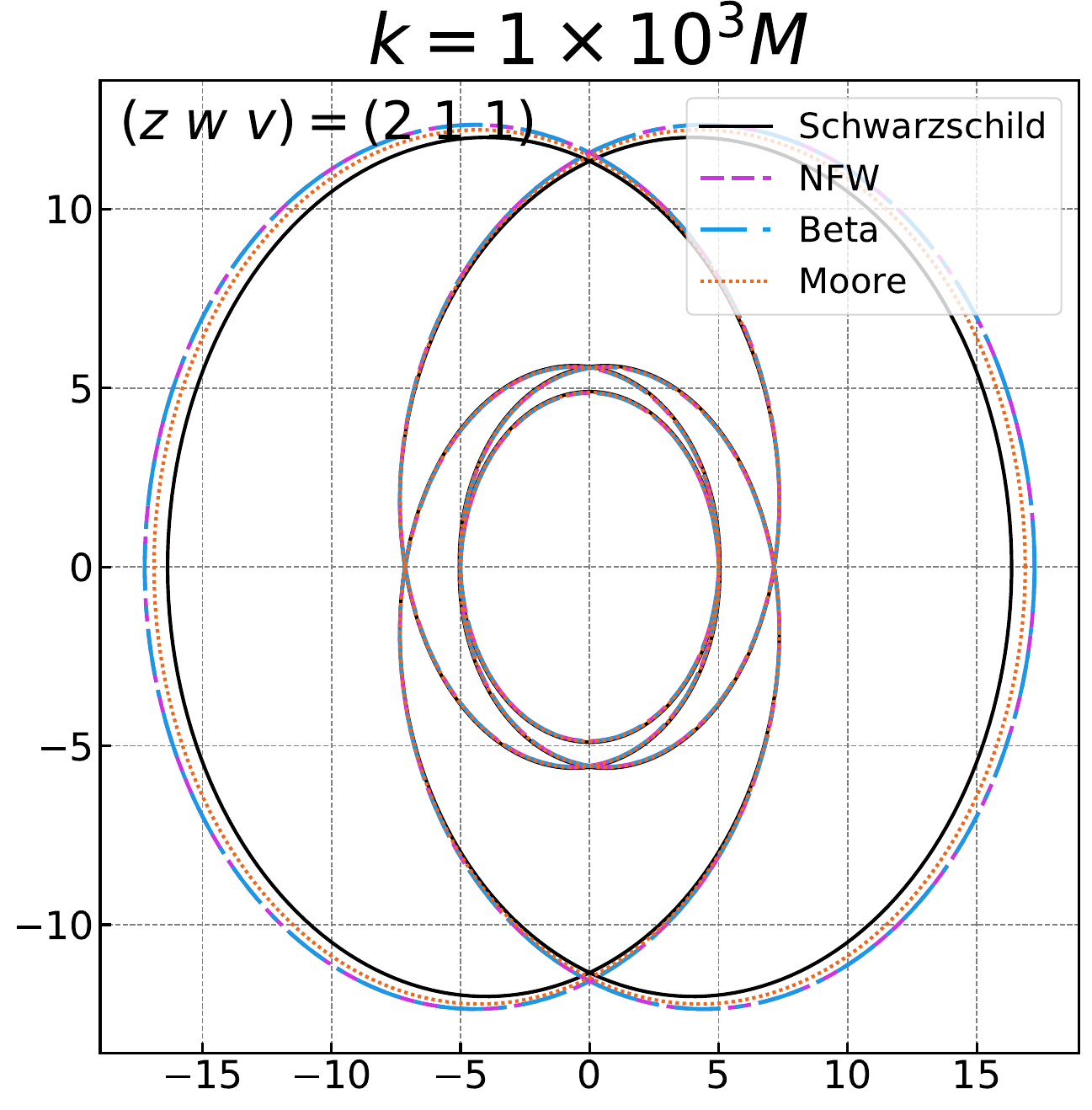}
	\includegraphics[width=0.23\linewidth]{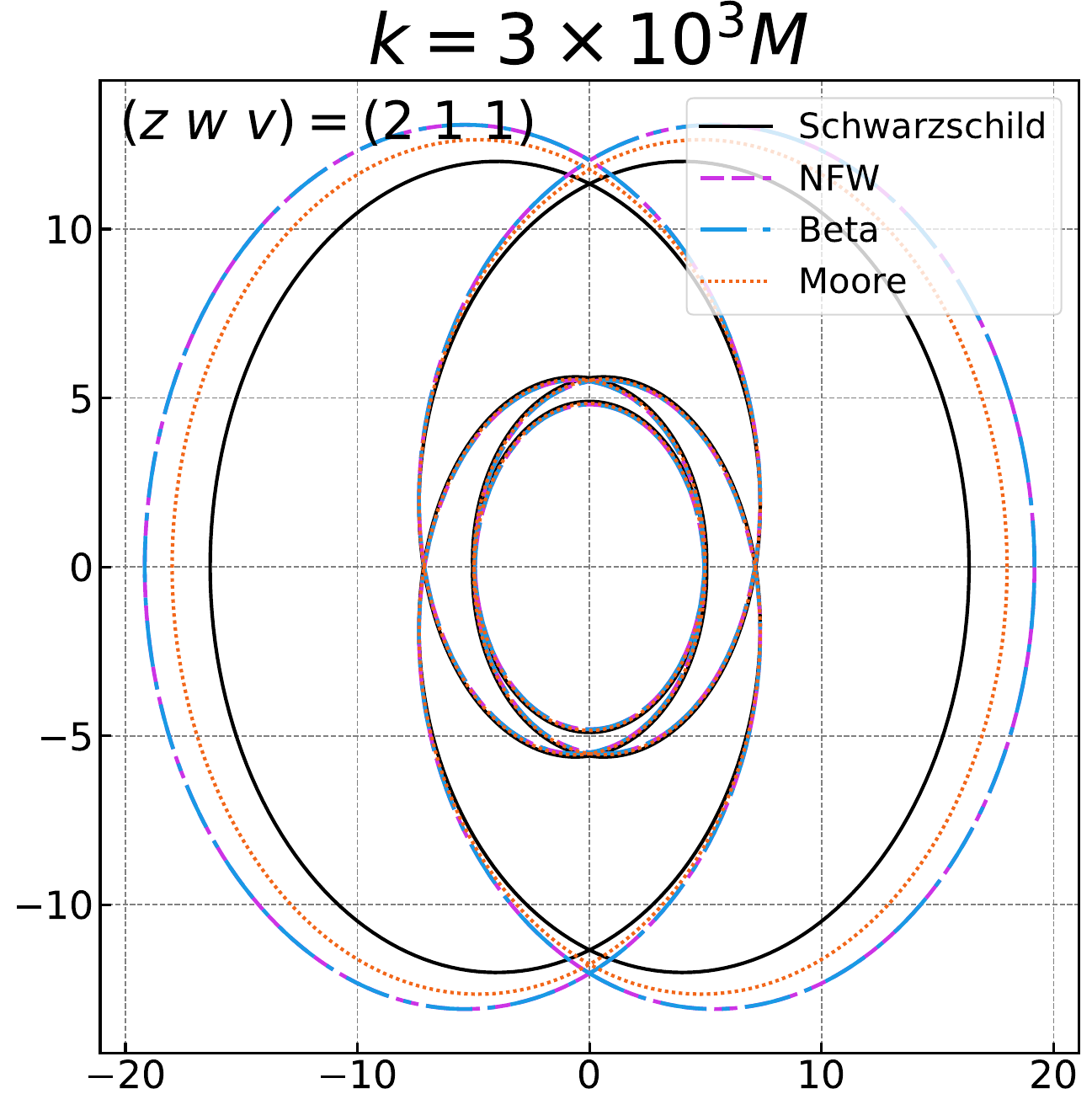}
	\includegraphics[width=0.23\linewidth]{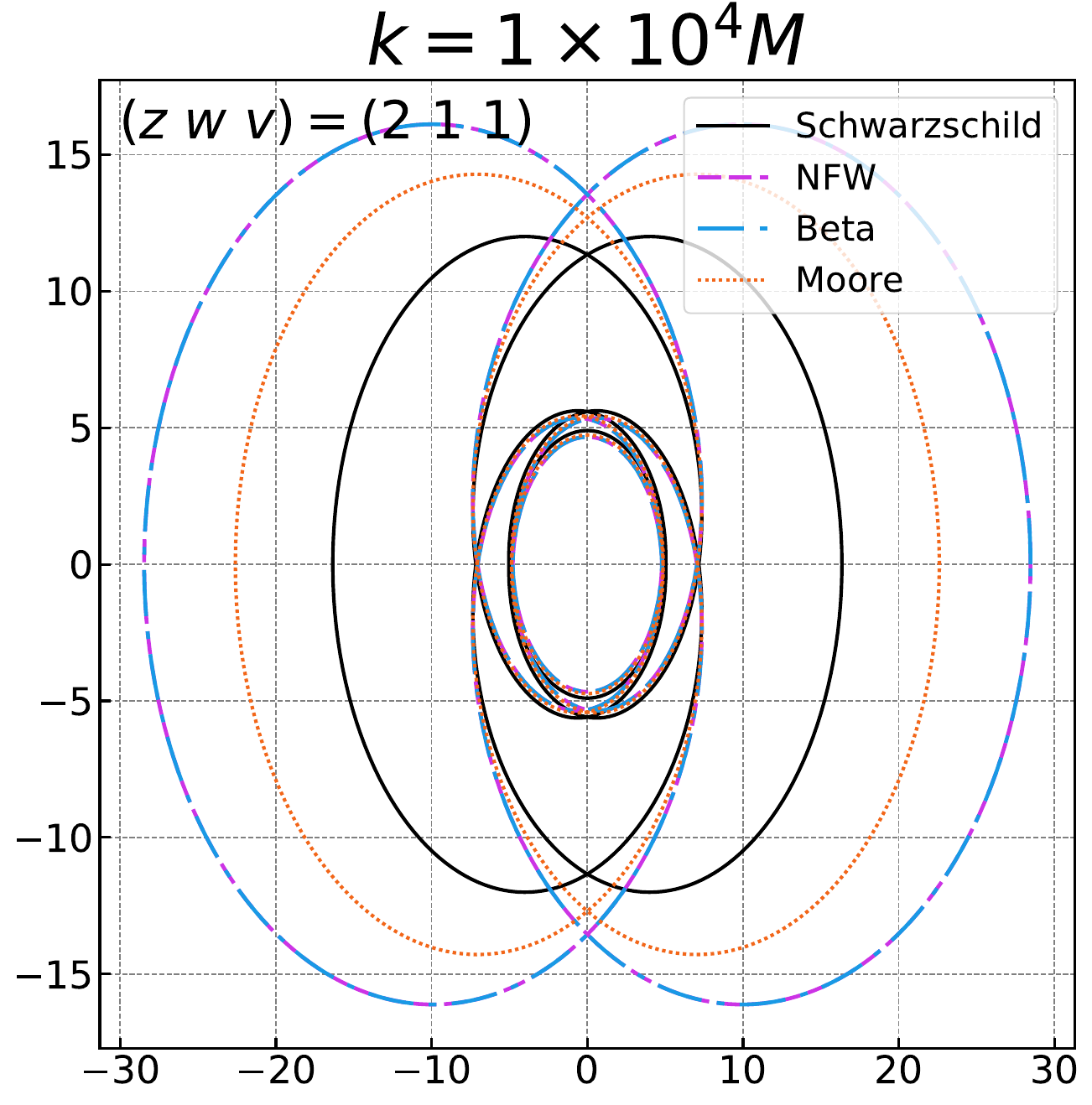}
	\includegraphics[width=0.23\linewidth]{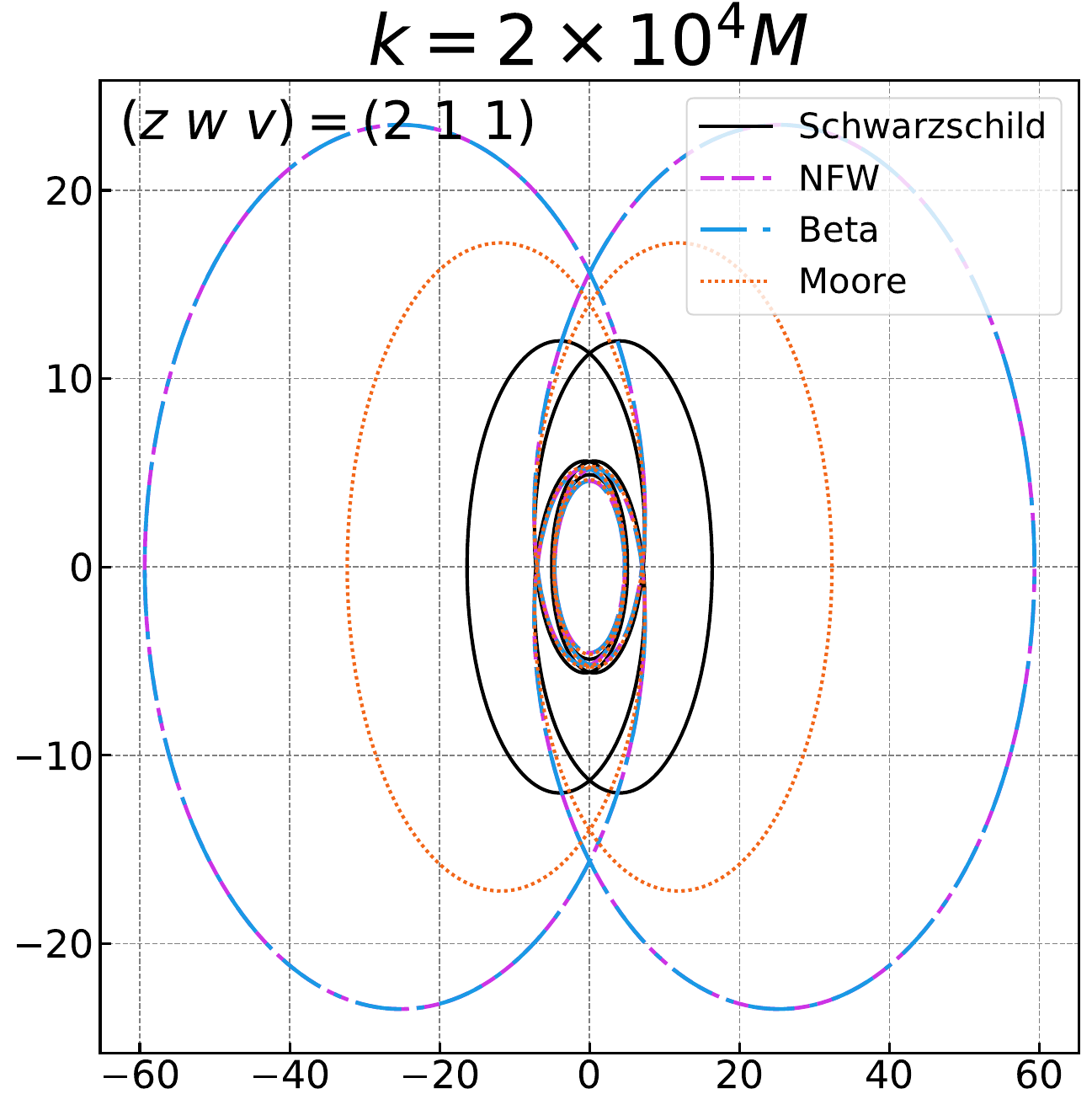}
	\end{subfigure}
	
	\vspace{0.4cm}
	
	\begin{subfigure}{\textwidth}
	\includegraphics[width=0.23\linewidth]{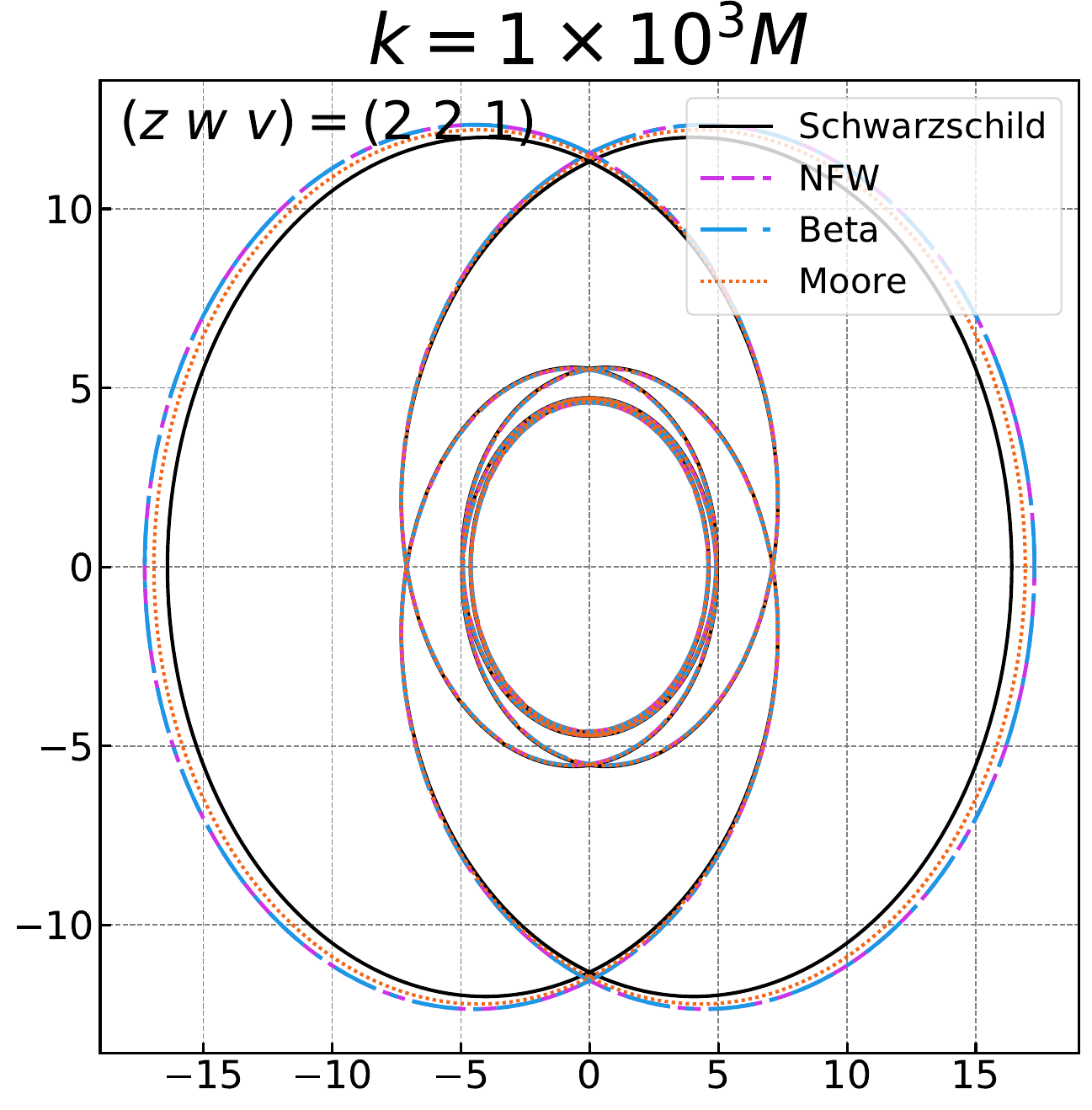}
	\includegraphics[width=0.23\linewidth]{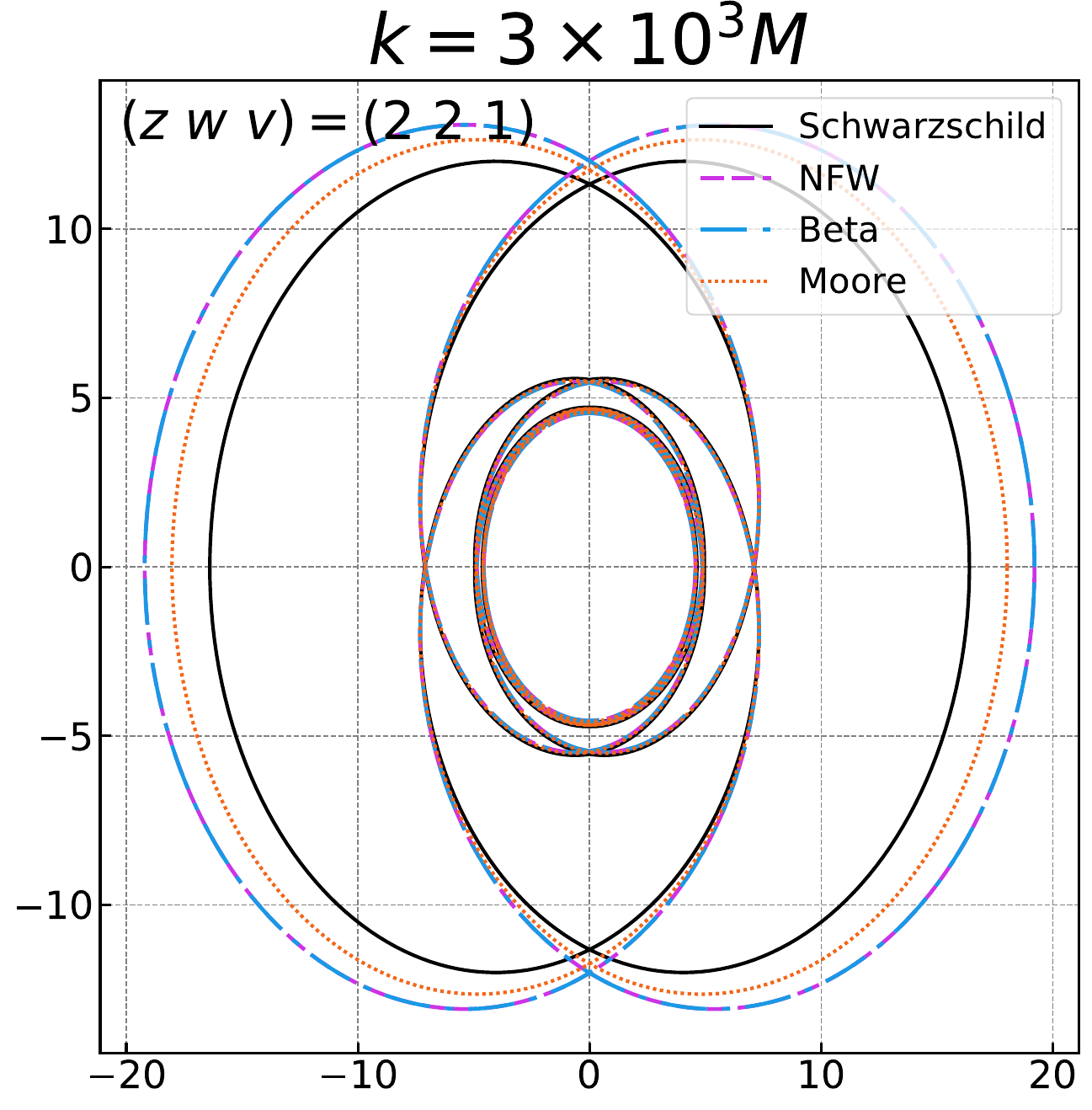}
	\includegraphics[width=0.23\linewidth]{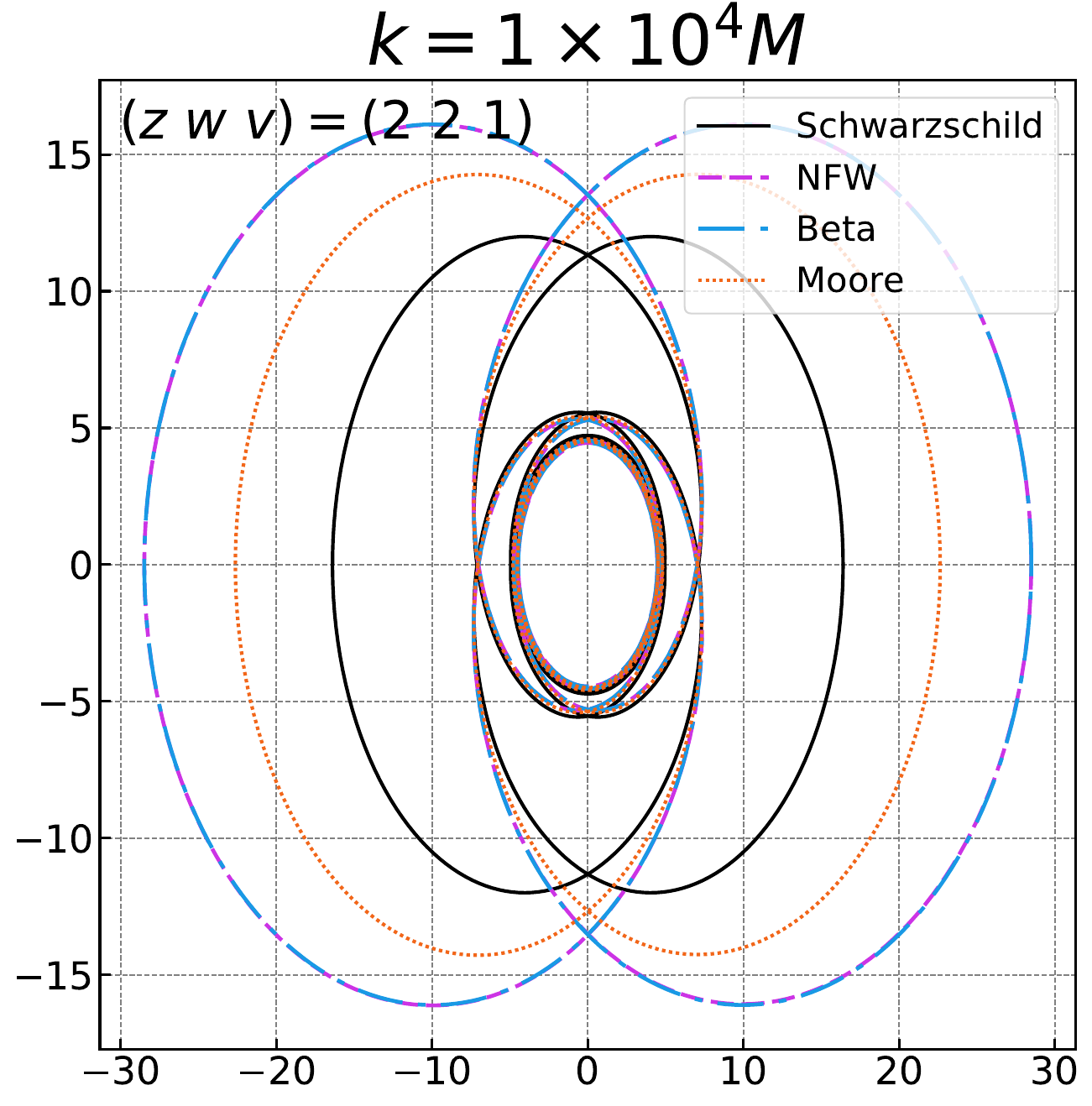}
	\includegraphics[width=0.23\linewidth]{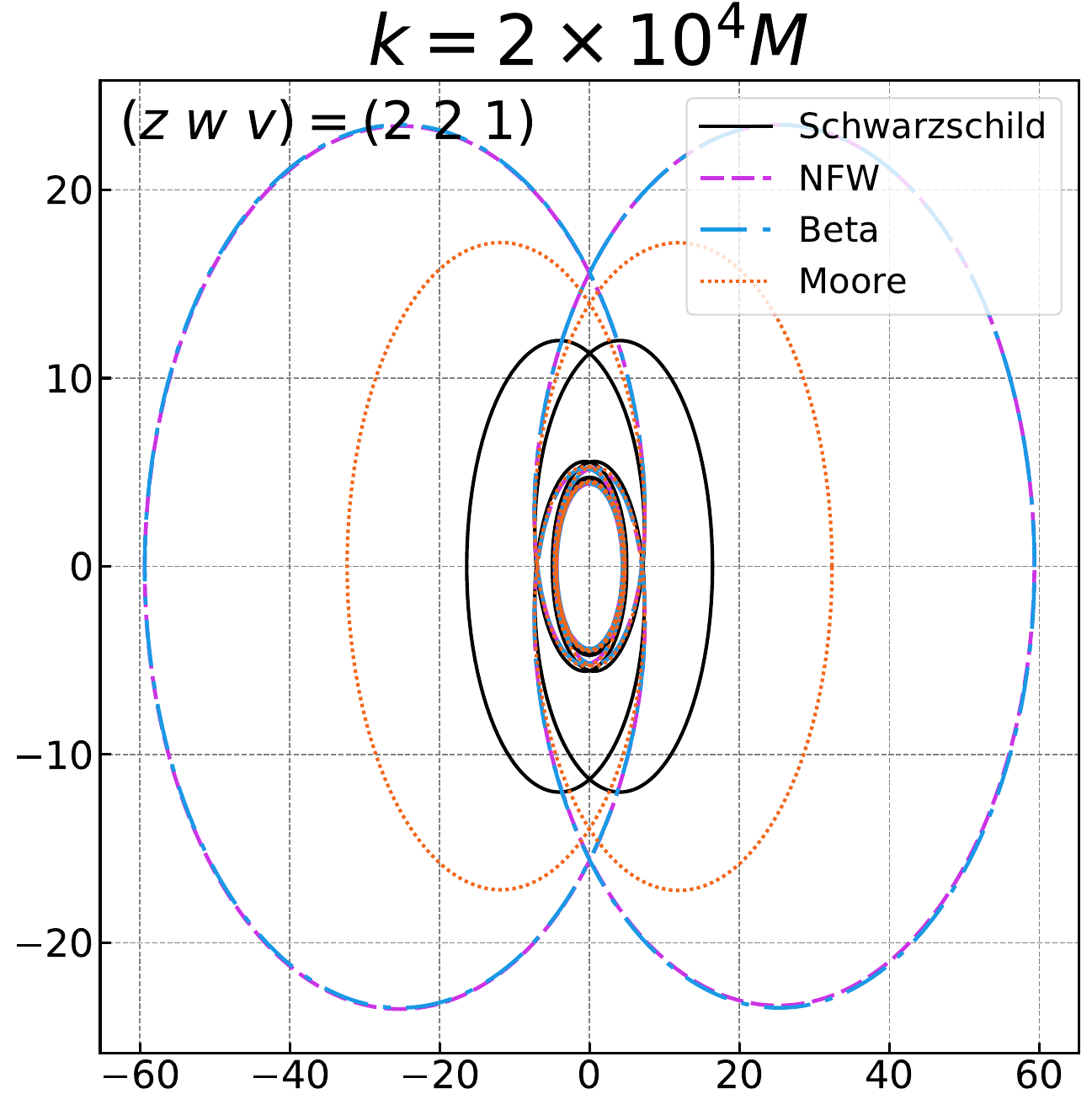}
	\end{subfigure}
	
	\caption{Periodic orbits around black holes embedded in different dark matter halos with selected energy $E=0.96$. The several panels of this figure plot the periodic orbits for different configurations $(z~w~v)$, where the dark matter mass $k$ ranges from $1 \times 10^3\,M \sim 2 \times 10^4\,M$, while the dark matter halo characteristic radius is selected at $h = 10^7\,M$.}
	\label{dif_E=E0_k}
\end{figure}

To perform our analysis, we set the energy to $E=0.96$ to study how dark matter mass affects the precession angle. Other choices of energy values lead to the same conclusions, which are not presented in the present work.  The relationship between the precession angle \(q\) and the angular momentum \(L\) for different dark matter masses \(k\) is shown in Figure~\ref{dif_k_E=E0_q_models}. The numerical results indicate that for any given precession angle $q$, increasing the dark matter mass $k$ requires higher angular momentum $L$. This is because the dark matter mass alters the effective potential of test particles, which increases the angular momentum needed to achieve a particular orbital precession. Fig.~\ref{dif_E=E0_k_q} provides detailed precession characteristics for each model at fixed dark matter masses. An important observation is that the precession angle curves for the NFW and Beta models overlap completely, even at large dark matter masses. For the same precession angle $q$, both NFW and Beta models require higher angular momentum $L$ than the Moore model.

Based on this precession angle analysis, we present the corresponding periodic orbits (shown in Fig.~\ref{dif_E=E0_k}); their angular momentum values are listed in Table~\ref{tab:k_E=E0_L}. The overall patterns and characteristics in orbital shape with different orbital configurations $(z~w~v)$ follow those observed in the fixed angular momentum case (Fig.~\ref{dif_k}). Besides, the periodic orbits calculated in the NFW and Beta models are relatively more extended than those in the Moore model. The orbital trajectories predicted by the NFW and Beta models remain identical for all dark matter masses $k$, which aligns with their coinciding precession angle curves in Fig.~\ref{dif_E=E0_k_q}. The influence of dark matter becomes increasingly evident as $k$ grows: periodic orbits around SMBH in the presence of dark matter halos diverge more substantially from those around a Schwarzschild black hole  with a increasement in the apoapsis distance. These results are consistent with the conclusions in Section~\ref{s3_1}, where we analyzed periodic orbits using the angular momentum $L = L_{\text{ISCO}} + \varepsilon(L_{\text{MBO}} - L_{\text{ISCO}})$.

\section{The Parameter $E_{(z~w~v)}$ and $L_{(z~w~v)}$ for Periodic Orbits Around Black Holes}\label{a2}

\begin{table}[b]
	\tiny
	\centering
	\begin{tabular}{lcccccc}
		\toprule
		\multicolumn{6}{c}{Energy Values} \\ \cmidrule(lr){1-6}
		\multicolumn{6}{c}{$k = 1 \times10^3M$} \\ \midrule
		Model & $E_{(1~1~0)}$ &$ E_{(1~2~0)}$ & $E_{1~3~0)}$ & $E_{(2~1~1)}$ & $E_{(2~2~1)}$ \\ 
		Schwarzschild & 0.965425 & 0.968383 & 0.968442 & 0.968026 & 0.968434 \\
		NFW & 0.964752 & 0.967716 & 0.967775 & 0.967360 & 0.967767 \\
		Beta & 0.964752 & 0.967716 & 0.967775 & 0.967360 & 0.967767 \\
		Moore & 0.965019 & 0.967981 & 0.968040 & 0.967625 & 0.968032 \\ \midrule
		\multicolumn{6}{c}{$k = 3 \times10^3M$} \\ \midrule
		Model & $E_{(1~1~0)}$ &$ E_{(1~2~0)}$ & $E_{1~3~0)}$ & $E_{(2~1~1)}$ & $E_{(2~2~1)}$ \\ 
		Schwarzschild & 0.965425 & 0.968383 & 0.968442 & 0.968026 & 0.968434 \\
		NFW  & 0.963405 & 0.966385 & 0.966443 & 0.966030 & 0.966436 \\
		Beta & 0.962058 & 0.965056 & 0.965114 & 0.964702 & 0.965106 \\
		Moore & 0.963395 & 0.966375 & 0.966433 & 0.966020 & 0.966426 \\ \midrule
		\multicolumn{6}{c}{$k = 1 \times10^4M$} \\ \midrule
		Model & $E_{(1~1~0)}$ &$ E_{(1~2~0)}$ & $E_{1~3~0)}$ & $E_{(2~1~1)}$ & $E_{(2~2~1)}$ \\ 
		Schwarzschild & 0.965425 & 0.968383 & 0.968442 & 0.968026 & 0.968434 \\
		NFW & 0.958698 & 0.961746 & 0.961802 & 0.961392 & 0.961795 \\
		Beta & 0.958698 & 0.961746 & 0.961802 & 0.961392 & 0.961795 \\
		Moore & 0.961367 & 0.964373 & 0.964430 & 0.964019 & 0.964423 \\ \midrule
		\multicolumn{6}{c}{$k = 2 \times10^{4}M$} \\ \midrule
		Model & $E_{(1~1~0)}$ &$ E_{(1~2~0)}$ & $E_{1~3~0)}$ & $E_{(2~1~1)}$ & $E_{(2~2~1)}$ \\ 
		Schwarzschild & 0.965425 & 0.968383 & 0.968442 & 0.968026 & 0.968434 \\
		NFW & 0.955001 & 0.958103 & 0.958153 & 0.957759 & 0.958147 \\
		Beta & 0.952002 & 0.955179 & 0.955233 & 0.954819 & 0.955227 \\
		Moore & 0.957319 & 0.960388 & 0.960443 & 0.960033 & 0.960437 \\ \bottomrule
	\end{tabular}
	\caption{Energy values $E_{(z~w~v)}$ of periodic orbits around SMBH calculated in different dark matter models, where angular momentum is fixed as $L=L_{\text{ISCO}} + \varepsilon (L_{\text{MBO}} - L_{\text{ISCO}})$ and dark matter halo scale is $h=10^7M$.}
	\label{tab:k_L=L0_E}
\end{table}	

\begin{table}[h]
	\tiny
	\centering
	\begin{tabular}{lcccccc}
		\toprule
		\multicolumn{6}{c}{Energy Values} \\ \cmidrule(lr){1-6}
		\multicolumn{6}{c}{$h = 10^7M$} \\ \midrule
		Model & $E_{(1~1~0)}$ &$ E_{(1~2~0)}$ & $E_{1~3~0)}$ & $E_{(2~1~1)}$ & $E_{(2~2~1)}$ \\ 
		Schwarzschild & 0.965425 & 0.968383 & 0.968442 & 0.968026 & 0.968434 \\
		NFW & 0.958697 & 0.961746 & 0.961802 & 0.961392 & 0.961795 \\
		Beta & 0.958698 & 0.961746 & 0.961802 & 0.961392 & 0.961795 \\
		Moore & 0.961367 & 0.964373 & 0.964430 & 0.964019 & 0.964423 \\ \midrule
		\multicolumn{6}{c}{$h = 10^8M$} \\ \midrule
		Model & $E_{(1~1~0)}$ &$ E_{(1~2~0)}$ & $E_{1~3~0)}$ & $E_{(2~1~1)}$ & $E_{(2~2~1)}$ \\ 
		Schwarzschild & 0.965425 & 0.968383 & 0.968442 & 0.968026 & 0.968434 \\
		NFW & 0.964752 & 0.967716 & 0.967775 & 0.967360 & 0.967767 \\
		Beta & 0.964752 & 0.967716 & 0.967775 & 0.967360 & 0.967767 \\
		Moore & 0.965018 & 0.967980 & 0.968039 & 0.967624 & 0.968031 \\ \midrule
		\multicolumn{6}{c}{$h = 10^9M$} \\ \midrule
		Model & $E_{(1~1~0)}$ &$ E_{(1~2~0)}$ & $E_{1~3~0)}$ & $E_{(2~1~1)}$ & $E_{(2~2~1)}$ \\ 
		Schwarzschild & 0.965425 & 0.968383 & 0.968442 & 0.968026 & 0.968434 \\
		NFW & 0.965358 & 0.968316 & 0.968375 & 0.967960 & 0.968368 \\
		Beta & 0.965358 & 0.968316 & 0.968375 & 0.967960 & 0.968368 \\
		Moore & 0.965385 & 0.968342 & 0.968402 & 0.967986 & 0.968394 \\ \midrule
		\multicolumn{6}{c}{$h = 10^{10}M$} \\ \midrule
		Model & $E_{(1~1~0)}$ &$ E_{(1~2~0)}$ & $E_{1~3~0)}$ & $E_{(2~1~1)}$ & $E_{(2~2~1)}$ \\ 
		Schwarzschild & 0.965425 & 0.968383 & 0.968442 & 0.968026 & 0.968434 \\
		NFW & 0.965419 & 0.968376 & 0.968435 & 0.968020 & 0.968428 \\
		Beta & 0.965419 & 0.968376 & 0.968435 & 0.968020 & 0.968428 \\
		Moore & 0.965421 & 0.968379 & 0.968438 & 0.968022 & 0.968430 \\ \bottomrule
	\end{tabular}
	\caption{$E _{(z~w~v)}$ of periodic orbits around SMBH obtained in different dark matter models, where angular momentum is fixed as $L=L_{\text{ISCO}} + \varepsilon (L_{\text{MBO}} - L_{\text{ISCO}})$ and dark matter mass is  \( k = 10^4M \).}
	\label{tab:h_L=L0_E}
\end{table}

\begin{table}[h]
	\tiny
	\centering
	\begin{tabular}{lcccccc}
		\toprule
		\multicolumn{6}{c}{Angular Momentum Values} \\ \cmidrule(lr){1-6}
		\multicolumn{6}{c}{$k = 1 \times10^3M$} \\ \midrule
		Model & $L_{(1~1~0)}$ & $L_{(1~2~0)}$ & $L_{(1~3~0)}$ & $L_{(2~1~1)}$ & $L_{(2~2~1)}$ \\
		Schwarzschild & 3.683589 & 3.653406 & 3.652581 & 3.657596 & 3.652701 \\
		NFW & 3.704068 & 3.674259 & 3.673477 & 3.678331 & 3.673589 \\
		Beta & 3.704068 & 3.674259 & 3.673477 & 3.678331 & 3.673589 \\
		Moore & 3.695949 & 3.665997 & 3.665198 & 3.670114 & 3.665311 \\
		\midrule
		\multicolumn{6}{c}{$k = 3 \times10^3M$} \\ \midrule
		Model & $L_{(1~1~0)}$ & $L_{(1~2~0)}$ & $L_{(1~3~0)}$ & $L_{(2~1~1)}$ & $L_{(2~2~1)}$ \\
		Schwarzschild & 3.683589 & 3.653406 & 3.652581 & 3.657596 & 3.652701 \\
		NFW & 3.745094 & 3.715932 & 3.715221 & 3.719798 & 3.715319 \\
		Beta & 3.745094 & 3.715932 & 3.715221 & 3.719798 & 3.715319 \\
		Moore & 3.720693 & 3.691164 & 3.690413 & 3.695147 & 3.690518 \\
		\midrule
		\multicolumn{6}{c}{$k = 1 \times10^4M$} \\ \midrule
		Model & $L_{(1~1~0)}$ & $L_{(1~2~0)}$ & $L_{(1~3~0)}$ & $L_{(2~1~1)}$ & $L_{(2~2~1)}$ \\
		Schwarzschild & 3.683589 & 3.653406 & 3.652581 & 3.657596 & 3.652701 \\
		NFW & 3.889576 & 3.861881 & 3.861333 & 3.865253 & 3.861401 \\
		Beta & 3.889576 & 3.861881 & 3.861333 & 3.865253 & 3.861401 \\
		Moore & 3.807581 & 3.779198 & 3.778571 & 3.782813 & 3.778654 \\
		\midrule
		\multicolumn{6}{c}{$k = 2 \times10^{4}M$} \\ \midrule
		Model & $L_{(1~1~0)}$ & $L_{(1~2~0)}$ & $L_{(1~3~0)}$ & $L_{(2~1~1)}$ & $L_{(2~2~1)}$ \\
		Schwarzschild & 3.683589 & 3.653406 & 3.652581 & 3.657596 & 3.652701 \\
		NFW & 4.099288 & 4.072265 & 4.071828 & 4.075284 & 4.071878 \\
		Beta & 4.099288 & 4.072266 & 4.071828 & 4.075284 & 4.071878 \\
		Moore & 3.932690 & 3.905278 & 3.904760 & 3.908547 & 3.904824 
		\\ \bottomrule
	\end{tabular}
	\caption{Angular Momentumy values $L _{(z~w~v)}$ of periodic orbits around SMBH obtained in different dark matter models when $E$ is selected as $E=0.96$ and dark matter halo scale is \( h = 10^7M \).}
	\label{tab:k_E=E0_L}
\end{table}

As established in Section~\ref{s3}, the characteristics of a periodic orbit, identified by its zoom-whirl-vertex numbers $(z~w~v)$, are determined by its energy $E_{(z~w~v)}$ and angular momentum $L_{(z~w~v)}$. In Sections~\ref{s3_1} and~\ref{s3_2}, we systematically analyze the influence of the dark matter halo by varying the mass parameter $k$ (from $1\times 10^3M \sim 2 \times 10^4 M$) and the scale parameter $h$ (from $10^7 M \sim 10^{10} M$), while keeping the angular momentum fixed at $L = L_{\text{ISCO}} + \varepsilon(L_{\text{MBO}} - L_{\text{ISCO}})$ with $\varepsilon = 0.5$ selected as a representative case to present numerical results on periodic orbits. The corresponding energy values $E_{(z~w~v)}$ for the periodic orbits shown in Figs.~\ref{dif_k} and~\ref{dif_h} are listed in Tables~\ref{tab:k_L=L0_E} and~\ref{tab:h_L=L0_E}, respectively. Furthermore, an alternative analysis is conducted in Appendix~\ref{a2} with a selected energy of $E = 0.96$. For this case, Table~\ref{tab:k_E=E0_L} provides the angular momentum values $L_{(z~w~v)}$ for the periodic orbits plotted in Fig.~\ref{dif_E=E0_k} for various dark matter models and different dark matter mass parameter $k$.

\section{The Stability of Periodic Orbits Under the Long-Term Orbital Evolution}\label{a3} 

In the main text, the periodic orbits and their gravitational wave signatures are studied assuming short-term observation, in which neither the dynamical evolution of the orbits nor dissipative effects are taken into consideration, and the orbital energy and angular momentum are treated as conserved quantities. However, in any realistic astrophysical EMRI system, the orbits of small celestial bodies are dynamical rather than static. There exist a number of mechanisms that could change the energy and angular momentum of bound orbits. For instance, the continuous emission of gravitational waves leads to a reduction in energy and angular momentum. The accumulated change in energy and angular momentum would break the periodicity of orbits. Therefore, it would be interesting to see how long these periodic orbits can retain their periodicity under long-term observation. This would significantly influence the observational detectability of periodic orbits in realistic EMRI systems. In this appendix, we provide a quantitative analysis of the long-term orbital evolution of these periodic orbits, with gravitational radiation effects incorporated into the orbital dynamics.

The adiabatic approximation provides a simple framework to study the dynamical evolution of orbits in EMRIs, especially when the orbital evolution timescale is significantly longer than the orbital period of the system~\cite{Hughes:1999bq,Hughes:2001jr,Glampedakis:2002ya}. In this approximation, the long-term evolution of the EMRI system is governed by the slow and continuous variation of the orbital energy $E$ and angular momentum $L$. The trajectory of the small celestial object can therefore be approximated as a sequence of instantaneous geodesic orbits. In particular, the loss of orbital energy and angular momentum caused by gravitational radiation can be evaluated using the standard quadrupole emission formulas:
\begin{subequations}
\begin{align}
	\left( \frac{dE}{dt} \right)_{\text{GW}} &= -\frac{1}{5} \dddot{\mathcal{I}}^{jk} \dddot{\mathcal{I}}^{jk}, \\
	\left( \frac{dL_i}{dt} \right)_{\text{GW}} &= -\frac{2}{5} \epsilon_{ijk} \ddot{\mathcal{I}}^{jl} \dddot{\mathcal{I}}^{kl},
\end{align}
\end{subequations}
where $\mathcal{I}^{jk}$ denotes the symmetric-traceless part of the system's mass quadrupole moment. By substituting the small celestial body's orbits into the quadrupole moment expression, one can calculate the instantaneous energy and angular momentum fluxes. Averaging these fluxes over one orbital period $T$ yields the secular rates of change for the orbital parameters, which can be explicitly expressed as:
\begin{subequations}
\begin{align}
	\left\langle \frac{dE}{dt} \right\rangle_{\text{GW}} &= \frac{1}{T} \int_{0}^{T} \left( \frac{dE}{dt} \right)_{\text{GW}} dt, \\
	\left\langle \frac{dL}{dt} \right\rangle_{\text{GW}} &= \frac{1}{T} \int_{0}^{T} \left( \frac{dL}{dt} \right)_{\text{GW}} dt.
\end{align}
\end{subequations}
The dynamic evolution of the system is accomplished by updating the energy and angular momentum at each time step ($E(t_{i}) \to E(t_{i+1})$, $L(t_{i}) \to L(t_{i+1})$). 

However, deriving the analytical time-averaged formula for EMRIs in the presence of dark matte halos is highly challenging. Instead, one can approximate this evolution using the conventional Peters-Mathews formula, and view the dark matter effect as a perturbation effect (which changes the orbital eccentricity and semi-latus rectum). In this way, the rates of change for the orbital parameters $(E, L)$ are calculated through \cite{Peters:1963ux,Blanchet:2013haa}
\begin{subequations}
\begin{align}
	\left\langle \frac{dE}{dt} \right\rangle_{\text{GW}} & \approx -\frac{32}{5} \frac{\mu M^3}{a^5} \big(1-e^2\big)^{\frac{3}{2}} \bigg( 1 + \frac{73}{24} e^2 + \frac{37}{96} e^4 \bigg)
	, \\
	\left\langle \frac{dL}{dt} \right\rangle_{\text{GW}} & \approx -\frac{32}{5} \frac{\mu M^{5/2}}{a^{7/2}} \big(1-e^2\big)^{\frac{3}{2}} \bigg( 1 + \frac{7}{8}e^2 \bigg) , 
\end{align}
\end{subequations}
with $\mu=\frac{mM}{m+M}$ to be the reduced mass of system \footnote{In our notation, the parameters $E$ and $L$ are defined to be the orbital energy and angular momentum per unit mass. Therefore, the Peters-Mathews formula presented here drops a factor of $mu$, compared with the formula given in literature.}. A periodic orbit characterized by $(z~w~v)$ can be perfectly decomposed into $z$ distinct leaves (see Fig. \ref{( z w v )}), and each individual leaf exhibits a precessing trajectory that can be parameterized by
\begin{equation}
	r = \frac{p}{1+e\cdot\cos[\chi(\phi)]} ,
\end{equation}
where $0 \le \chi(\phi) \le 2\pi$ is a Keplerian-like parameterization for the precessing trajectory. The corresponding orbital eccentricity and semi-latus rectum can be defined as 
\begin{equation}
e \equiv \frac{r_{\text{max}}-r_{\text{min}}}{r_{\text{max}}+r_{\text{min}}} , 
\ \ \ \ \ \ 
p \equiv \frac{2r_{\text{max}}r_{\text{min}}}{r_{\text{max}}+r_{\text{min}}} .  \label{eccentricity}
\end{equation}
For EMRIs in dark matter environments, the time evolution of orbital energy and angular momentum can be calculated using the following scheme. Starting from a periodic orbit with a specific orbital configuration $(z~w~v)$, its orbital energy and angular momentum are taken as the initial values $E_{0}$, $L_{0}$ for the evaluation. One can extract the maximal and minimal radius $r_{\text{max}}$ and $r_{\text{min}}$ for this periodic orbit to determine the initial orbital eccentricity $e_{0}$ and semi-latus rectum $p_{0}$. These quantities are used to compute the time-averaged flux $\left\langle \frac{dE}{dt} \right\rangle$ and $\left\langle \frac{dL}{dt} \right\rangle$ at the initial step, with the orbital period $T$ of periodic orbits obtained numerically. At any subsequent time step $t_{i}$, one can numerically solve the orbit's trajectory using the instantaneous orbital parameters $E(t_{i})$ and $L(t_{i})$ (the orbit is periodic at the initial step but is not necessary to be periodic in subsequent steps). From this trajectory, the orbital eccentricity $e(t_{i})$ and semi-latus rectum $p(t_{i})$ can be extracted using Eq. (\ref{eccentricity}) (the dark matter effects have been incorporated into these parameters), which are necessary for evaluating the time-averaged fluxes at this time step. Furthermore, the period $T$ is not a constant during the time evolution process, and it must be calculated numerically at each step. Particularly, for non-periodic orbits that emerged in subsequent steps, $T$ is defined as the time interval for a celestial body moving from one apoapsis to the next apoapsis. The dynamical evolution of the system is accomplished by updating the energy and angular momentum at each time step 
\begin{subequations}
\begin{align}
	E(t_{i+1}) &= E(t_{i}+T_{i}) = E(t_{i}) + T_{i} \cdot \left\langle \frac{dE}{dt} \right\rangle_{\text{GW}}, \\ 
	L(t_{i+1}) &= L(t_{i}+T_{i}) = L(t_{i}) + T_{i} \cdot \left\langle \frac{dL}{dt} \right\rangle_{\text{GW}}.
\end{align}
\end{subequations}
By following this procedure, the entire time evolution is carried out by applying a sufficiently large number of steps.

\begin{figure}
	\centering
	
	\begin{minipage}{\linewidth}
		\centering
		\includegraphics[width=0.495\textwidth]{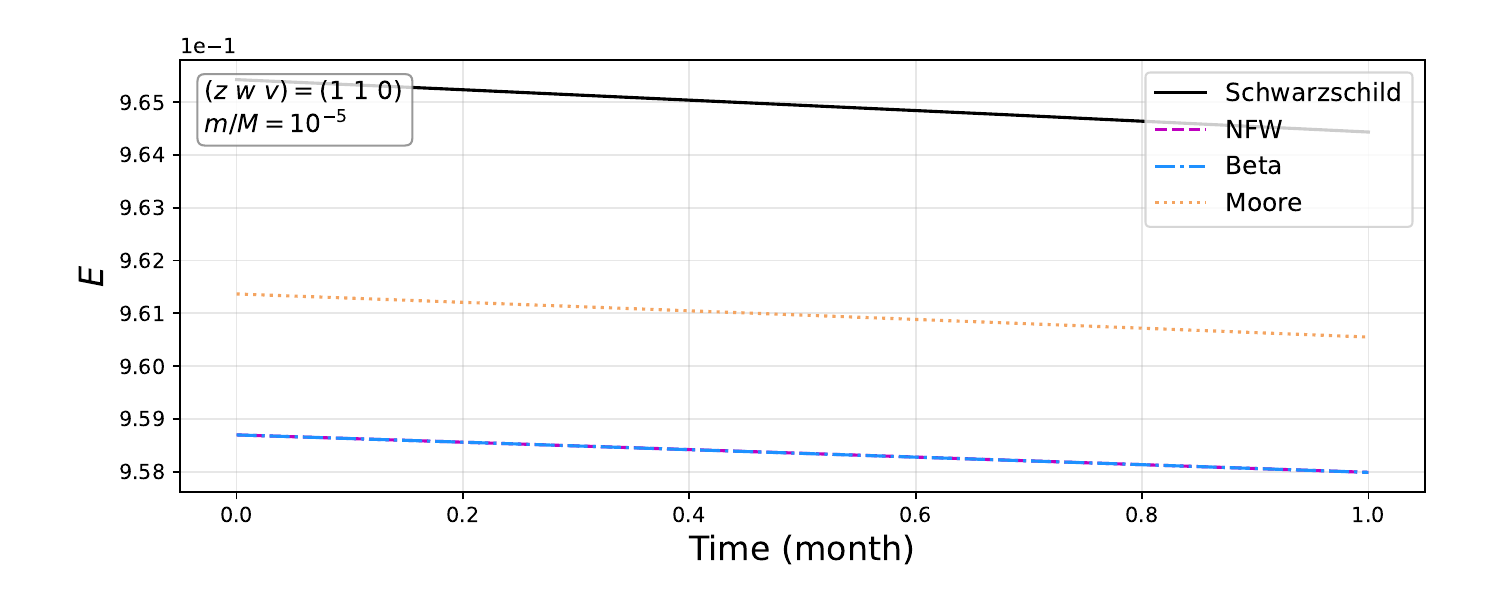}
		\includegraphics[width=0.495\textwidth]{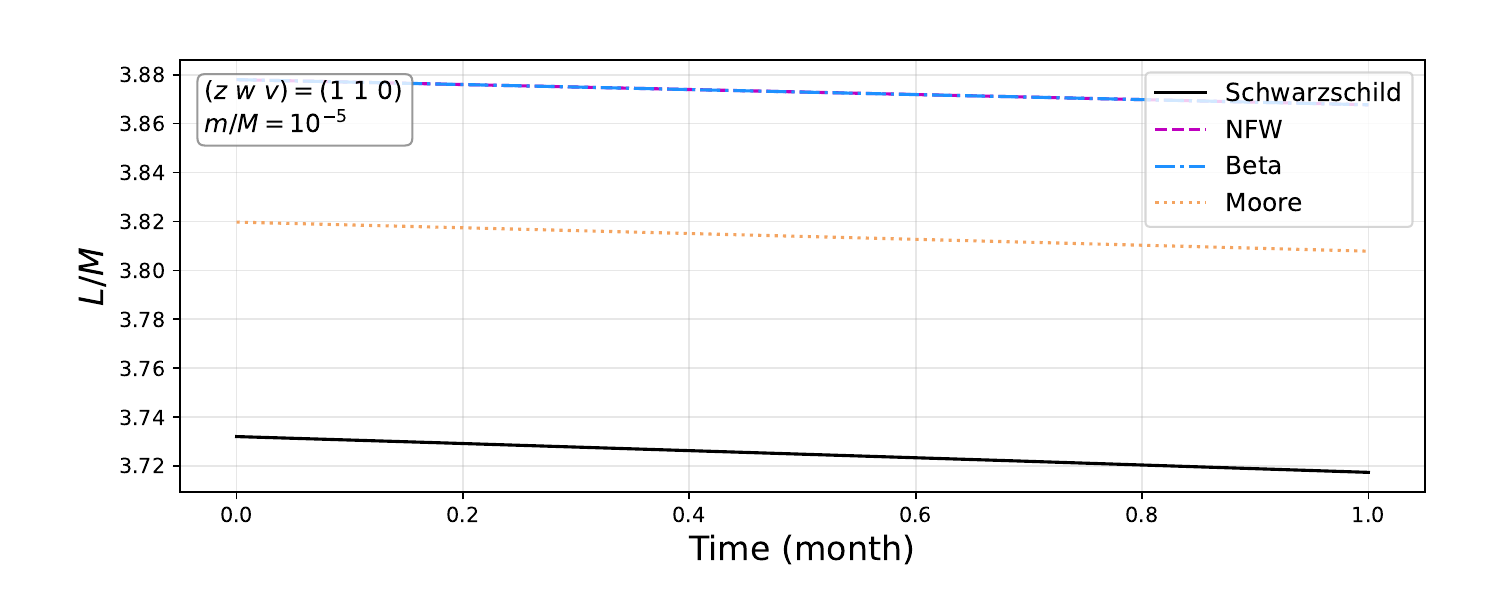}
	\end{minipage}
	
	\vspace{0.3cm}
	
	\begin{minipage}{\linewidth}
		\centering
		\includegraphics[width=0.495\textwidth]{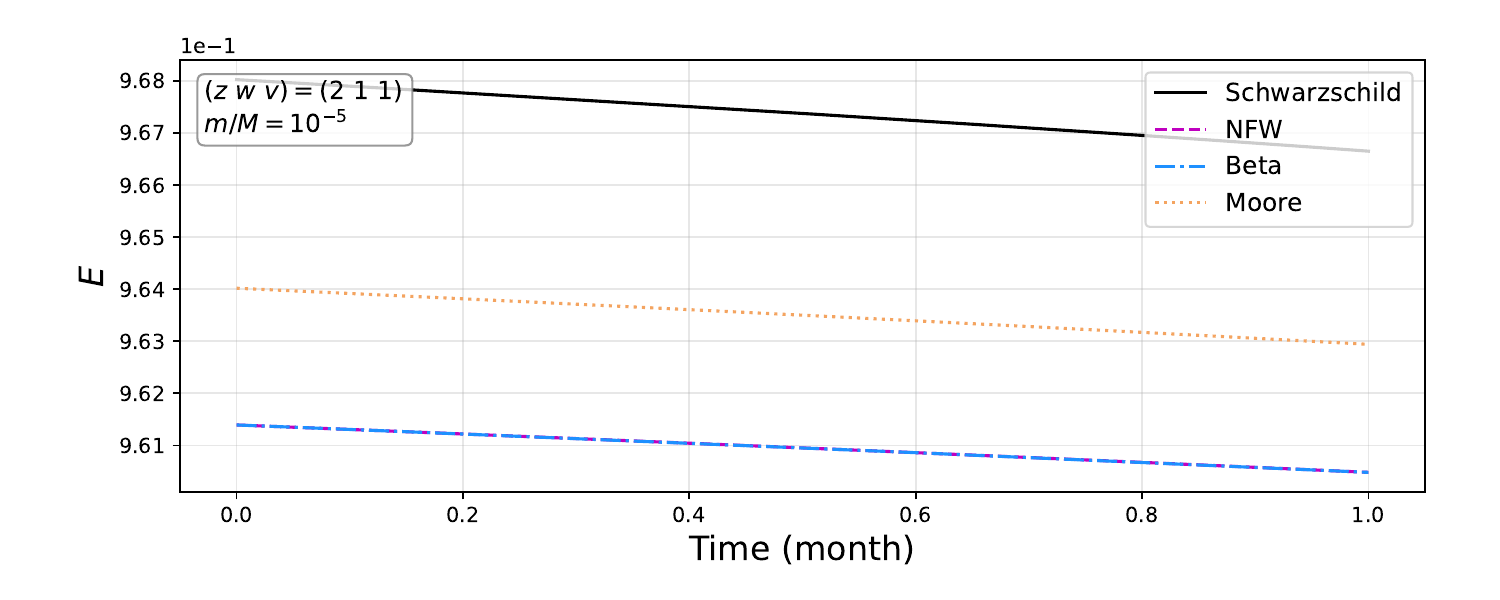}
		\includegraphics[width=0.495\textwidth]{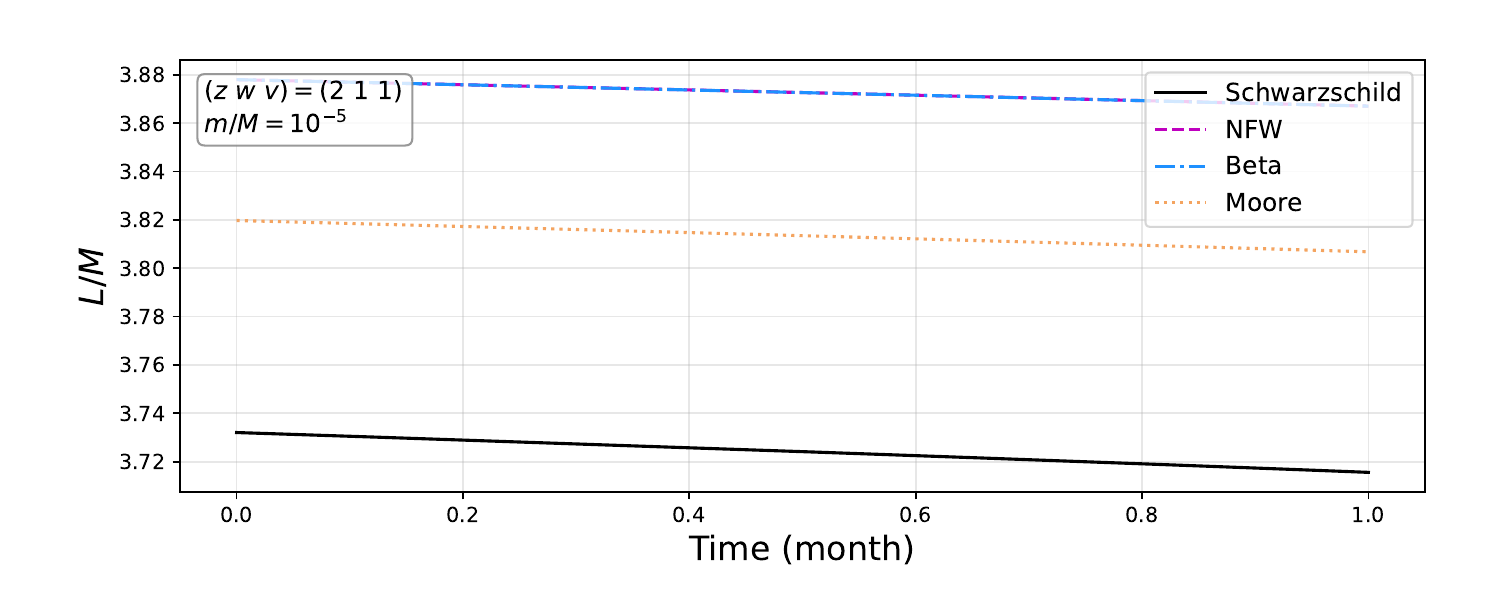}
	\end{minipage}
	
	\vspace{0.3cm}
	
	\begin{minipage}{\linewidth}
		\centering
		\includegraphics[width=0.495\textwidth]{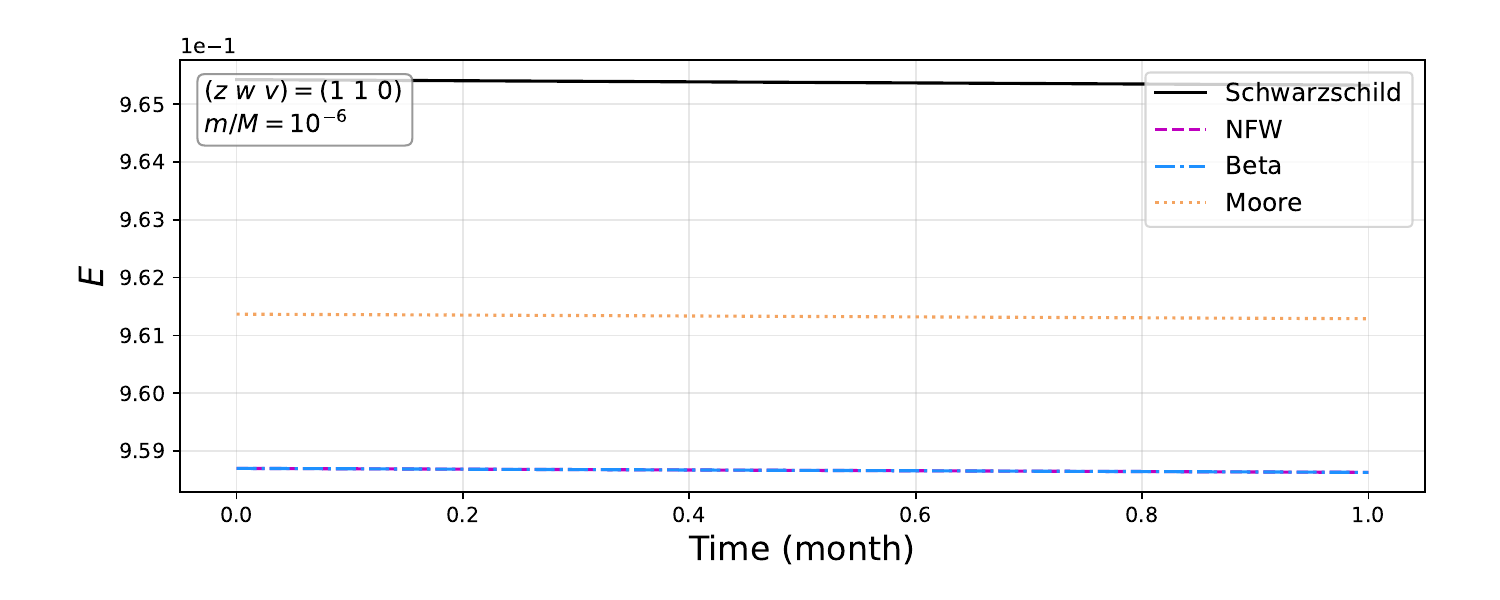}
		\includegraphics[width=0.495\textwidth]{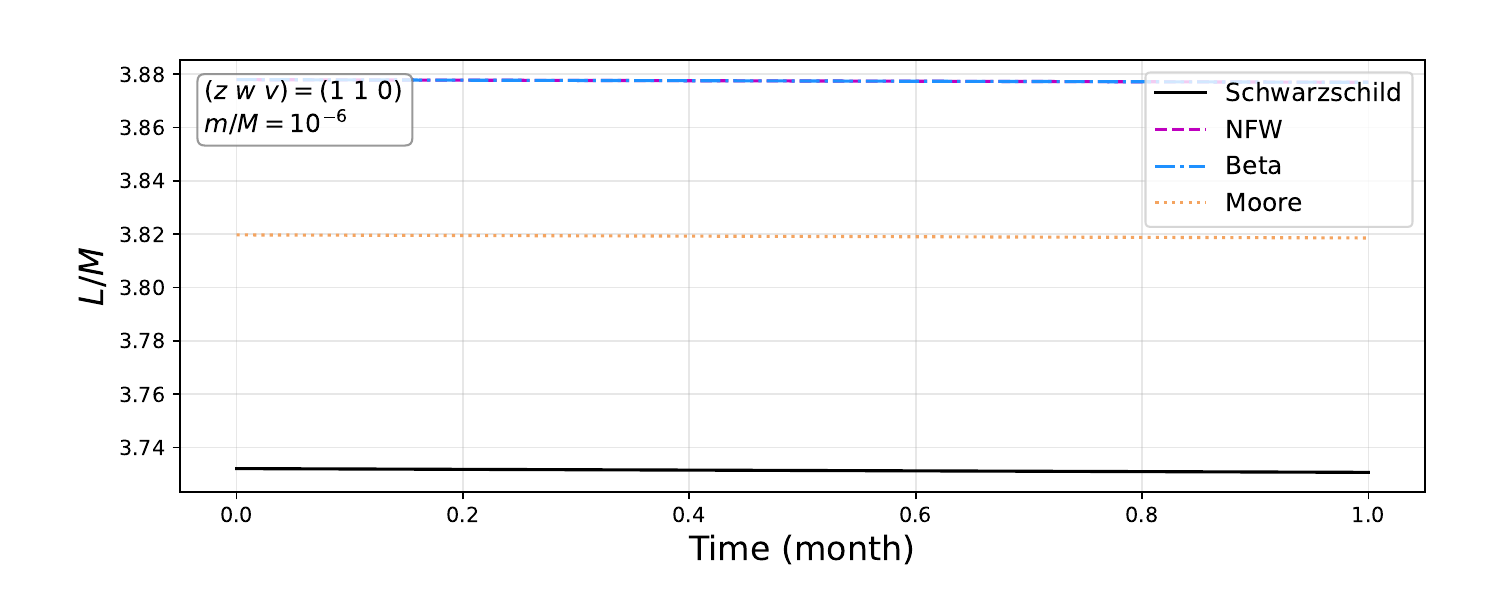}
	\end{minipage}
	
	\vspace{0.3cm}
	
	\begin{minipage}{\linewidth}
		\centering
		\includegraphics[width=0.495\textwidth]{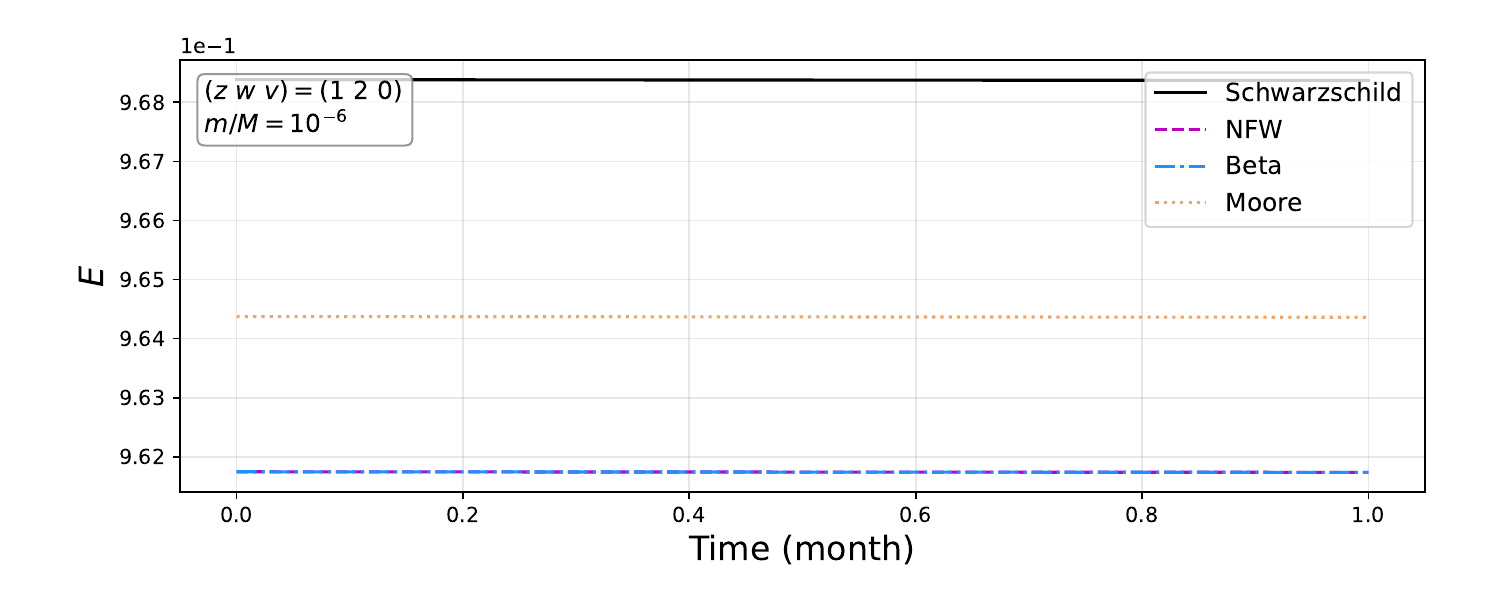}
		\includegraphics[width=0.495\textwidth]{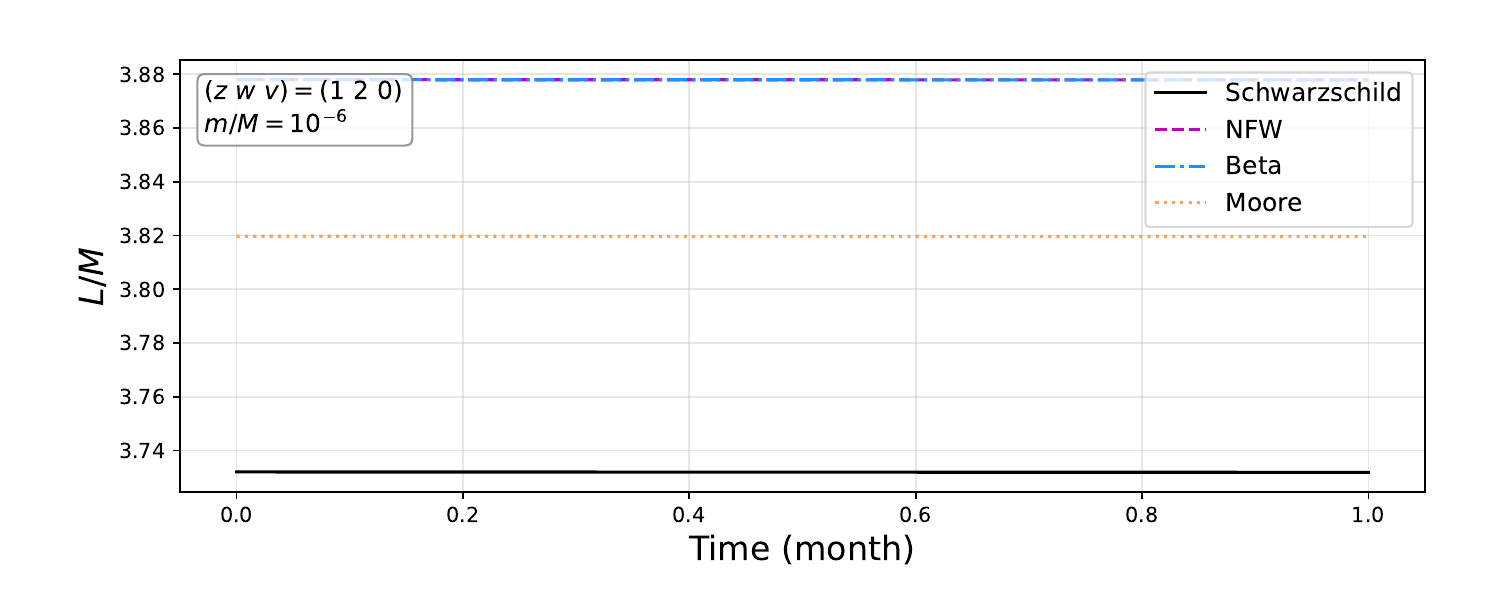}
	\end{minipage}
	
	\vspace{0.3cm}
	
	\begin{minipage}{\linewidth}
		\centering
		\includegraphics[width=0.495\textwidth]{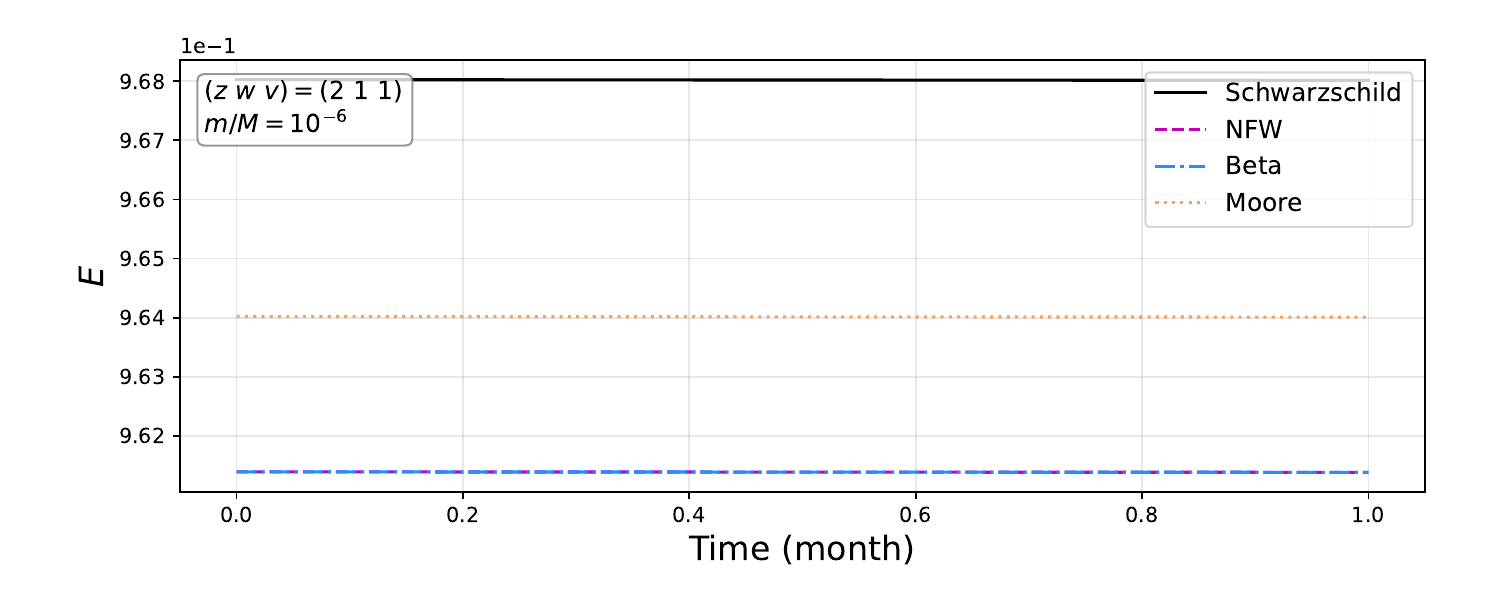}
		\includegraphics[width=0.495\textwidth]{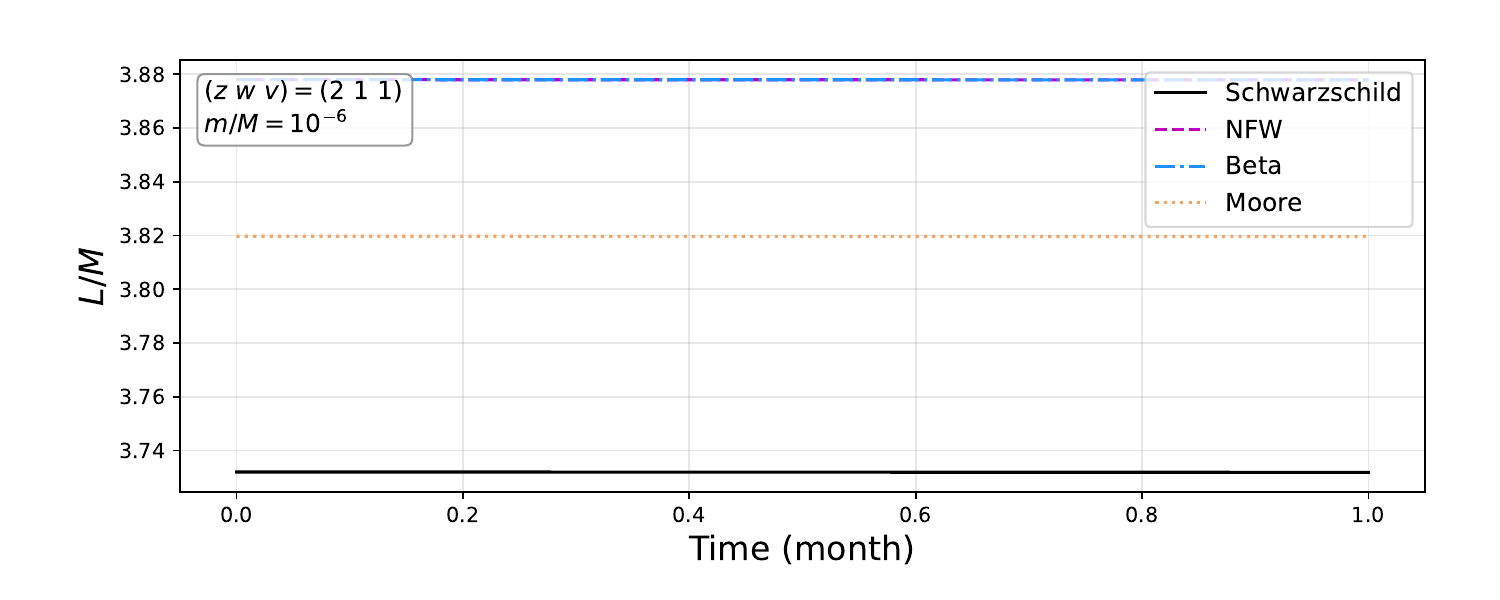}
	\end{minipage}
	
	\caption{Time evolution of orbital energy $E$ (left column) and angular momentum $L$ (right column) over a one-month period under the influence of gravitational radiation reaction, with a dark matter halo mass $k=10^4 M$ and characteristic radius $h=10^7 M$, with the small celestial body's mass fixed at $m=10 M_{\odot}$. First Row: the evolution of orbital parameters for the (1~1~0) configuration with a mass ratio $m/M=10^{-5}$. Second Row: the evolution of orbital parameters for the (2~1~1) configuration with a mass ratio $m/M=10^{-5}$. Third row: the evolution of orbital parameters for the (1~1~0) configuration with a mass ratio $m/M=10^{-6}$. Fourth row: the evolution of orbital parameters for the (1~2~0) configuration with a mass ratio $m/M=10^{-6}$. Fifth row: the evolution of orbital parameters for the (2~1~1) configuration with a mass ratio $m/M=10^{-6}$.}
	\label{fig:EL_evolution}
\end{figure}

To show the quantitative results on the long-term variation of orbital energy and angular momentum, we present their time evolution over a one-month timescale in Fig.~\ref{fig:EL_evolution} for several distinct orbital configurations: (1~1~0), (1~2~0) and (2~1~1). These results are calculated for two selected typical mass ratios: $m/M=10^{-5}$ and $m/M=10^{-6}$. The initial values $(E_0, L_0)$ are set according to their respective periodic orbit configurations, where we adopt $L_0 = L_{\mathrm{ISCO}} + \varepsilon(L_{\mathrm{MBO}} - L_{\mathrm{ISCO}})$ with $\varepsilon = 1/2$, and $E_0$ is taken from Table~\ref{tab:k_L=L0_E}. Over the one-month timescale, a steady, monotonic decrease in both $E$ and $L$ is clearly evident in all cases, which simply indicates that the gravitational rediation acts as a continuous dissipation mechanism for the long-term orbital evolution. From Fig.~\ref{fig:EL_evolution}, we notice that the loss of energy and angular momentum becomes more pronounced when periodic orbital configuration $(z~w~v)$ becomes more complex and the corresponding precession parameter $q=w+\frac{v}{z}$ increases. 
Furthermore, the numerical results in Fig.~\ref{fig:EL_evolution} demonstrate that the changes in $E$ and $L$ proceed more slowly for the more extreme mass ratio $m/M=10^{-6}$ than those for $m/M=10^{-5}$. In particular, for the periodic orbits with orbital configuration (1~1~0) in an EMRI system with $m/M=10^{-6}$, the orbital energy and angular momentum remain nearly constant over a one-month observational timescale.

\begin{figure}
	\centering
	
	\begin{minipage}{\linewidth}
		\centering
		\includegraphics[width=0.3\textwidth,height=0.25\textwidth]{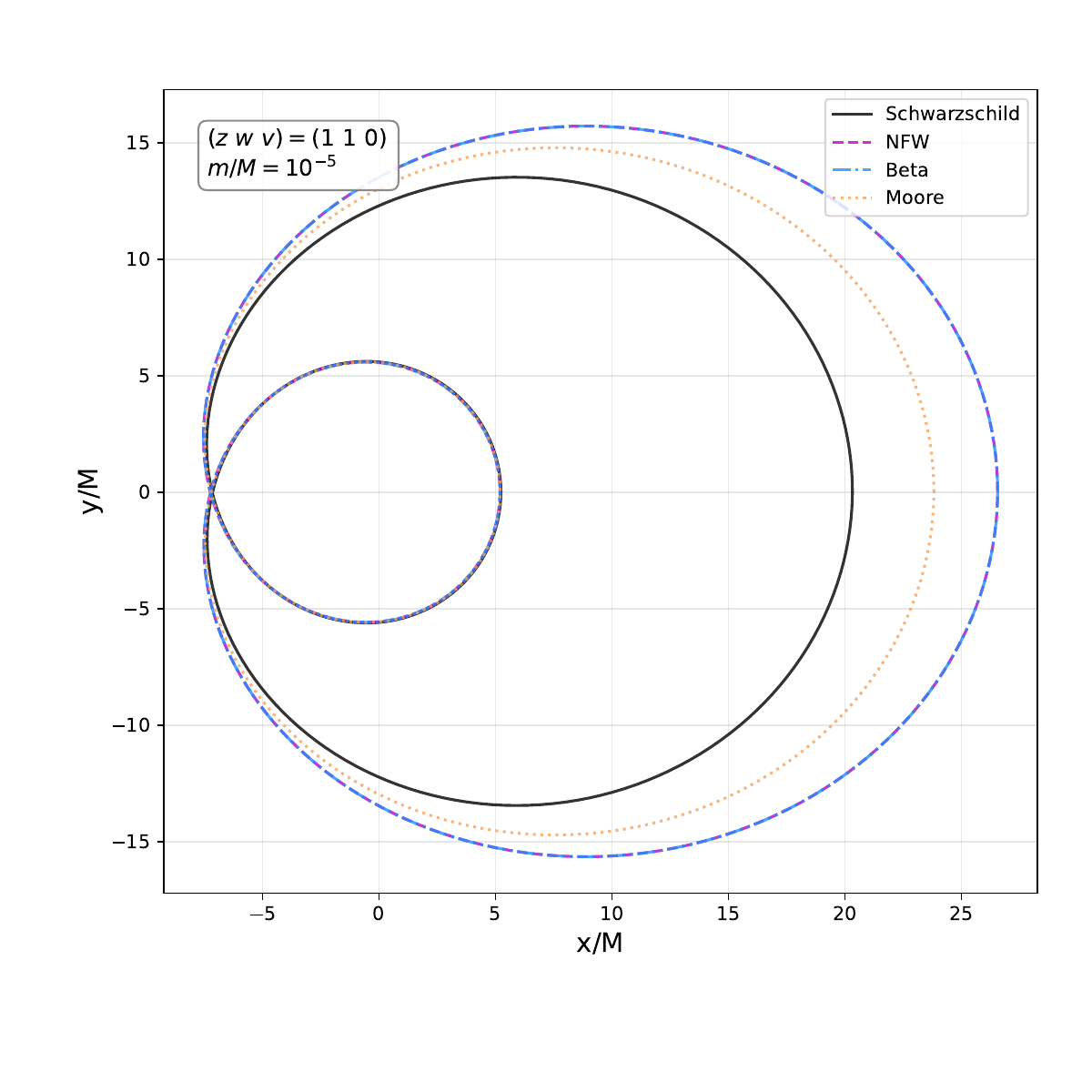}
		\includegraphics[width=0.3\textwidth,height=0.25\textwidth]{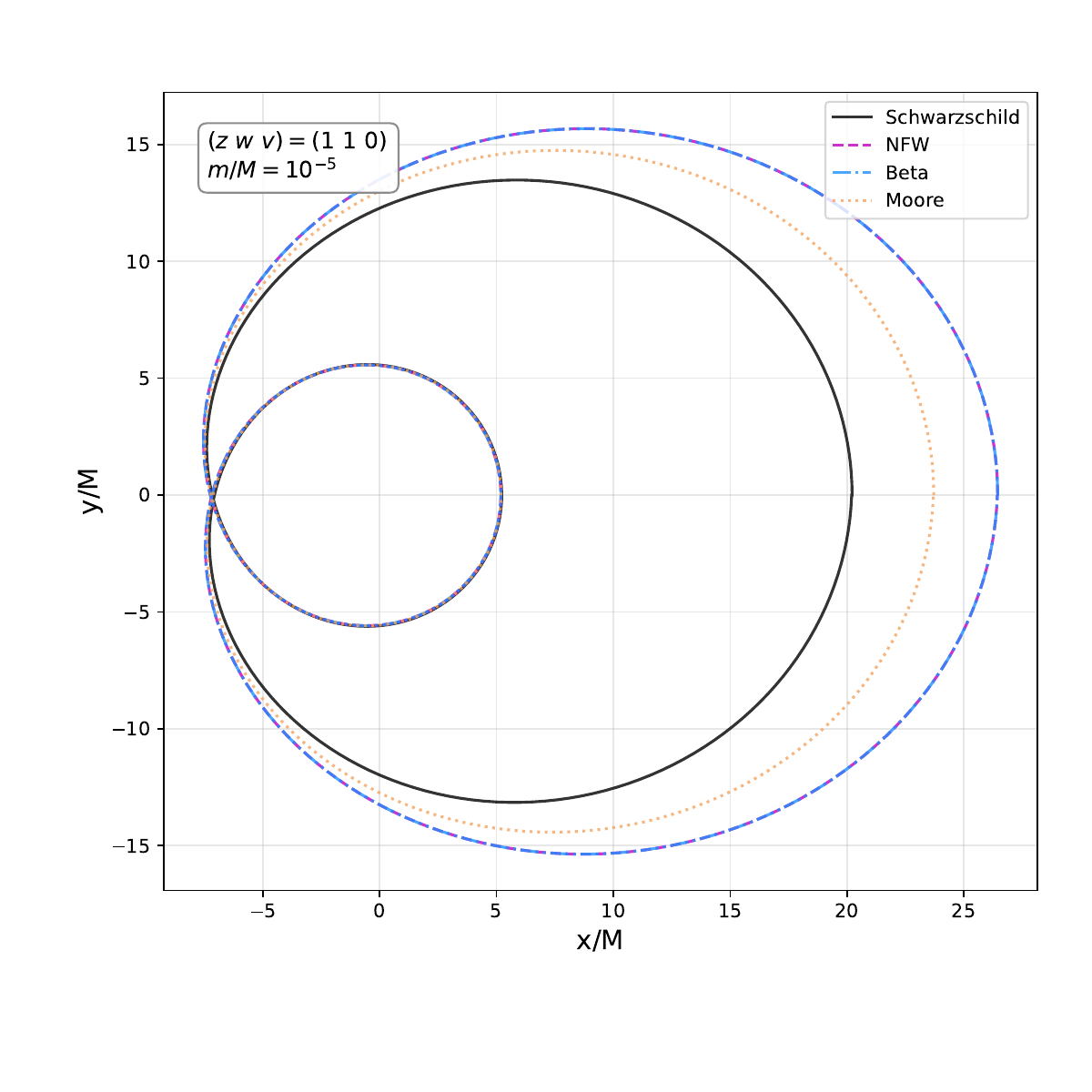}
		\includegraphics[width=0.3\textwidth,height=0.25\textwidth]{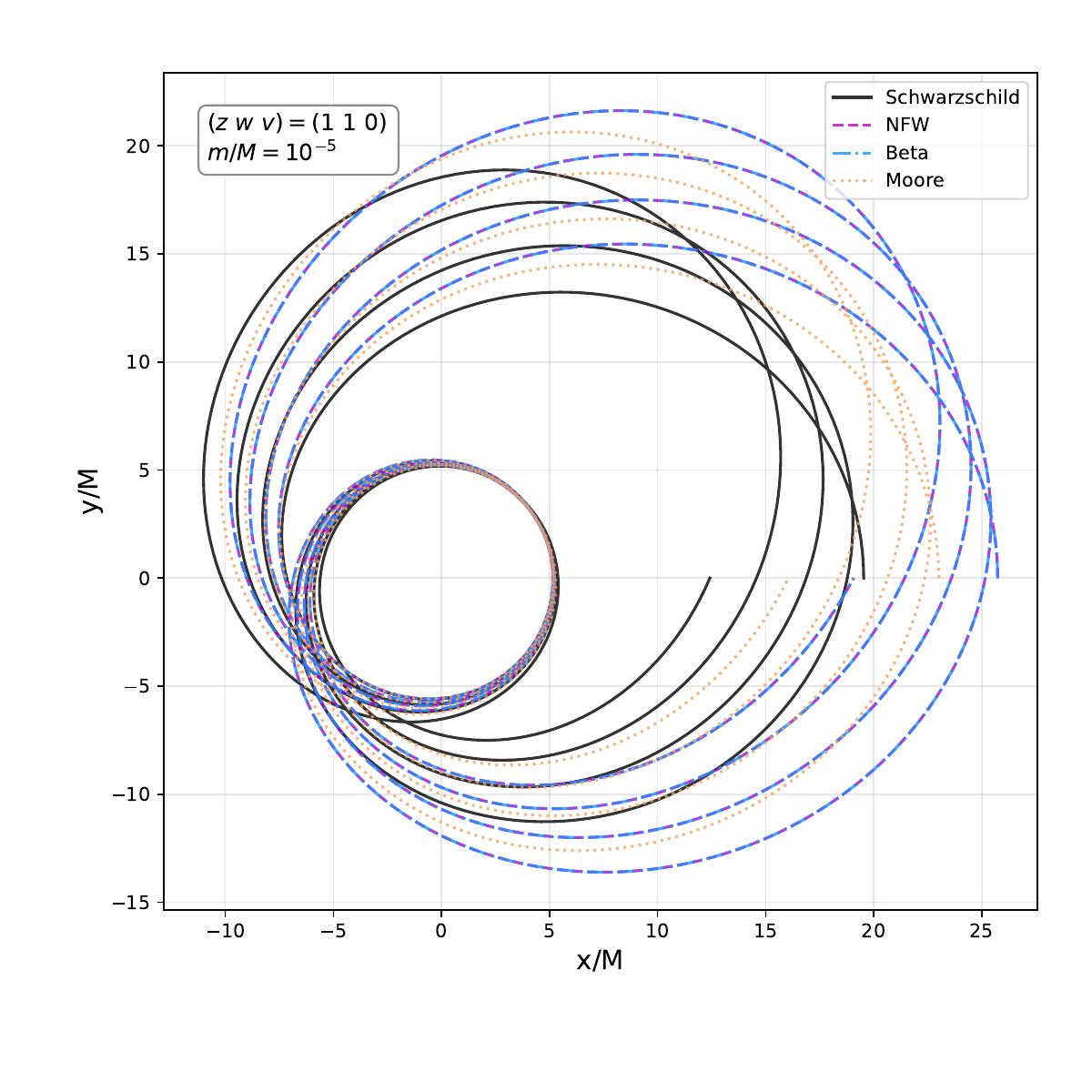}
	\end{minipage}
	
	
	
	\begin{minipage}{\linewidth}
		\centering
		\includegraphics[width=0.3\textwidth,height=0.25\textwidth]{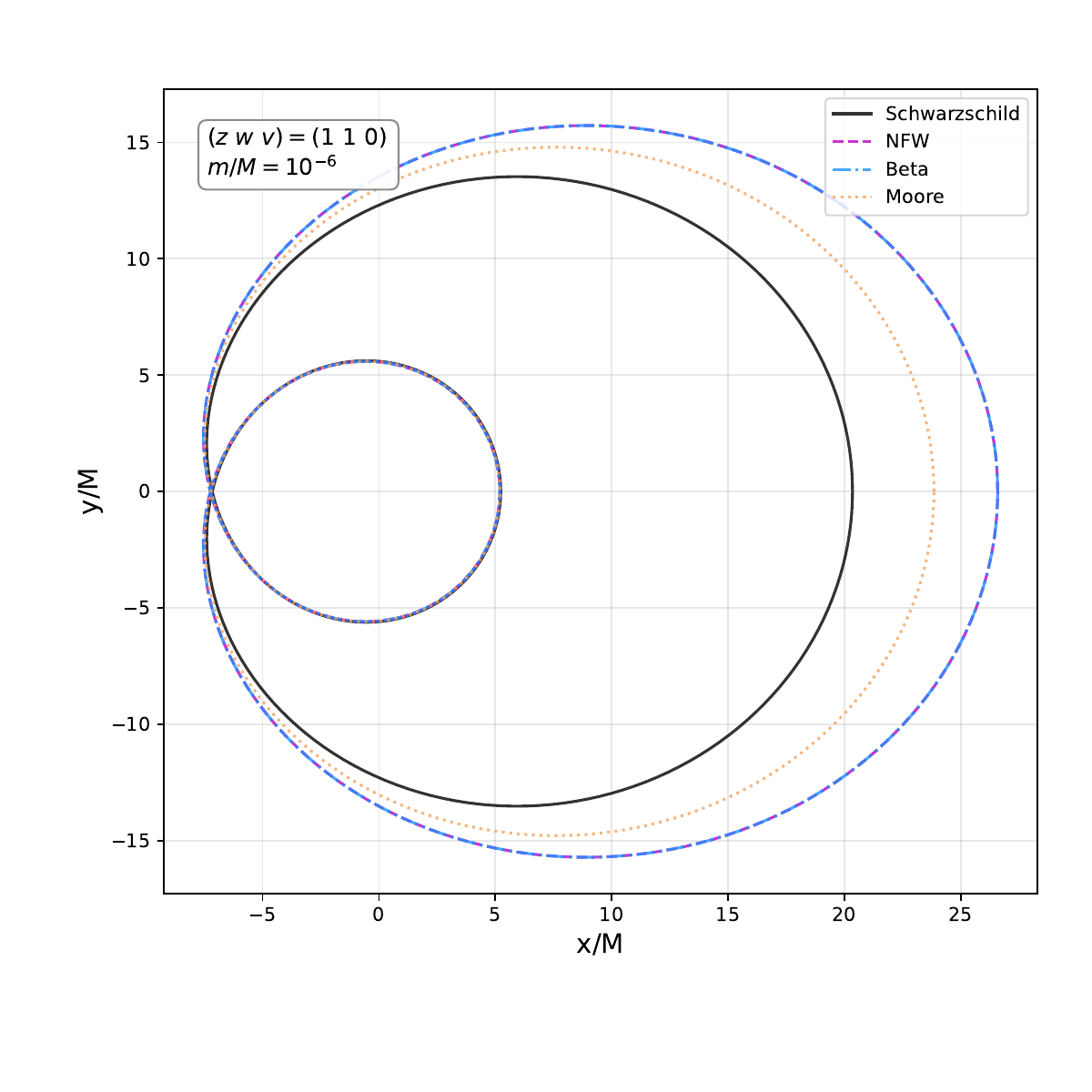}
		\includegraphics[width=0.3\textwidth,height=0.25\textwidth]{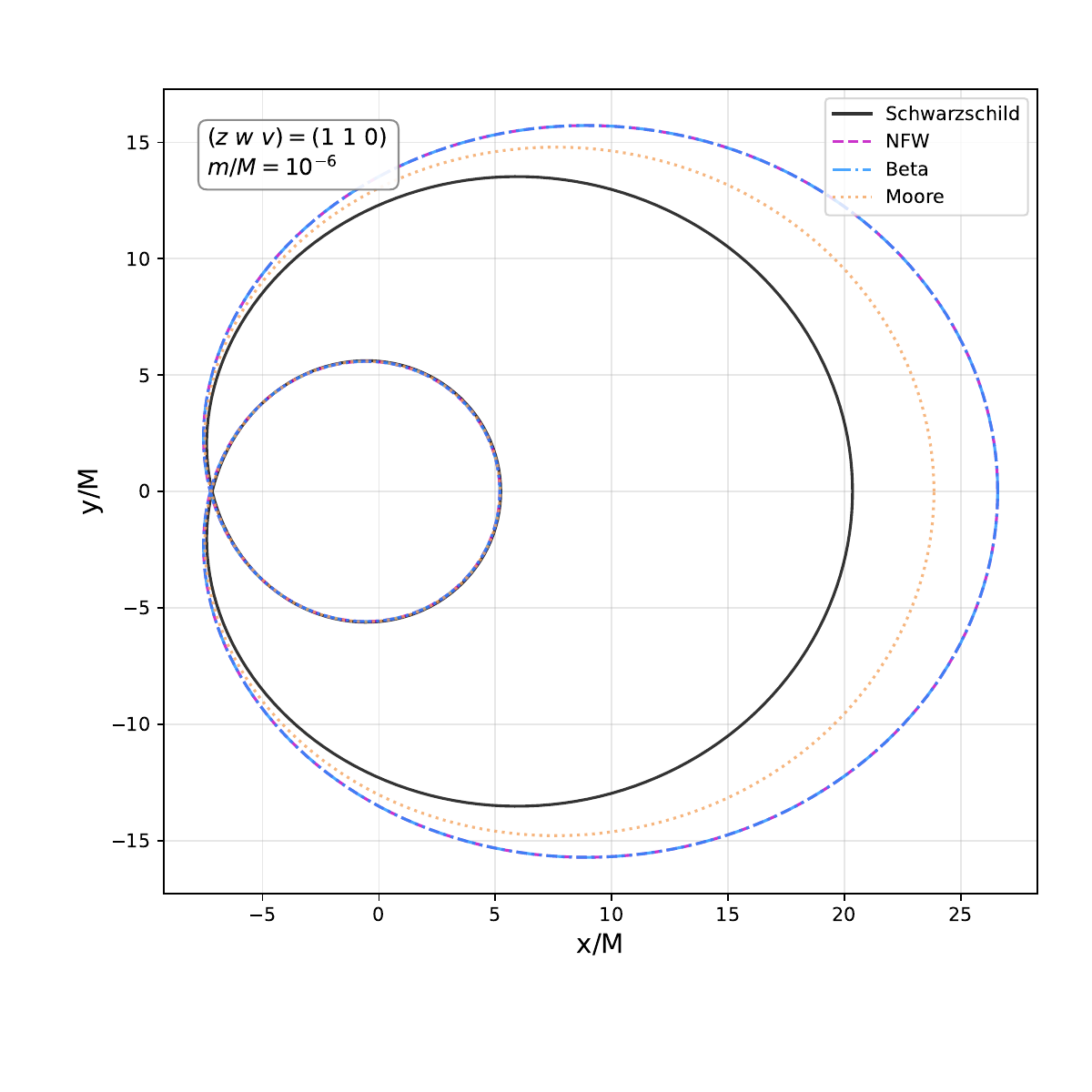}
		\includegraphics[width=0.3\textwidth,height=0.25\textwidth]{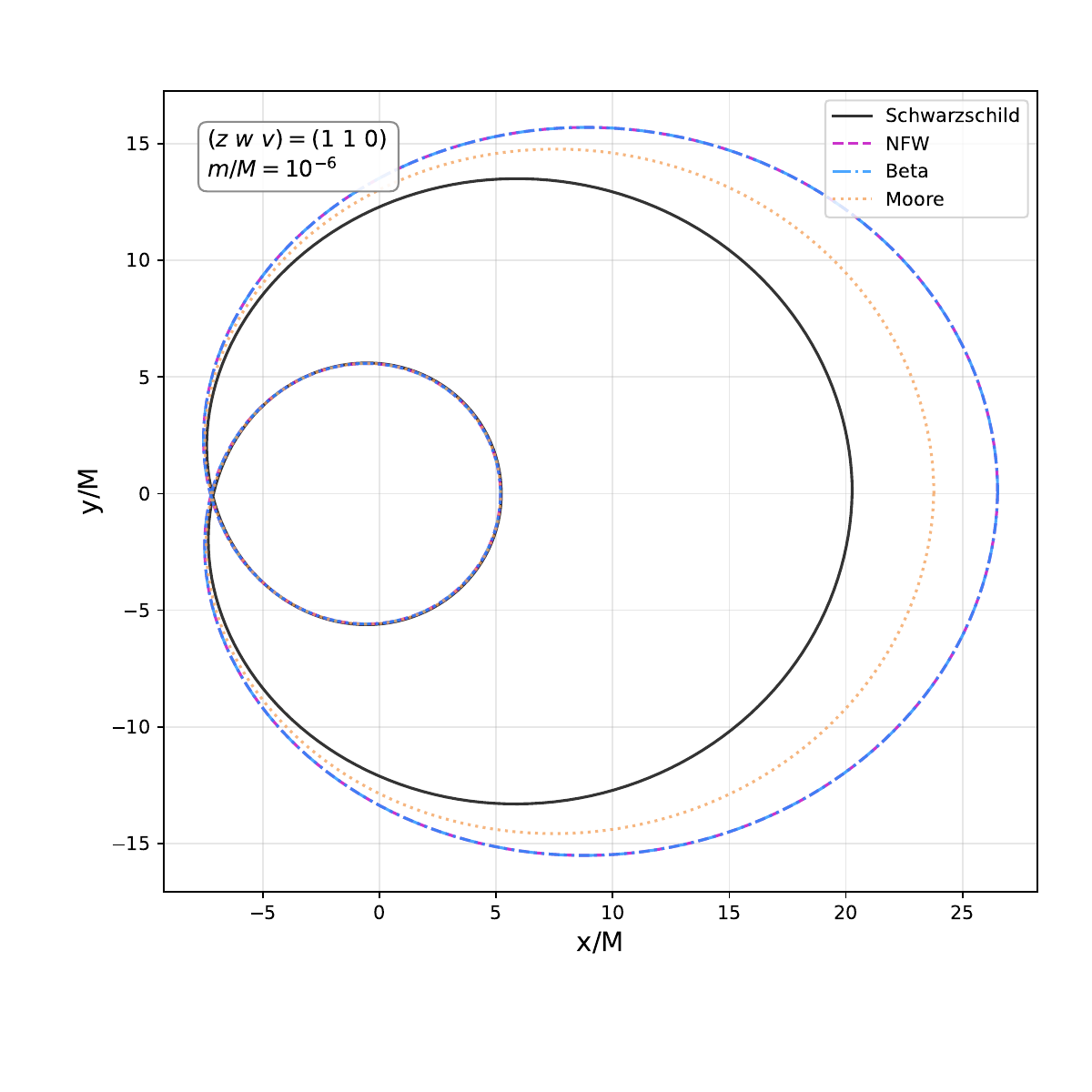}
	\end{minipage}
	
	
	\begin{minipage}{\linewidth}
		\centering
		\includegraphics[width=0.3\textwidth,height=0.25\textwidth]{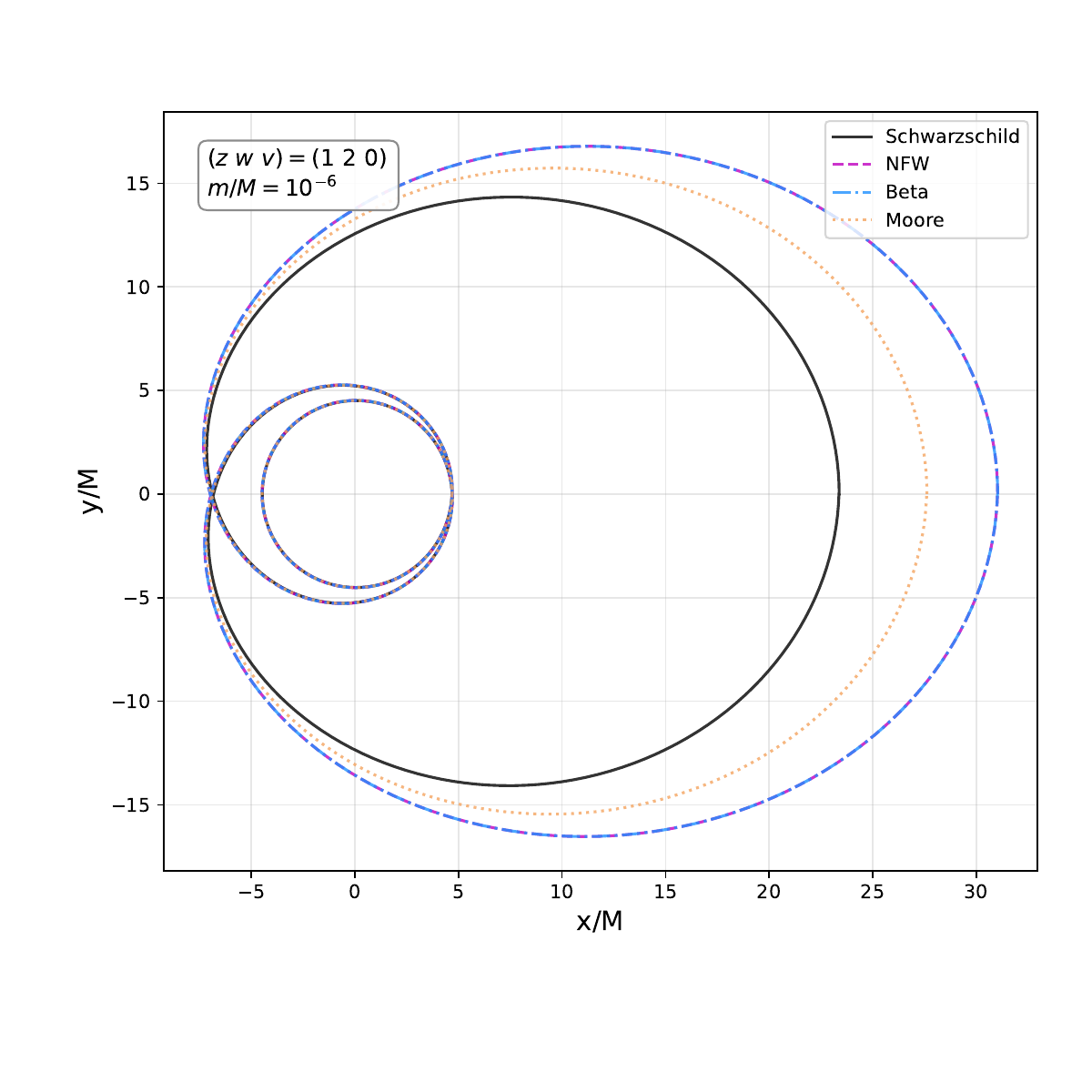}
		\includegraphics[width=0.3\textwidth,height=0.25\textwidth]{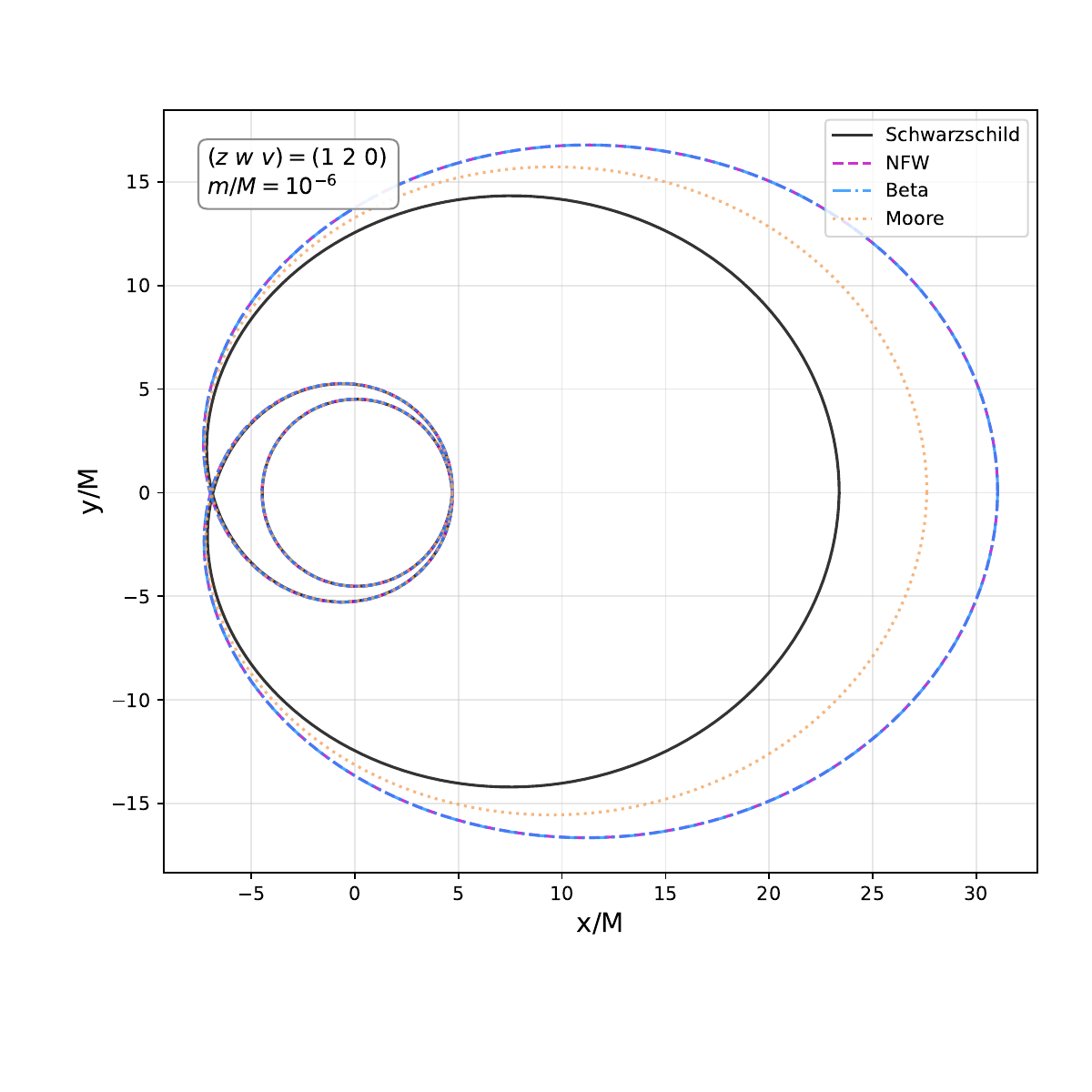}
		\includegraphics[width=0.3\textwidth,height=0.25\textwidth]{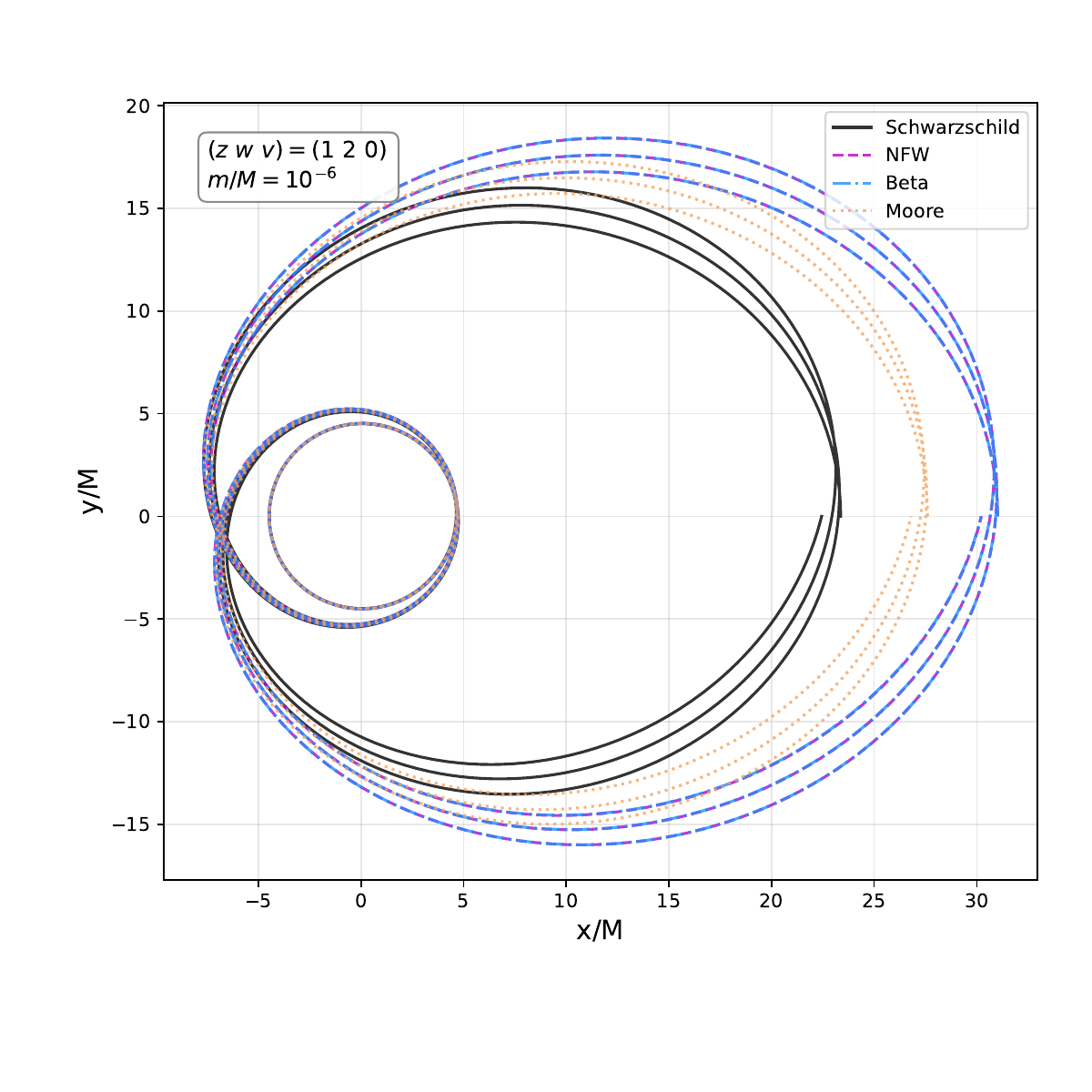}
	\end{minipage}
	
	\begin{minipage}{\linewidth}
		\centering
		\includegraphics[width=0.3\textwidth,height=0.25\textwidth]{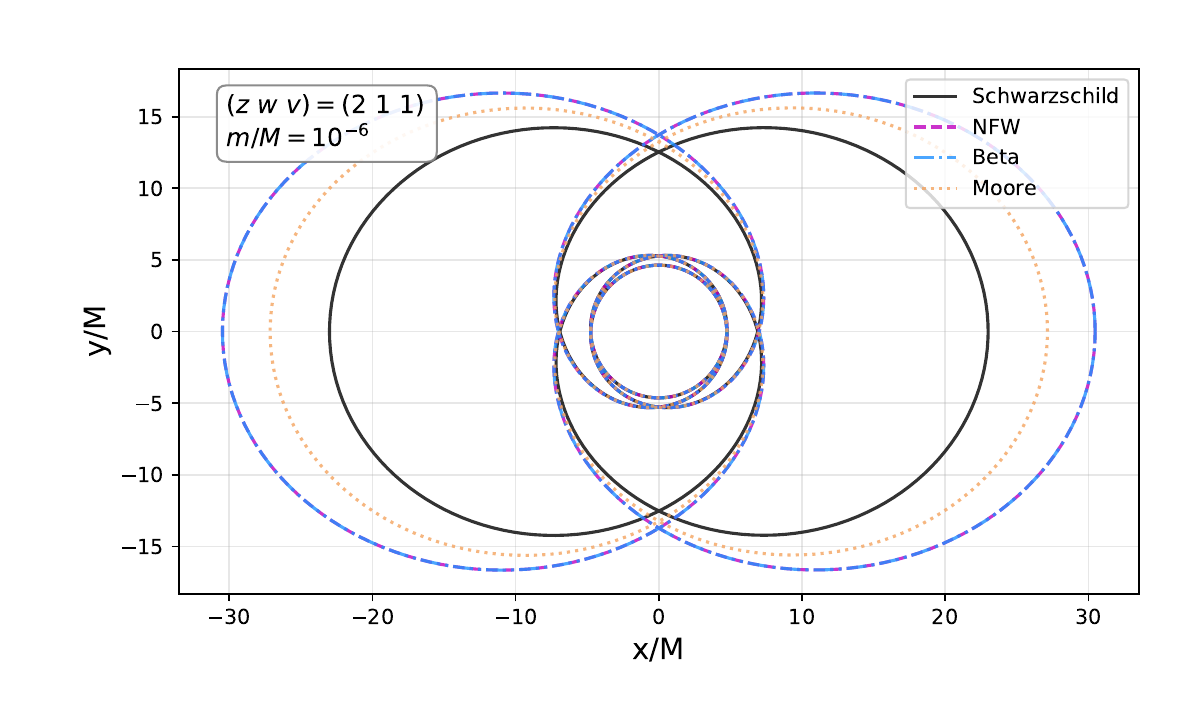}
		\includegraphics[width=0.3\textwidth,height=0.25\textwidth]{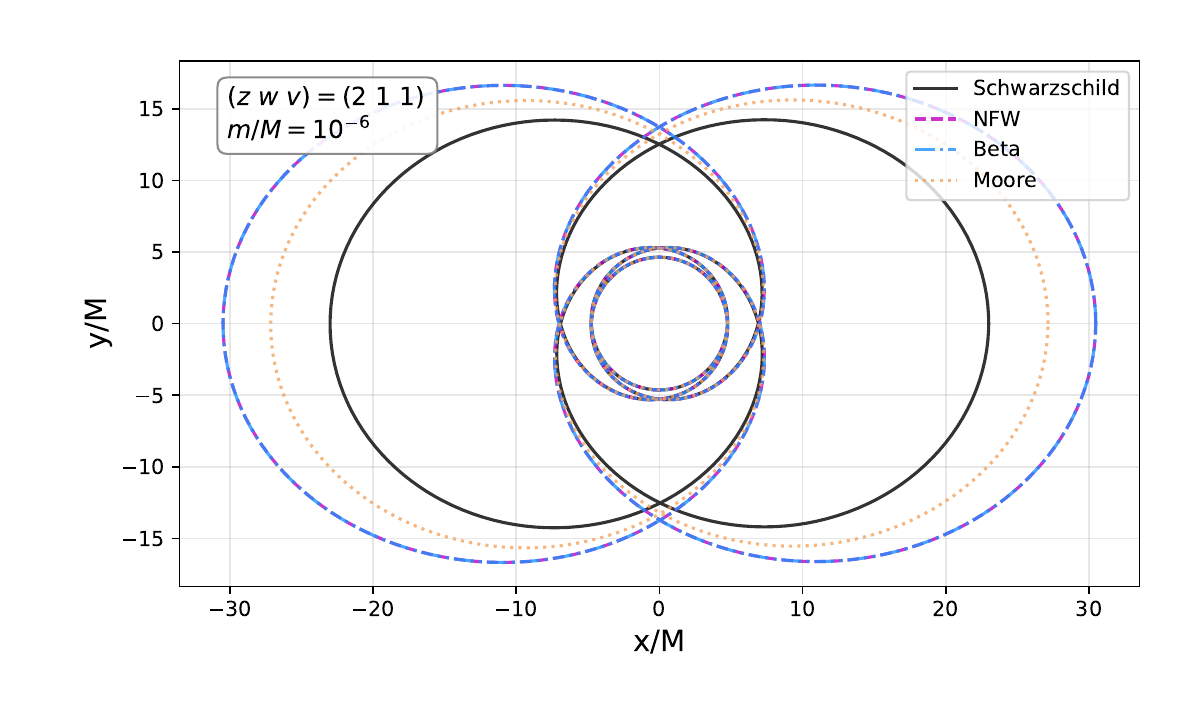}
		\includegraphics[width=0.3\textwidth,height=0.25\textwidth]{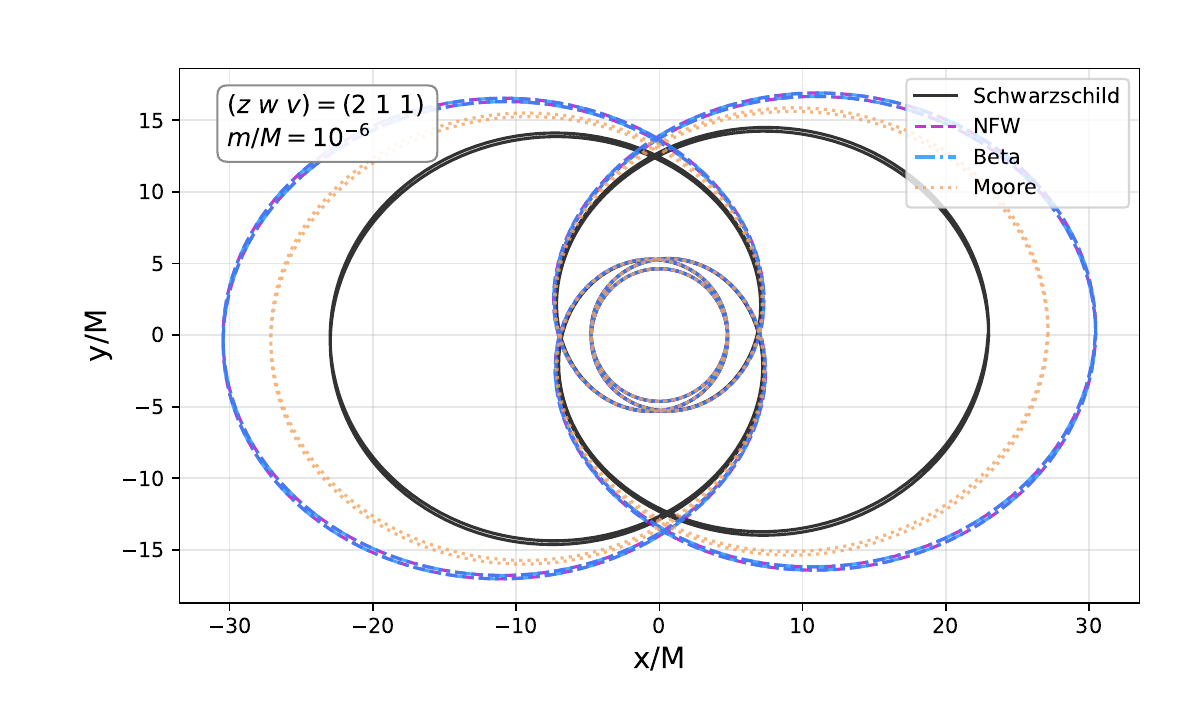}
	\end{minipage}
	
	\caption*{ \centering \ \ \ 
		$t=1$ day  \ \ \ \ \ \ \ \ \ \ \ \ \ \ \ \ \ \ \ \ \ \ \ \ \ \ 
		$t=5$ days \ \ \ \ \ \ \ \ \ \ \ \ \ \ \ \ \ \ \ \ \ \ \ \ \ \ 
		$t=1$ month
	}
	
	\caption{Stability of periodic orbits over specific timescales of orbital evolution under the influence of gravitational radiation reaction. The dark matter halo parameters are set to a mass of $k=10^4 M$ and a characteristic radius of $h=10^7 M$, with the small celestial body's mass fixed at $m=10 M_{\odot}$. The left, middle, and right columns present the orbital configurations over the observation windows of one day ($t=1$~day), five days ($t=5$~days), and one month ($t=1$~month), respectively. The first row illustrates the stability of periodic orbit for $(1~1~0)$ configuration with a mass ratio $m/M=10^{-5}$. Second row shows the stability of periodic orbit for $(1~1~0)$ configuration with a smaller mass ratio $m/M=10^{-6}$. Third row shows the stability for $(1~2~0)$ configuration with a mass ratio $m/M=10^{-6}$. The last row illustrates the stability for $(2~1~0)$ configuration with a mass ratio $m/M=10^{-6}$.}
	\label{fig:long_term_evolution}
\end{figure}

In principle, the more slowly $E$ and $L$ vary, the less easily these periodic orbits are perturbed from their initial states (where the precession parameter $q$ is rational), and the longer their periodicity can be maintained. Consequently, Fig.~\ref{fig:EL_evolution} suggests that the simplest orbital configuration $(1~1~0)$ combined with a smaller mass ratio ($m/M=10^{-6}$) can preserve its periodicity for the longest duration during dynamical evolution. To further verify this conclusion and assess the long-term stability of these periodic orbits, we numerically calculate their trajectories over specific timescales of orbital evolution, and the results are displayed in Fig.~\ref{fig:long_term_evolution}. This figure shows the orbital trajectories obtained within the Beta, NFW, and Moore dark matter models for two characteristic mass ratios, $m/M=10^{-5}$ and $m/M=10^{-6}$, over different observation timescales \footnote{For the cases of mass ratio $m/M=10^{-5}$, the periodicity of orbits for $(1~2~0)$ and $(2~1~1)$ configurations maintains a shorter timescale (less than 5 days) compared with the $(1~1~1)$ configuration, and their results are not presented in Fig.~\ref{fig:long_term_evolution}.}. The left column demonstrates that all examined periodic orbits remain highly stable over a one-day timescale. Since a typical EMRI detection involves tracking a large number of orbital cycles within a one-day observation window, this short-term stability confirms that the instantaneous waveform can be accurately modeled using the methods in Sections \ref{s4} and \ref{s5}. However, when the observation is extended to one month (right column), the impact of the mass ratio and orbital configuration $(z~w~v)$ becomes evident. For a relatively large mass ratio of $m/M = 10^{-5}$, the first row of Fig.~\ref{fig:long_term_evolution} clearly indicates that the periodicity of the $(1~1~0)$ configuration can be maintained for more than five days but less than a month. When the mass ratio becomes $m/M = 10^{-6}$, the dissipation of energy and angular momentum occurs much more slowly, thereby strengthening the stability of the periodic orbits. Specifically, the second row shows that the periodicity of the $(1~1~0)$ configuration can be maintained for more than a month, while the third and fourth rows demonstrate that the $(1~2~0)$ and $(2~1~1)$ configurations can sustain their periodicity for nearly a month. The stability of periodic orbits in systems with smaller mass ratios provides reliable evidence that the periodic signature can be preserved over a relatively long observation timescale, enhancing the probability of tracking these orbits in realistic astrophysical EMRIs.

\acknowledgments

The authors thank Yu-Chen Zhou for helpful discussion and comments on the numerical schemes. This research was funded by the Natural Science Foundation of China (Grant No. 12175212), Natural Science Foundation of Chongqing Municipality (Grant No. CSTB2022NSCQ-MSX0932), the Scientific and Technological Research Program of Chongqing Municipal Education Commission (Grant No. KJQN202201126), the Scientific Research Program of Chongqing Municipal Science and Technology Bureau (the Chongqing “zhitongche” program for doctors, Grant No. CSTB2022BSXM-JCX0100).

\paragraph{Note added.} This is also a good position for notes added after the paper has been written.


\bibliographystyle{JHEP}   
\bibliography{ref2}



\end{document}